\documentclass[galaxies,article,accept,pdftex,moreauthors]{Definitions/mdpi}
\firstpage{1} 
\pubvolume{1}
\issuenum{1}
\articlenumber{0}
\pubyear{2025}
\copyrightyear{2025}
\datereceived{28 May 2025} 
\daterevised{22 July 202} 
\dateaccepted{24 July 2025 } 
\datepublished{ } 
\usepackage[toc]{appendix}  
\hreflink{https://doi.org/} 
\Title{AGN contribution on the morphological parameters of their host galaxies up to intermediate redshifts of z\,$\sim$\,2}

\TitleCitation{AGN contribution on morphological parameters}

\Author{Tilahun Getachew-Woreta, $^{1, 2, 3, *}$\orcidA{}, Mirjana Povi\'c $^{1, 4, 5}$, Jaime Perea $^{4}$,  Isabel Marquez $^{4}$, Josefa Masegosa $^{4}$, Antoine Mahoro $^{6, 7}$, and Shimeles Terefe Mengistue $^{1, 2, 8}$}
\address{%
$^{1}$ \quad  Astronomy and Astrophysics Department, Entoto Observatory and Research Center (EORC), Space  	Science and Geospatial Institute (SSGI), P.O. Box 33679, Addis Ababa, Ethiopia\\ 
$^{2}$ \quad  Addis Ababa University (AAU), P.O. Box 1176, Addis Ababa, Ethiopia\\
$^{3}$ \quad Physics Department, Bule Hora University (BHU), P.O. Box 144, Bule Hora, Ethiopia\\
$^{4}$ \quad Instituto de Astrof\'isica de Andaluc\'ia (IAA-CSIC), 18008, Granada, Spain\\
 $^{5}$ \quad Physics Department, Faculty of Science, Mbarara University of Science and Technology (MUST), P.O. Box 1410, Mbarara, Uganda\\
 $^{6}$ \quad South African Astronomical Observatory (SAAO), P.O. Box 9 Observatory, Cape Town, South Africa\\
 $^{7}$ \quad Southern African Large Telescope (SALT), P.O. Box 9, Observatory, Cape Town 7935, South Africa\\
 $^{8}$ \quad Physics Department, College of Natural Sciences, Jimma University, P.O. Box 378, Jimma, Ethiopia
}

\corres{{tilahun85@gmail.com}}

\abstract{The presence of Active Galaxy Nuclei (AGN) can affect the morphological classification of galaxies. This work aims to determine how the contribution of AGN affects the most used morphological parameters down to the redshift of z\,$\sim$\,2  in COSMOS-like conditions. We use a sample of $>$\,2000 local non-active galaxies, with a well-known visual morphological classification, and add an AGN as an unresolved component that contributes to the total galaxy flux with 5\%-75\%. We moved all the galaxies to lower magnitudes (higher redshifts) to map the conditions in the COSMOS field, and we measured six morphological parameters. The greatest impact on morphology occurs when considering the combined effect of magnitude, redshift and AGN, with spiral galaxies being the most affected. In general, all the concentration parameters change significantly if the AGN contribution is $>$\,25\% and the magnitude $>$\,23. We find that the GINI coefficient is the most stable in terms of AGN and magnitude/redshift, followed by the M20, Conselice-Bershady (CCON), and finally the Abraham (CABR) concentration indexes. We find that, when using morphological parameters, the combination of CABR, CCON and asymmetry is the most effective in classifying active galaxies at high-redshift, followed by the combination of CABR and GINI.}


\begin{document}


\section{Introduction}
The study of galaxy properties over cosmic time is crucial for understanding galaxy evolution. One of such properties is the morphology, which is intertwined with properties like stellar mass, colour, luminosity, star formation, gas and dust content, nuclear activity, and environment, and is fundamental in constraining models of galaxy formation and evolution  ({e.g.,} \citep{Gabor2009, Povic2009a, Povic2009b,  Povic2013a, Povic2013b, Con2014, Amado2019, Dimauro2022, Mei2023}). The techniques of morphological classification of galaxies can be broadly separated into 4 groups:\\
\indent (1) Visual classification is the traditional method, and it is still commonly used to perform morphological classification of galaxies \cite[e.g.,][]{
	Lintott2008, Nair2010, Buta2013, Simmons2017, Willett2017, Mahoro2019, Zy2021}. It is time-consuming when dealing with a large sample of galaxies, as well as subjective, but it is one of the most reliable methods for classifying nearby and well-resolved objects. It is therefore often used to test the reliability of other classification methods.\\
\indent (2) Parametric methods assume an analytical model to fit the surface brightness of galaxies using a set of mathematical, predefined functions \citep[e.g.,][]{Sersic1963, Peng2002, Peng2010, Simard2011, Jimenez2012, Buta2013, Nadolny2021}. These methods provide detailed information on the different components of galaxies (e.g., bulge and disk), and the measurement of the S\'ersic index. They have been commonly used in large surveys to obtain a broad classification into early- and late-types, although classification using the S\'ersic index can sometimes be difficult (e.g., for late-type spirals and irregular galaxies, galaxies with central light peaks, etc.) \citep[e.g.,][]{Vika2014, Nadolny2021, Cutler2022}, including an additional difficulty related to the inclusion of spiral arms in the decomposition \citep[e.g.,][]{Marchuk2024, Chugunov2024}.\\
\indent (3) The non-parametric approach relies on measuring a set of carefully chosen observables rather than assuming any pre-existing analytic model. Non-parametric methods have also been widely used to measure various parameters of galaxies corresponding to morphological types such as light distribution and galaxy shape \cite[e.g.,][]{Abraham1994, Abraham1996, Con2000, Con2003, HC2008, Povic2013a, Mahoro2019, Nersesian2023, Mukundan2024}. 
Refs. \cite{Abraham1994, Abraham1996} introduced the non-parametric technique by defining two observables, the Abraham concentration index and asymmetry. Subsequently, other parameters were introduced, such as smoothness \citep[][]{Con2000, Con2003}, Gini coefficient \citep[][]{Abraham2003, Lotz2004}, Conselice-Bershady concentration index \citep{Bersh2000}, and moment of light \citep[][]{Lotz2004}. The full summary of these parameters can be found \linebreak{} in \cite{Povic2015,  Amado2019, Getachew2022}. Morphological classifications for a large sample of galaxies at both low and high redshift using non-parametric techniques have been widely used to classify galaxies into early- and late-types in large surveys, often in combination with machine learning \citep[e.g.,][]{HC2008, Povic2009a, Povic2012, Povic2013a, Povic2015, Pintos2016, Amado2019, Getachew2022, Nersesian2023}. \\
\indent (4) (4) Finally, machine learning methods for the morphological classification of galaxies, both supervised (e.g., support vector machine (SVM), random forest, decision trees, logistic regression) and unsupervised (e.g., principal component analysis, k-means clustering), and, in particular, the (convolutional) neural networks, have been extensively used over recent years (e.g., \citep{Martin2020, deDiego2020, Cheng2020, Cavanagh2021, Vavilova2021, deDiego2021, Sazonov2021, DominguezSanchez2022, Urrutia2023, Cakir2024}); the reliability and success rate of classification are often judged by how well they agree with the visual classification.

The presence of AGN can affect the morphological classification of active galaxies, and inconsistencies have been found in previous studies. Some studies concluded that AGN host galaxies tend to reside in spheroids or bulge-dominated galaxies \citep[e.g.,][]{Kauffmann2003, Pierce2007, Pierce2010, Povic2012, Dubois2016}, while others found a higher fraction of later types \citep[][]{Choi2009, Gabor2009, Mahoro2019}. It has also been shown that the morphology of AGN host galaxies may depend on the wavelength used for AGN detection. Radio-detected AGN are mainly hosted by red sequence galaxies (e.g., \citep{Hickox2009}), {while X-ray-detected AGN are found in greater numbers in the green valley} (e.g., \citep{Povic2012}). Therefore, it is important to carry out further studies on the morphologies of active galaxies; for this, it is also essential to study how different morphological parameters, commonly used in morphological classification, are affected by the contribution of AGN.

\indent Even though numerous studies found that AGN contributions may also affect the morphological classification of their host galaxies, in the best-case scenario, limited contributions are made (see, for example, \citep[][]{Gabor2009, Cardamone2010, Pierce2010, Trump2015}).   
At present, the most detailed study on the impact of AGN on morphological parameters and the classification at z\,$\sim$\,0 is  \cite{Getachew2022}. In this work, we studied how different contributions of AGN to the total optical light, covering a wide range of AGN fractions between 5\% and 75\% to consider both type-1 and type-2 AGN, affect the six commonly used morphological parameters: Abraham concentration index, Gini coefficient, Conselice-Bershady concentration index, M20 moment of light, asymmetry index, and smoothness. We found that most morphological parameters are affected by the presence of an AGN when its contribution to the total galaxy light is above 25\%, with late-type galaxies being more affected than early-types. However, an exhaustive analysis of the relationship between the commonly used morphological parameters and the effect of the AGN on the classification of galaxies at higher redshift, where the impact of survey depth and resolution on the classification of galaxies is significant \citep{Povic2015} is still missing. 

In this work, we artificially added 5\%\,-\,75\% AGN contribution to the centre of non-active local galaxies with known morphology, and we moved them to the redshift and magnitude distributions corresponding to a sample of galaxies from the COSMOS \citep{Scoville2010} survey, using five magnitude and redshift ranges. We then studied how the contribution of the AGN has an impact on the derived morphological parameters up to the intermediate redshift of z\,$\sim$\,2.\\
\indent The structure of the paper is as follows: The data are described in Section~\ref{sec:data1}, together with a summary of the sample selection. Section~\ref{sec:3} defines the methodology used. The analysis and results are presented in Section~\ref{sec:4}. The results are discussed in Section~\ref{sec5}, while conclusions are drawn in Section~\ref{6}.
We assume a cosmological model with $\Omega_{\Lambda} = 0.7$, $\Omega_{M} = 0.3$, and $H_{0}$ = 70\,kms$^{-1}$Mpc$^{-1}$. All magnitudes given in this paper are in the $AB$ system described by \cite{Oke2015}.

\section{Data and sample selection}\label{sec:data1}
In this work, we use two data sets corresponding to the local universe (at z\,$\sim$\,0) and intermediate redshift (up to z\,$\sim$\,2).
The local sample of galaxies is the same as that described in \cite{Getachew2022}, and therefore, we provide only a summary here. We use a total sample of 2301 local, non-AGN galaxies, randomly selected from the initial sample of $\sim$\,14000 galaxies with available visual morphological classification (T-type) from \cite{Nair2010} (see their Table 1). Galaxies were selected as non-AGN using the BPT-NII diagram \citep{Baldwin1981} and emission line ratios using fluxes from the MPA-JHU SDSS DR7 catalogue \endnote{\url{https://www.mpa.mpa-garching.mpg.de/SDSS/} (accessed on 31 October 2008)}.

The final local sample used in this work contains 2251 galaxies (after excluding mergers, interactions, peculiar galaxies, and images affected by nearby stars), of which 471 (21\%) are early-types (elliptical and lenticular, with T-type\,$\le$\,0), 891 (40\%) early-spirals (0\,$<$\, T-type\,$\le$\,4) and 889 (39\%) late-spirals (4\,$<$\,T-type\,$\le$\,8). We do not have strong edge-on galaxies in our sample, and therefore, although the effect of inclination is not taken into account in this study, we do not expect it to significantly affect our main results. Fig~\ref{fig:Mag_Redshift} shows the g-band magnitude and redshift distributions for the sample of selected local non-AGN galaxies. 
The selected galaxies are, in general, well separated between ellipticals and spirals when using different morphological diagrams, as shown in \cite{Getachew2022} (Figures 12--15 for the original sample, top left plots). Finally, all selected galaxies are nearby and extended, with the PSF size much smaller than the galaxy size, as shown in \cite{Getachew2022} (see their Figure 4 and Table 1). The impact of AGN on galaxy sizes was already tested in our previous work by \cite{Getachew2022} (their Section 4.1 and Figure 6), showing a negligible impact on the effective radius of $<$\,10\% up to the AGN contributions of 25\%, and of $\sim$\,20\%, 40\%, and 60\% for AGN contributions of 25\%, 50\%, and 75\%, respectively. The impact of magnitude/redshift on galaxy sizes was already tested in previous works, including the work of \cite{Mosleh2013} where the authors found a negligible effect on sizes up to redshift of z\,$\sim$\,1.

At intermediate redshift, we use as a reference the Cosmic Evolution Survey {\citep[COSMOS;][]{Scoville2007}. We selected COSMOS since it is one of the deepest surveys with multiwavelength data available, and it is widely used in galaxy formation and evolution studies. It is the Hubble Space Telescope's (HST) largest project to date, covering an area of 2 deg$^{2}$ and reaching a depth of I(F814W)\,=\,27.2 mag. In this work, we use the photometric images and catalogue in the F814W band from the HST-ACS survey. The data reduction, images, and photometric catalogue are all described in \cite{Koekemoer2007, Leauthaud2007}. For photometric redshifts, we used the catalogue by \cite{Ilbert2009}, where the redshifts were measured using the \textit{Le Phare} code and photometric information from 30 broad, intermediate, and narrow-band filters from UV, optical, NIR, and MIR bands. For more details on the images and catalogues used in this work see \cite{Povic2015}. 
	
	\begin{figure}[h!!!]
	\begin{center}
	\includegraphics[height=1.85in, width=4.4in]{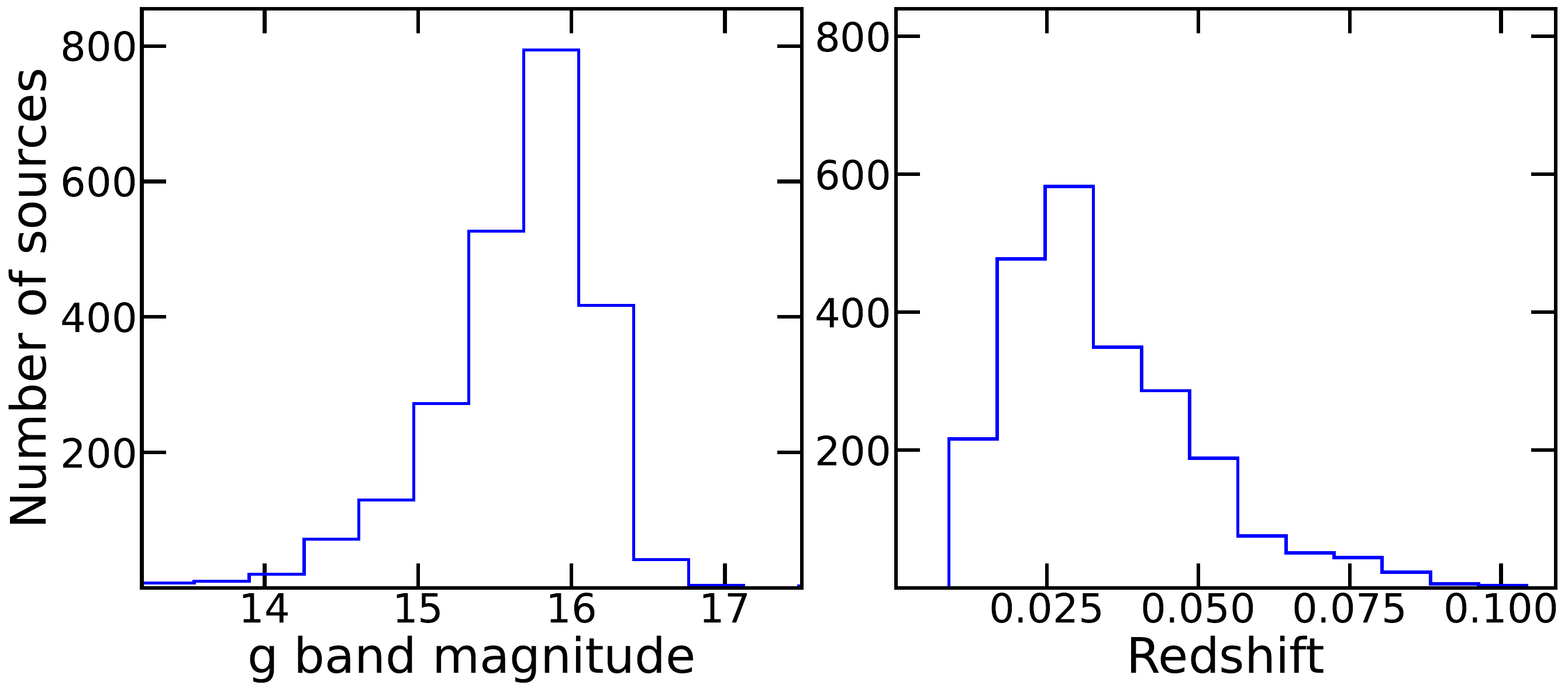}
	\includegraphics[height=1.85in, width=4.36in]{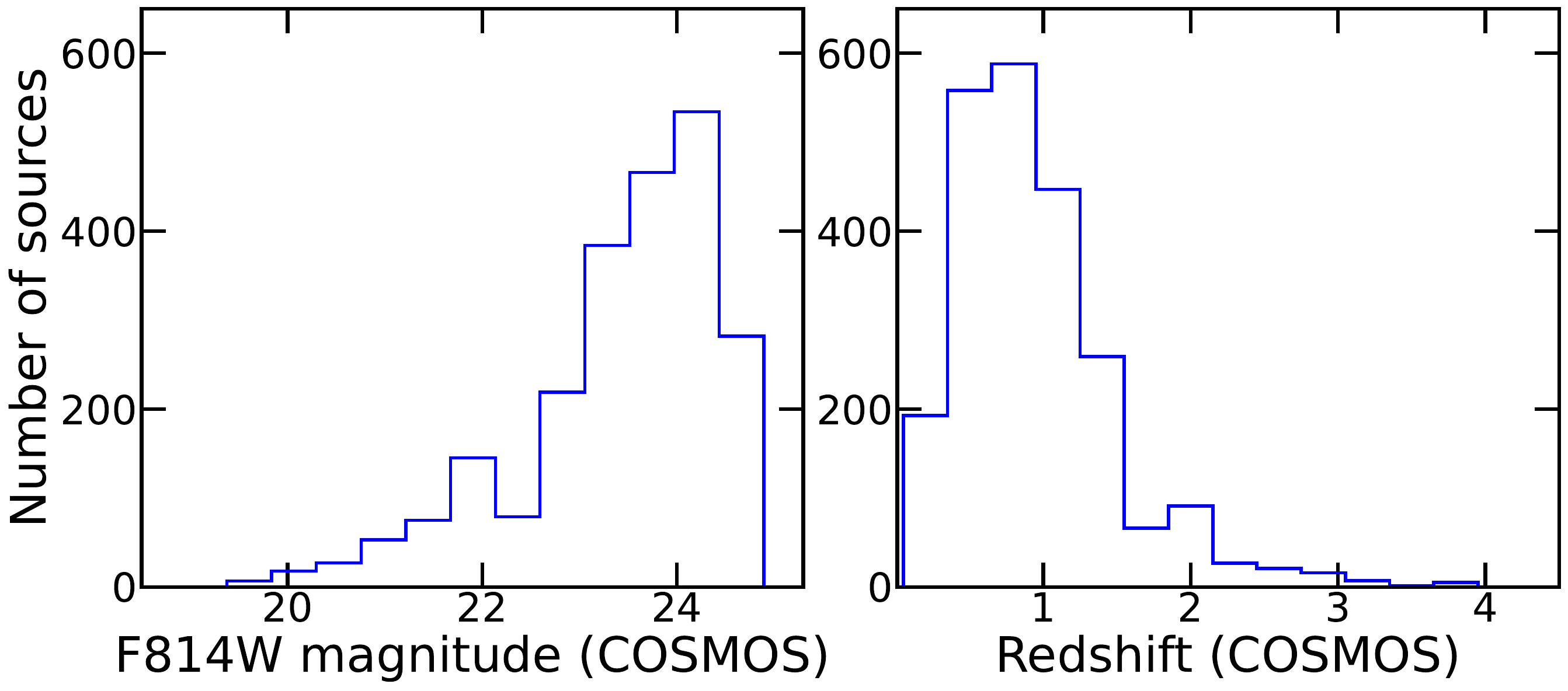}
	\caption{Top: The g-band magnitude (left) and redshift (right) distributions for the selected sample of non-active local galaxies. Bottom: The F814W magnitude (left) and redshift (right) of the used COSMOS sample.}
	\label{fig:Mag_Redshift}
	\end{center}
	\end{figure}
	
	\section{Methodology}\label{sec:3}
	Non-parametric methods of morphological classification have been widely used to separate galaxies with different structures, especially when dealing with large datasets and faint sources at higher redshift \citep[e.g., see][and references therein]{Povic2015}. In this work, we have measured and analysed the following six morphological parameters commonly used in the literature and non-parametric morphological classification: Abraham concentration index \citep[hereafter CABR;][]{Abraham1996}, as the ratio between the flux at 30\% of the Petrosian radius and the total flux of the galaxy; Gini coefficient \citep[hereafter GINI;][]{Abraham2003, Lotz2004}, that gives the distribution of
	light among the pixels associated with a galaxy, taking into account the number of pixels, the flux in each pixel, and the mean flux; Conselice-Bershady concentration index \citep[hereafter CCON;][]{Bersh2000}, measured as the logarithm of the ratio of the circular radii containing 80\% and 20\% of
	the total flux; $M_{20}$ moment of light \citep[hereafter M20;][]{Lotz2004}, which depends on the spatial distribution of the
	light in a galaxy relative to its centre, summed over the 20\% brightest pixels and normalised by the total second-order moment of light; asymmetry index \citep[hereafter ASYM;][]{Abraham1994}, which gives the level of asymmetry in the shape of a galaxy by comparing an object flux in every pixel using an original image and the rotated one by 180\,deg, and taking into account the background; and smoothness or clumpiness \citep[hereafter SMOOTH;][]{Con2000}, which is sensitive to the small-scale structures in galaxies and takes into account the original galaxy image smoothed with a boxcar of a given width and subtracted from the original image. For more details on how all these parameters are defined and measured, see any of the previous studies by \cite{Povic2015,Pintos2016,Amado2019}, or \cite{Getachew2022}.
	We used the galaxy Support Vector Machine (galSVM) code \citep{HC2008} to measure all six parameters. The galSVM is a freely available code written in IDL that uses the libSVM library \citep{CL2011}, also freely available, which helps with the multi-parametric analysis and allows for an automatic morphological classification of galaxies by using the SVM. 
	
	\subsection{Morphological parameters at intermediate redshifts}\label{sec:cosmoscon}
	Using galSVM, we first randomly scaled in brightness a local sample of 2251 non-AGN galaxies to match the magnitude and redshift distributions of galaxies in the COSMOS field. In this way, when scaling the galaxy in flux, the surface brightness dimming was directly taken into account. All local galaxies have been re-sampled with a pixel scale of 0.03 arcsec/pix corresponding to COSMOS and convolved with their point spread function (PSF) obtained from the SDSS survey for each galaxy to match the same spatial resolution. In addition, the galSVM code is optimised to classify faint sources.
	Our previous studies showed that the use of several magnitude limits tends to give a better classification, rather than mapping the entire COSMOS sample of galaxies at once \citep[see][]{Povic2013a, Pintos2016}. Furthermore, since in this work we want to analyse the effect of AGN on the morphological parameters when dealing with different brightness and redshift of galaxies, we decided to perform the study considering five magnitude limits (and corresponding redshift limits) such as: F814W\,$\le$\,21 (z\,$\lesssim$\,0.7), F814W\,$\le$\,22 (z\,$\lesssim$\,0.9), F814W\,$\le$\,23 (z\,$\lesssim$\,1.1), F814W\,$\le$\,24 (z\,$\lesssim$\,1.6), and F814W\,$\le$\,25 (z\,$\lesssim$\,2). Figure~\ref{sec:Mag21_25} shows these distributions in COSMOS using the HST/ACS F814W photometric band. All 2251 local non-AGN galaxies have been simulated, as explained above, to match the five selected magnitude and redshift distributions. \\
	\indent For each magnitude (and corresponding redshift) limit, we then dropped the simulated galaxies into the real background of the COSMOS image. Special care is taken to place all simulated galaxies in empty background regions to avoid superposition with COSMOS sources. For doing that, galSVM uses the corresponding SExtractor \citep[][]{Bertin1996} segmentation images in COSMOS to find sky regions free of sources. In addition, to ensure that each simulated galaxy will be dropped in an empty background region, we only allow 10 galaxies per image as the maximum number of objects per galSVM run. This process is repeated until all simulated galaxies have been dropped. With all this, we expect to reproduce the noise of COSMOS images by placing the local galaxies in a real background of high-redshift sources.\\
	\indent Finally, we measured the six morphological parameters (CABR, GINI, CCON, M20, ASYM, and SMOOTH) of all simulated galaxies. Depending on the F814W band used in COSMOS and the redshift to which the galaxy was randomly shifted, we measured the morphological parameters of all simulated galaxies using the corresponding SDSS rest-frame bands and their central wavelengths, to avoid K-correction effects (for more information see \cite{HC2008}). With this, we generated new images that show how the galaxies would appear at higher redshift (z\,$<$\,2). We repeated this process five times, for five magnitude (and therefore redshift) limits, each time measuring all six morphological parameters. In \cite{Povic2015}, it has been shown how images of galaxies at fainter magnitudes and higher redshifts can be affected (see their Figures 3--5). For further information on the methodology, see the galSVM reference paper, \cite{HC2008}.
	\begin{figure}[h!!!]
		{\includegraphics[height= 1.35in, width=1.32in]{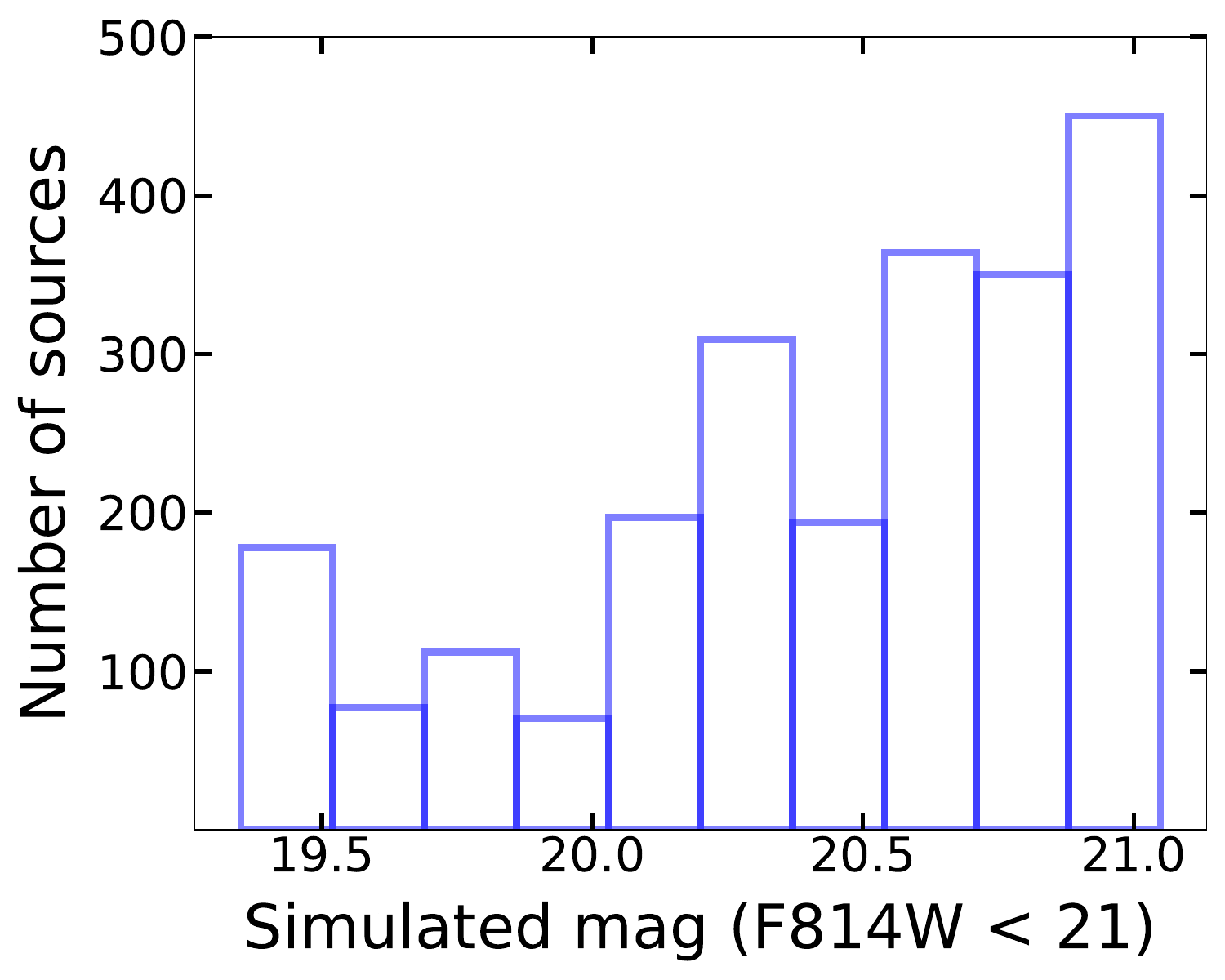}}
		{\includegraphics[height= 1.35in, width=1.32in]{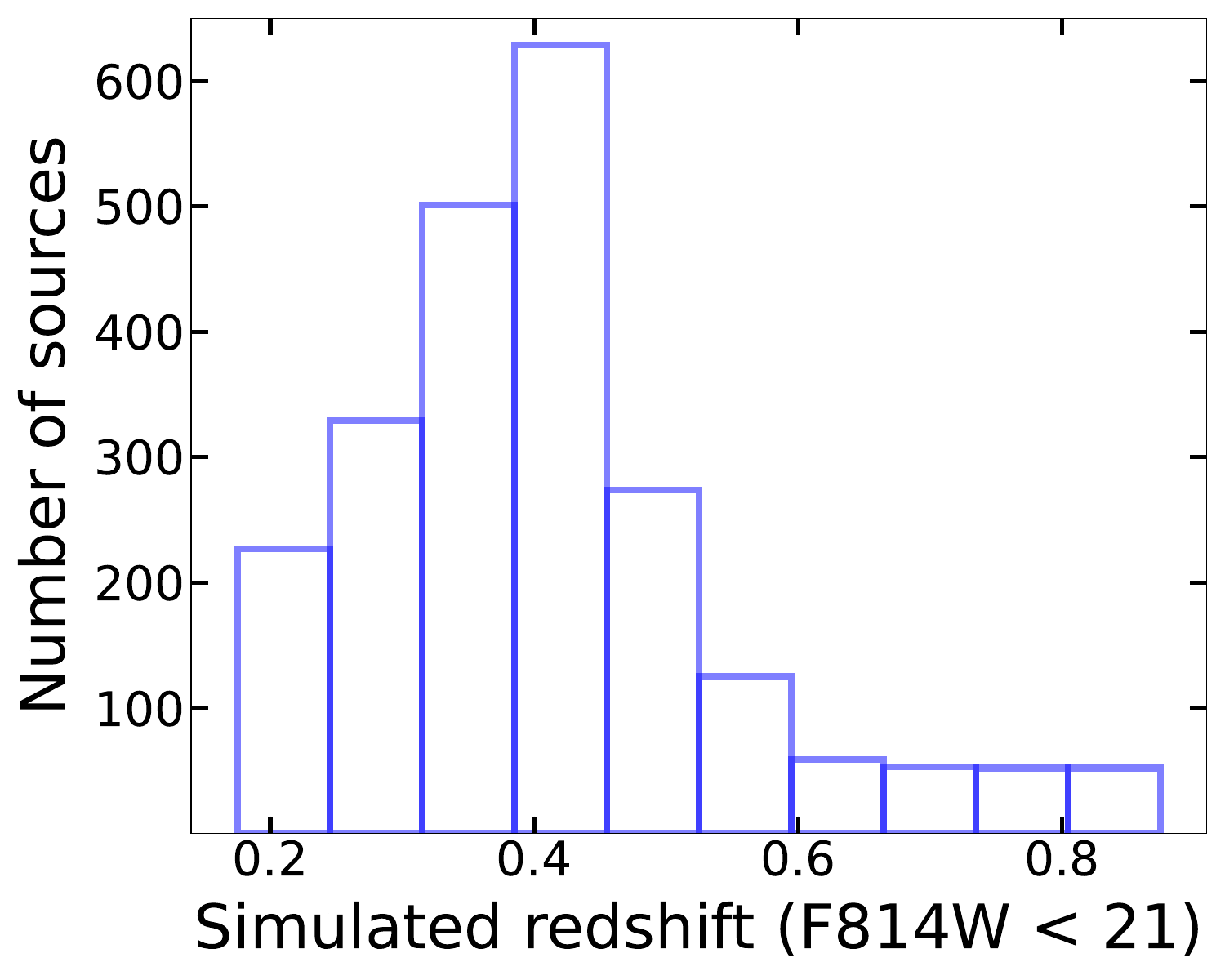}}
		{\includegraphics[height= 1.35in, width=1.32in]{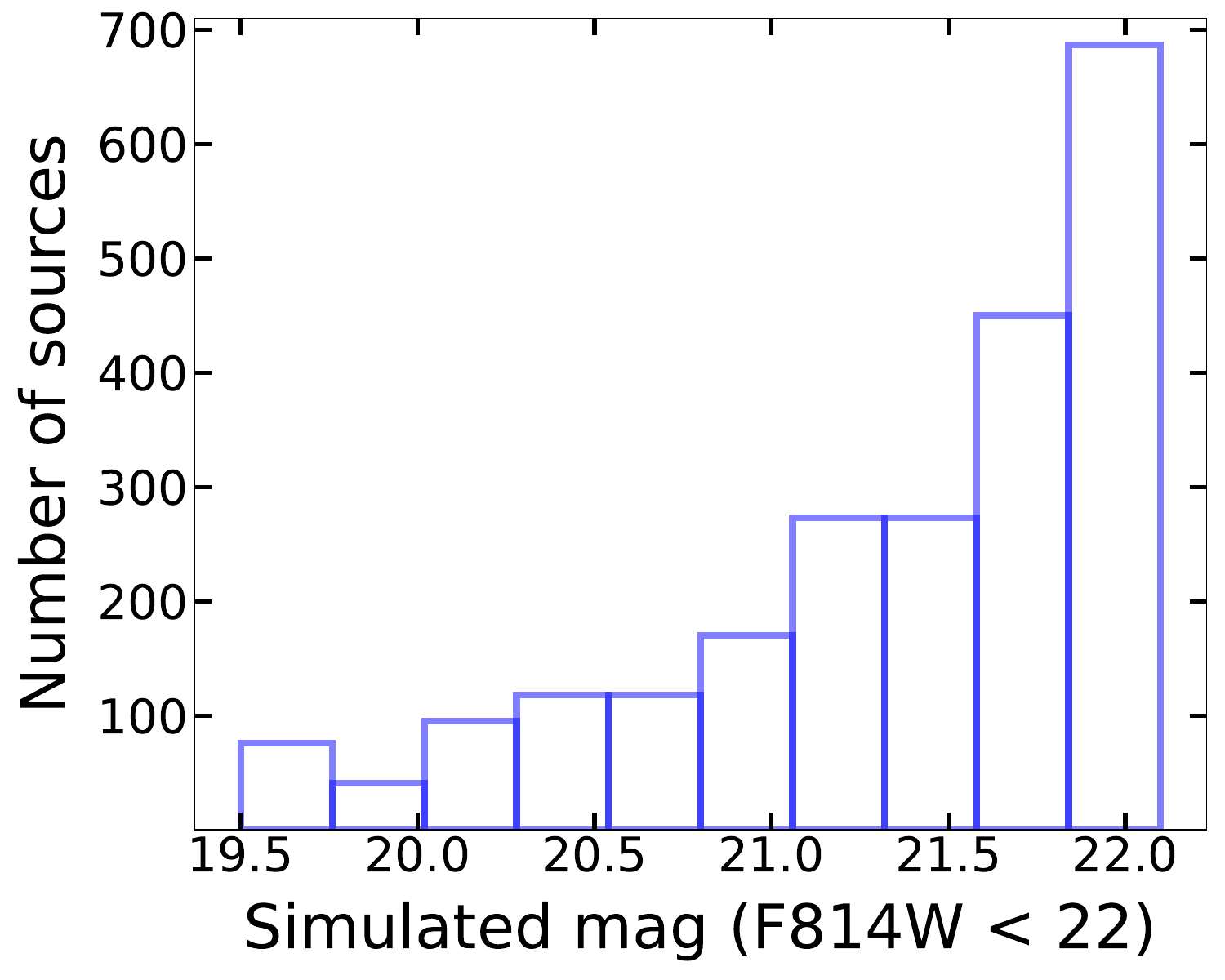}}
		{\includegraphics[height= 1.35in, width=1.32in]{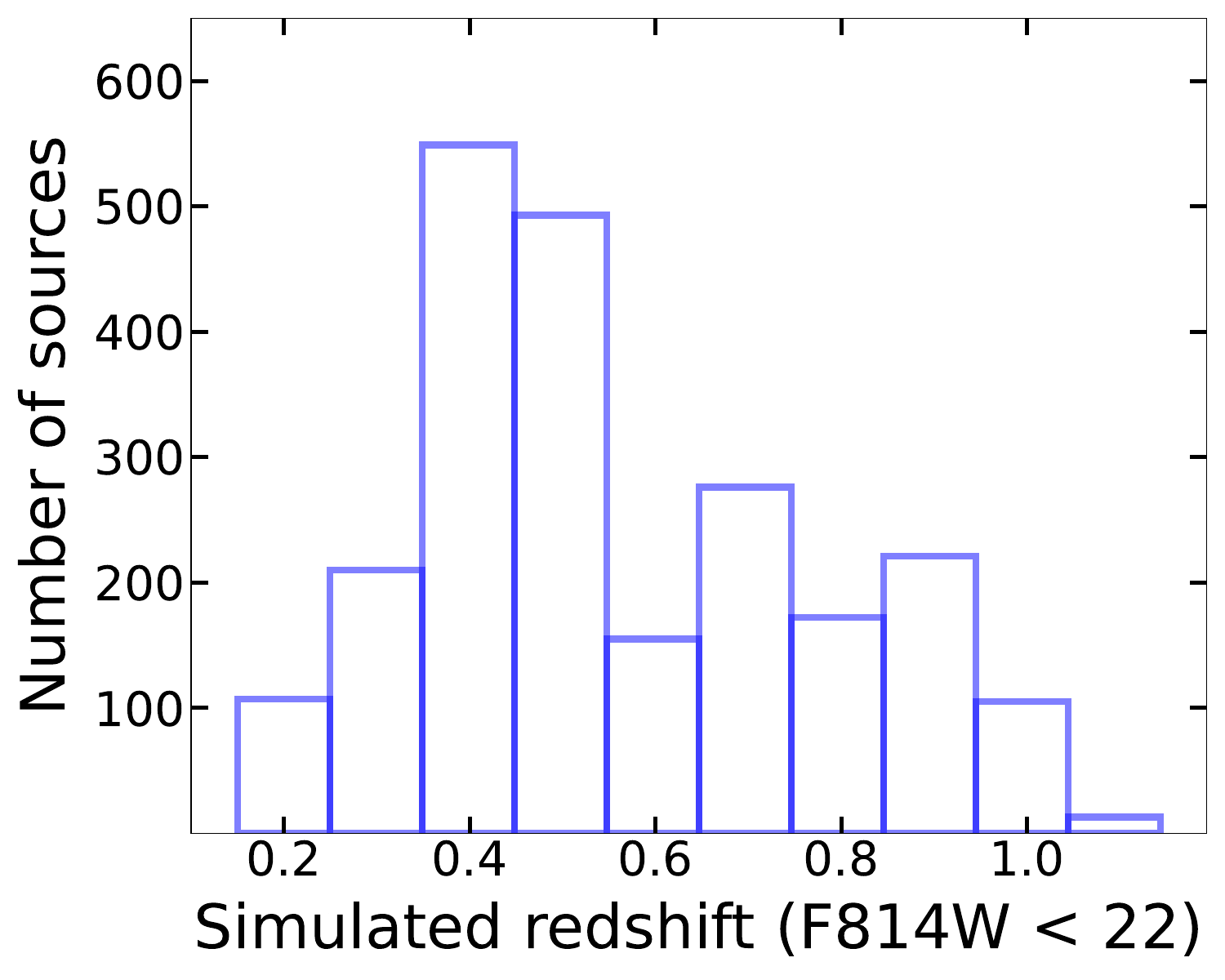}}
		{\includegraphics[height= 1.35in, width=1.32in]{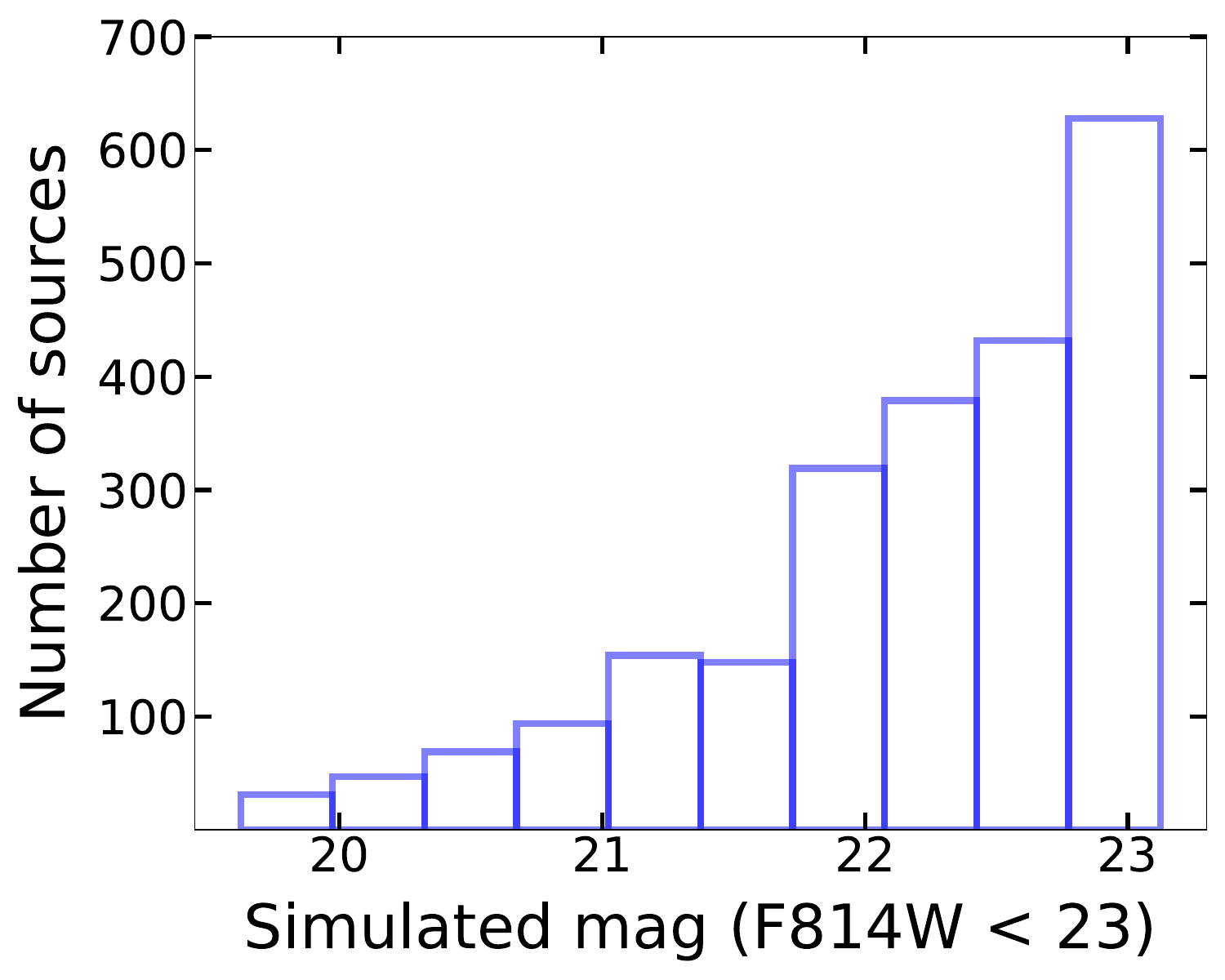}}
		{\includegraphics[height= 1.35in, width=1.32in]{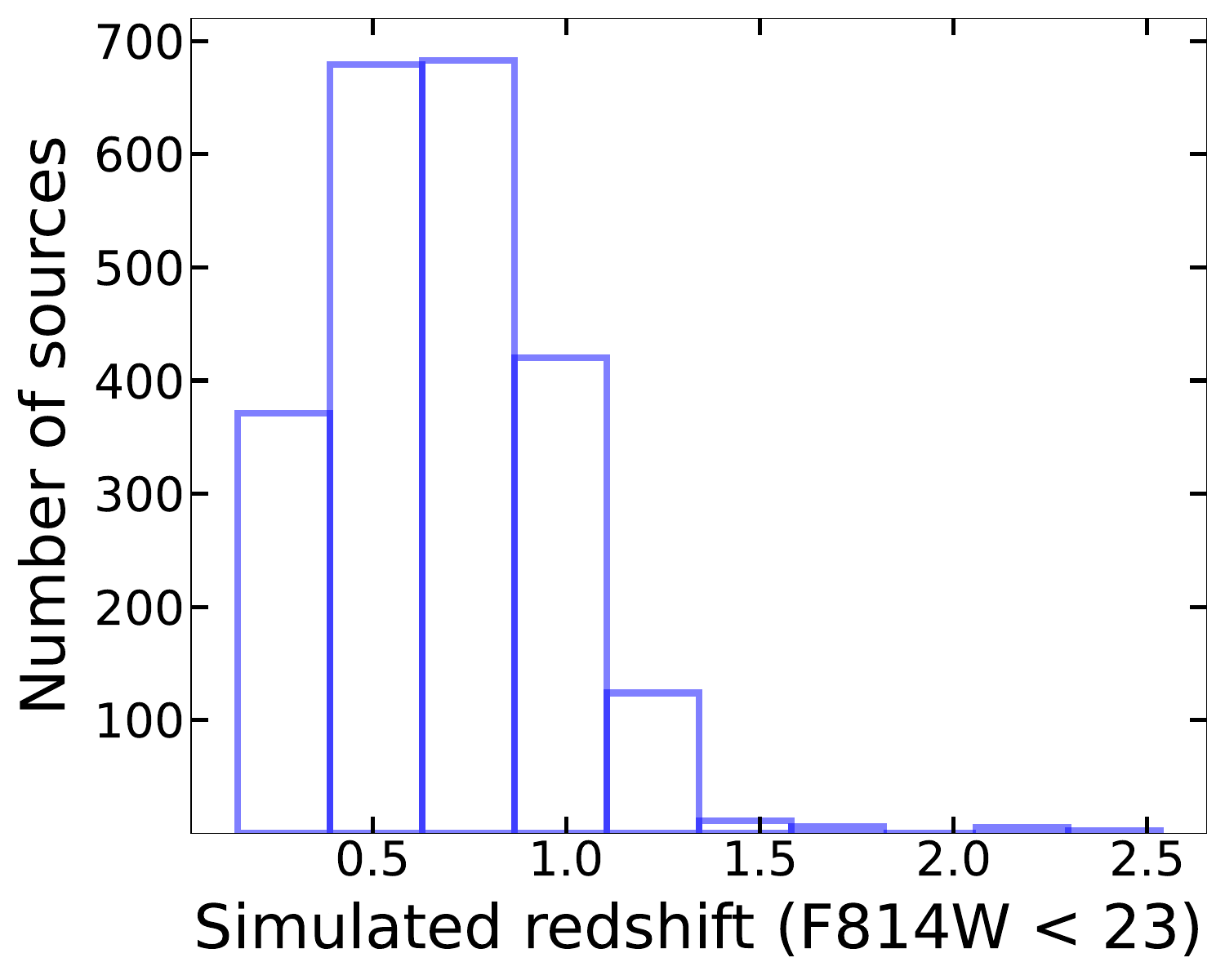}}
		{\includegraphics[height= 1.35in, width=1.32in]{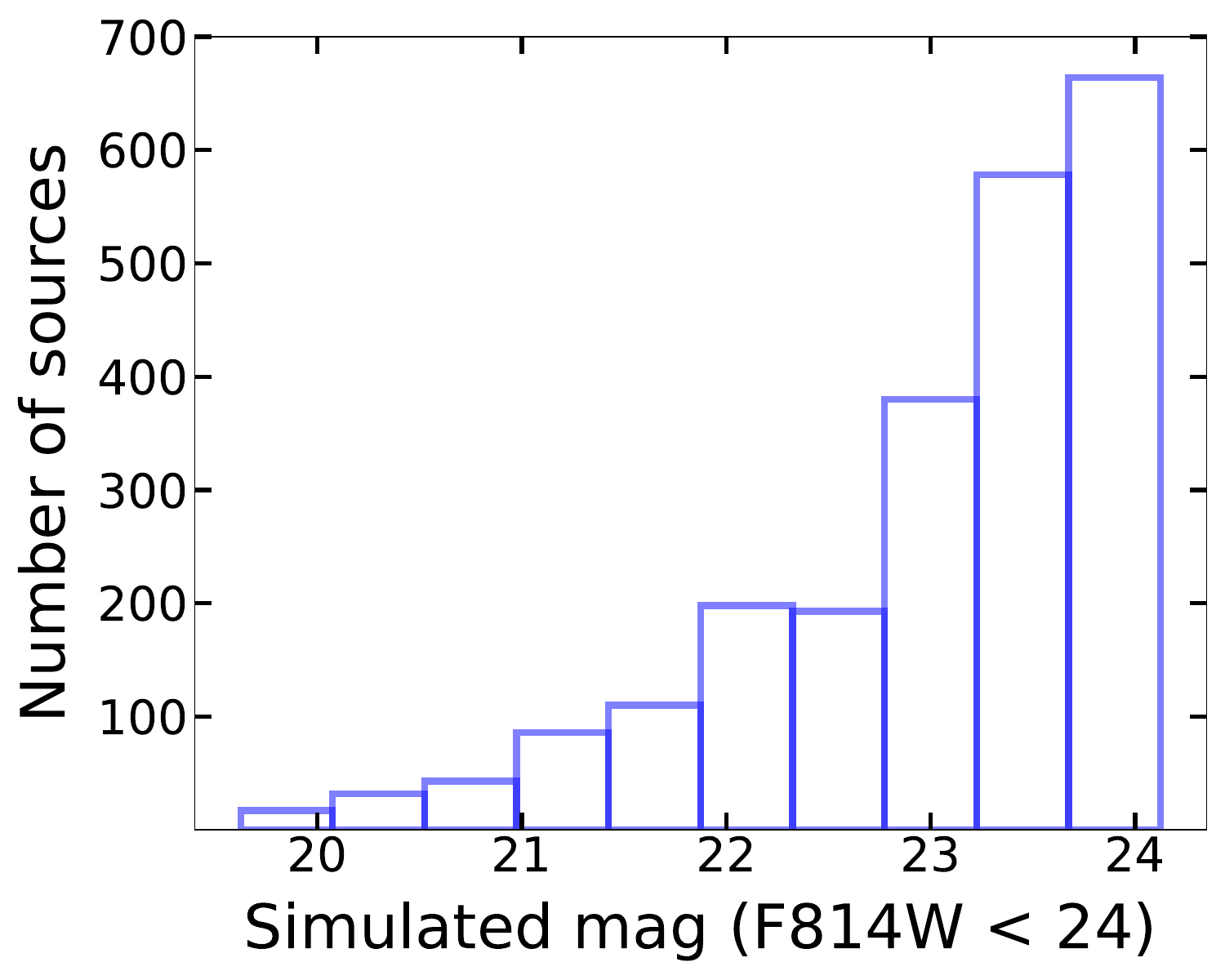}}
		{\includegraphics[height= 1.35in, width=1.32in]{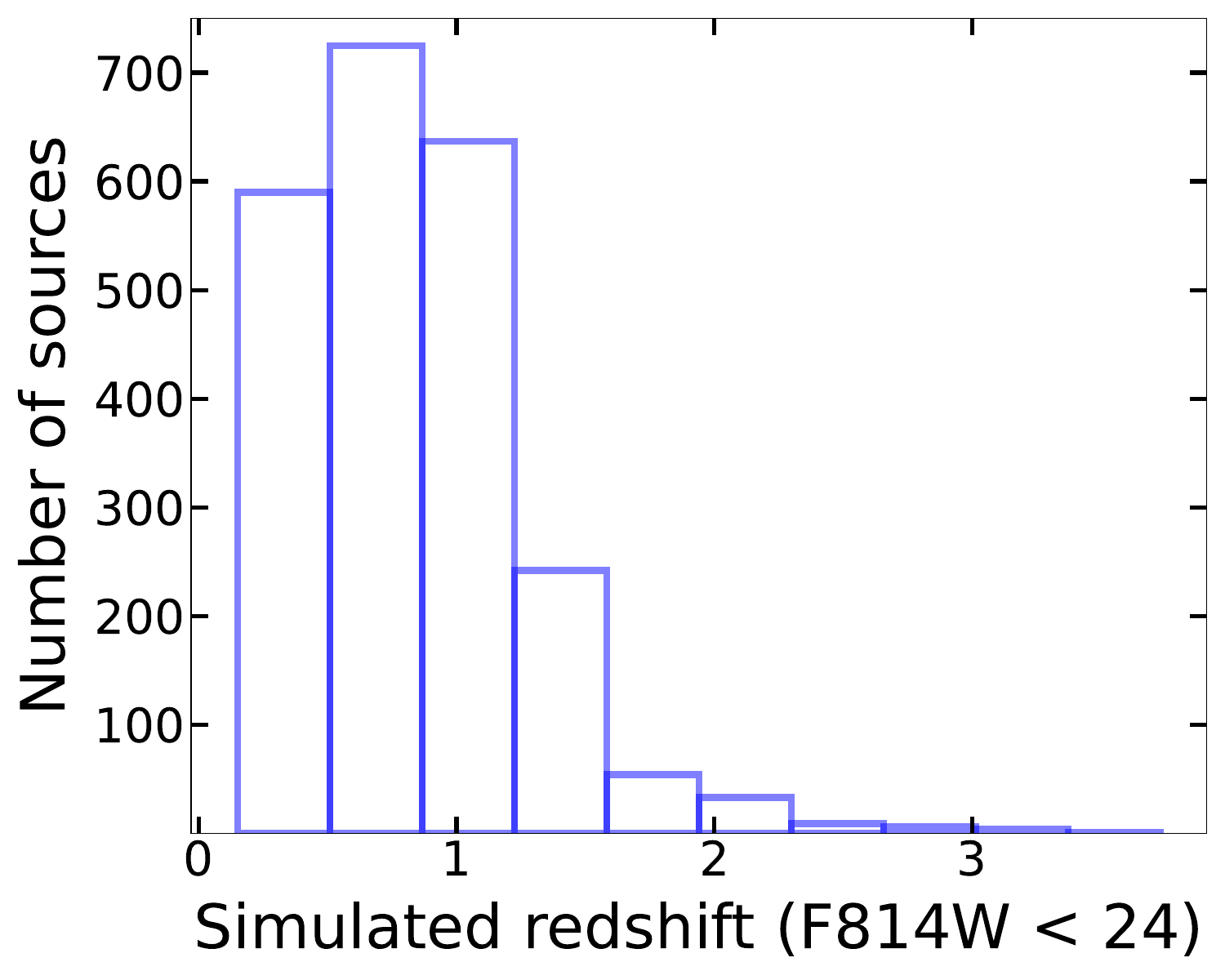}}
		\centering
		{\includegraphics[height= 1.35in, width=1.32in]{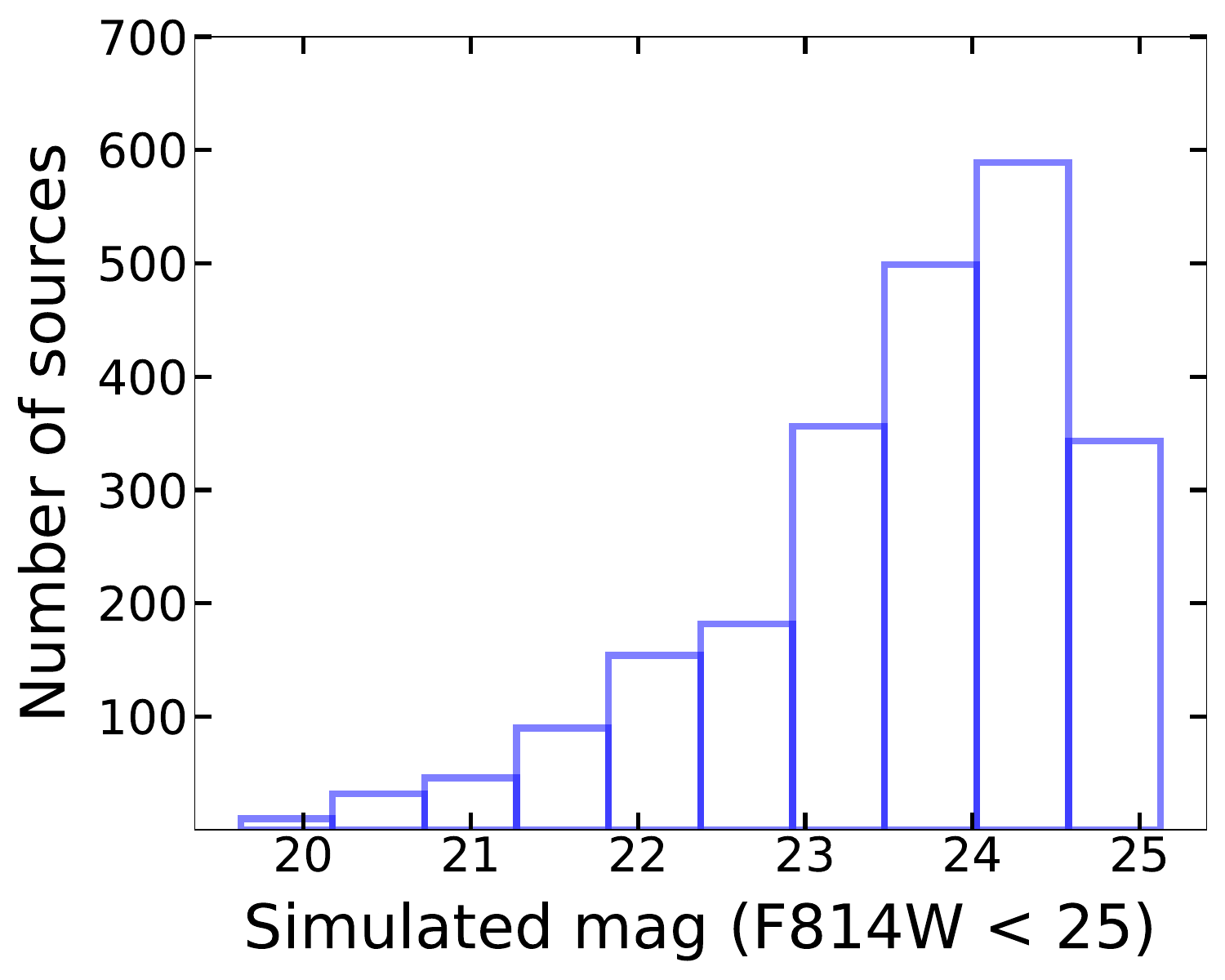}}
		{\includegraphics[height=1.35in, width=1.32in]{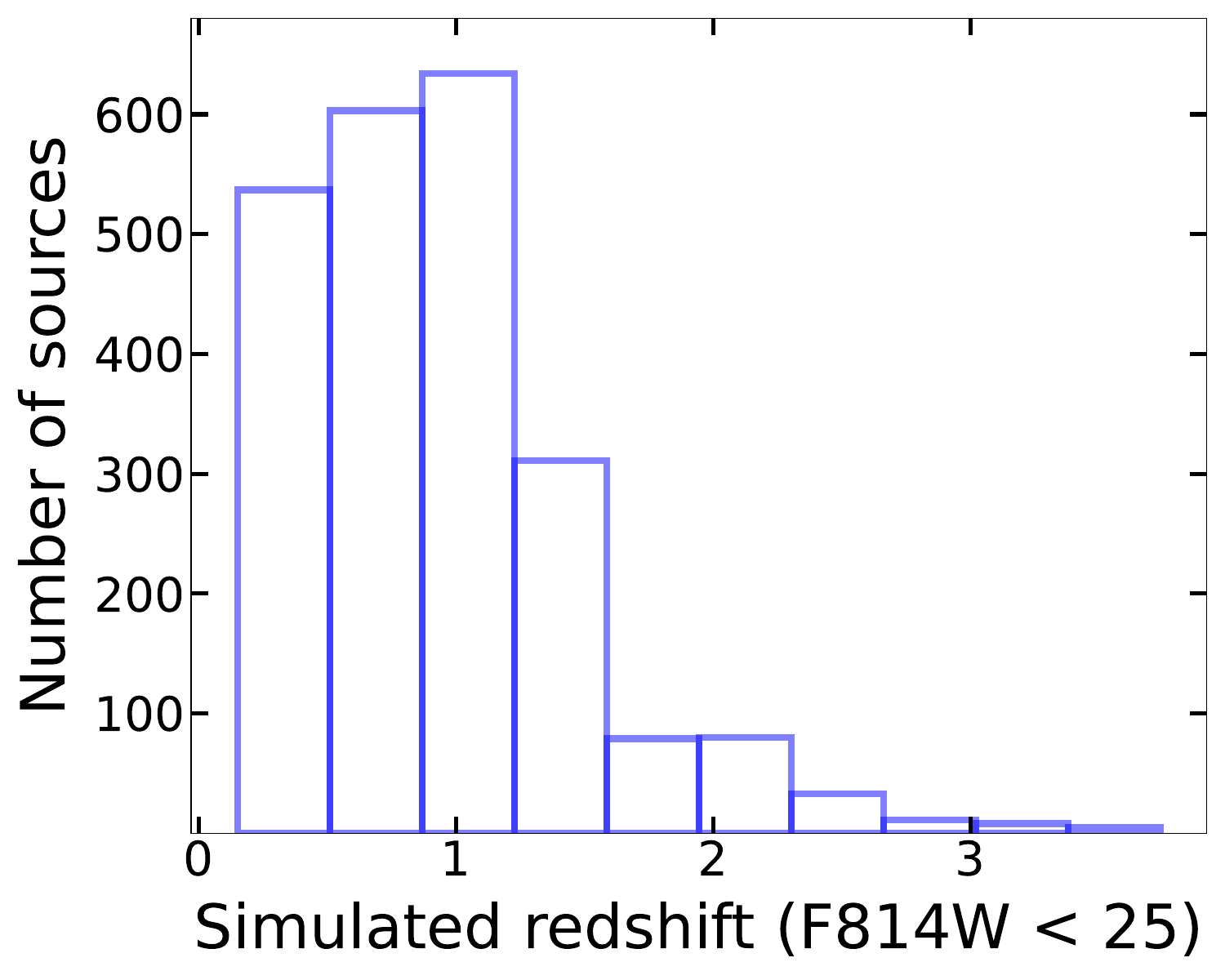}}
		\caption{The simulated magnitude (left plot per magnitude limit) and redshift (right plot per magnitude limit) distributions of the non-AGN local galaxies after moving them to higher redshift to represent the distribution of galaxies in the COSMOS field at F814W\,$\le$\,21 (top left plots), F814W\,$\le$\,22 (top right plots), F814W\,$\le$\,23 (middle left plots), F814W\,$\le$\,24 (middle right plots), and F814W\,$\le$\,25 (bottom plots).}
		\label{sec:Mag21_25}
	\end{figure}
	
	\subsection{Morphological parameters at intermediate redshift with added AGN}\label{sec:methodAGNcont}
	As explained in \cite{Getachew2022}, to measure the effect of AGN on morphological parameters, we used the above sample of 2251 local non-AGN galaxies. We simulated them by adding at their centres different fractions of AGN contribution to the total galaxy flux. For each galaxy, the PSF image obtained from the SDSS survey as described in \cite{Povic2013a} was used to simulate the AGN contribution, while the PSF was modelled using the Moffat function \citep{Moffat1969}. For each galaxy, new images have been generated by adding five different AGN contributions of 5\%, 10\%, 25\%, 50\% and 75\% to the total flux of the galaxy, which was measured with SExtractor. All additional information is given in \cite{Getachew2022} and their Figure 4. By selecting 5\%-75\% of AGN contribution, we wanted to cover the wide range of AGN fractions, considering both type-1 and type-2 AGN. This is supported by previous studies. For example, in the local universe \cite{Mountrichas2024}, using the SDSS database, obtained AGN fractions to the host total luminosity between 5\% and 80\%. The advent of the JWST also allowed to test this fraction at higher redshift. In fact, Ref.\cite{Durodola2025} quotes AGN fractions between 20\% and 70\% at z\,$>$\,4.
	
	After obtaining for each galaxy these five images with different AGN fractions, we followed the steps given in Section~\ref{sec:cosmoscon}. We first simulated all images to match higher redshift (and fainter magnitude) distributions of COSMOS, and then we dropped the simulated images into the real COSMOS background to map its resolution, and we measured six morphological parameters. Finally, we obtained the morphological parameters for the five different AGN contributions to the total flux (from 5\% to 75\%) and intermediate redshift corresponding to the five simulated magnitude limits (from $\le$\,21 to $\le$\,25). 
	
	\section{Analysis and results}\label{sec:4}
	This section describes the analysis carried out. Figure~\ref{analysis} helps to visualise the comparisons made between morphological parameters measured under different conditions, at z\,$\sim$\,0 and in COSMOS-like conditions, with and without added AGN, {described in \linebreak{}Sections \ref{sec4.1}--\ref{sec4.3}. The impact of AGN and magnitude/redshift on the morphological diagrams is presented in Section \ref{sec4.4}}.
	
	\begin{figure}
		\begin{center}
			{\includegraphics[height=2.5in, width=4in]{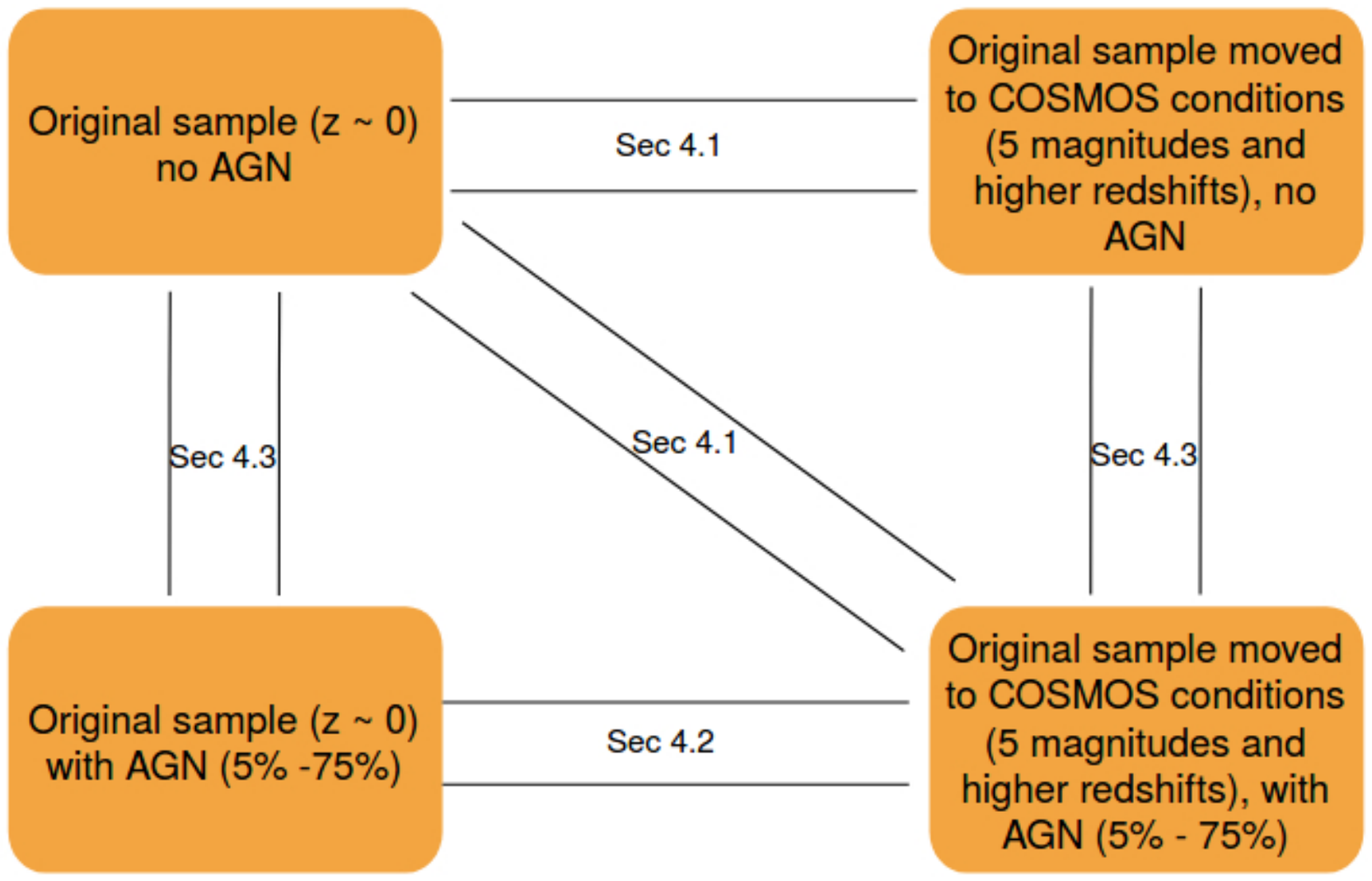}}
			\caption{Summary of analysis and comparisons carried out in Sections~\ref{sec4.1}, \ref{sec4.2}, and \ref{sec4.3}.}
			\label{analysis}
		\end{center}
	\end{figure}
	
	\subsection{Impact of magnitude, redshift, and AGN on
		morphological parameters}\label{sec4.1}
	In this section, we want to analyse how the morphological classification of galaxies can be affected by the magnitude, redshift, and AGN contribution, when all are combined and when non-parametric methods are used. To do so, we analysed the impact of the magnitude, redshift, and AGN on six morphological parameters by comparing the local non-AGN galaxies without any AGN added, with the same sample in COSMOS-like conditions without and with added AGN. These analyses are summarised in Appendix~\ref{appendixA}. \\
	\indent The detailed comparison of the distributions of CABR, GINI, CCON, M20, ASYM, and SMOOTH parameters measured for real and simulated galaxies can be seen in Figures~\ref{fig4.1:part1}\,-\,\ref{fig4.6:part1}, respectively. For each parameter, we performed all comparisons for a total sample (in orange) of galaxies (regardless of their morphological type), and for three morphological types (early in red, early-spirals in blue, and late-spirals in green), as defined in Section~\ref{sec:data1}, to test to what extent the different morphological types are sensitive to the magnitude, redshift and AGN contribution.
	In all figures, the filled (coloured) histograms show the reference distribution for all comparisons, in this case of the original sample at z\,$\sim$\,0 without any AGN added. The open histograms show the simulated galaxies in COSMOS-like conditions, moved to higher redshift and fainter magnitudes from 21 to 25 (rows from top to bottom), with no added AGN (first columns) and with 5 added AGN contributions of 5\%, 10\%, 25\%, 50\%, and 75\% (from the second to the last columns, respectively).\\
	\indent In all comparisons, we measured the variance (Var) between the parameters obtained for a real non-simulated sample and in COSMOS-like conditions without and with AGN added, to quantify the level of change and to what extent the parameters are affected by the magnitude, redshift and AGN contribution. The variance is measured as:
	\begin{equation}\label{eqn1}
		Var[\%] = 100/\textit{N}tot * \sum_{\textit{i}}^{}(abs((\textit{P}_{sim_\textit{i}} - \textit{P}_{orig_\textit{i}})/ \textit{\textit{P}}_{orig_\textit{i}})),
	\end{equation}
	where $\textit{P}_{sim_\textit{i}}$ and $\textit{P}_{orig_\textit{i}}$  are the values of the corresponding simulated and original parameters of individual galaxies, respectively, while $\textit{N}tot$ is the total number of galaxies for which the parameter has been measured. In line with \cite{Getachew2022} and \citep{Pierce2010}, we consider that parameters are significantly affected by AGN contribution and/or magnitude/redshift when Var is $\ge$\, 20\%. Figure~\ref{fig_var:part1} shows the variance in the CABR, GINI, CCON, and M20 parameters (from top to bottom), of early (left, red), early-spiral (centre, blue), and late-spiral (right, green) galaxies between the original sample at z\,$\sim$\,0 with no AGN added and in COSMOS-like conditions with AGN contribution between 0\% and 75\% (from left to right in each subplot) and high-redshift magnitudes (\textit{mag\_hz}) from $\textit{mag\_hz}<$\,21 to $\textit{mag\_hz}<$\,25 (from top to bottom in each subplot). The measured Var values are also indicated in each graph in Appendix~\ref{appendixA}. Note that equation~\ref{eqn1} has difficulties when dealing with very low parameter values close to 0, as is often the case for the ASYM and SMOOTH parameters. For these two parameters, the variances are not included in the figures, but they are discussed below and in Section~\ref{sec5}.   
	
	\begin{figure*}
		\begin{center}
			\includegraphics[height= 4.0in, width=5.5in]{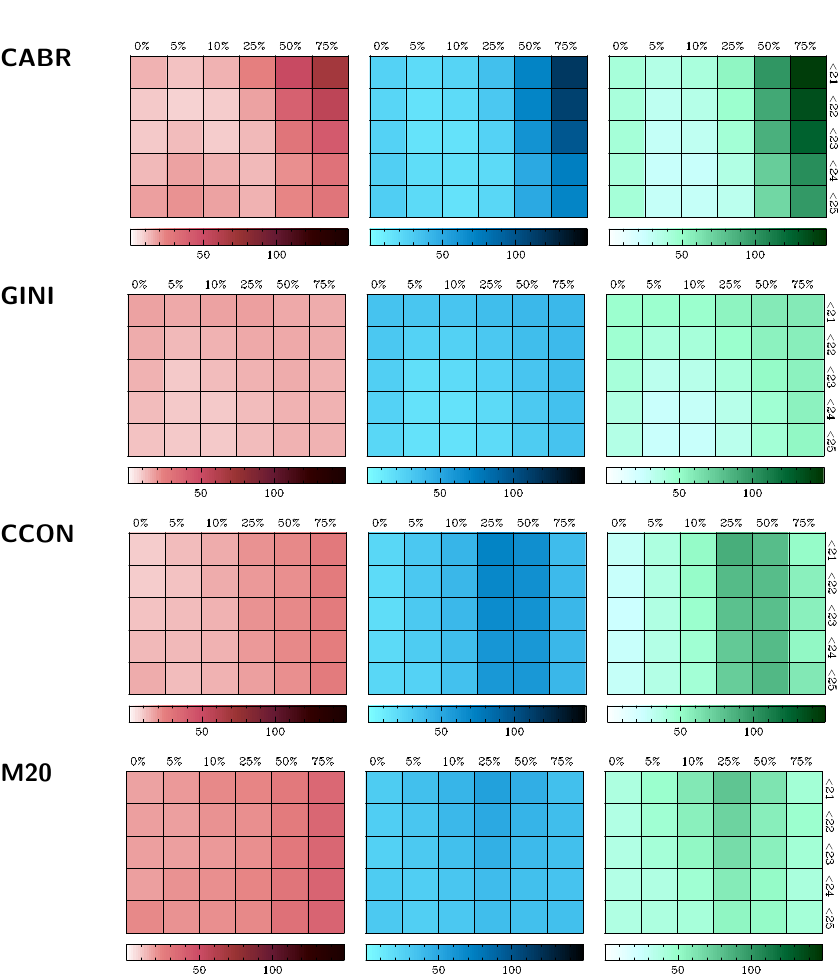}
			\caption{Change (Var) in CABR, GINI, CCON, and M20 parameters (from top to bottom) when comparing the original sample at z\,$\sim$\,0 with no AGN added and the simulated sample at COSMOS-like conditions without AGN and with five AGN contributions added from 5\% to 75\% (from left to right in each panel), at $\textit{mag\_hz}<$\,21, $\textit{mag\_hz}<$\,22, $\textit{mag\_hz}<$\,23, $mag\_hz\textit{mag\_hz}<$\,24, and $\textit{mag\_hz}<$\,25 (from top to bottom rows in each panel). The comparison is given for early-type (left, in red), early-spiral (centre, in blue), and late-spiral (right, in green) galaxies. For the ASYM and SMOOTH parameters, the variances are not included in the figure, but they are discussed in the text (see Sections~\ref{sec4.1} and \ref{sec5}).}
			\label{fig_var:part1}
		\end{center}
	\end{figure*}
	
	In general, the number of sources of all morphological parameters is decreasing from brighter to fainter magnitudes and from lower to higher AGN added contributions, as it becomes more difficult to measure morphological parameters. In the case of the concentration parameters (CABR, GINI, CCON, and M20), in general, $>$\,90\% of the sources have well-measured parameters, except for the faintest magnitudes of $\textit{mag\_hz}<$\,24 and $\textit{mag\_hz}<$\,25, and for the largest added AGN contributions of 50\% and 75\%, when the number of sources with measured parameters drops to 60-70\% of the total sample, with spirals being more affected than early-type galaxies. On the other hand, the measurements of the ASYM and SMOOTH parameters are more complicated. When considering the total sample at z\,$\sim$\,0 with no AGN added, we only obtained reliable values for 75\% and 45\% of the sources, respectively. These numbers are as low as 50\% and 20\%, respectively, under COSMOS-like conditions when considering the faintest magnitudes and the largest AGN contributions.\\
	\indent For comparison, we show the impact of magnitude and redshift without adding any AGN contribution. This is
	represented in the first columns ofFigure~\ref{fig_var:part1}, and corresponds to the case of normal galaxies in COSMOS-like conditions. We can see that only the early-type galaxies are not affected in COSMOS-like conditions, showing for all the parameters Var\,$\lesssim$\,20\%, while spiral galaxies (both early and late) are significantly affected in the case of all 5 magnitude/redshift limits, showing the change in all parameters above 20\%. The results obtained here are in agreement with those of \cite{Povic2015}, where the effect of resolution and survey depth has been studied in three different surveys, including COSMOS. In both works, the same trends were obtained for all parameters with similar variances for all morphological types. In this work, we can also see that the morphological parameters of normal galaxies are more affected by magnitude and redshift in COSMOS-like conditions than active galaxies, as discussed in Section~\ref{sec5}. \\
	\indent Here we briefly discuss the impact of magnitude, redshift, and AGN, all combined, on the six morphological parameters, as shown in Figure~\ref{fig_var:part1} and Figures~\ref{fig4.1:part1}-\ref{fig4.6:part1}:
	\begin{itemize}
		\item[$\bullet$] CABR. In Figure~\ref{fig_var:part1} (top panels), it can be seen that this parameter is the most affected compared to other morphological parameters. On average, when considering the total sample of galaxies, early-types, and early-spirals, the variance in the parameter is significant in all magnitude and redshift conditions when the contribution of the AGN is equal to or larger than 25\%, 50\%, and 75\%, respectively. For late-spirals, the CABR parameter undergoes significant changes above 20\% in all magnitudes, redshifts, and AGN conditions. Figure~\ref{fig4.1:part1} shows that the distribution of CABR after the simulations can move in both directions, towards higher and lower values compared to the non-simulated sample. Taking into account the definition of CABR, as described in Section~\ref{sec:3}, this can be interpreted that when the parameter is shifted towards higher values the contribution of the AGN, increasing the flux within the inner radius, has the dominant effect compared to the impact of the magnitude and redshift, which is dimming the light in the galaxy, in particular in the outer regions. Similarly, when it is shifted towards lower values, it means that the magnitude and redshift have more impact on the CABR than the added AGN itself. We can observe that for 50\% and 75\% of the AGN added, the AGN contribution plays a dominant role independently of the magnitude and redshift. This means that if more light is added to the central part of the galaxy, according to the CABR definition, its values will increase. For 5\%-25\% added AGN, AGN plays the dominant role up to \textit{mag\_hz}$<$\,22, as the central light of the galaxy within 30\% of the galaxy radius is less affected than the rest of the galaxy. At fainter magnitude limits (23 and above), the impact of the magnitude and redshift on the CABR becomes a dominant effect, affecting light in the entire galaxy.
		\item[$\bullet$] GINI. This is the least affected parameter, with the lowest values of variance, as can be seen in Figures~\ref{fig_var:part1} (middle top panels). Even so, the total sample of galaxies and early-spirals will be significantly affected when the AGN contribution is equal to or larger than 25\%, regardless of the magnitude and redshift limits. Again, late-spirals will be significantly affected in all cases, similarly to the CABR. On the other hand, the GINI parameter for early-types in all cases will suffer variations below 20\%.
		The tendency of GINI to shift towards higher or lower values compared to the original sample is similar to that of CABR, but with smaller shifts. This is due to the difference in the definition of GINI and CABR, in which, although both are concentration parameters, CABR is much more sensitive to the concentration of light in the central part of the galaxy than GINI.
		\item[$\bullet$] CCON. This parameter will undergo significant changes of more than 20\%, in all cases of the total sample, due to the impact magnitude, redshift, and AGN on spiral galaxies, both early and late, as shown in Figures~\ref{fig_var:part1} (middle bottom panels). The early-types will be significantly affected only when the added contribution from the AGN is greater than 25\%. In general, the change in the CCON parameter for the simulated sample will be towards higher values. However, for 50\% and 75\% of added AGN in the early-types and early-spirals (75\% of AGN added for late-spirals), because the definition of CCON is the radius ratio and not the flux ratio, as in the case of CABR, the distribution of the simulated sample will start to change the trend towards lower concentrations in the simulated sample, compared to the lower AGN contributions. This effect is more pronounced in early-type galaxies, which, in general, have higher central concentrations and CCON values and are therefore more affected.
		\item[$\bullet$] M20. This parameter will also undergo significant changes of more than 20\%, in all cases of the total sample, early- and late-spirals, as shown in Figure~\ref{fig_var:part1} (bottom panels). In early-type galaxies, the variations of M20 are around 20\% up to the added contribution of AGN of 25\%. For AGN contributions above this value, M20 is more affected, with variations exceeding 30\%.
		\item[$\bullet$] ASYM. We found similar distributions of ASYM between simulated and original samples in most cases of magnitude, redshift and AGN (see Figure~\ref{fig4.5:part1}). However, this parameter is very sensitive to noise, as has already been studied and discussed in \cite{Povic2015}. Similar was reported in other studies \citep[e.g.,][]{Zhao2022, Nersesian2023, Sazonova2024}. Therefore, we do not suggest its use except in combination with other morphological parameters.
		\item[$\bullet$] SMOOTH. We find larger changes in the distribution of this parameter in early-type galaxies than in spiral galaxies, but towards lower values, as can be seen in Figure~\ref{fig4.6:part1}. This parameter is not convenient to use due to the very small number of sources for which it can be measured, as mentioned above. 
	\end{itemize}
	
	\subsection{Impact of magnitude and redshift on the morphological parameters of active galaxies}\label{sec4.2}
	We compared the measurements of six morphological parameters after adding an AGN contribution of 5\% to 75\% to the total flux, using the original sample at z\,$\sim$\,0 and simulated sample in COSMOS-like conditions. This corresponds to the case of measurement and comparison of the morphological parameters of active galaxies at z\,$\sim$\,0 and higher redshifts, and it can help us to better understand the effect of magnitude and redshift on different morphological parameters when classifying AGN host galaxies. Overall, among the three types of analysis conducted in sections~\ref{sec4.1}-\ref{sec4.3}, here we find the lowest changes in morphological parameters. These analysis are represented in Appendix~\ref{appendixB}.\\
	\indent Figures~\ref{fig4.1:part2}-\ref{fig4.6:part2} show the distribution of CABR, GINI, CCON, M20, ASYM, and SMOOTH, respectively, of simulated active galaxies at z\,$\sim$\,0 (filled and coloured histograms) and under COSMOS-like conditions (open histograms), or the direct impact of magnitude and redshift on the AGN host-galaxy classification. The different panels show the same analysis for different magnitudes and redshifts (top to bottom) and AGN contributions (left to right), as indicated in each panel. The different colours again examine different types of galaxies as described in Section~\ref{sec4.1}.\\
	\indent The variances have been measured similarly as in Section~\ref{sec4.1}, where now $P_{sim_i}$ and $P_{orig_i}$ are the values of the morphological parameters of individual galaxies, corresponding to the simulated active galaxies in COSMOS-like conditions and simulated active galaxies at z\,$\sim$\,0, respectively. The obtained variances are shown in Figure~\ref{fig_var:part2}, while the obtained results are briefly summarised here for each parameter and are discussed more in Section~\ref{sec5}:
	\begin{itemize}
		\item[$\bullet$] CABR. Up to 50\% of the AGN added, the effect of magnitude and redshift is up to $\sim$\,20\% for all morphological types (with slightly higher values of up to 30\% for late-spirals), as can be seen in Figures~\ref{fig_var:part2} (top panels). The change in variance, towards lower values of CABR, increases with magnitude, as expected (except for the AGN contribution of 75\%). The highest measured CABR values are obtained at \textit{mag\_hz}\,<\,21, when the outskirts of a galaxy are the ones mainly affected, leading to higher CABR values. On the other hand, at \textit{mag\_hz}\,<\,25 the brightness of the entire galaxy is affected, giving generally lower CABR values than at z\,$\sim$\,0. The most significant impact of magnitude and redshift on active galaxies is observed for 75\% of the AGN added contribution, being around 20\% for early-type galaxies, 20-35\% for early-spirals, and 30\%-40\% for late-spirals.
		\item[$\bullet$] GINI. Again, this is the most stable of all parameters, as can be seen in Figure~\ref{fig_var:part2} (middle top panels). In this case, for active galaxies with an AGN contribution of 25\%-75\%, the impact of the magnitude and redshift on GINI is $\lesssim$\,20\% for all morphological types (except when the AGN contribution is 25\% for late-spirals and \textit{mag\_hz}\,$<$\,23, then the variance is up to 30\%). In early-type galaxies, the difference is insignificant for less luminous active galaxies, with AGN contributions of 5\% and 10\%. In the case of spirals, at magnitudes of $\leq$\,21 and $\leq$\,22 that mainly affect the disc, active galaxies at higher redshift will show higher GINI values with a variance of $\sim$\,15\%-30\% for early-spirals and $\sim$\,30\%-40\% for late-spirals. As the magnitude limit increases, which affects the brightness of the entire galaxy, the variance will decrease in both early and late spiral galaxies.
		\item[$\bullet$] CCON. The impact of the magnitude and redshift on the CCON concentration index in early-type active galaxies will be insignificant ($<$\,20\%), up to an AGN contribution of 25\%. These differences become significant (20\%-35\%) for a 50\% and 75\% AGN contribution (see Figure~\ref{fig_var:part2}, middle bottom panels). For the spiral active galaxies, the variance is significant in all cases, varying between 20\%-30\% and 20\%-40\% in the cases of early- and late-spiral galaxies, respectively. In all cases, as the magnitude limit increases, affecting a larger and larger fraction of the galaxy light, the CCON parameter of galaxies in COSMOS-like conditions will move towards lower values.
		\item[$\bullet$] M20. The behaviour of this parameter at $z\sim 0$ and in COSMOS-like conditions is shown in Figure~\ref{fig_var:part2} (bottom panels) for all morphological types. In all cases, the impact of the magnitude and redshift is insignificant ($<$\,20\%) for all morphological types when the AGN contribution is up to 25\%. The variances become significant (20\%-36\%) at 50\% and 75\% AGN contribution.
		\item[$\bullet$] ASYM and SMOOTH. The results and conclusions of these two parameters, shown in Figures~\ref{fig4.5:part2} and \ref{fig4.6:part2}, respectively, are similar to those discussed in the previous section.
	\end{itemize}
	
	\begin{figure*}
		\begin{center}
			{\includegraphics[height=4in, width=5.5in] {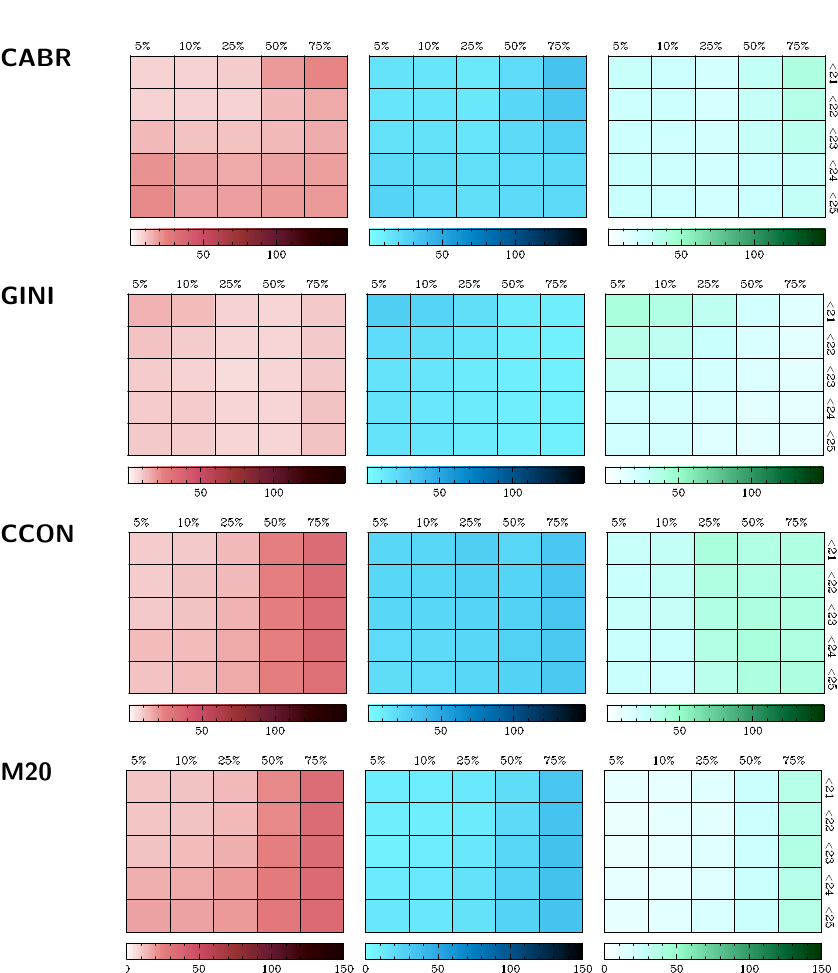}}
			\caption{Same as in Figure~\ref{fig_var:part1}, but with the variance measured when comparing samples at z\,$\sim$\,0 and under COSMOS-like conditions with AGN added in both cases.}
			\label{fig_var:part2}
		\end{center}
	\end{figure*}

	\subsection{Impact of AGN on the morphological parameters of
		intermediate-redshift galaxies in COSMOS-like
		conditions}\label{sec4.3}
	Finally, we compared the six morphological parameters obtained in COSMOS-like conditions before and after adding the AGN contribution in the centre of galaxies, from 5\%\,-\,75\%. 
	With this analysis, we can better understand the impact of AGN on morphological parameters at higher redshifts, in COSMOS-like conditions. These analyses are represented in Appendix~\ref{appendixC}. \\
	\indent Figures~\ref{fig4.1:part3}-\ref{fig4.6:part3} show the distribution of CABR, GINI, CCON, M20, ASYM, and SMOOTH, respectively, of simulated non-AGN galaxies in COSMOS-like conditions (filled and coloured histograms) and simulated galaxies with added AGN contribution in COSMOS-like conditions (open histograms), or the direct impact of AGN on the classification of its host-galaxy at higher redshift. The different panels show the same analysis for different AGN contributions (top to bottom) and magnitudes and redshifts (left to right), as indicated in each panel. The different colours again examine different types of galaxies as described in Section~\ref{sec4.1}. In addition, we use for comparison the results of \cite{Getachew2022} (GW22, first column in all Figures of Appendix~\ref{appendixC}), where the same analysis has been carried out at z\,$\sim$\,0.\\
	\indent We again measured the variance of the morphological parameters similar to \linebreak{} Equation~\ref{eqn1}, using in this case, the values of the parameters of individual galaxies in COSMOS-like conditions without ($\textit{P}_{orig_\textit{i}}$) and with ($\textit{P}_{sim_i\textit{}}$) added AGN contribution. The variance values are shown in Figure~\ref{fig_var:part3} for all conditions and the three morphological types, similar to Figures~\ref{fig_var:part1} and \ref{fig_var:part2}.
	In all GW22 comparison columns (the first columns in all Figures), the variance is measured between the parameters of individual galaxies obtained at z\,$\sim$\,0 without ($P_{orig_i}$) and with ($\textit{P}_{sim_\textit{i}}$) AGN added, as in \cite{Getachew2022}. As shown in our previous work, at z\,$\sim$\,0 for all parameters, the effect of AGN becomes significant, and in most cases $\geq$\,20\%, when the AGN contribution to the total flux is equal to or greater than 25\%. We get the same results in this work as at z\,$\sim$\,0.
	
	Please note that the meaning of the rows and columns in Figures of Appendix~\ref{appendixC} and in Figure~\ref{fig_var:part3} is intentionally different, to mark different types of analysis.
	
	\begin{figure*}[h!!]
		\begin{center}
			{\includegraphics[height= 4in, width=5.5in]{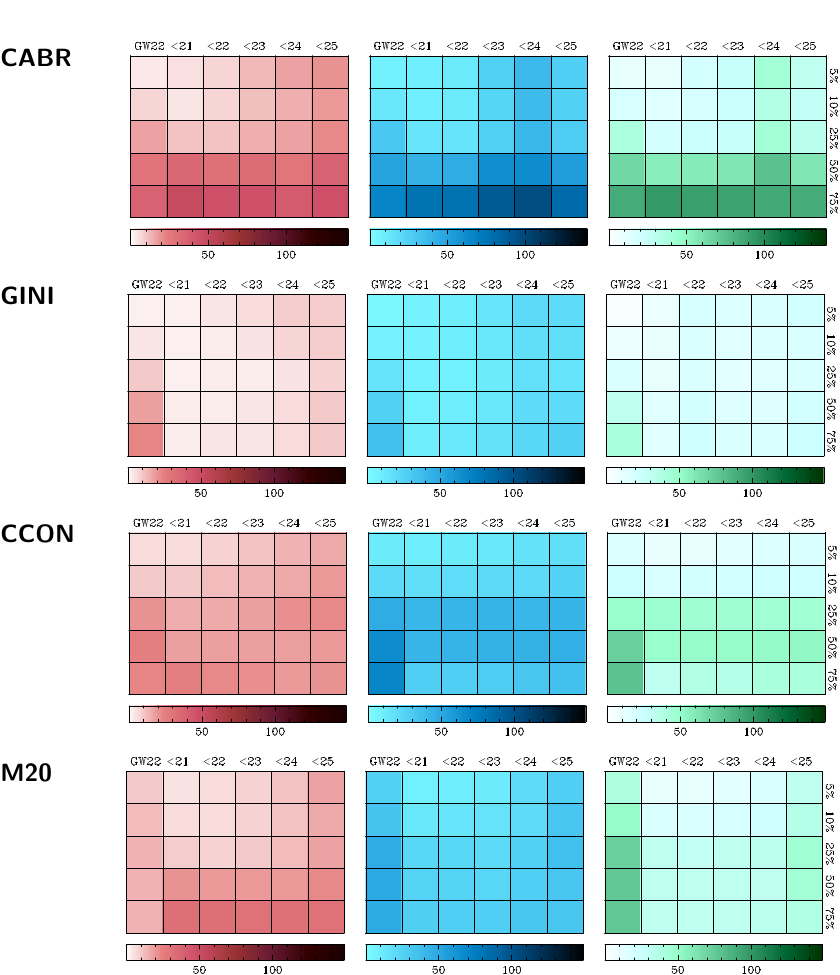}}
			\caption{Same as in Figure~\ref{fig_var:part1}, but with the variance measured when comparing samples in COSMOS-like conditions without and with AGN added. The first column (GW22) shows the same comparison, but at z\,$\sim$\,0, as measured in \cite{Getachew2022}.}
			\label{fig_var:part3}
		\end{center}
	\end{figure*}
	
	Following the analysis in Section~\ref{sec4.1}, here we find the second largest variances in morphological parameters, as shown in Figure~\ref{fig_var:part3}. Our main findings on the effect of AGN on morphological parameters of galaxies in COSMOS-like conditions are as follows:
	\begin{itemize}
		\item[$\bullet$]CABR. In general, there are no significant differences up
		to the added AGN contribution of 25\%, in the case of the early-type and early-spiral galaxies, as shown in Figures~\ref{fig_var:part3} (top panels). For the larger contributions from the AGN, of 50\% and 75\%, the variance becomes significant (above 20\%). In the case of late-spirals, the CABR parameter will be significantly affected in almost all cases except for the lowest
		AGN contributions of 5\%-10\% and magnitude limits of \textit{mag\_hz}\,$<$\,22. In this analysis, the AGN contribution dominates, and therefore, in almost all cases, the values of CABR after adding the AGN at the centre of a galaxy tend to shift towards higher values, due to the definition of the CABR parameter.
		\item[$\bullet$] GINI. This is again the most stable parameter showing in almost all cases the variance of $<$\,20\%, regardless of morphological type, as seen in Figures~\ref{fig_var:part3} (top bottom panels). This means that, similar to what we found in \cite{Getachew2022} that AGN does not have a significant impact on GINI at z\,$\sim$\,0, is also true for galaxies at higher redshift, when comparing non-active and active galaxies in COSMOS-like conditions.
		\item[$\bullet$] \textls[-5]{CCON. For early-type galaxies, the change in variance is $\leq$\,20\% in almost all cases except in 75\% added AGN contribution, as shown in {Figure \ref{fig_var:part3} (middle bottom panels)}. In spiral galaxies, this parameter will undergo significant changes above 40\%, where the added AGN contribution is $>$25\%, and 20--25\% changes for 10\% added AGN. Only in the case of 5\% added AGN, the variance in CCON is insignificant. Once again, we are observing the complex changes in behavior of the CCON parameter, as in Sections \ref{sec4.1} and \ref{sec4.3}, and its change from higher values to lower values depending on the AGN contribution, compared to the reference sample in COSMOS-like conditions without AGN added. For all morphologies, up to the AGN contribution of 25\%, the simulated sample with AGN shows higher values than the reference sample without AGN added. When the AGN contribution is 50\% and 75\%, CCON will move to lower values at all magnitudes, as most of the flux is now accumulated within the very small radii. }
		\item[$\bullet$] M20. In early-type galaxies, this parameter will now go through significant changes only when the added AGN contribution is $>$\,50\%, as shown in Figures~\ref{fig_var:part3} (bottom panels). In spiral galaxies, the variance is significant for AGN added contribution of $\ge$\,25\% at all magnitudes, and at the highest magnitude limits of mag\_24\,$<$\,24 and mag\_24\,$<$\,25 when AGN contribution is 5\% and 10\%.
	\end{itemize}
	The ASYM and SMOOTH parameters are shown in Figures~\ref{fig4.5:part3} and \ref{fig4.6:part3}, respectively. They show a similar behaviour to that discussed in Section~\ref{sec4.1}.
	
	\subsection{Morphological diagrams}\label{sec4.4}
	In this section, we analysed some of the morphological diagrams commonly used in previous studies to distinguish among different types of galaxies \citep[e.g.,][]{Abraham1994, Abraham1996, Abraham2003, Con2000, Con2003, Lotz2004, Pierce2010, Povic2013a, Povic2015, Pintos2016, Tarsitano2018, Amado2019, Mahoro2019, Getachew2022, Nersesian2023}. The main objective is to understand the effect of magnitude, redshift, and AGN together on the morphological diagrams and the final morphological classification. We want to see to what extent commonly used morphological diagrams can be used in the classification of active galaxies at higher redshifts (and fainter magnitudes). \\
	\indent We analyse some of the most commonly used diagrams, such as CABR vs. CCON (see Figure~\ref{CABR_vs_CCON}), CABR vs. GINI (see Figure~\ref{CABR_vs_GINI}), GINI vs. M20 (see Figure~\ref{M20_vs_GINI}), and ASYM vs. CABR (see Figure~\ref{logASYM_vs_logCABR}) using the simulated sample in COSMOS-like conditions, with 5\%\,-\,75\% added AGN contribution, as indicated in the panels, and at z\,$\sim$\,0 with no AGN added (bottom right panel in all Figures). Each Figure has five plots (a-e), corresponding to 5 simulated magnitude limits (from \textit{mag\_hz}\,$<$\,21 to \textit{mag\_hz}\,$<$\,25). Each plot consists of 6 panels, which show the distribution of galaxies with added AGN of 5\% (top, left), 10\% (top, middle), 25\% (top, right), 50\% (bottom, left), 75\% (bottom, middle), and without AGN added (bottom, right). 
	The different coloured contours correspond to different types of galaxies, such as: early (red contours), early-spiral (blue), and late-spiral (green).
	
	\begin{figure}[!t]
		\subfloat[]{\includegraphics[height=1.7in, width=2.93in]{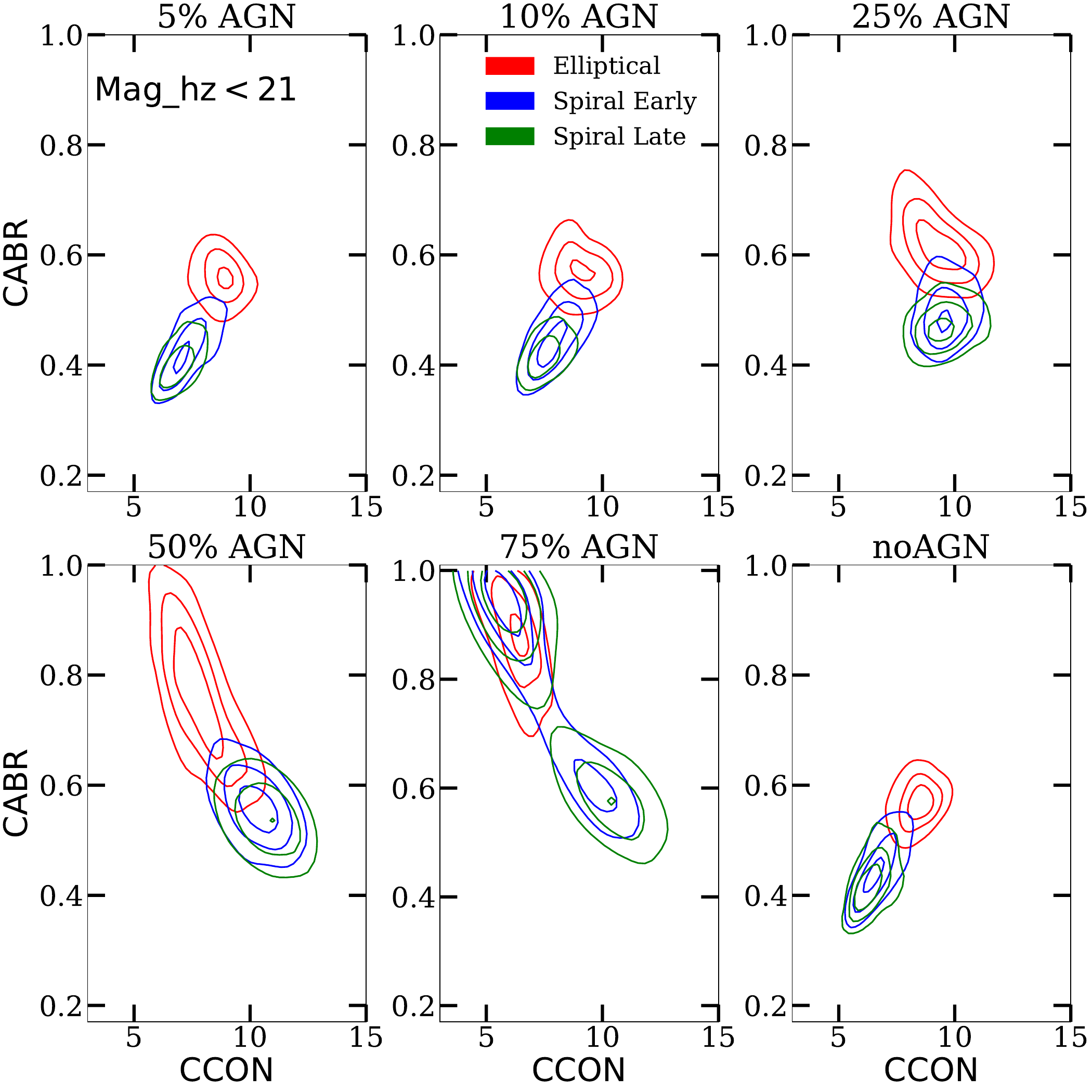}}
		\subfloat[]{\includegraphics[height=1.7in, width=2.93in]{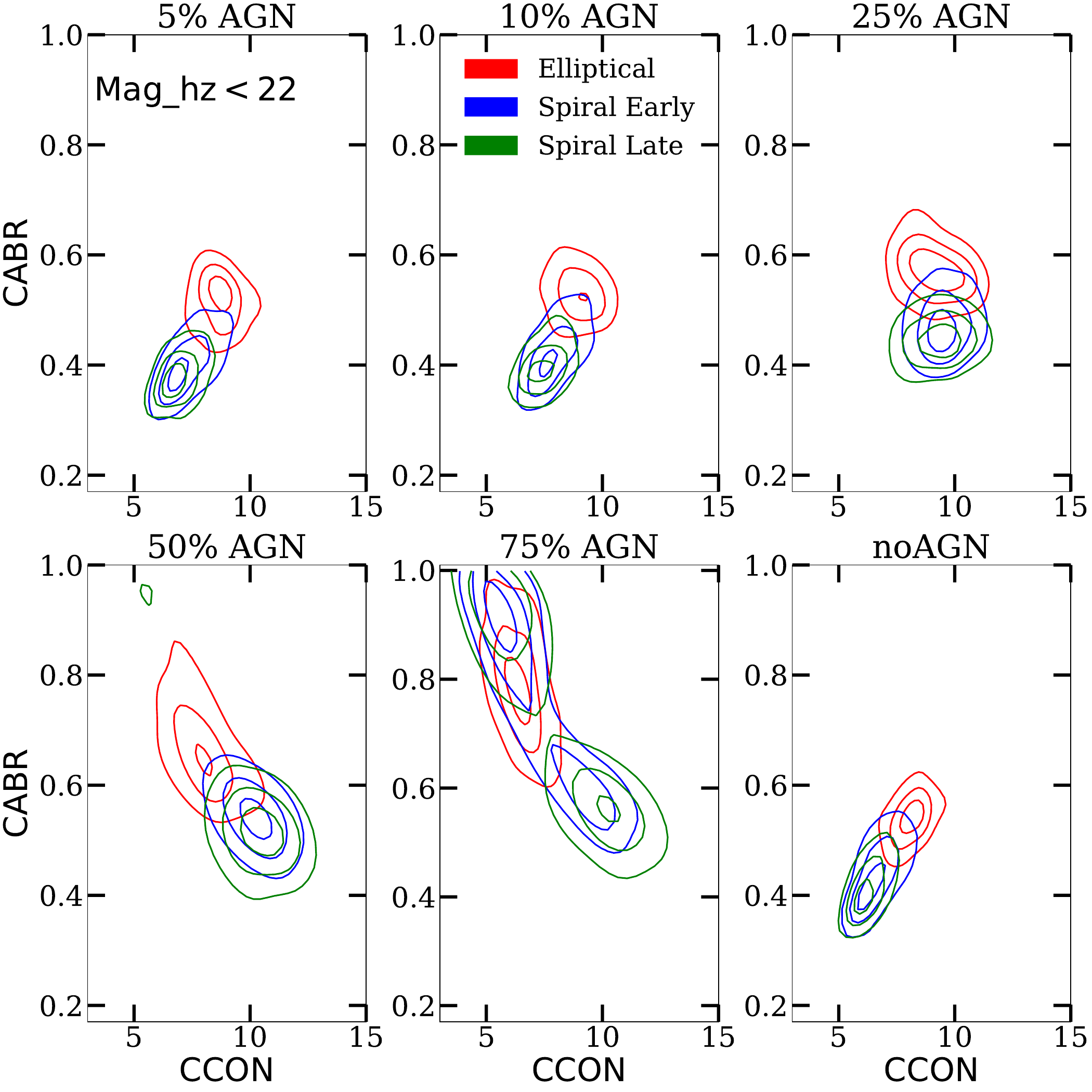}}
		\begin{center}
			\subfloat[]{\includegraphics[height=1.7in, width=2.93in]{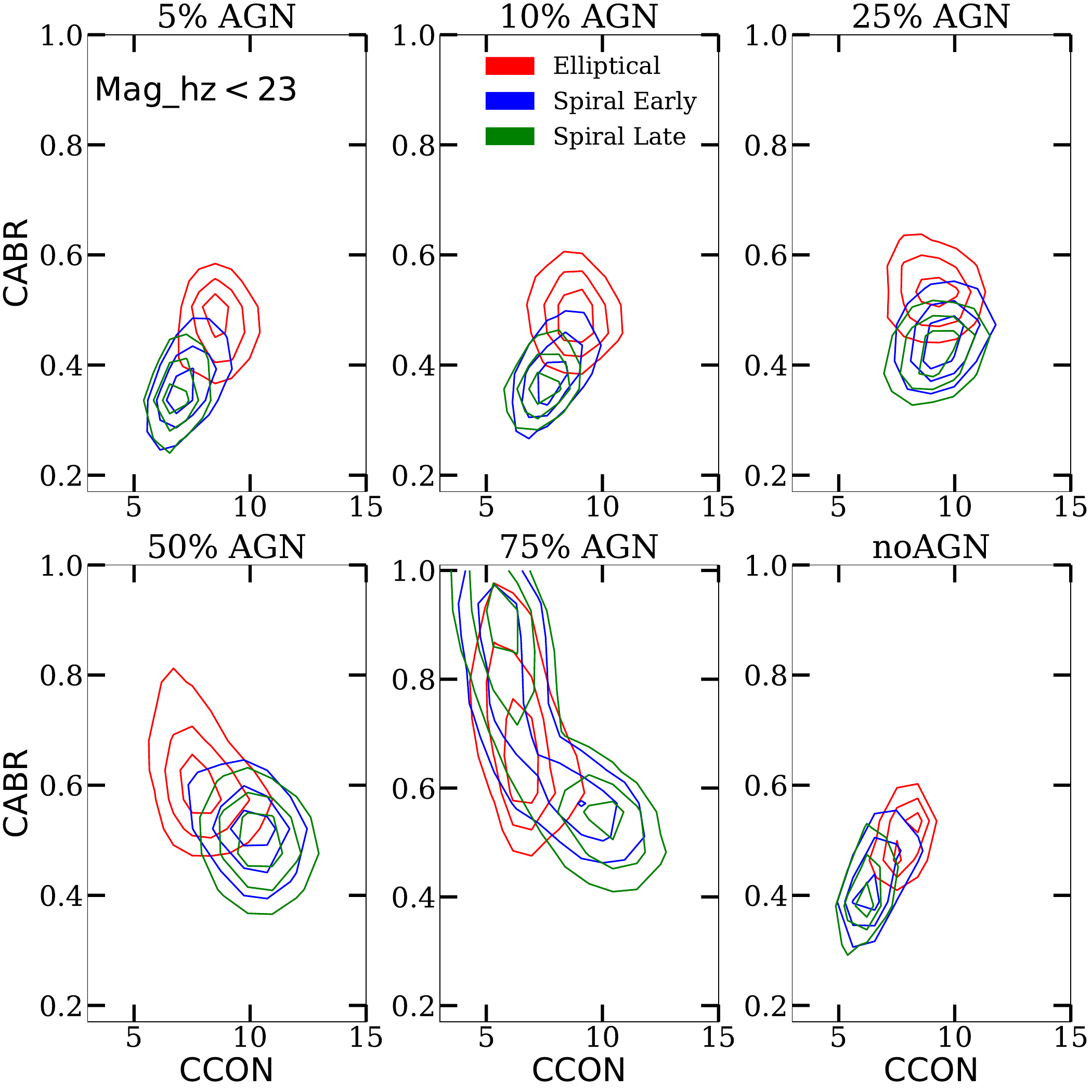}}
			\subfloat[]{\includegraphics[height=1.7in, width=2.93in]{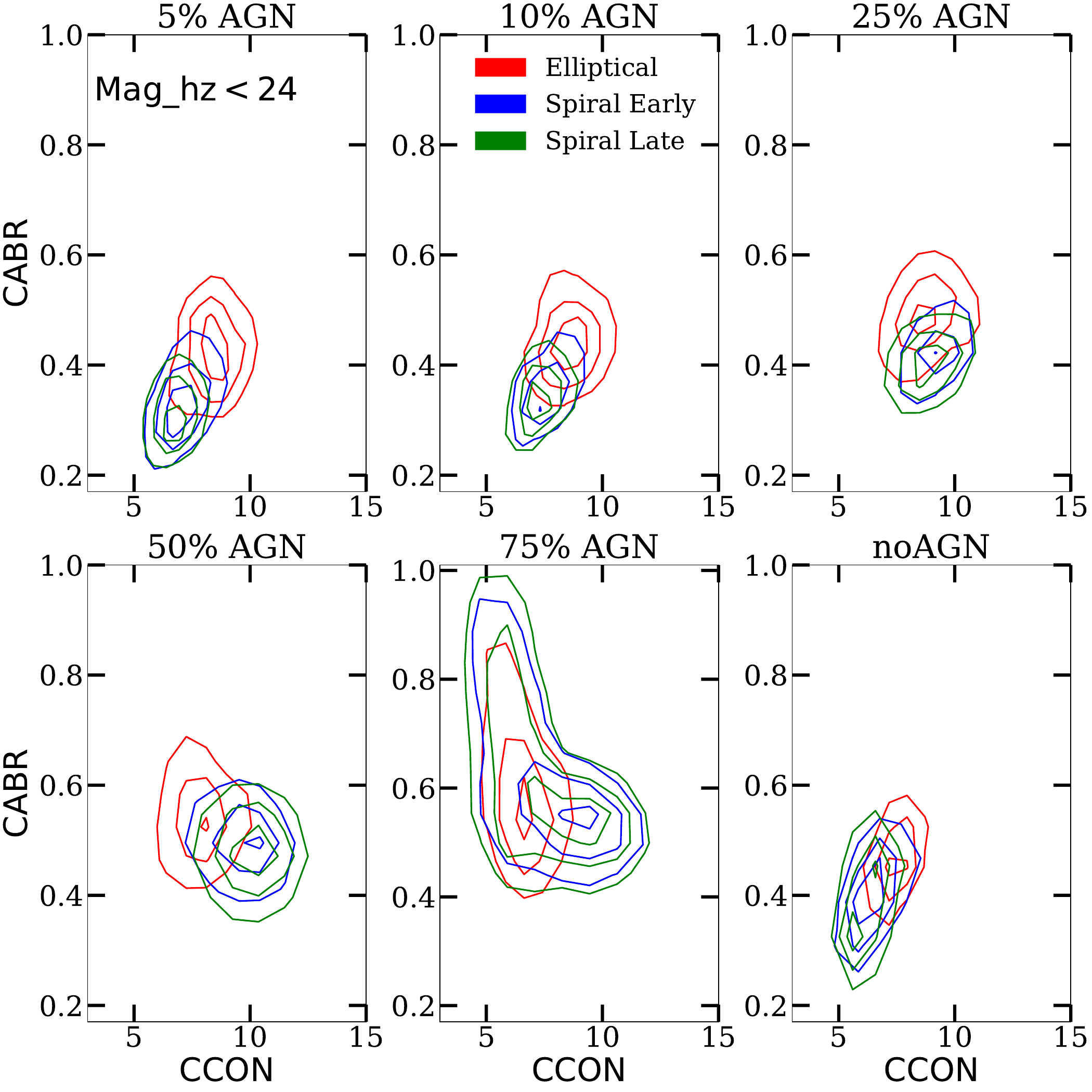}}
		\end{center}
		\begin{center}
			\subfloat[]
			{\includegraphics[height=1.7in, width=2.93in]{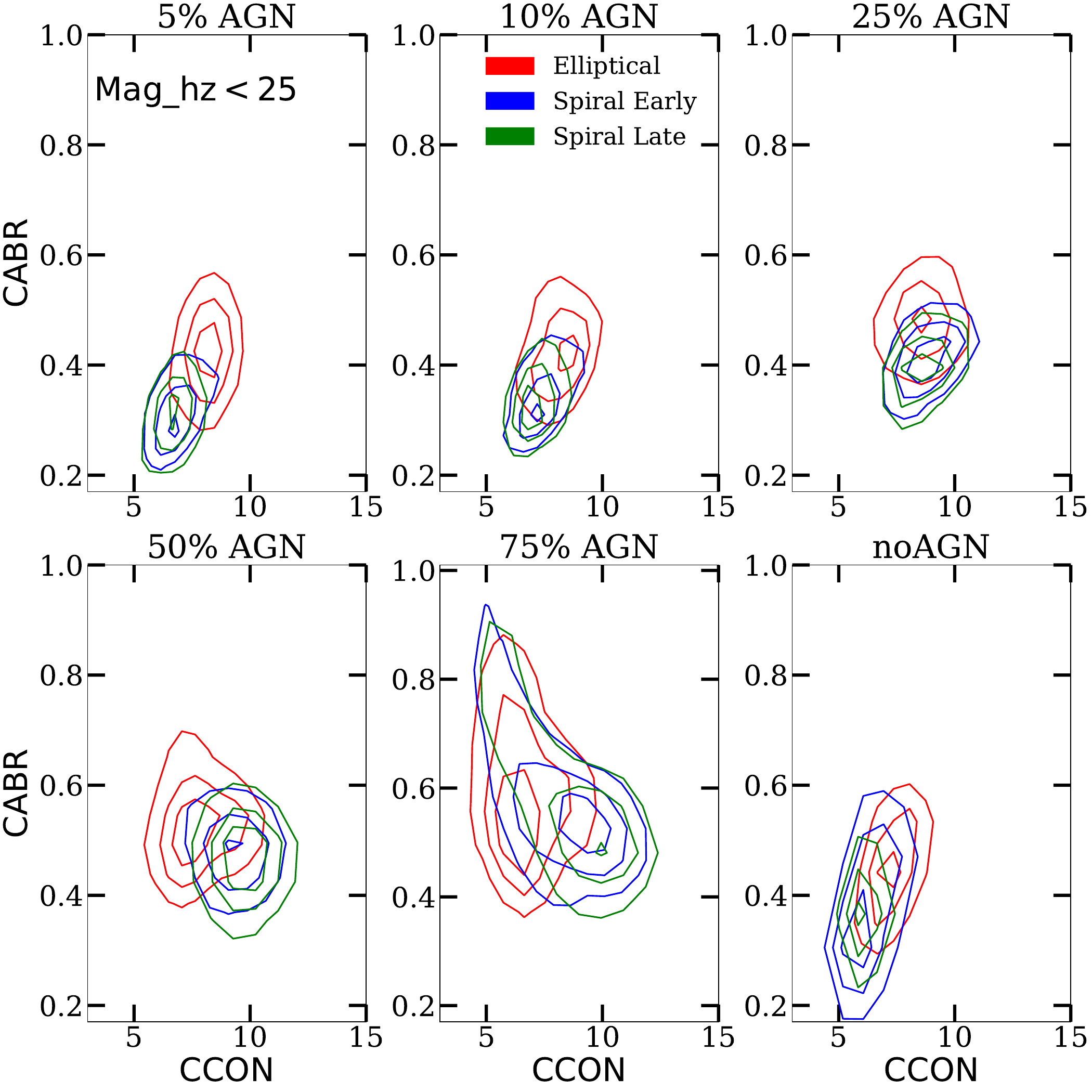}}
		\end{center}
		\caption{Relation between CABR and CCON morphological parameters for a simulated sample moved to fainter magnitudes of \textit{mag\_hz}\,$<$\,21 (a), \textit{mag\_hz}\,$<$\,22 (b), \textit{mag\_hz}\,$<$\,23 (c), \textit{mag\_hz}\,$<$\,24 (d), and \textit{mag\_hz}\,$<$\,25 (e). For each panel a\,-\,e, we present the distribution of early-type (red contours), early-spiral (blue contours), and late-spiral (green contours) galaxies, with parameters measured after adding AGN contribution of 5\% (top, left), 10\% (top, middle), 25\% (top, right), 50\% (bottom, left), 75\% (bottom, middle), and without AGN added (bottom, right).}
		\label{CABR_vs_CCON}
	\end{figure}
	
	\begin{figure}[!t]
		\subfloat[]{\includegraphics[height=1.7in, width=2.93in]{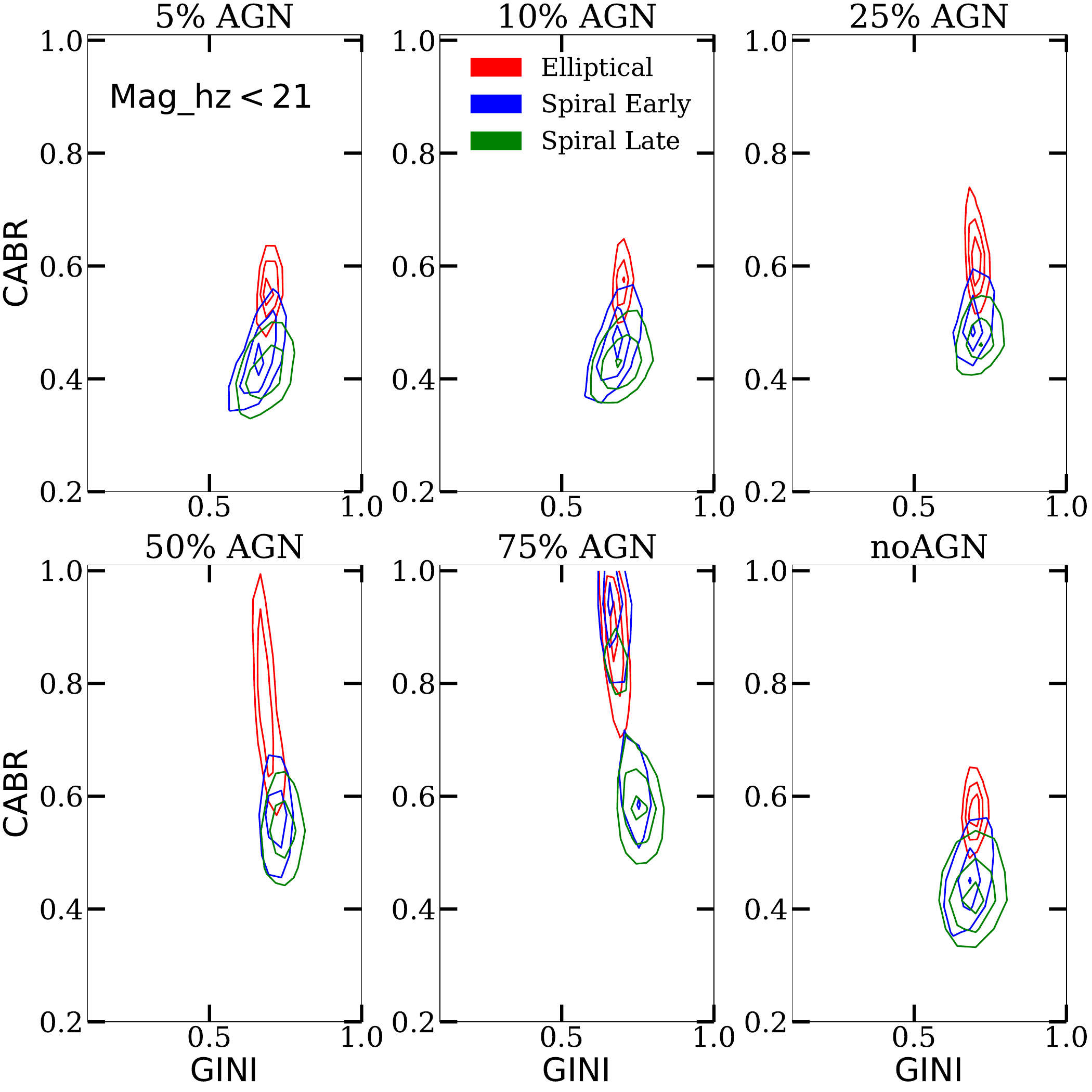}}
		\subfloat[]{\includegraphics[height=1.7in, width=2.93in]{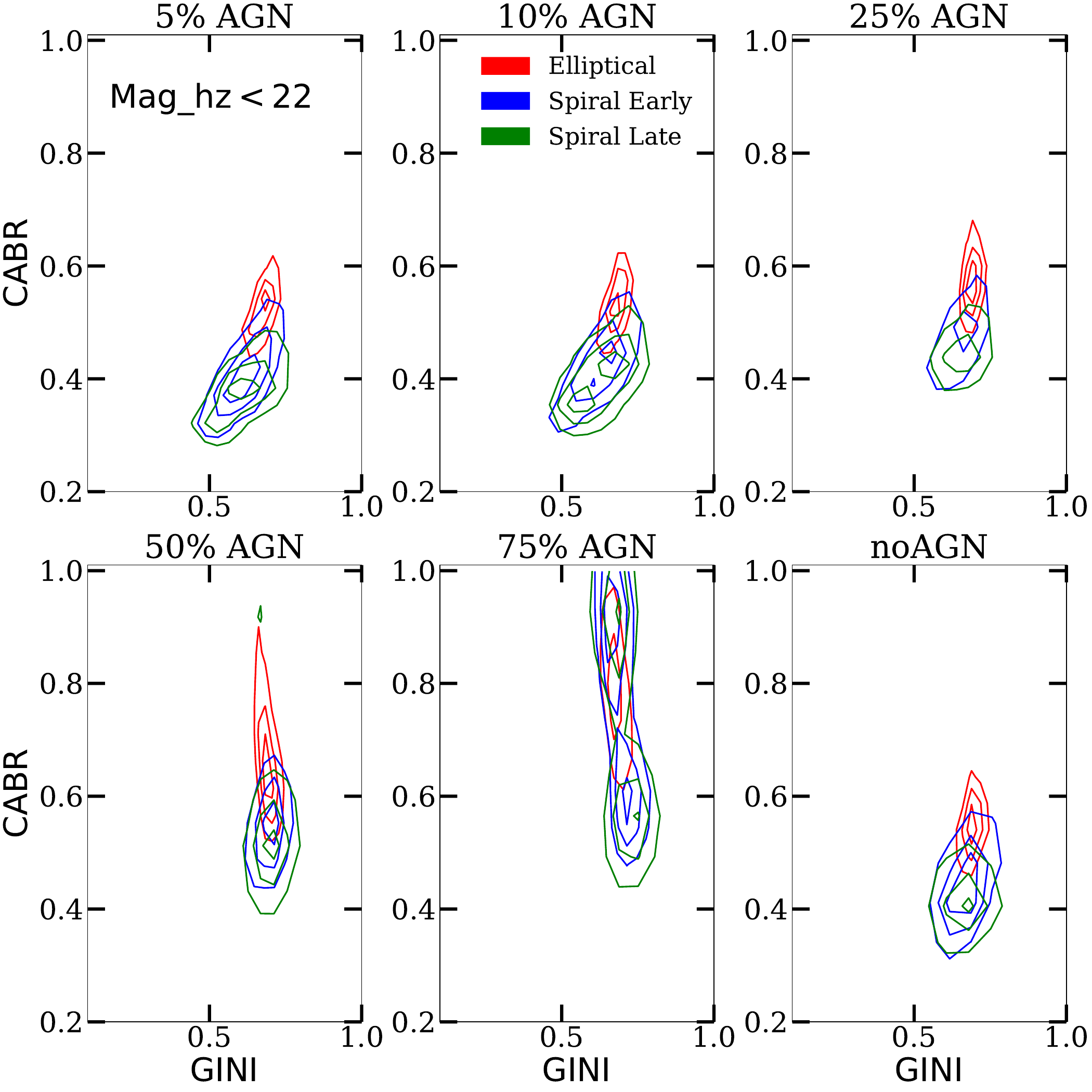}}
		\begin{center}
			\subfloat[]{\includegraphics[height=1.7in, width=2.93in]{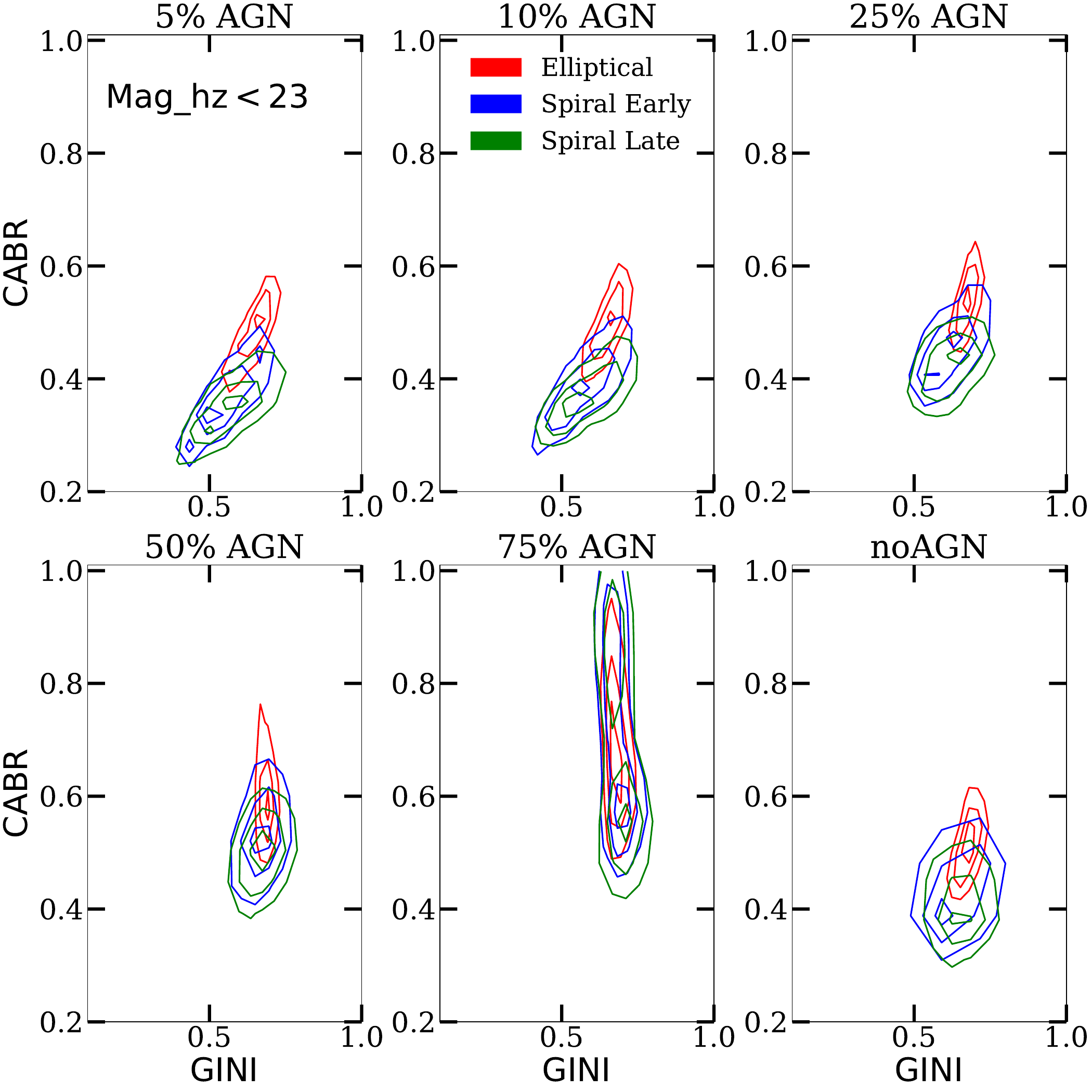}}
			\subfloat[]{\includegraphics[height=1.7in, width=2.93in]{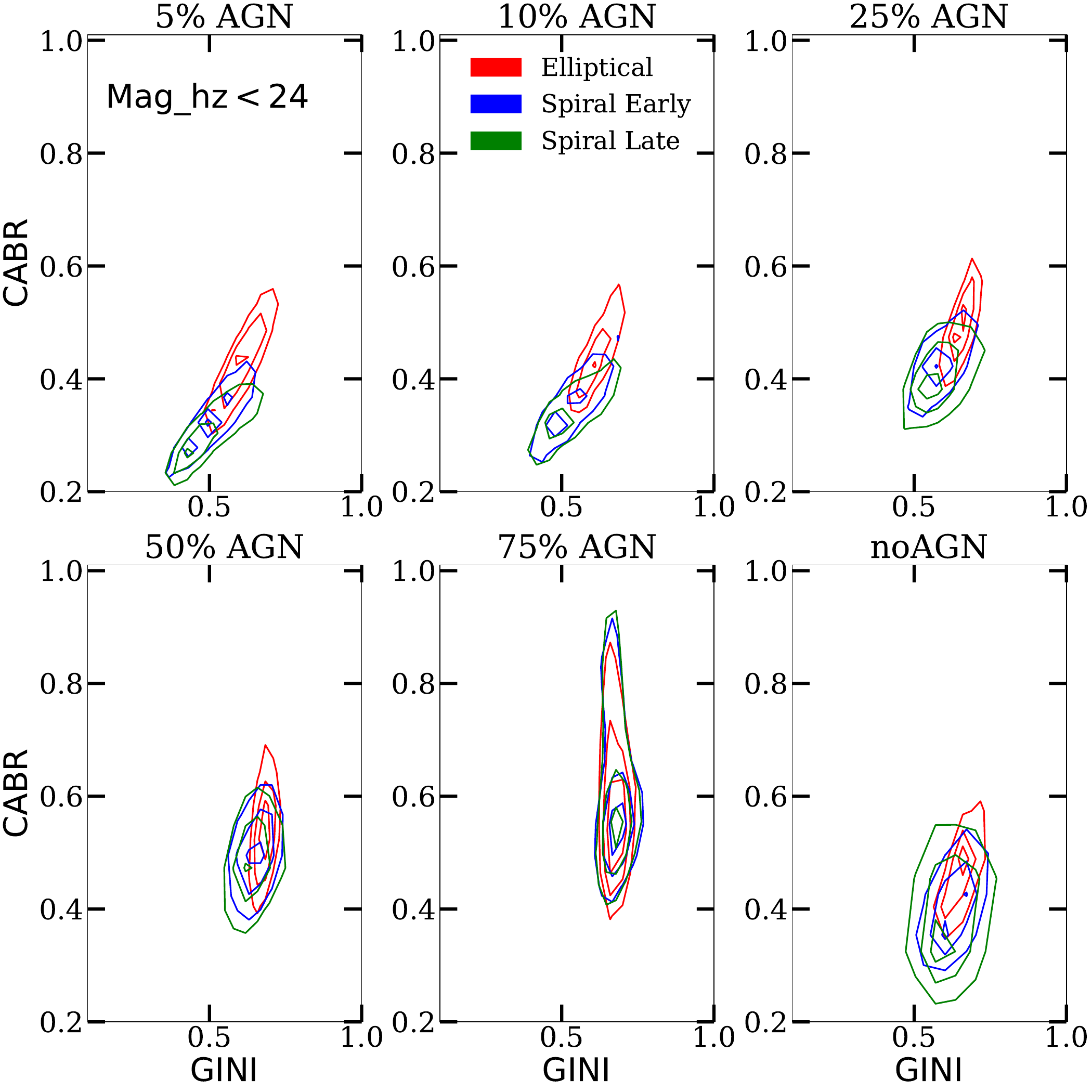}}
		\end{center}
		\begin{center}
			\subfloat[]
			{\includegraphics[height=1.7in, width=2.93in]{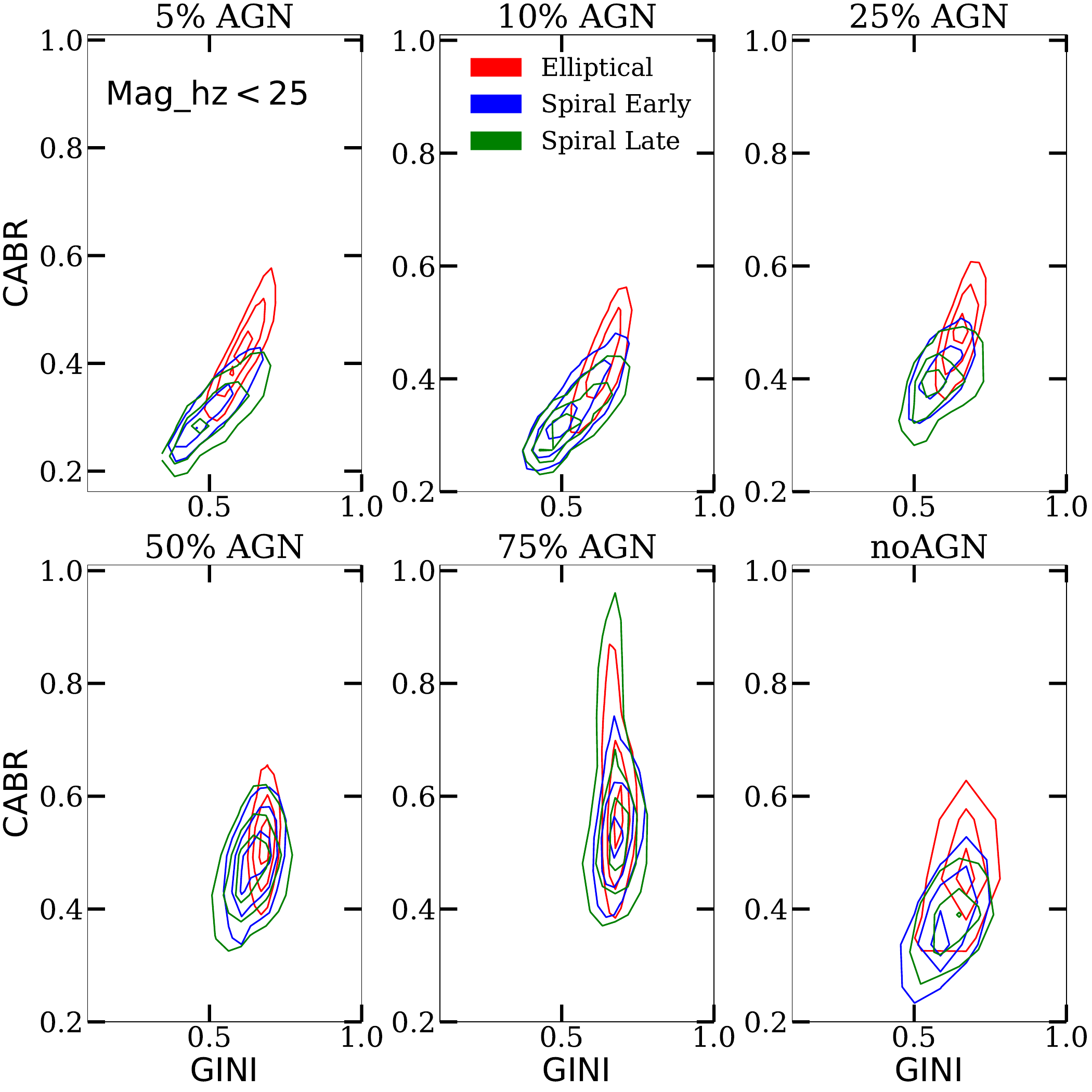}}
		\end{center}
		\caption{Same as Figure~\ref{CABR_vs_CCON}, but showing the relation between CABR and GINI morphological parameters for a simulated sample moved to fainter 
			magnitudes of \textit{mag\_hz}\,$<$\,21 (a), \textit{mag\_hz}\,$<$\,22 (b), \textit{mag\_hz}\,$<$\,23 (c), \textit{mag\_hz}\,$<$\,24 
			(d), and \textit{mag\_hz}\,$<$\,25 (e). For each panel (a--e), colours for morphological type are the same that in Figure~\ref{CABR_vs_CCON}.}
		\label{CABR_vs_GINI}
	\end{figure}
	
	\begin{figure}[!t]
		\subfloat[]{\includegraphics[height=1.7in, width=2.93in]{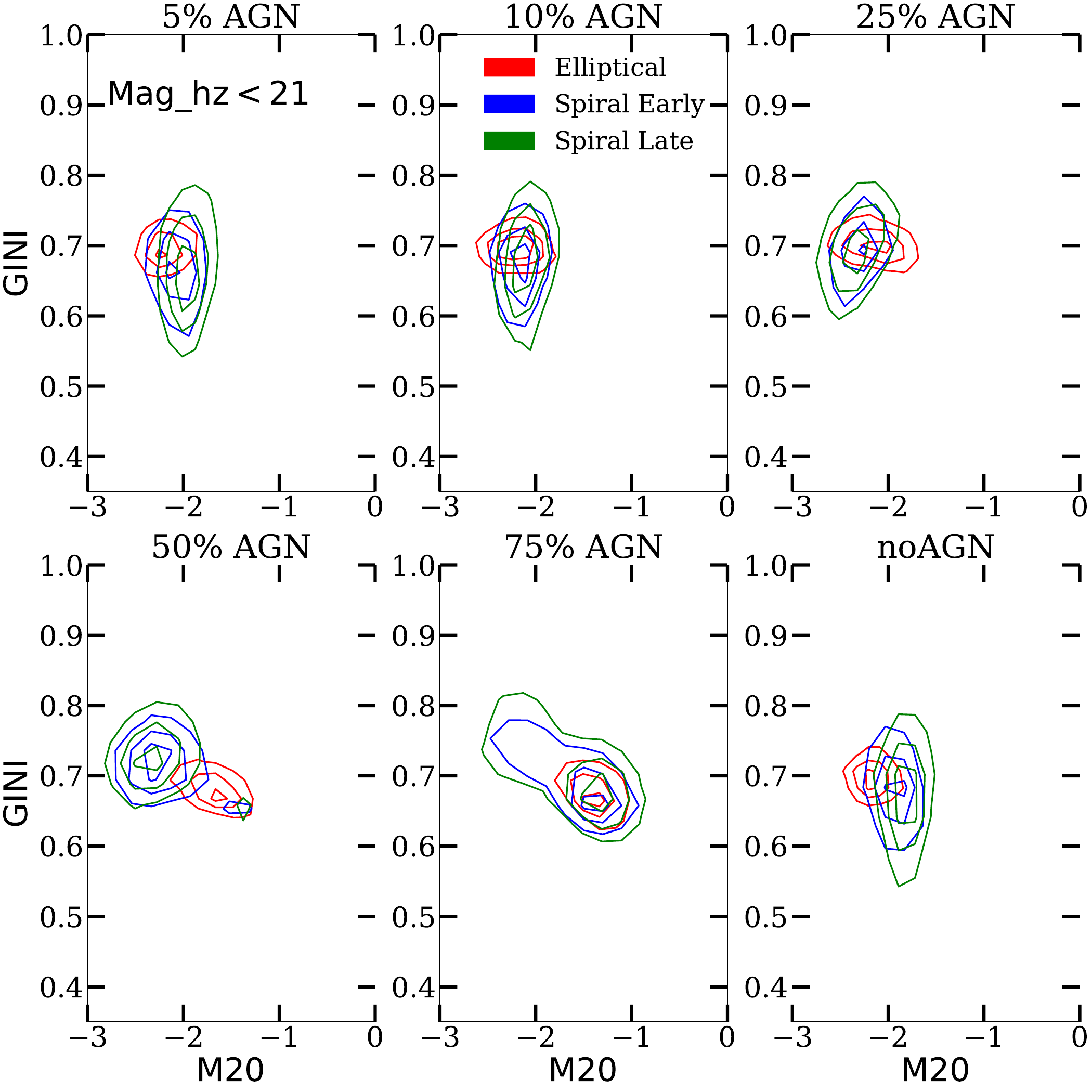}}
		\subfloat[]{\includegraphics[height=1.7in, width=2.93in]{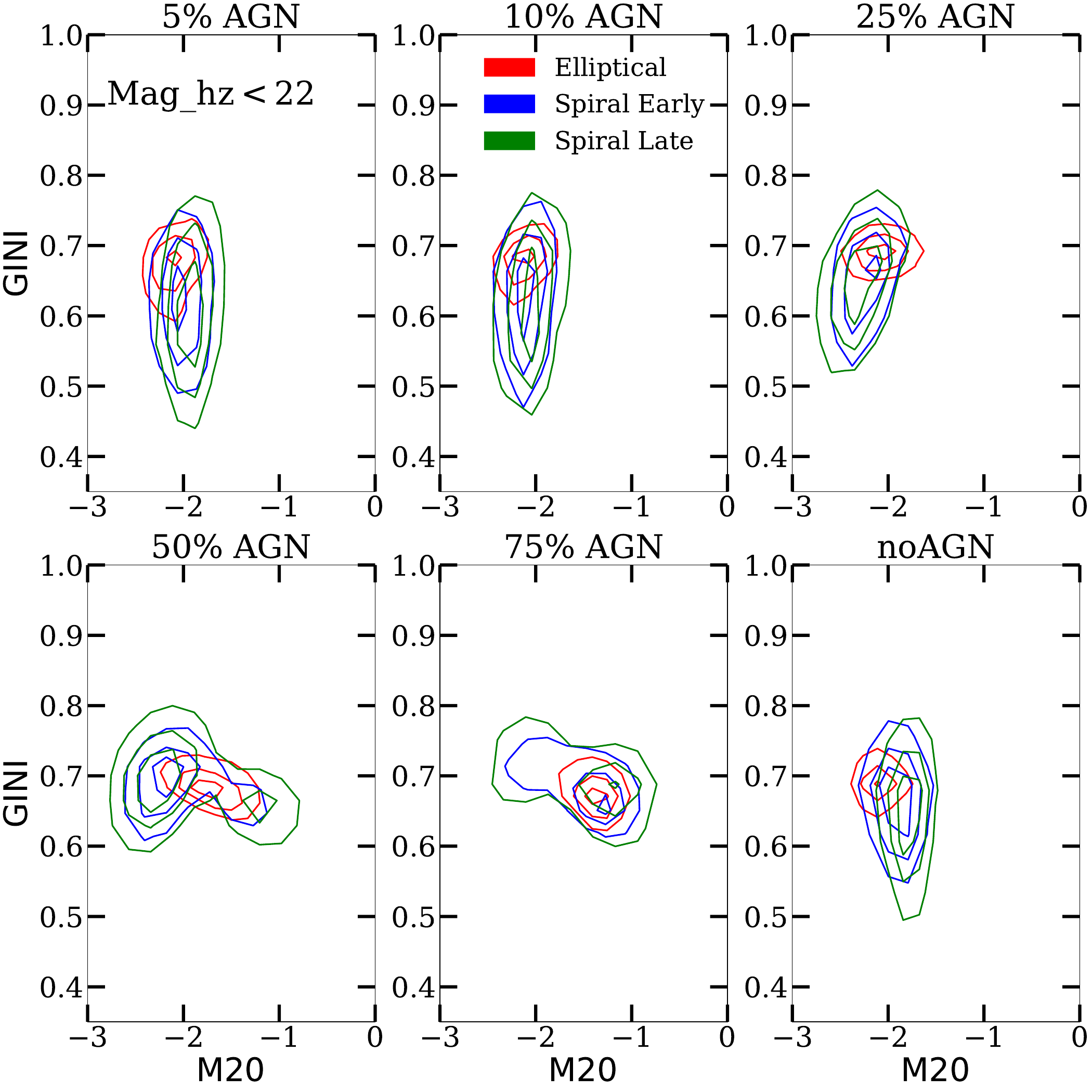}}
		\begin{center}
			\subfloat[]{\includegraphics[height=1.7in, width=2.93in]{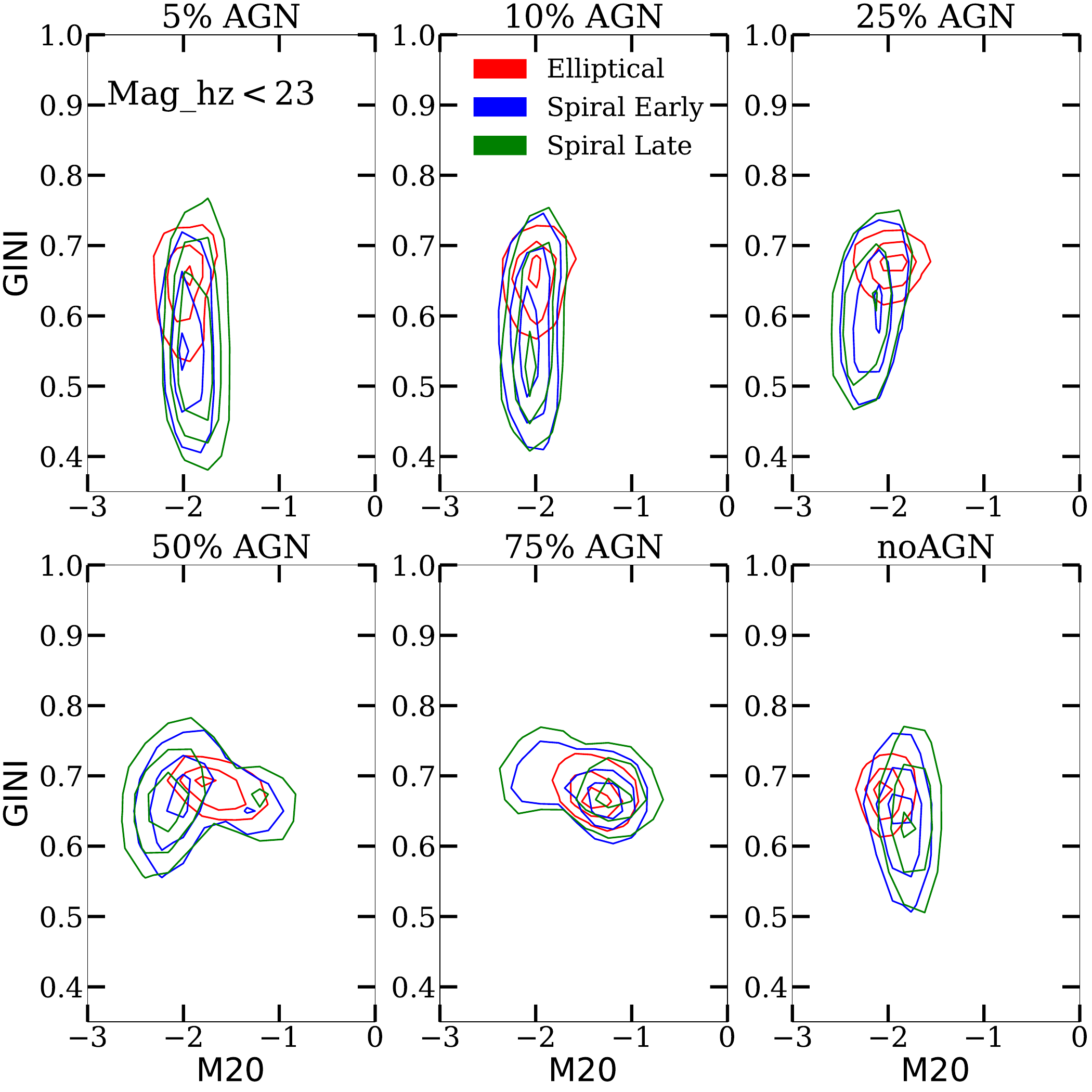}}
			\subfloat[]{\includegraphics[height=1.7in, width=2.93in]{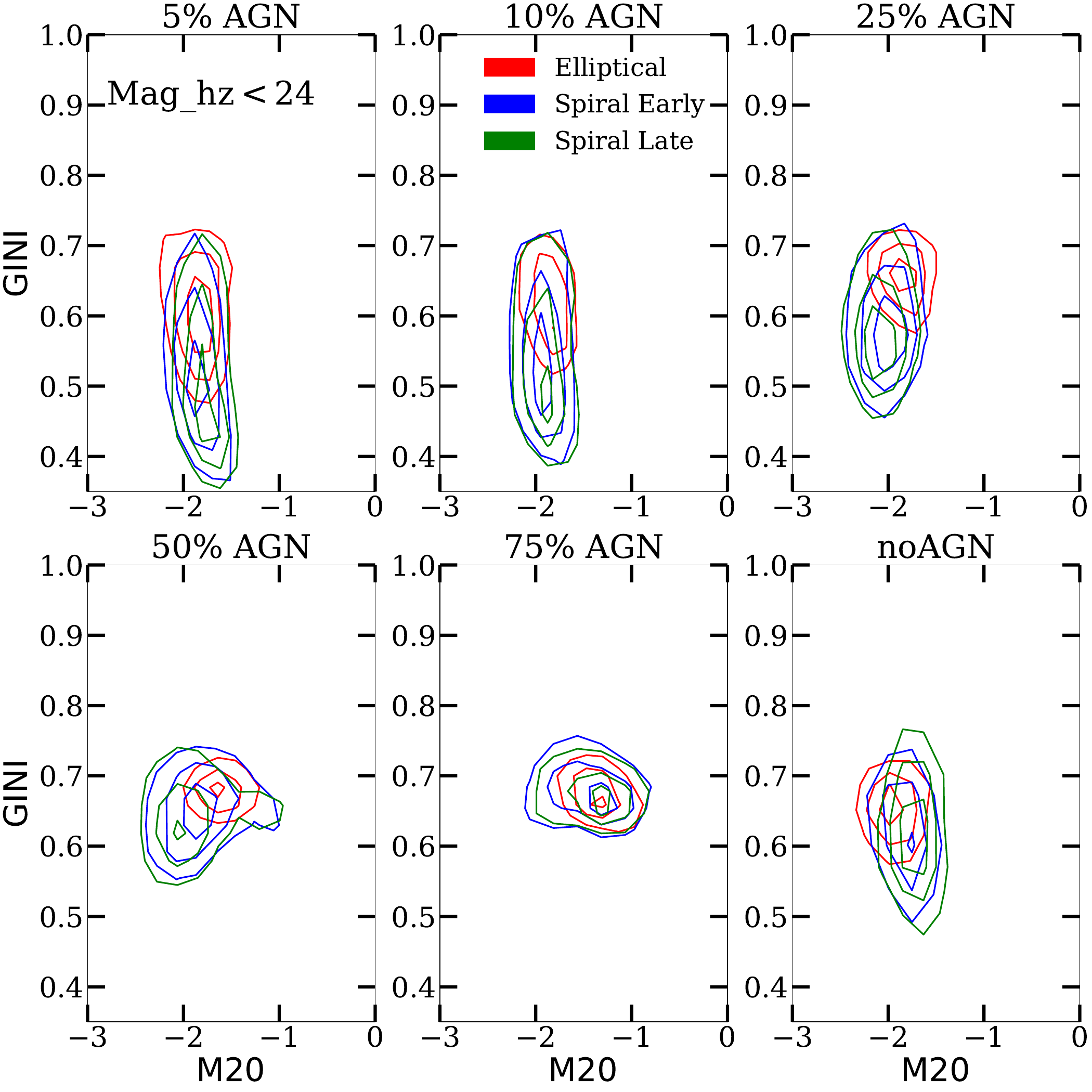}}
		\end{center}
		\begin{center}
			\subfloat[]
			{\includegraphics[height=1.7in, width=2.93in]{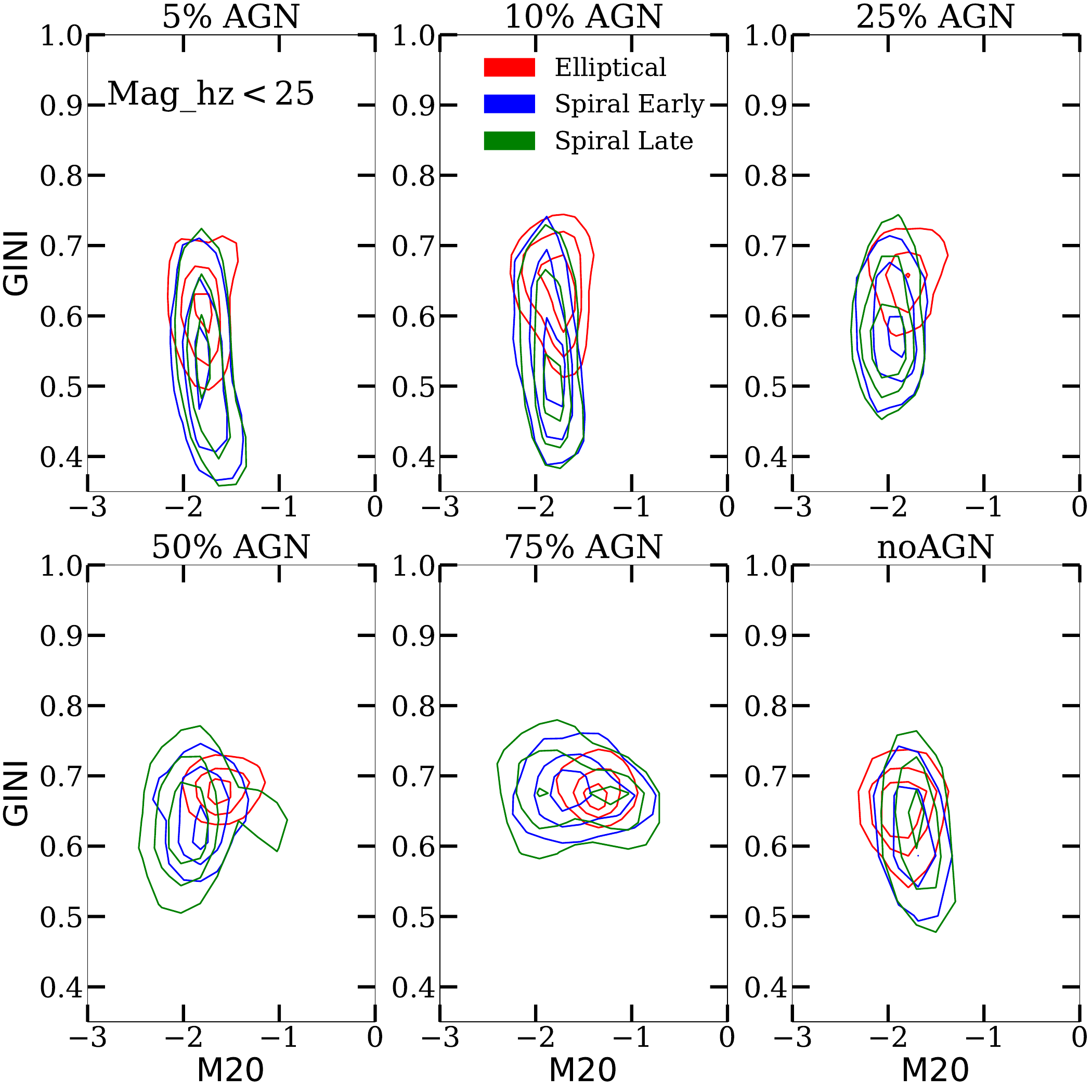}}
		\end{center}
		\caption{Same as Figure~\ref{CABR_vs_CCON}, but showing the relation between GINI and M20 morphological parameters for a simulated sample moved to fainter magnitudes of \textit{mag\_hz}\,$<$\,21 (a), \textit{mag\_hz}\,$<$\,22 (b), \textit{mag\_hz}\,$<$\,23 (c), \textit{mag\_hz}\,$<$\,24 (d), and \textit{mag\_hz}\,$<$\,25 (e). For each panel (a--e), colours for morphological type are the same that in Figure~\ref{CABR_vs_CCON}.}
		\label{M20_vs_GINI}
	\end{figure}
	
	\begin{figure}[!t]
		\subfloat[]{\includegraphics[height=1.7in, width=2.93in]{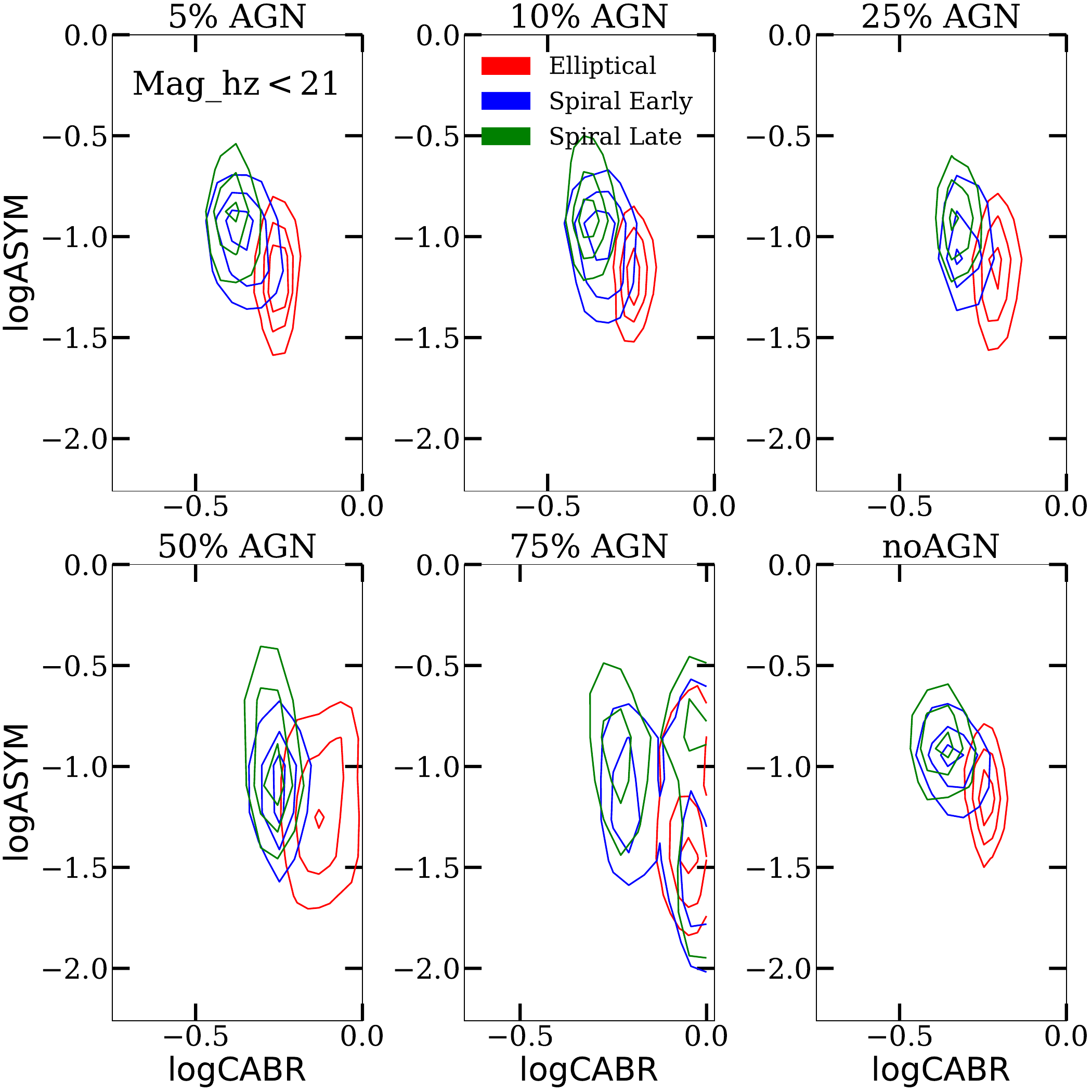}}
		\subfloat[]{\includegraphics[height=1.7in, width=2.93in]{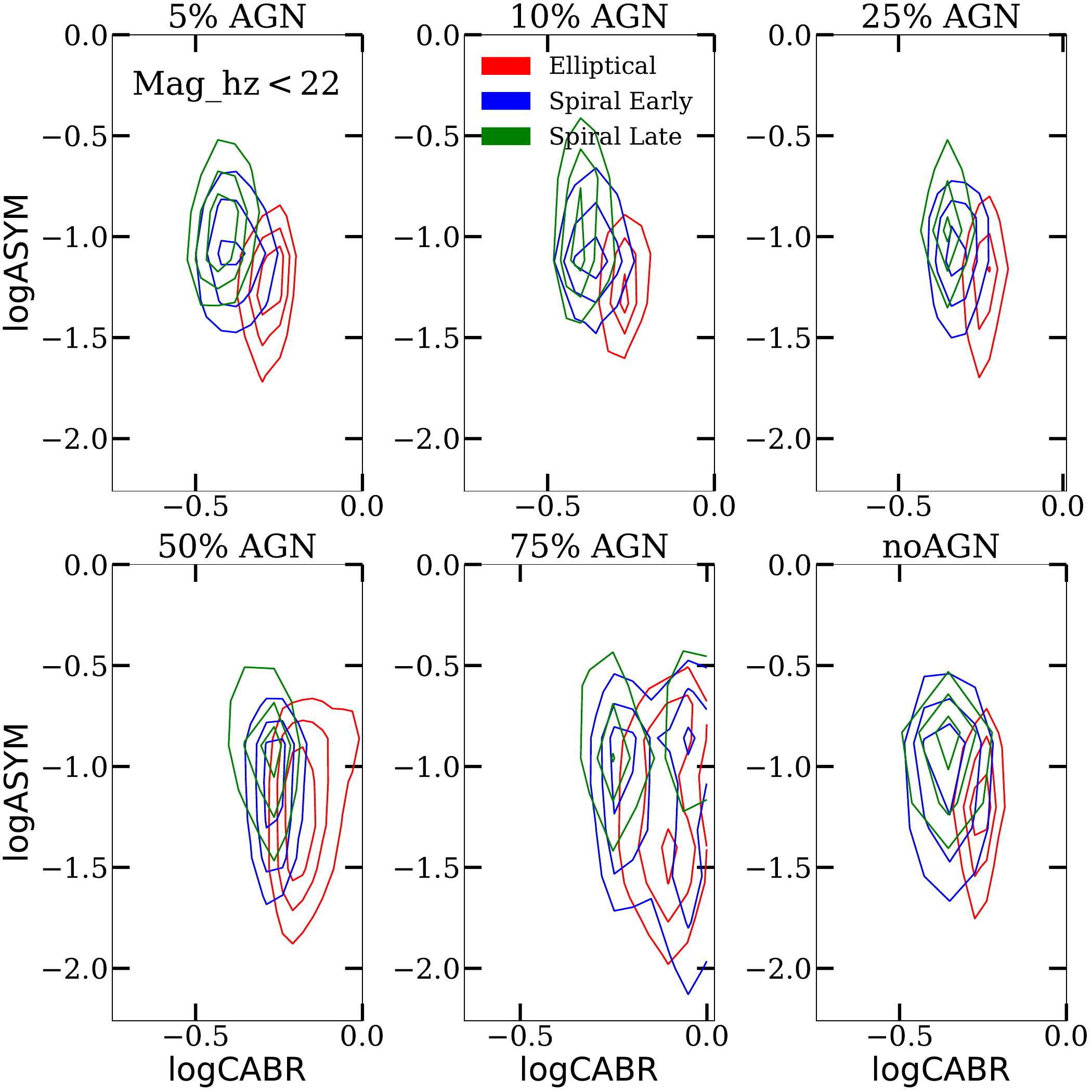}}
		\begin{center}
			\subfloat[]{\includegraphics[height=1.7in, width=2.93in]{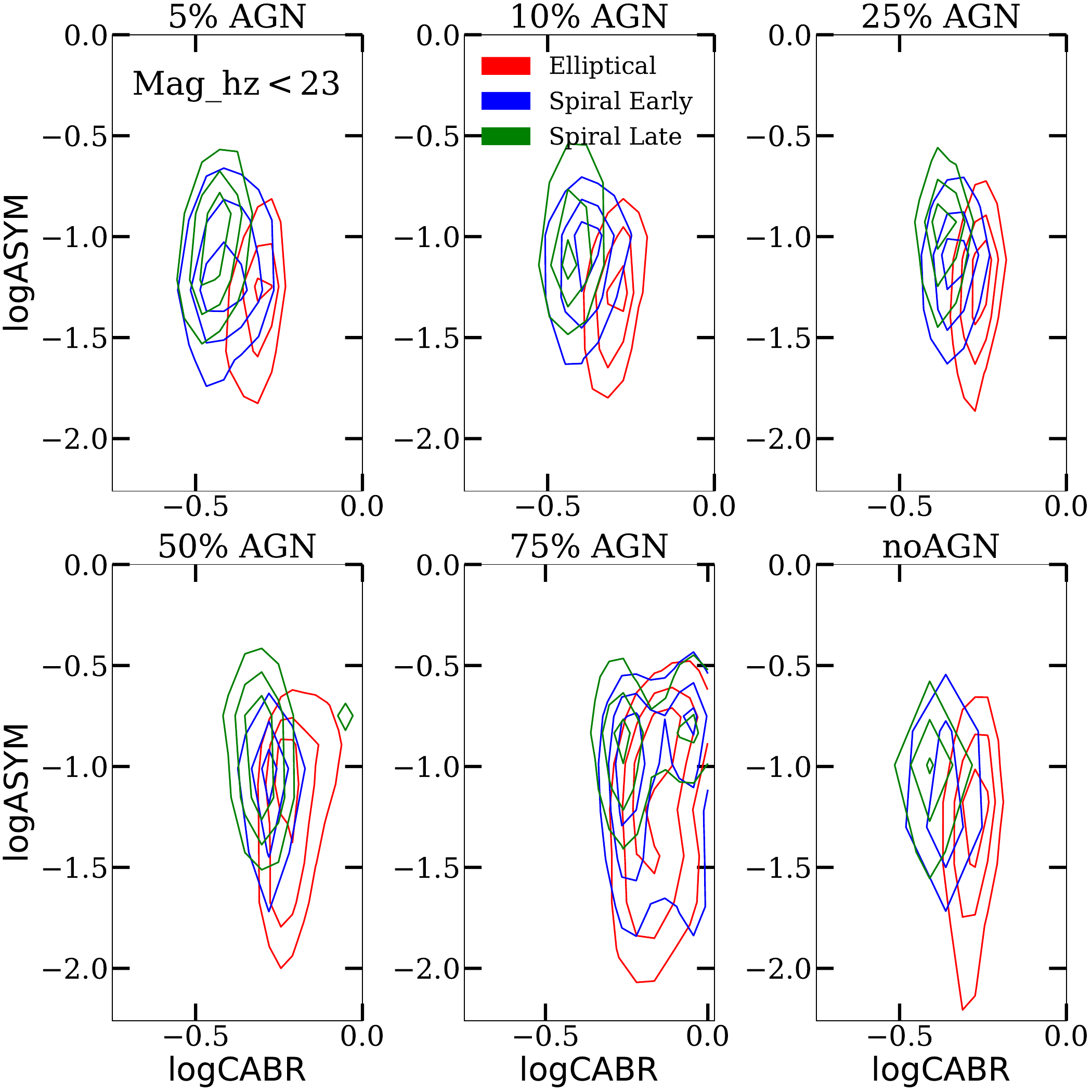}}
			\subfloat[]{\includegraphics[height=1.7in, width=2.93in]{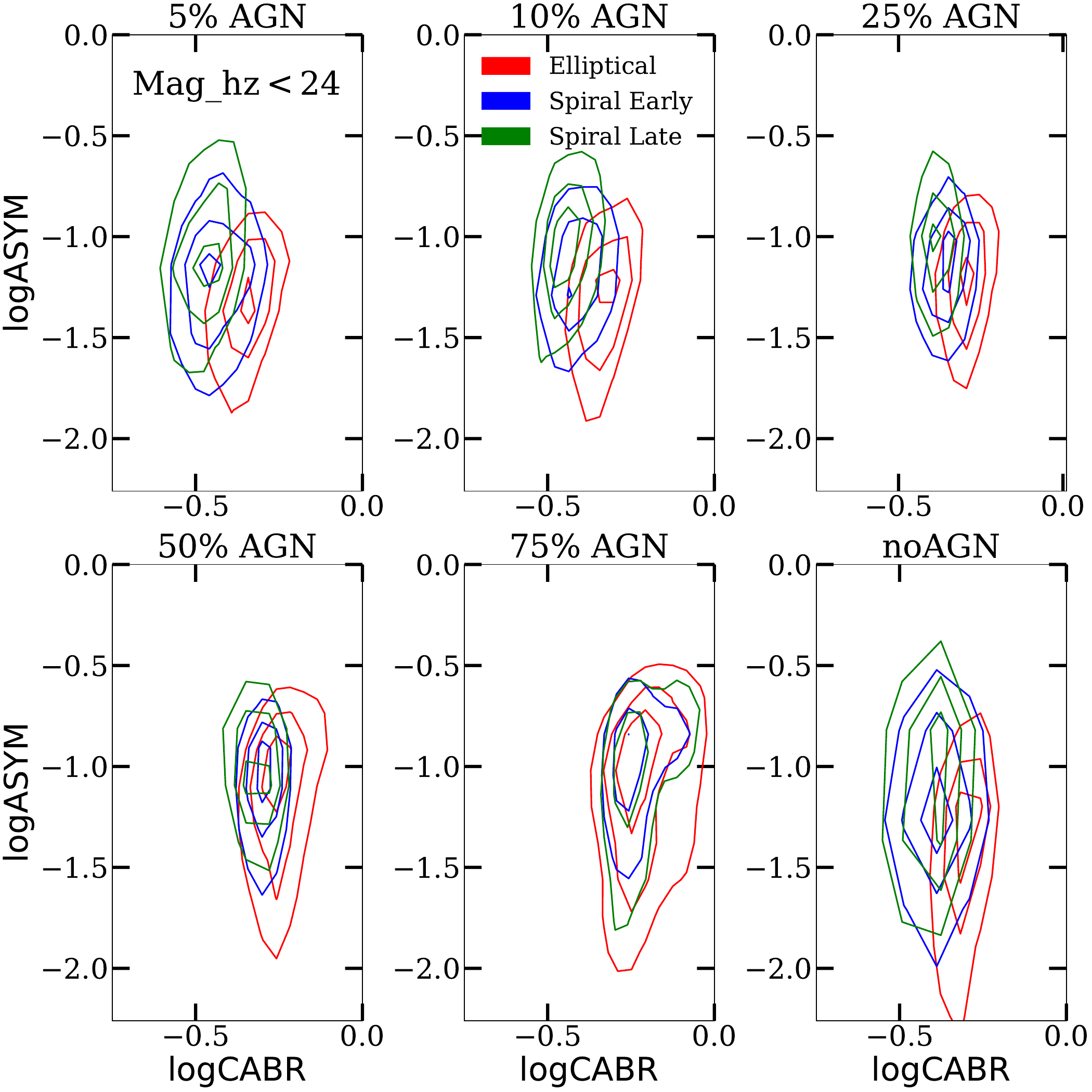}}
		\end{center}
		\begin{center}
			\subfloat[]
			{\includegraphics[height=1.7in, width=2.93in]{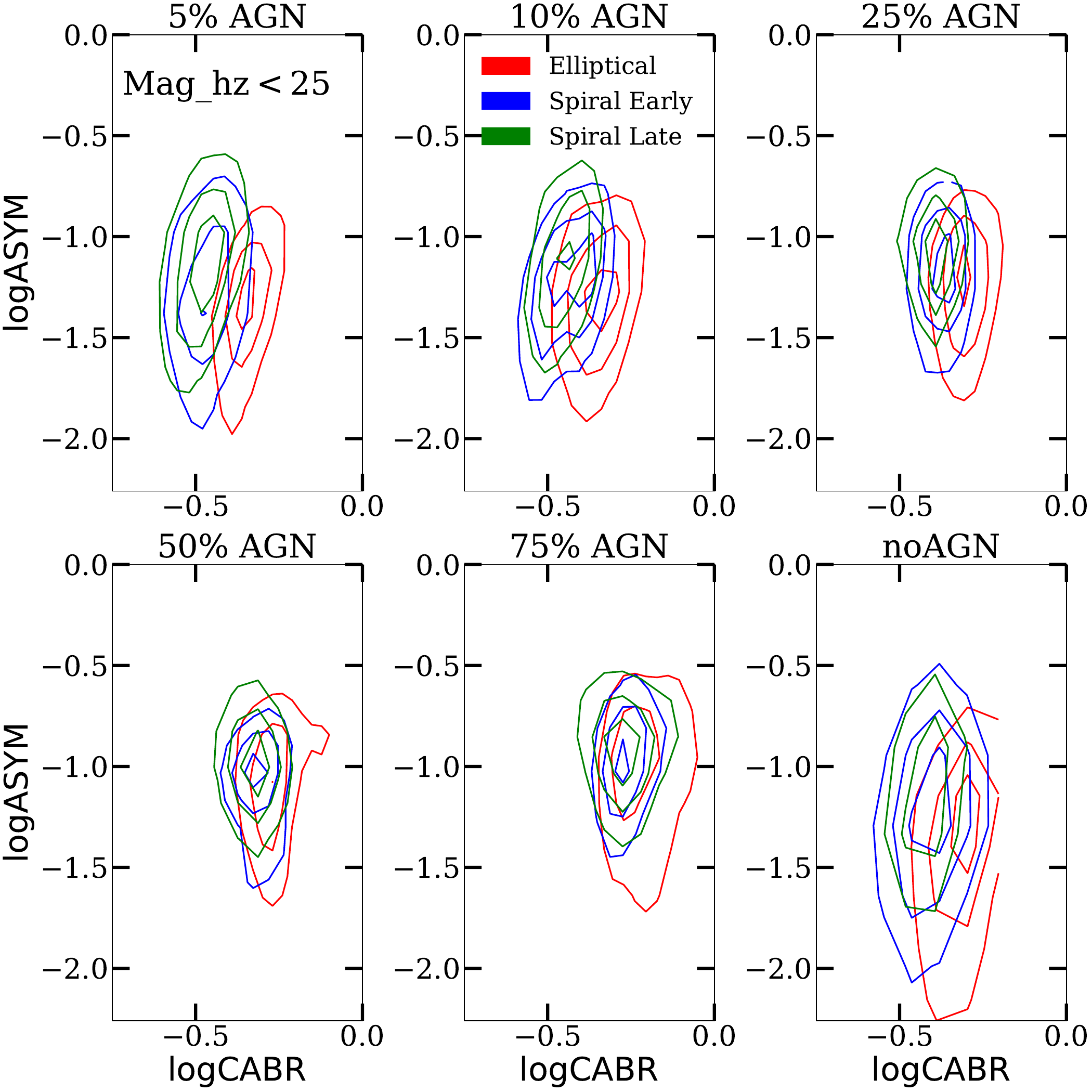}}
		\end{center}
		\caption{Same as Figure~\ref{CABR_vs_CCON}, but showing the relation between logASYM and logCABR. morphological parameters for a simulated sample moved to fainter magnitudes of \textit{mag\_hz}\,$<$\,21 (a), \textit{mag\_hz}\,$<$\,22 (b), \textit{mag\_hz}\,$<$\,23 (c), \textit{mag\_hz}\,$<$\,24 (d), and \textit{mag\_hz}\,$<$\,25 (e). For each panel (a--e), colours for morphological type are the same that in Figure~\ref{CABR_vs_CCON}.}
		\label{logASYM_vs_logCABR}
	\end{figure}
	
	\section{Discussion}\label{sec5}
	In this section, we discuss the results obtained in this work, presented in Sections~\ref{sec4.1}--\ref{sec4.4}. We present the results obtained for the six morphological parameters and the analysed morphological diagrams, for the total sample, early-type galaxies, early-spirals, and late-spirals. 
	
	Of the three types of analyses carried out in Sections~\ref{sec4.1}--\ref{sec4.3}, we find that the greatest impact on morphological parameters occurs when considering the effect of magnitude, redshift, and AGN together, discussed in Section~\ref{sec4.1}. Next, we measure the largest variations of the morphological parameters in the third case study, when AGN is added under COSMOS-like conditions, as explained in Section~\ref{sec4.3}. Finally, the lowest variances were measured in the second case analysed in Section~\ref{sec4.2} when comparing simulated active galaxies under local-like (at z\,$\sim$\,0) and COSMOS-like conditions. This indicates that, in general, the AGN influences the morphological properties of galaxies more than the 
	magnitude/redshift effect. In addition, we find that in most cases significant changes in the morphological parameters start to occur with an AGN contribution $>$\,25\% and a magnitude \textit{mag\_hz}\,$\ge$\,23 (z\,$>$\,1) in COSMOS-like conditions. This is an overall observation, but the numbers change slightly depending on the conditions analysed and the morphological type of the galaxy, as shown above in Section~\ref{sec:4}. We also find that in the case of all three types of analysis, all morphological parameters of spiral galaxies are more affected than in early-type galaxies. Furthermore, late-spirals are more affected by magnitude, redshift, and AGN contribution than early-spirals. This is not surprising and has already been reported in \cite{Getachew2022}. Spiral galaxies, in particular late-spirals, have weaker bulges, and introducing an added AGN as a central component will affect all morphological parameters. In addition, the light in the disc will be more affected by the magnitude and redshift effect than in the case of early-type galaxies, which will lead to a larger change in the morphological parameters. Difficulties in the classification of late-type galaxies compared to early-type galaxies have already been observed in the past when classifying a large sample of galaxies (e.g., \citep[][]{Pintos2016, Amado2019}), and when studying the effect of resolution and survey depth on the morphological classification of galaxies (e.g., \citep[][]{Povic2015}, and references therein). Similarly, we find that the morphological parameters in normal galaxies are more affected by the magnitude and redshift than in active galaxies when comparing galaxies in local-like (at z\,$\sim$\,0) and COSMOS-like conditions. Again, this is not surprising, since AGN host galaxies are brighter and have concentrated light in the central part of the galaxy, they will be, by definition, less sensitive to the effect of magnitude and redshift, especially when it comes to concentration parameters.  \\
	
	\indent Of all the parameters, GINI is the most stable on the magnitude and redshift effects and the contribution of AGN in all three cases analysed and in the case of both early- and late-type galaxies (for more details on variances, see Section~\ref{sec:4}). We found the same result at z\,$\sim$\,0 in our previous study by \cite{Getachew2022}. 
	The GINI coefficient emerges as the most robust morphological parameter, likely due to its inherent insensitivity to variations in spatial resolution and flux distribution. Unlike other concentration or asymmetry indices, which depend on the precise spatial distribution of light, the Gini coefficient measures only the difference in pixel flux values. This makes it less susceptible to redshift-induced PSF smoothing or AGN-driven central point sources. Being normalised by the total flux, it remains stable even when a bright AGN contaminates the nuclear region, as long as the light distribution of the underlying galaxy remains relatively unchanged. This parameter was commonly used in previous studies to separate early-type and spiral galaxies \linebreak{}(e.g., \citep{Povic2013a, Pintos2016, Amado2019, Mahoro2019}), to trace sub-structures in galaxies (e.g., \citep{Lisker2008}), mergers and interacting galaxies (e.g., \citep{Lotz2004, Lotz2008}), strong-lensed galaxies (e.g., \citep{Florian2016b}). Some studies suggest that it can trace the overall structure of a galaxy better than any other morphological parameter (e.g., \citep{Zamojski2007}). On the other hand, GINI may depend strongly on the signal-to-noise ratio and on the choice of the aperture within which it is measured (e.g., \cite{Lisker2008, Povic2015}). After GINI, the light concentration parameter that shows the lowest variance is M20, followed by CCON, and finally, CABR is the most affected. {The last two parameters are particularly sensitive to the central light distribution, especially CABR, and therefore any additional light added in the central part of the galaxy will influence them directly.} For ASYM, as in \cite{Getachew2022}, we do not find significant changes when adding the AGN contribution, as this parameter is not sensitive to the light in the central part of the galaxy. However, as already reported in \cite{Povic2013a, Povic2015} and \cite{Zhao2022}, this parameter is sensitive to noise and is measured for a smaller sample of galaxies (as mentioned in Section~\ref{sec:4}), and shall therefore be used in combination with other parameters to classify normal and active galaxies. Similar results have been seen previously in \cite{Sazonova2024}, suggesting that ASYM is extremely dependent on imaging properties, both resolution and depth. The SMOOTH parameter is the most difficult to measure in the case of both early-type and spiral galaxies. As shown in Section~\ref{sec:4}, this parameter is only measured in 20\%-45\% of the galaxies, depending on the magnitude and redshift limit and AGN added contribution. Therefore, we do not find this parameter efficient in classifying active galaxies at higher redshifts. This is in agreement with the previous studies \citep[e.g.,][]{Lotz2004, Nersesian2023} and also our previous work \cite{Povic2015}, where we studied the impact of magnitude and redshift on normal galaxies, suggesting that SMOOTH is especially sensitive to all, noise, spatial resolution, and data depth, in comparison to other parameters.\\
	\indent We observe complex behaviour of the parameters, shifting between higher and lower values in comparison to the reference sample as the magnitude and redshift limits and the AGN contribution change, as can be seen in Appendices~\ref{appendixA}, \ref{appendixB}, and \ref{appendixC}. In general, this depends on the definition of the parameters. The CABR parameter is the most sensitive to these changes. It is followed by the CCON concentration parameter, in particular at 50\% and 75\% of the added contribution of AGN when the distribution of the simulated samples changes completely, indicating that most of the total flux in the galaxy accumulates within a very small radius. This complex behaviour of the parameters has already been seen in \cite{Povic2015} for normal galaxies when studying only the impact of magnitude and redshift. However, when the AGN contribution is added, the behaviours become even more complex.\\
	
	\indent As already shown in \cite{Povic2015} and \cite{Getachew2022}, not necessarily a larger variation means that a parameter is less suitable for morphological classification, as commonly various parameters are used in classification, often simultaneously (e.g.,\cite{Povic2012, Povic2013a, Pintos2016, Amado2019, Aguilar2025}). We also see the same in this work. We analysed several morphological diagrams to evaluate how redshift, magnitude, and AGN contribution together influence the classification of galaxies. When using the CABR vs. CCON diagram, most early- and late-type galaxies can be separated up to the AGN contribution of 50\% if the \textit{mag\_hz} is $<$\,22, as seen in Figure~\ref{CABR_vs_CCON}. A stronger mixing starts at \textit{mag\_hz}\,$<$\,23. Above that, galaxies can be distinguished between early- and late-types only for AGN contributions of 5\%-10\%. If the AGN contribution is 75\% and above, there is a strong mix of types at all magnitudes, and galaxies can not differ from each other. Similar trends are observed in the CABR vs. GINI morphological diagram shown in Figure~\ref{CABR_vs_GINI}, but with slightly worse separations in all cases, including cases without any added AGN contribution. Although not shown in this paper, we find the worst separation between early- and late-type galaxies when using the GINI vs. CCON diagram, starting from lower AGN contributions of 10\% at all magnitude limits. GINI vs. M20 diagram has been commonly used in the literature to identify galaxy mergers and peculiar (disturbed) galaxies \cite[][]{Lotz2004, Lotz2008, Hung2014, Yao2023}, in addition to classifying early- and late-types (e.g., \citep[][]{Nersesian2023, Zaazou2025}, and references therein). . However, in this work, we do not find this diagram to be effective in separating early- and late-type galaxies when dealing with AGN hosts at higher redshift in COSMOS-like conditions, as shown in Figure \ref{M20_vs_GINI}. Under each case of magnitude and redshift limit and independently of the AGN contribution, we find that early- and late-type galaxies occupy similar areas in the diagram. Therefore, we do not recommend using this diagram alone when dealing with morphological classification of active galaxies in COSMOS-like conditions. {This is not entirely surprising, and is in line with the results of \cite{Povic2015}, where in three surveys analyzed with different survey depths, this diagram proved to be ineffective in separating the different morphological types.} Finally, CABR vs. ASYM diagrams (see Figure \ref{logASYM_vs_logCABR}) show a reasonable separation between early- and late-type galaxies up to \textit{mag\_hz}\,$<$ 23 and AGN contribution of 25\%. At fainter magnitude limits and higher AGN contributions of 50\% and 75\%, there is a higher contamination between morphological types. This diagram is commonly used in previous studies to classify both normal and active early- and late-type galaxies (e.g., \citep{Povic2012, Povic2013a, Mahoro2019, Amado2019, Zhao2022, Xu2025}). In addition, the ASYM parameter was recently used in combination with the CCON parameter to classify post-starburst galaxies \citep{Himoto2023}. To summarize, we find CABR vs. CCON and CABR vs. ASYM to be the most efficient in the classification of AGN host galaxies at higher redshift in COSMOS-like conditions, in particular up to \textit{mag\_hz}\,$<$ 22 and AGN contribution of $\leq$25\%, followed by the CABR vs. GINI diagram. We do not recommend using the GINI vs. M20 diagram to classify active galaxies in COSMOS-like conditions.

	\section{Conclusions}\label{6}
	In this work, we studied how the AGN can affect morphological parameters of galaxies under COSMOS-like conditions when dealing with fainter magnitudes and an intermediate redshift of z\,$\sim$\,2. The main objective of this study is to quantify the impact of magnitude, redshift, and AGN on six commonly used morphological parameters, Abraham concentration index (CABR), Gini coefficient (GINI), Conselice-Bershady concentration index (CCON), M20 moment of light (M20), asymmetry (ASYM) and smoothness (SMOOTH) in COSMOS-like conditions. We add 5\%\,-\,75\% AGN contribution to the centre of local galaxies from the SDSS survey with known morphology and move them to intermediate redshifts and fainter magnitudes corresponding to a sample of galaxies in the COSMOS field. 
	
	We performed three types of analysis on the total sample, early-type, early-spiral, and late-spiral galaxies, studying: 1) the combined impact of magnitude, redshift, and AGN on morphological parameters, 2) the impact of magnitude and redshift on the morphological parameters of active galaxies, and 3) the impact of AGN on the morphological parameters of intermediate-redshift galaxies in COSMOS-like conditions. In all analyses, we quantified the variance in morphological parameters for different magnitude/redshift conditions and AGN contribution. Our main findings are as follows: 
	\begin{itemize}
		\item The greatest impact on morphological parameters is when considering the combined effect of magnitude, redshift and AGN. Combined, these effects affect morphological parameters of spiral galaxies more than early-type galaxies.
		\item The impact of the magnitude/redshift will affect the morphological parameters of active galaxies less than those of normal galaxies in COSMOS-like conditions. In general, in most cases, for an AGN contribution to the total flux of up to 50\%, the impact of magnitude/redshift on the morphological parameters of active galaxies is $<$\,20\%, being higher at fainter magnitudes. 
		\item If the AGN contribution to the total galaxy flux is $\geq$\,50\%, the AGN effect dominates the impact on all morphological parameters compared to the effect of magnitude/redshift in the case of all morphological types. For  AGN contributions of $<$\,50\%, the AGN will have a dominant effect up to $g <$\,23, while at fainter magnitudes the impact of the magnitude/redshift will play a more important role in the change of morphological parameters than an AGN. The impact of AGN in combination with magnitude/redshift on morphological parameters under COSMOS-like conditions is particularly complex for spiral galaxies, compared to early-type galaxies, which are, in general, less affected.  
		\item In general, in COSMOS-like conditions up to intermediate redshift of z\,$\sim$\,2 all four concentration parameters will undergo significant changes of $>$\,20\% for $>$\,25\% of the added AGN, similar to what we found previously at z\,$\sim$\,0, affecting spiral galaxies more than early-types and, in particular, late-spirals.
		\item We find that GINI is the most stable in terms of the effect of both AGN and magnitude/redshift, followed by the M20 moment of light, CCON, and finally CABR. ASYM and SMOOTH are very sensitive to the noise and image properties, and are measured for a much smaller fraction of the total sample. 
		\item Finally, we find CABR vs. CCON and CABR vs. ASYM to be the most efficient diagrams in the classification of AGN host galaxies at higher redshift in COSMOS-like conditions, in particular up to \textit{mag\_hz}\,$<$ 22 and AGN contribution of $\leq$\,25\%, followed by the CABR vs. GINI diagram. We do not recommend using the GINI vs. M20 diagram to classify active galaxies in COSMOS-like conditions.
	\end{itemize}
	While our analysis focuses on the detailed behaviour of morphological parameters under AGN contamination and magnitude/redshift effects, the broader implications are directly relevant to galaxy evolution studies, as galaxy structure is a key property for understanding the complete physics of galaxies and thus morphological classification is essential in galaxy studies. Specifically, with the conclusions and recommendations given above, our study defines the limits for the reliable use of different concentration and galaxy shape indices, and diagnostic diagrams to identify structural features in galaxies where resolution is limited and AGN activity is prevalent. These results are especially important for understanding the role of AGN feedback in galaxy morphology, the connection between AGN and their host galaxy, and the classification of active galaxies in current and future large deep-field and all-sky surveys, where PSF variation and flux contamination are inevitable. Furthermore, by covering the full range of AGN contamination possibilities (5\%-75\%), our results can be applied to both type-1 and type-2 AGN. This is especially important at higher redshift, where the fraction of AGN, and in particular type-1 AGN, is higher. Finally, by clarifying which parameters remain robust across all conditions analysed, our study improves the selection of reliable morphological diagnostics to trace the evolution of galaxies over cosmic time.
	
	%
	
	
	
	\vspace{1.5cm}
	\authorcontributions{T.G.-W. was responsible for the main analysis and paper edition. M.P. and J.M. participated in the analysis, discussion, and drafting of the paper. J.P. was responsible for the development of the code for the simulation of AGN contribution and statistical analysis. I.M. contributed to the development of Figures \ref{fig_var:part1}--\ref{fig_var:part3}. A.M. supported the coding in Python. S.T.M. participated in the general discussion. All authors have read and agreed to the published version of the manuscript.}
	
	\funding{TGW acknowledges the support from Bule Hora University under the Ministry of Science and Higher Education.
		
		TGW, MP, and ST acknowledge financial support from the Space Science and Geospatial Institute (SSGI) under the Ethiopian Ministry of Innovation and Technology (MInT).
		
		ST acknowledges the financial support from Jimma University.
		
		TGW, JM, MP, and IM acknowledge the support given through the grant CSIC I-COOP 2017, COOPA20168.
		
		AM acknowledges support from the National Research Foundation of South Africa and financial support from the Swedish International Development Cooperation Agency (SIDA) through the International Science Programme (ISP)-Uppsala University to the University of Rwanda through the Rwanda Astrophysics, Space and Climate Science Research Group (RASCSRG).
		
		MP, JM, JP, and IMP acknowledge the support from the Spanish Ministerio de Ciencia e Innovación-Agencia Estatal de Investigación through projects PID2019-106027GB-C41, PID2022-140871NB-C21, and AYA2016-76682C3-1-P, and the State Agency for Research of the Spanish MCIU through the Center of Excellence Severo Ochoa award to the Instituto de Astrofísica de Andalucía (CEX2021-001131-S funded by MCIN/AEI/10.13039/501100011033). }
	
	\dataavailability{In support of this study, no new data were generated or analysed. The data used in this article can be obtained from the public sources cited in the article (or references therein).} 
	
	
	%
	%
	
	\acknowledgments{We thank the anonymous referees for accepting to review this paper and for providing constructive and valuable comments that improved the manuscript. In this work, we made use of the Virtual Observatory Tool for OPerations on Catalogues And Tables (TOPCAT) and Python.}
	
	\conflictsofinterest{The authors declare no conflicts of interest.} 
	
	
	
	\abbreviations{Abbreviations}{
		The following abbreviations are used in this manuscript:
		\\
		
		\noindent 
		\begin{tabular}{@{}ll}
			ACS & Advanced Camera for Surveys\\
			AGN & Active Galactic Nuclei\\
			ASYM & Asymmetry index\\
			BPT & Baldwin-Phillips-Terlevich diagram\\
			CABR & Abraham concentration index\\
			CCON & Conselice-Bershady concentration index\\
			galSVM & galaxy Support Vector Machine\\
			GINI & GINI coefficient\\
			IRAF & Image Reduction and Analysis Facility\\
			LINER  & Low-Ionization Nuclear Emission-line Region\\
			mag & magnitude\\
			M20 &  M20 moment of light\\
			SDSS & Sloan Digital Sky Survey\\
			SVM  & Support Vector Machine\\
			TOPCAT & Tool for OPerations on Catalogues And Tables\\
		\end{tabular}
	}
	
	\appendixtitles{no} 
	\appendixstart
	\appendix
	\begin{appendix}
		\onecolumn
		\section[width=1.9\textwidth]{: Comparing the local sample (no AGN) with COSMOS-like samples (without and with AGN added)}
		\label{appendixA}
		\begin{figure}[H]
			\begin{center}
				\includegraphics[height= 1.90in, width=5.54in]{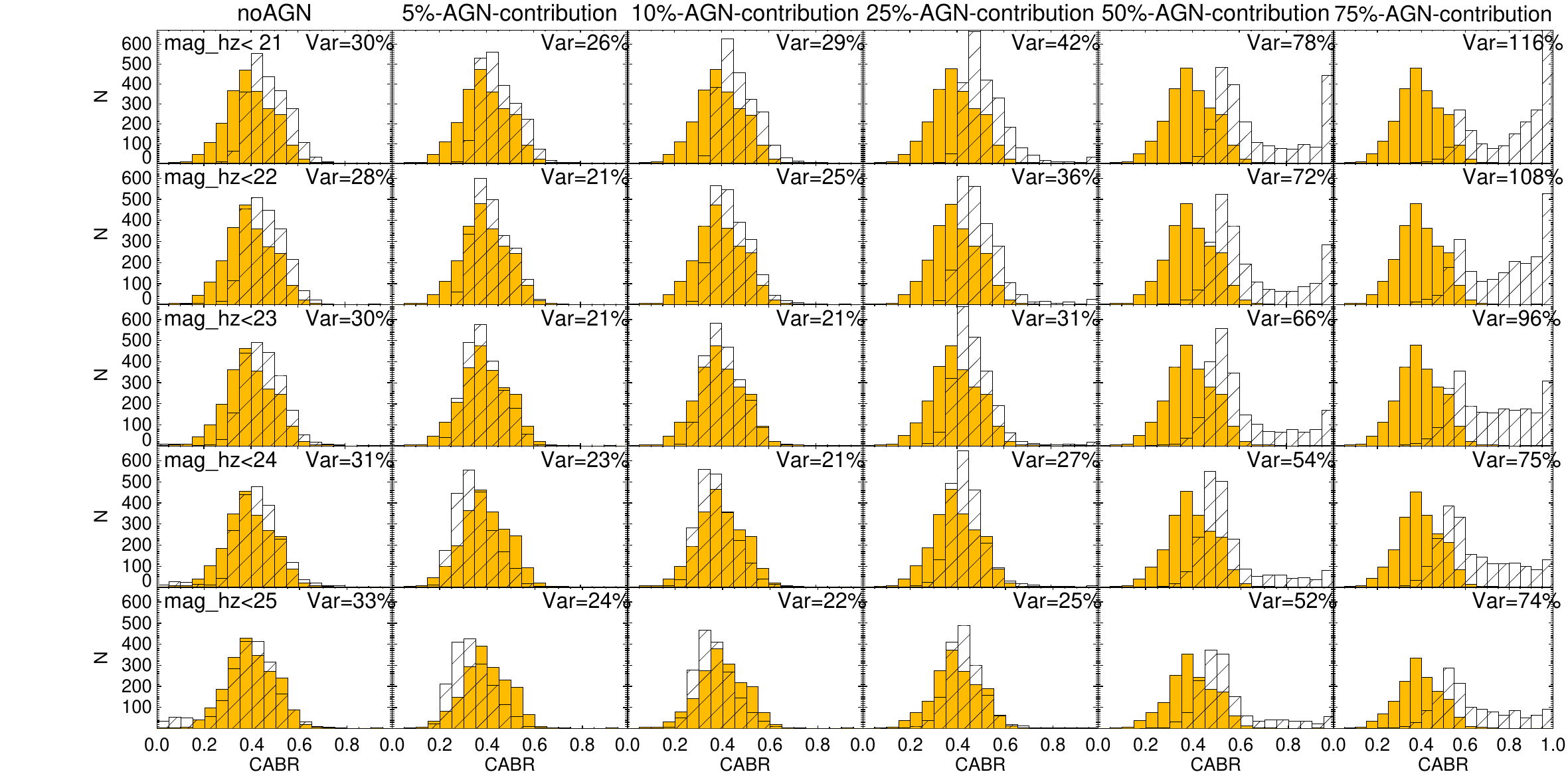}
				\includegraphics[height= 1.90in, width=5.54in]{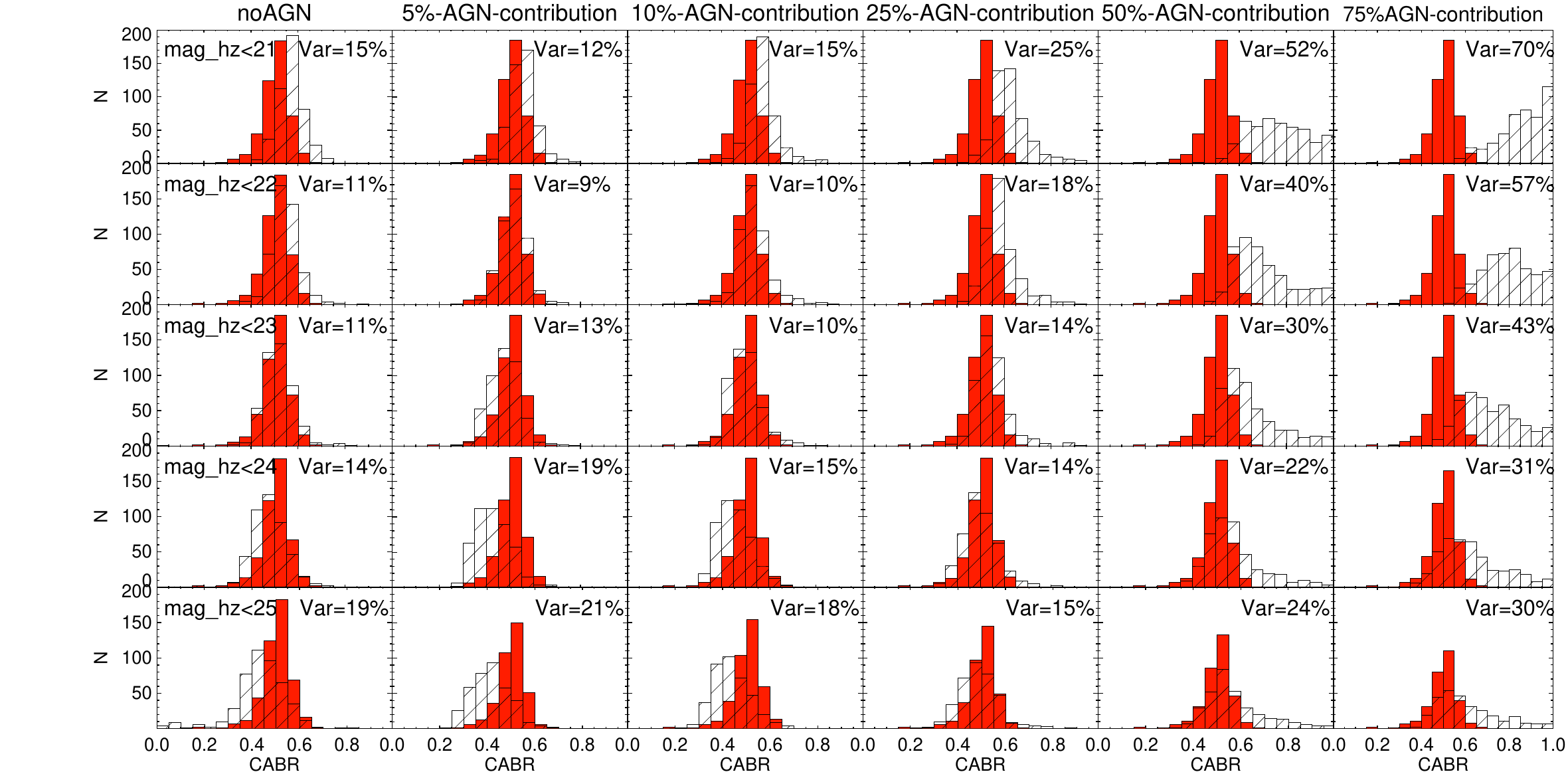}
				\includegraphics[height= 1.90in, width=5.54in]{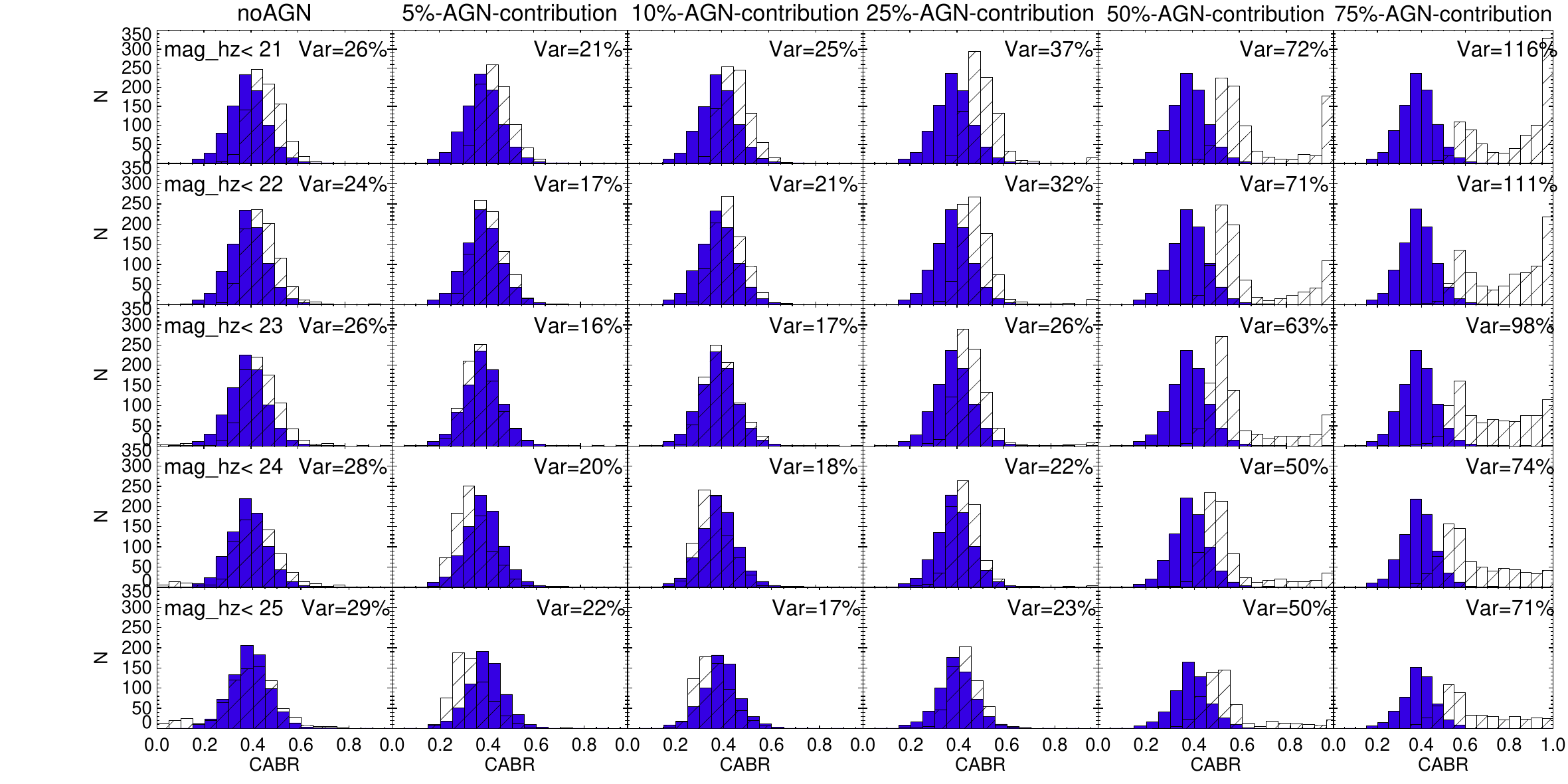}
				\includegraphics[height= 1.90in, width=5.54in]{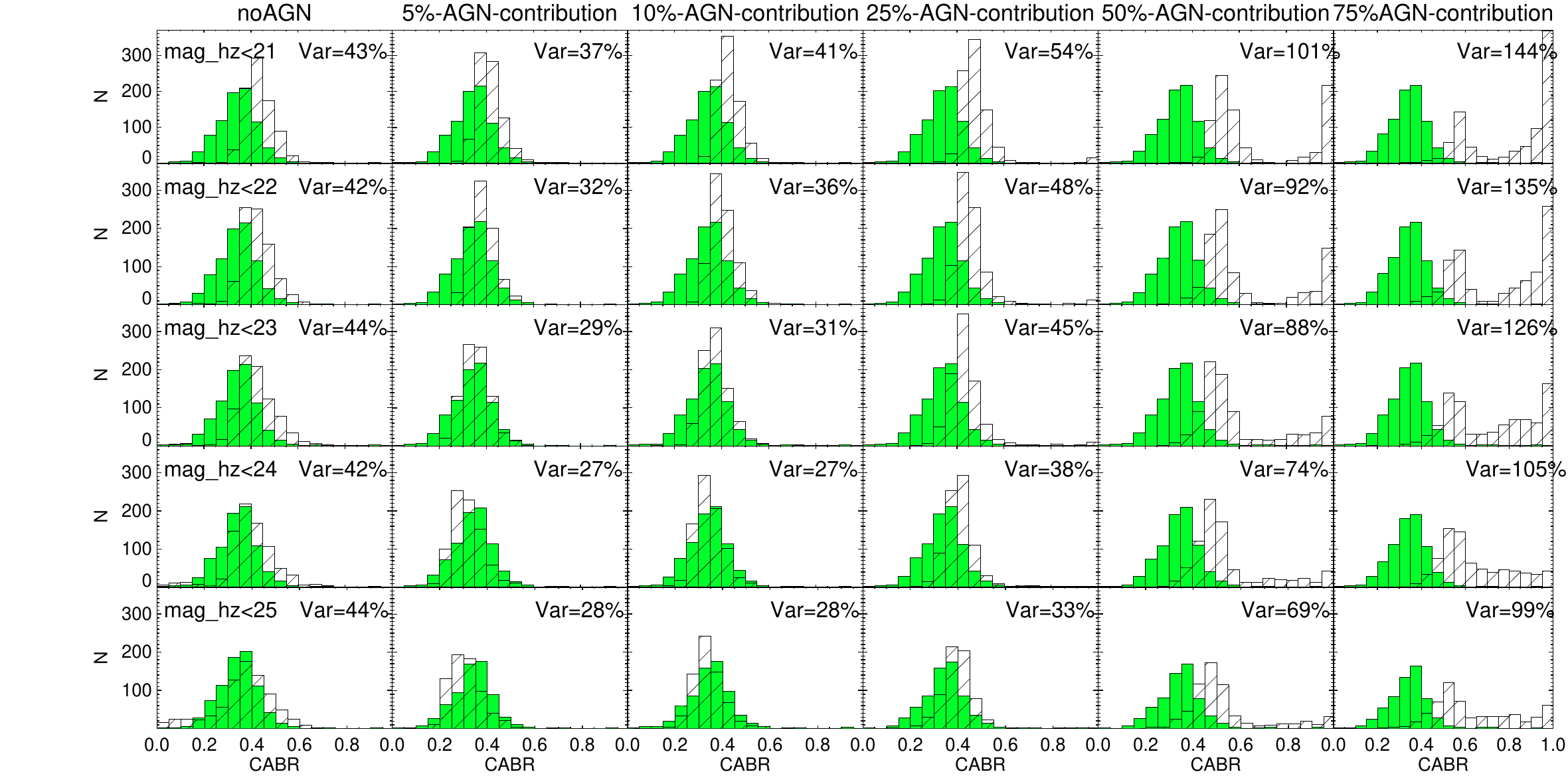}
				\caption{Comparison between the values of CABR for the original sample at z\,$\sim$\,0 without any added AGN contributions (filled histograms) and the simulated sample (opened histograms) moved to higher redshifts and fainter magnitudes from 21 to 25 (from top to bottom rows) without AGN (first columns) and for five AGN contributions added from 5\% to 75\% (second to last columns). From top to bottom, the distributions of the total sample (yellow colour), early-type (red), early-spiral (blue) and late-spiral  (green) galaxies are shown.}
				\label{fig4.1:part1}
			\end{center}
		\end{figure}
		
		\begin{figure*}[h!]
			\begin{center}
				{\includegraphics[height= 2.2in, width=5.54in]{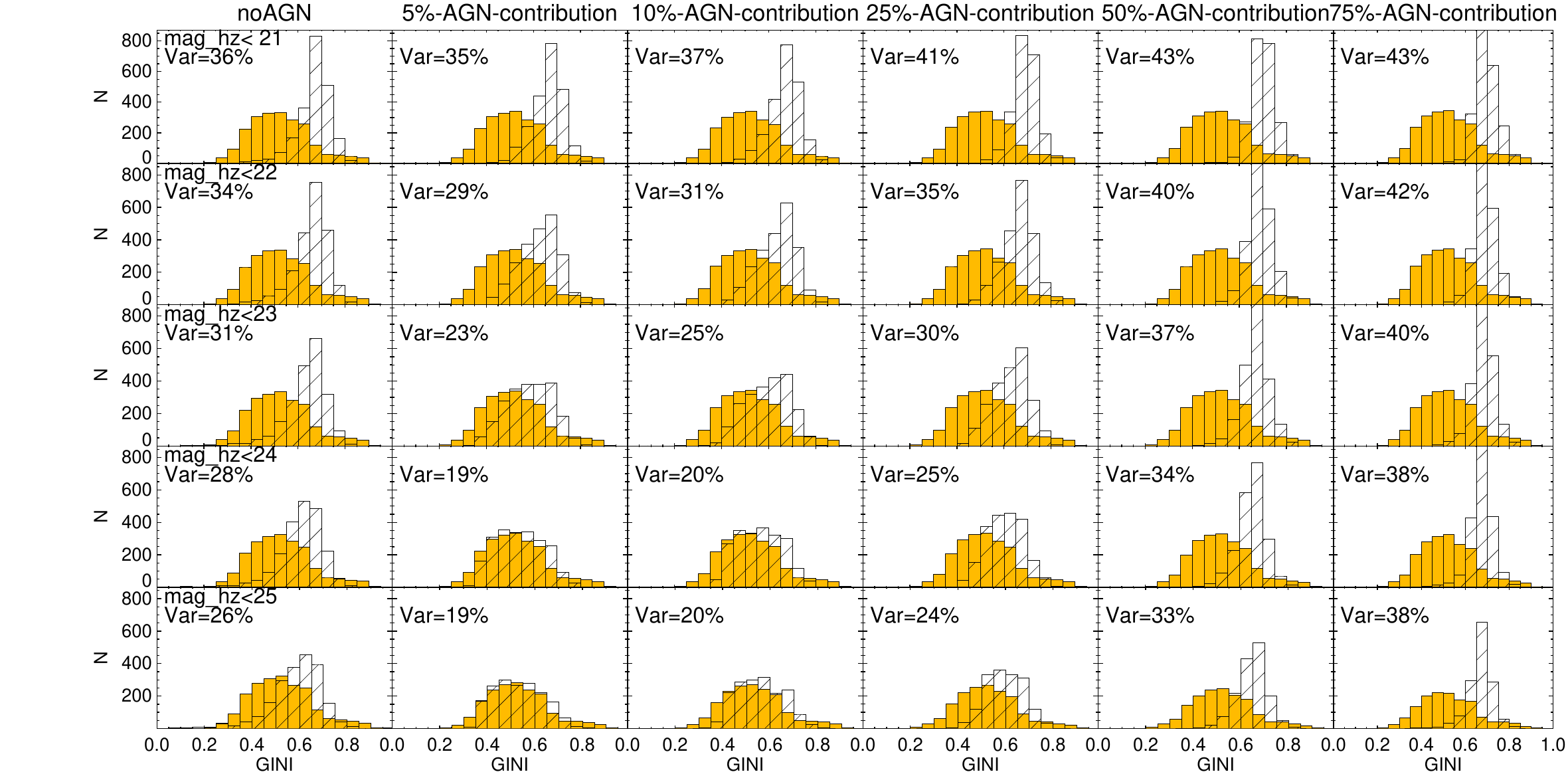}}
				{\includegraphics[height= 2.2in, width=5.54in]{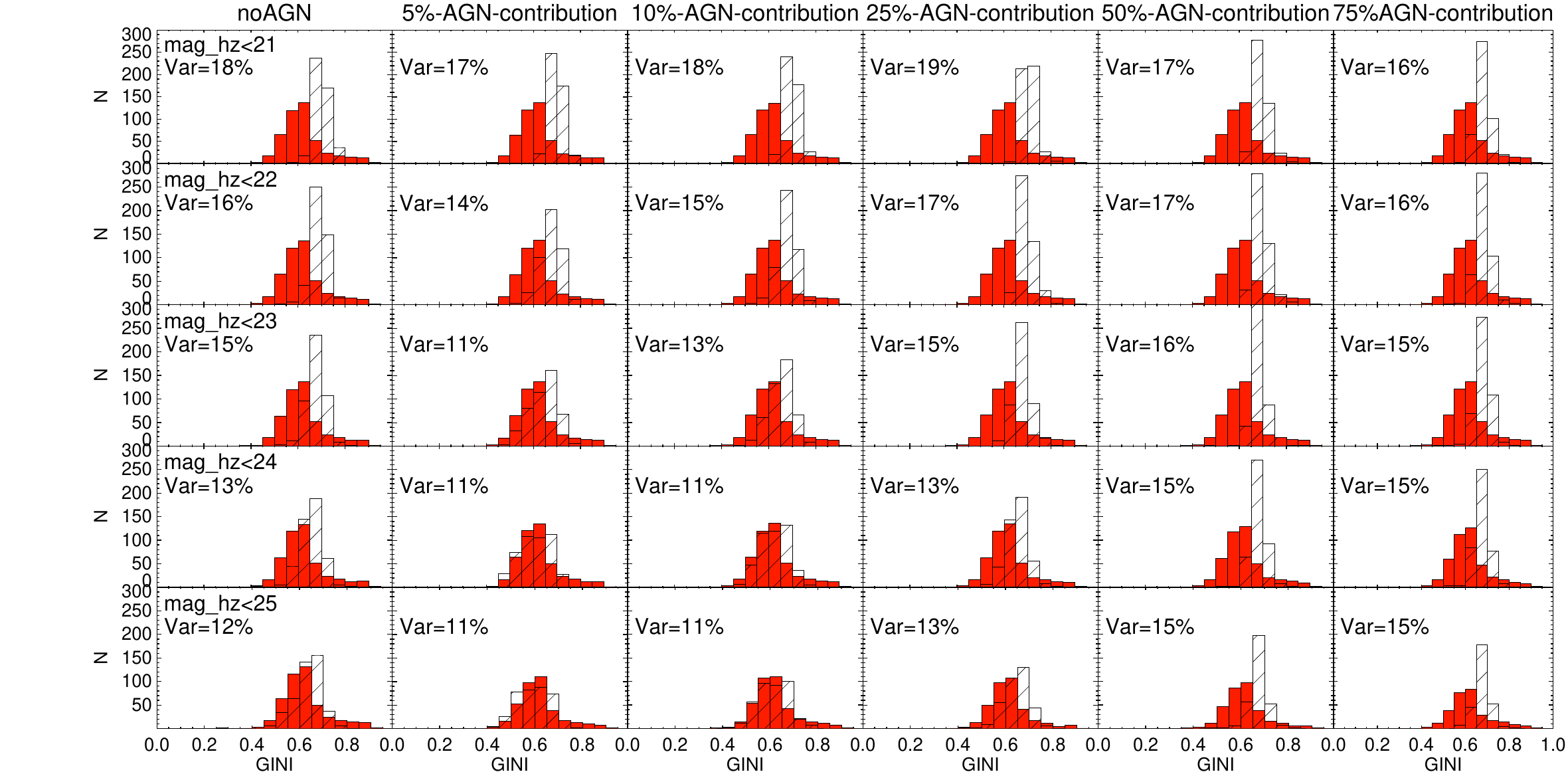}}
				{\includegraphics[height=2.2in,width=5.54in]{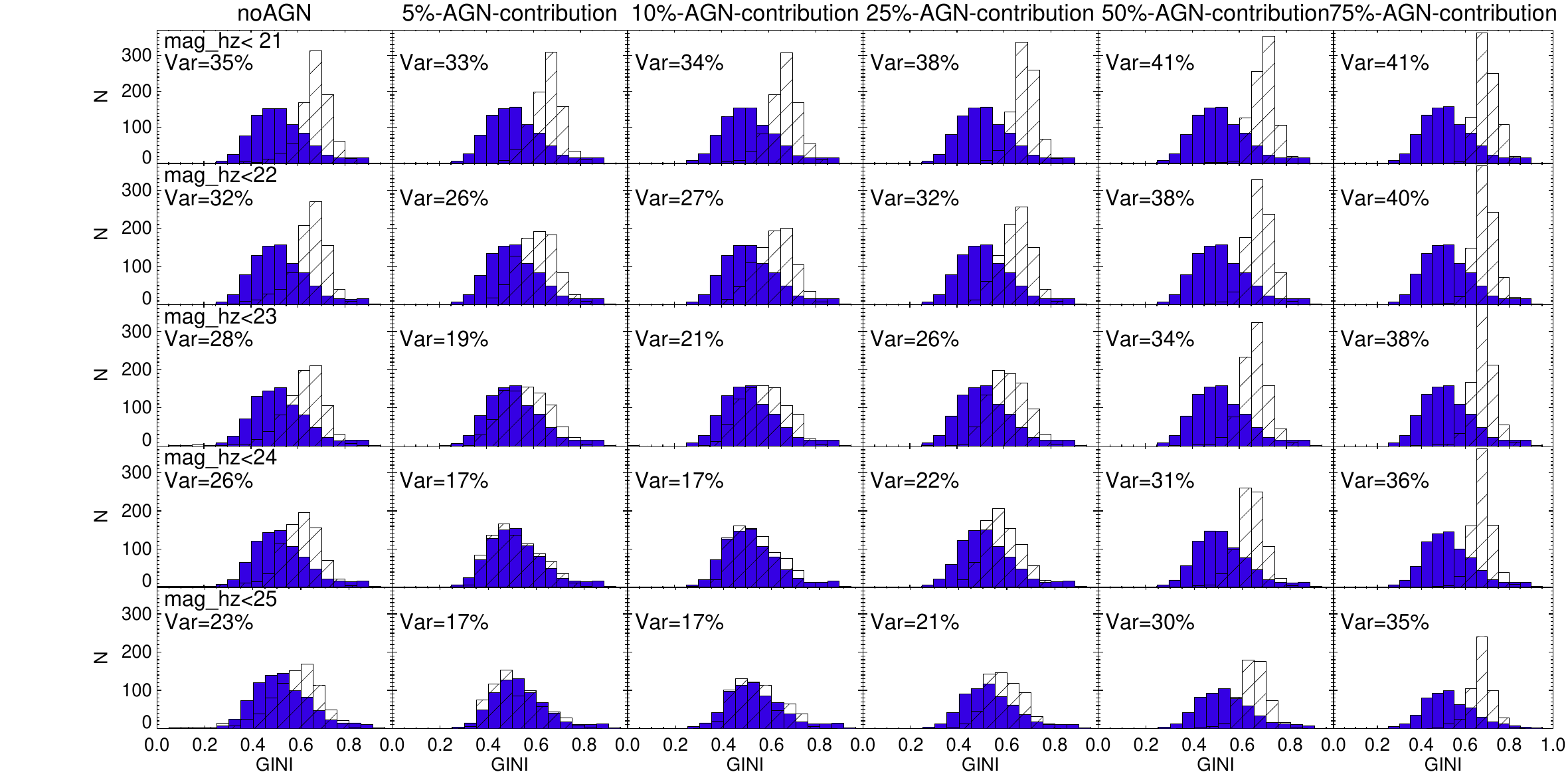}}
				\includegraphics[height=2.2in,width=5.54in]{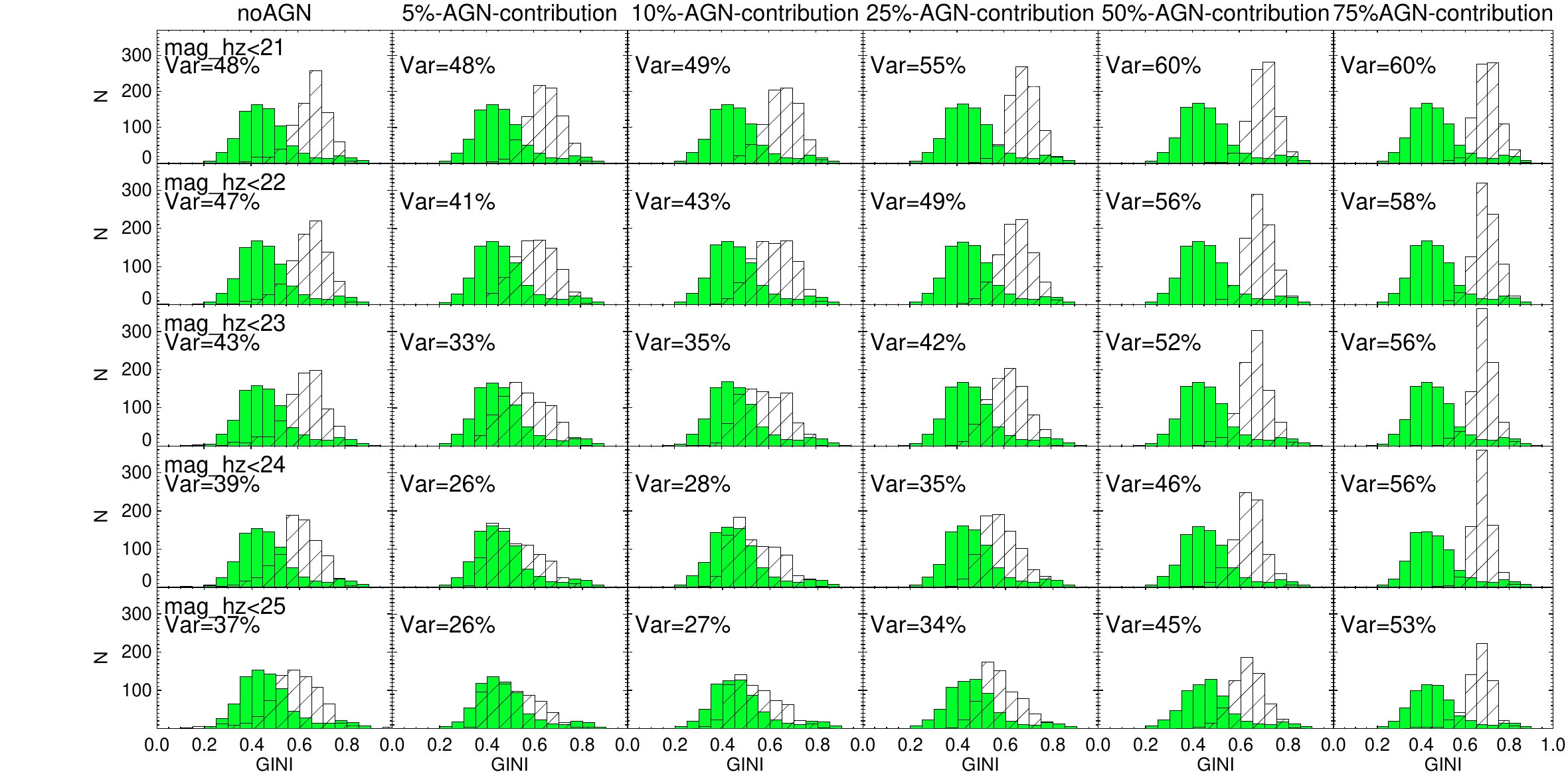}
				\caption{Same as Figure \ref{fig4.1:part1}, but for the GINI parameter.}
				\label{fig4.2:part1}
			\end{center}
		\end{figure*}
		
		\begin{figure*}[h!]
			\begin{center}
				{\includegraphics[height= 2.2in, width=5.54in]{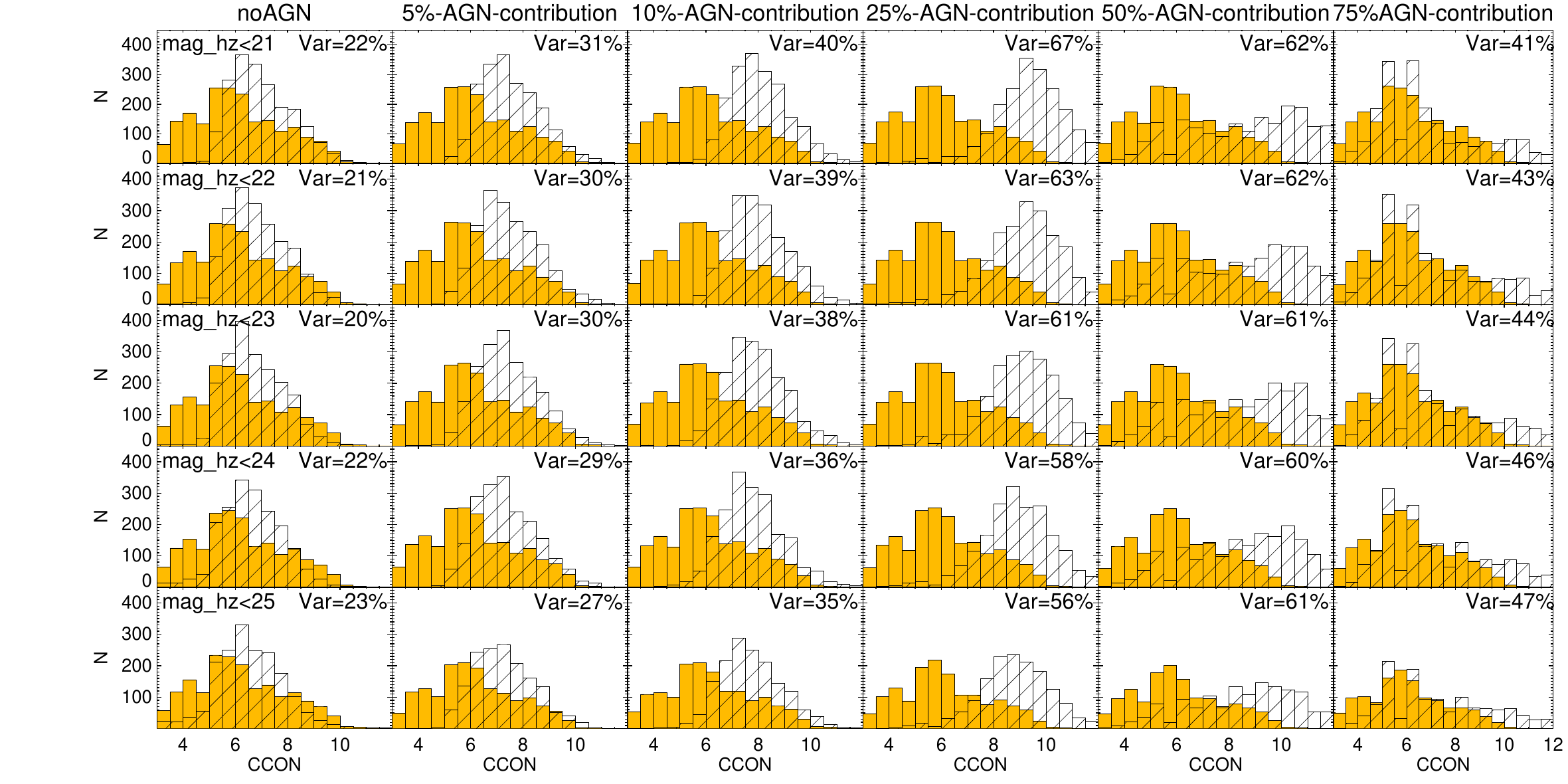}}
				{\includegraphics[height= 2.2in, width=5.54in]{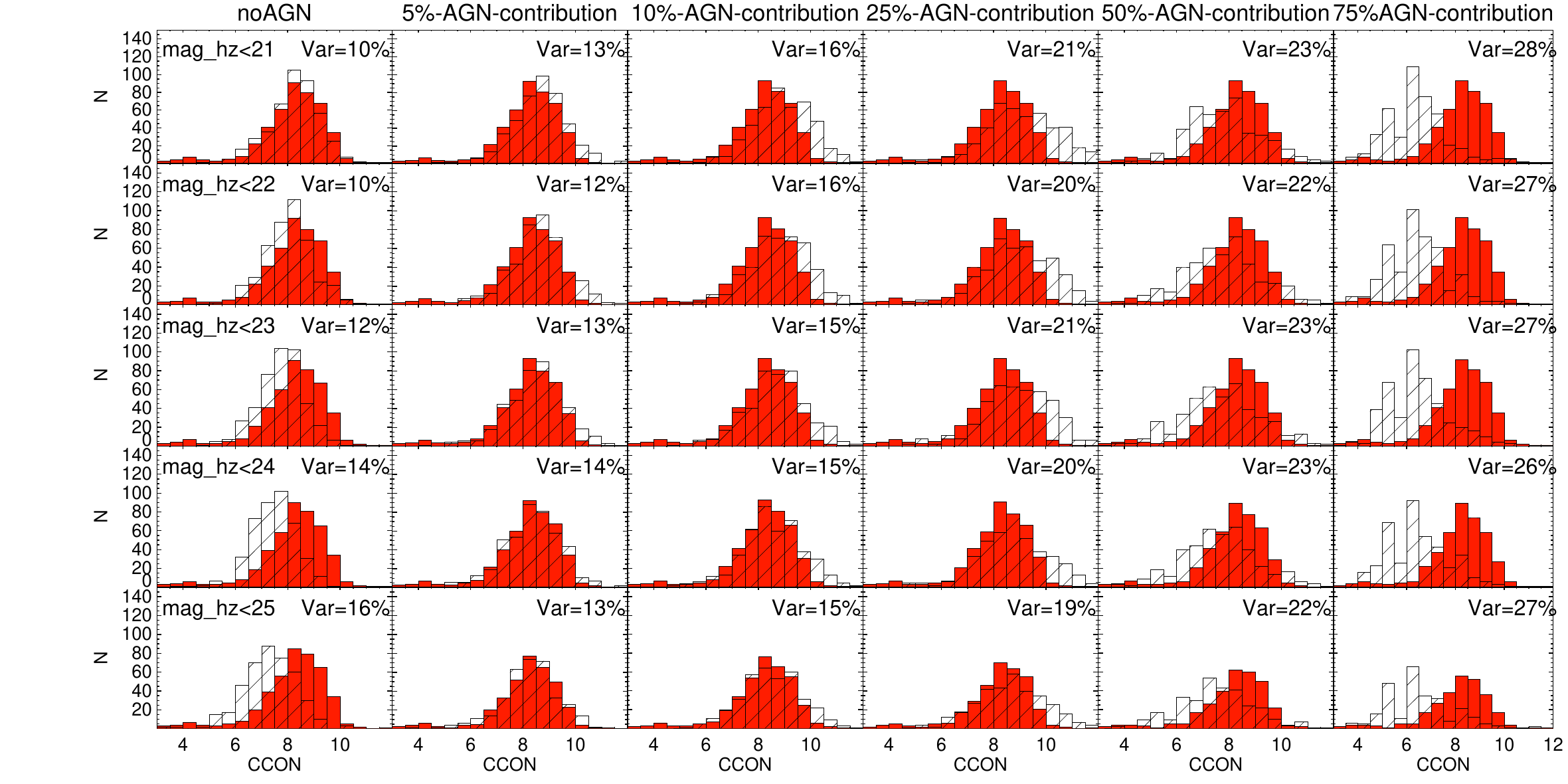}}
				{\includegraphics[height= 2.2in, width=5.54in]{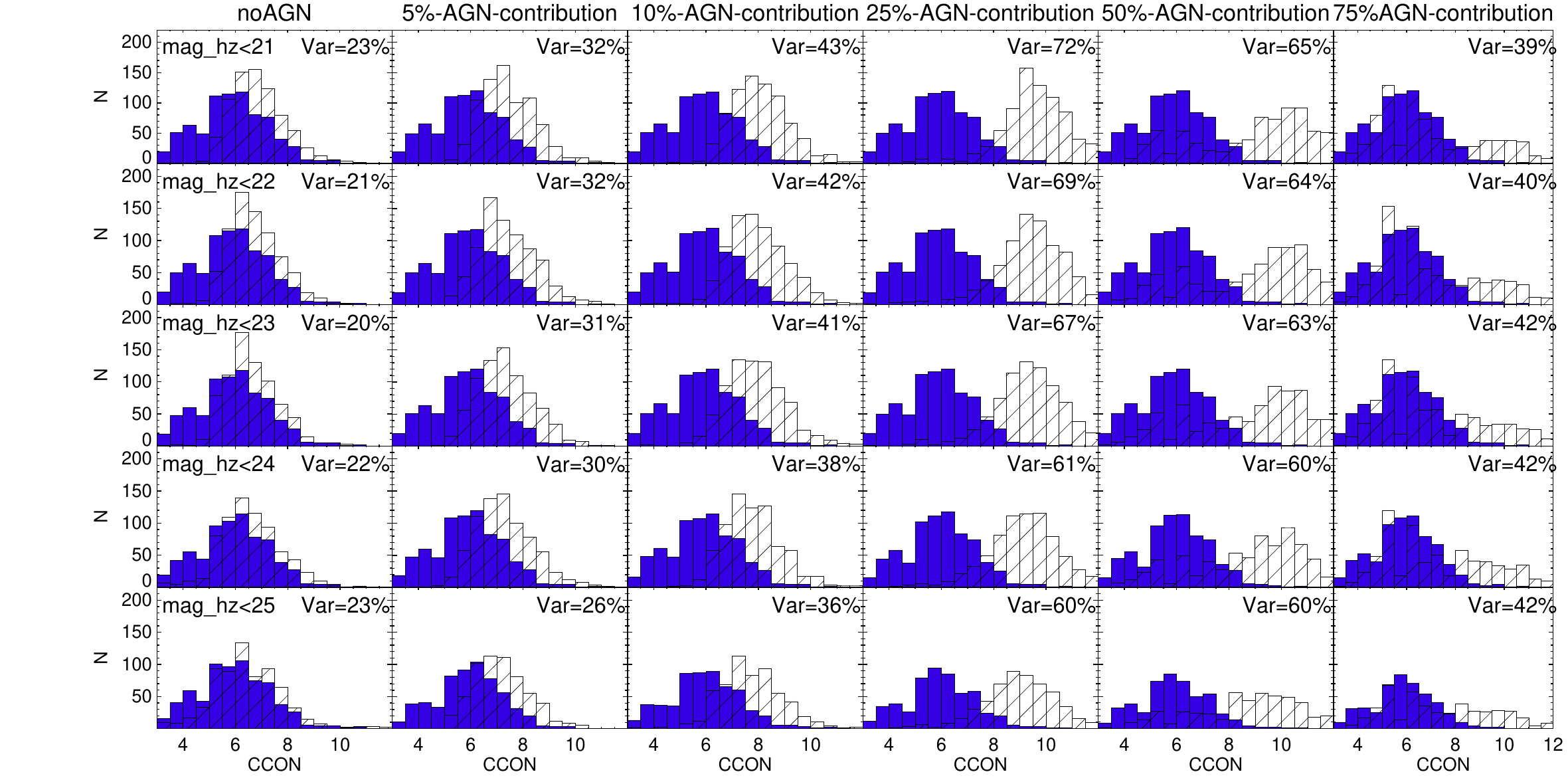}}
				{\includegraphics[height= 2.2in, width=5.54in]{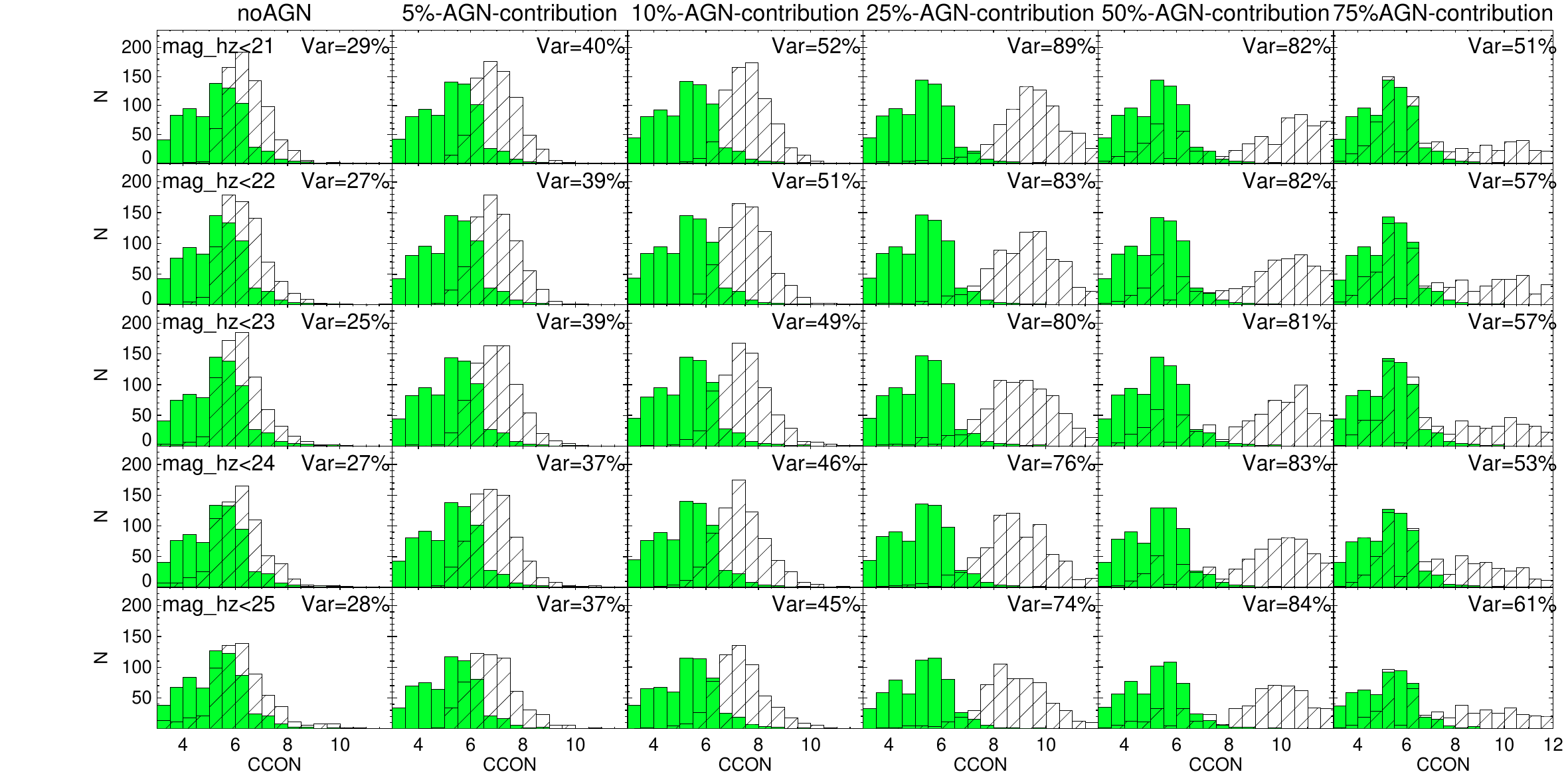}}
				\caption{Same as Figure \ref{fig4.1:part1}, but for the CCON parameter.}
				\label{fig4.3:part1}
			\end{center}
		\end{figure*}
		
		\begin{figure*}
			\begin{center}
				{\includegraphics[height= 2.2in, width=5.54in]{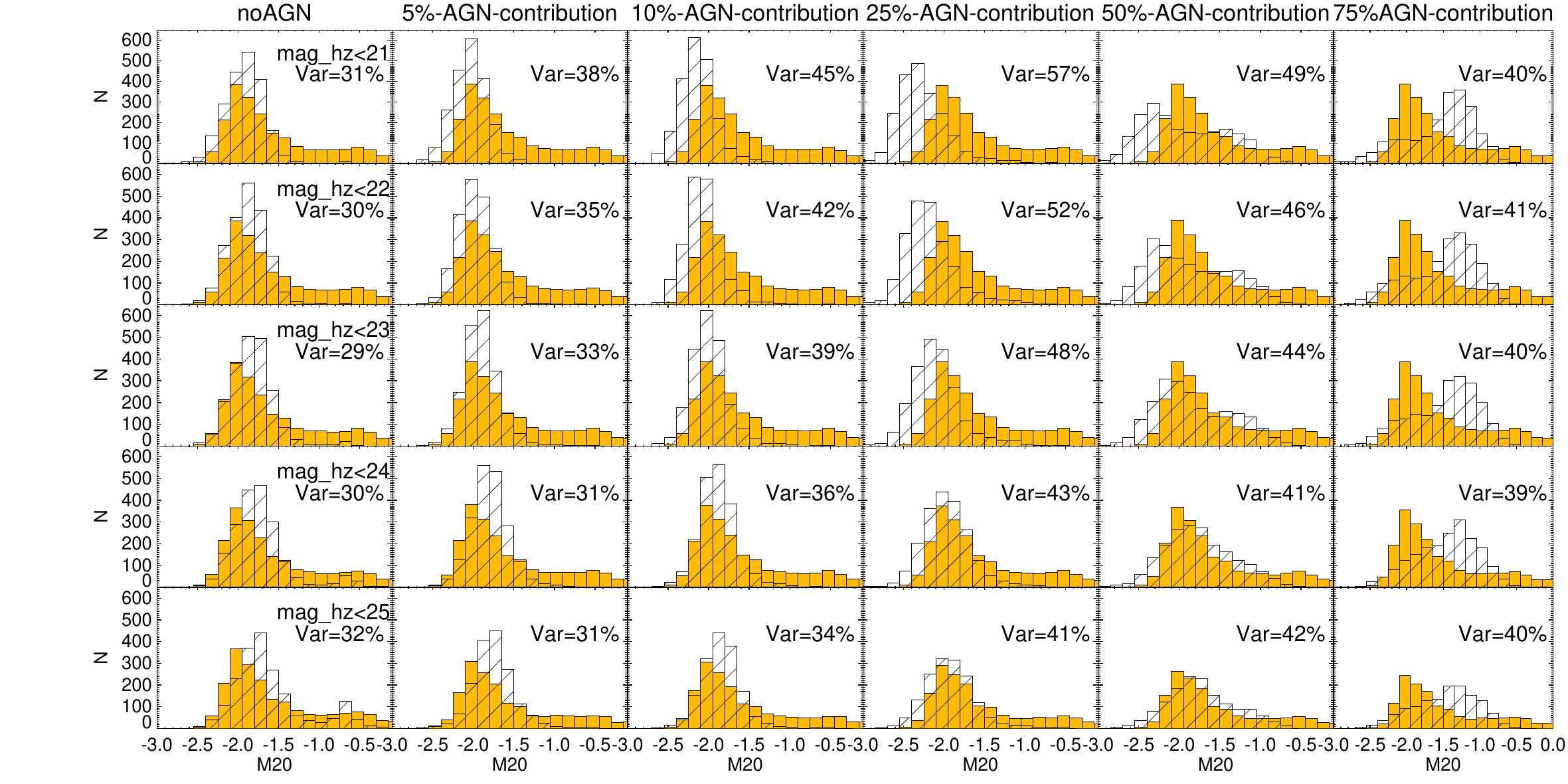}}
				{\includegraphics[height= 2.2in, width=5.54in]{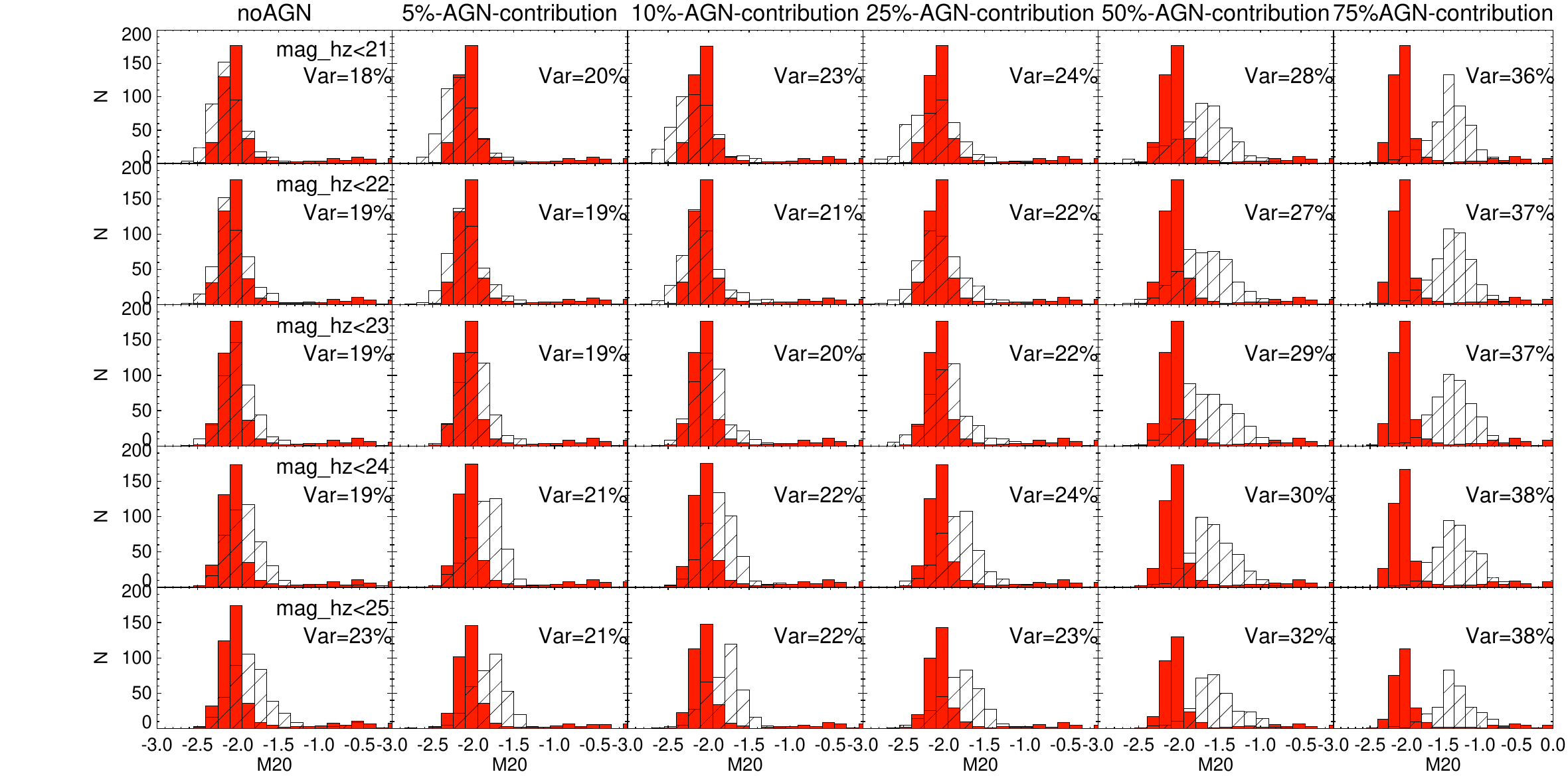}}
				{\includegraphics[height= 2.2in, width=5.54in]{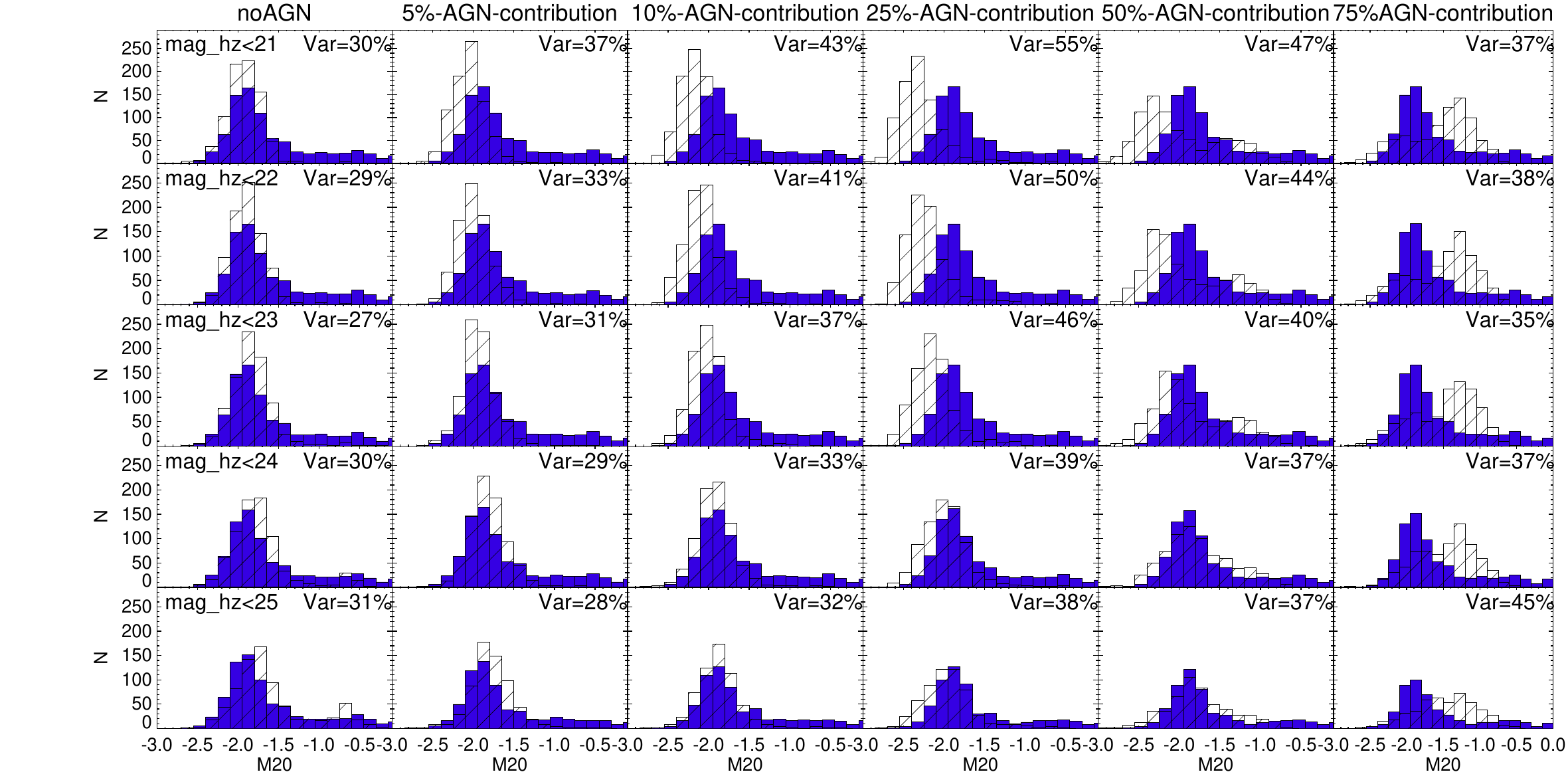}}
				{\includegraphics[height= 2.2in, width=5.54in]{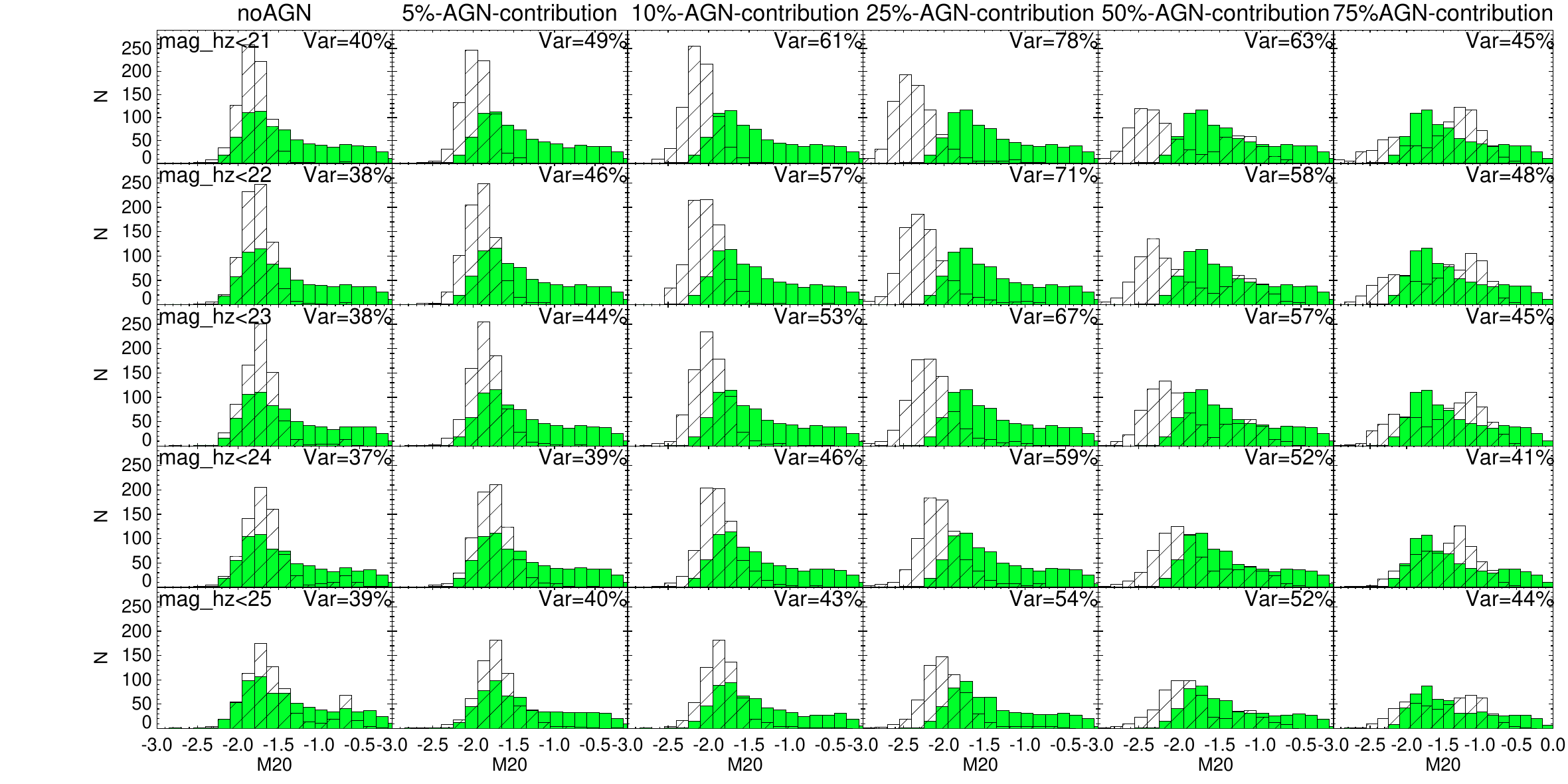}}
				\caption{Same as Figure \ref{fig4.1:part1}, but for the M20 moment of light.}
				\label{fig4.4:part1}
			\end{center}
		\end{figure*}
		
		\begin{figure*}
			\begin{center}
				{\includegraphics[height= 2.2in, width=5.54in]{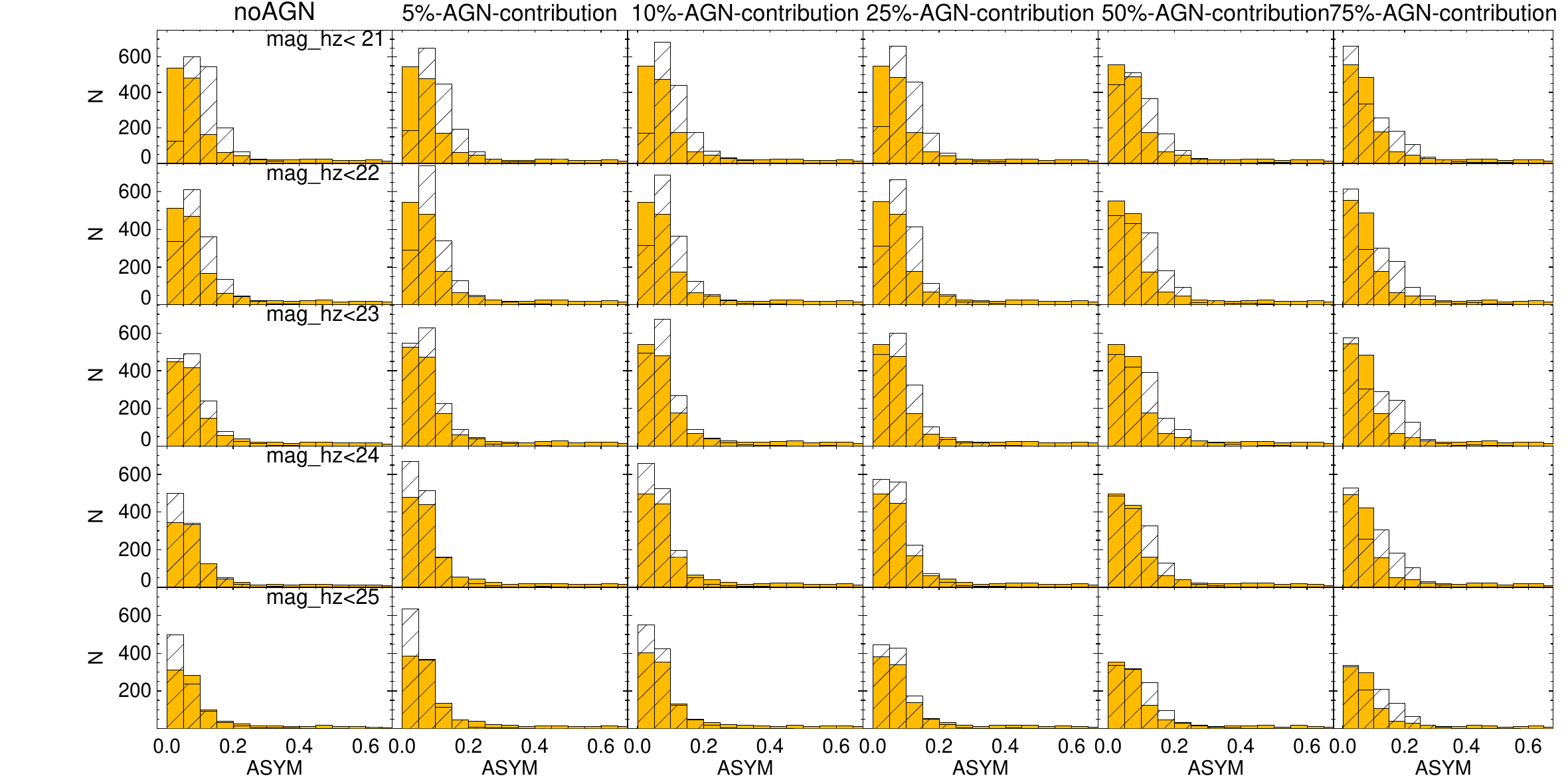}}
				{\includegraphics[height= 2.2in, width=5.54in]{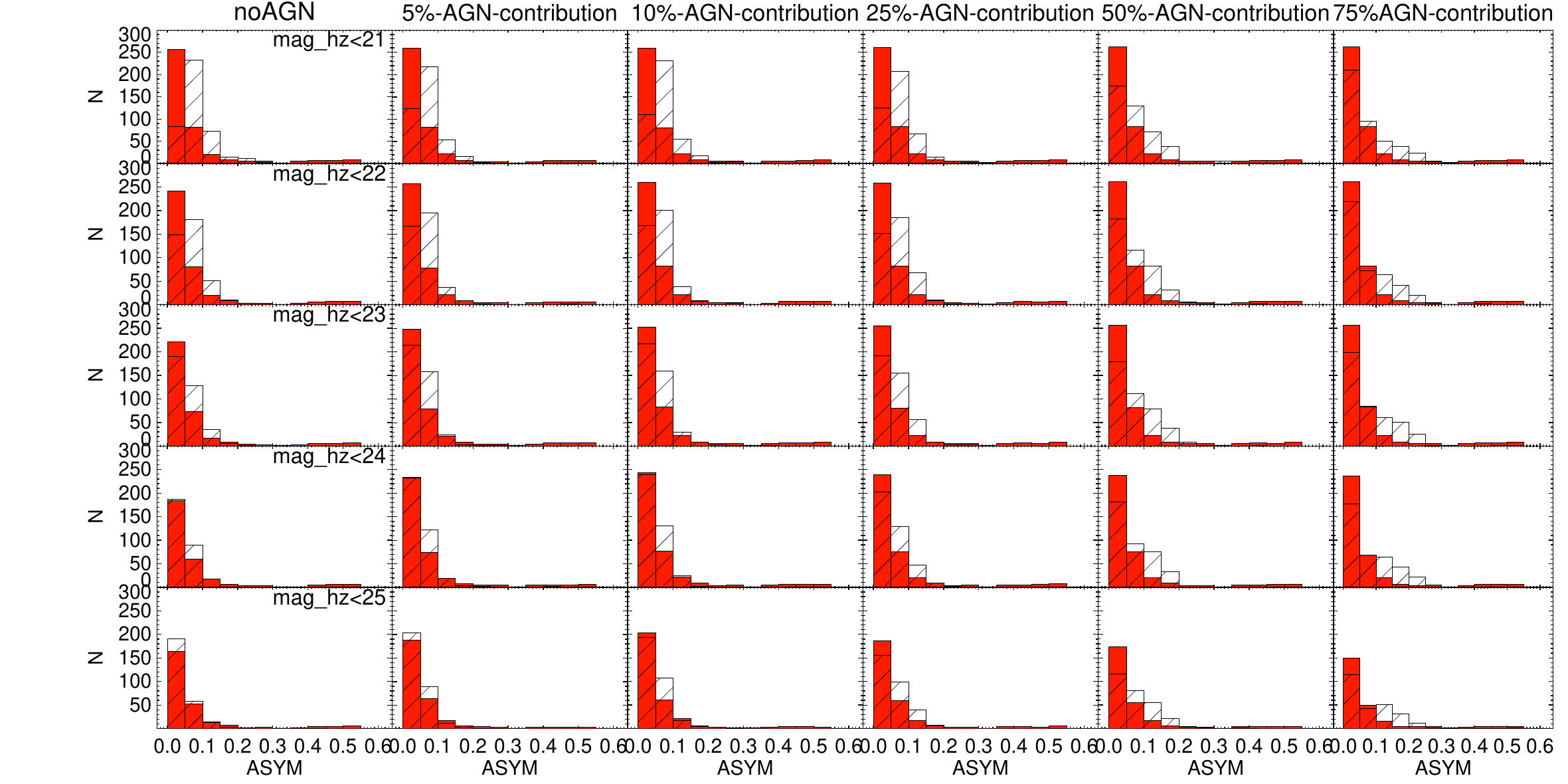}}
				{\includegraphics[height= 2.2in, width=5.54in]{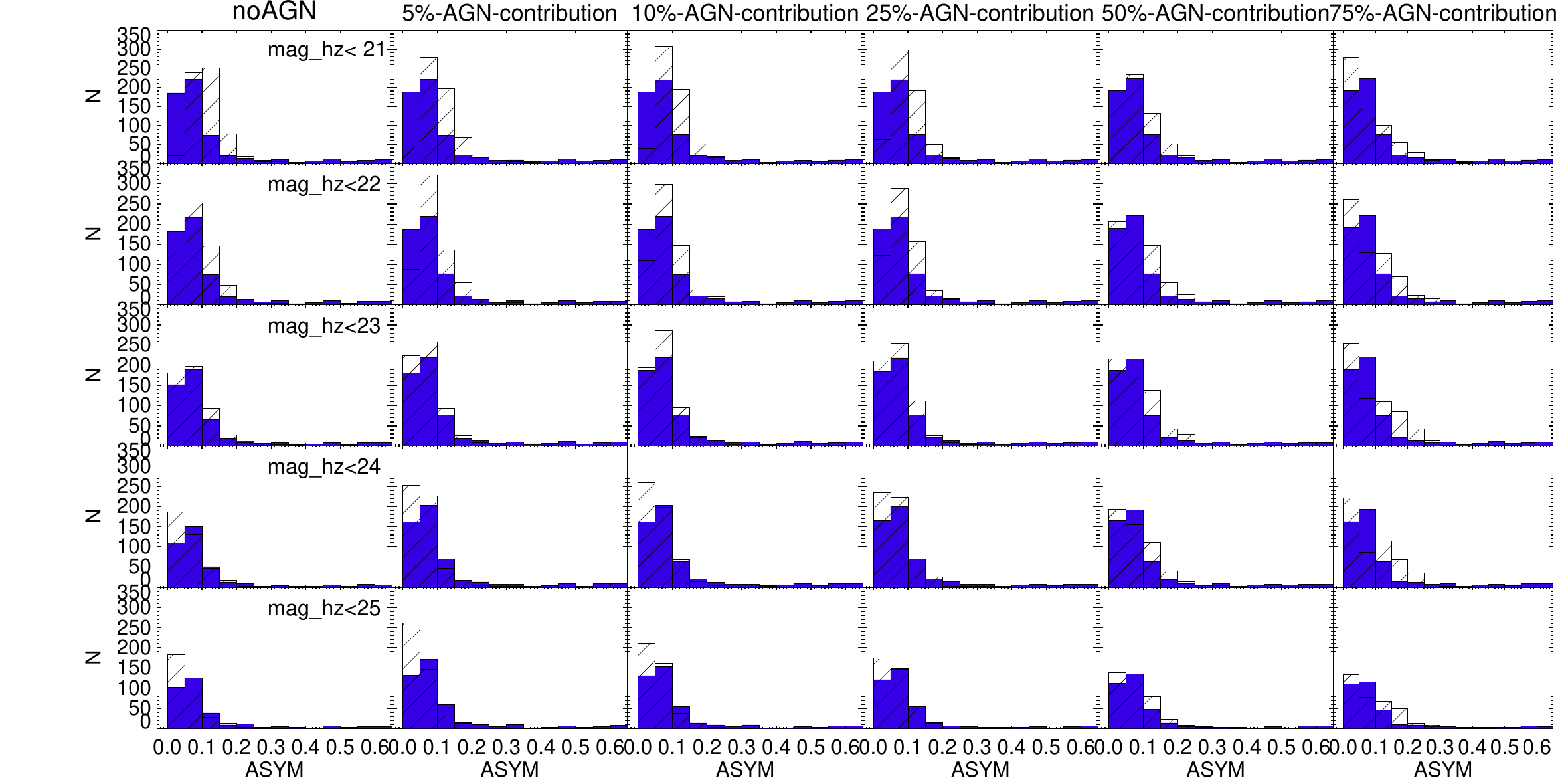}}
				{\includegraphics[height= 2.2in, width=5.54in]{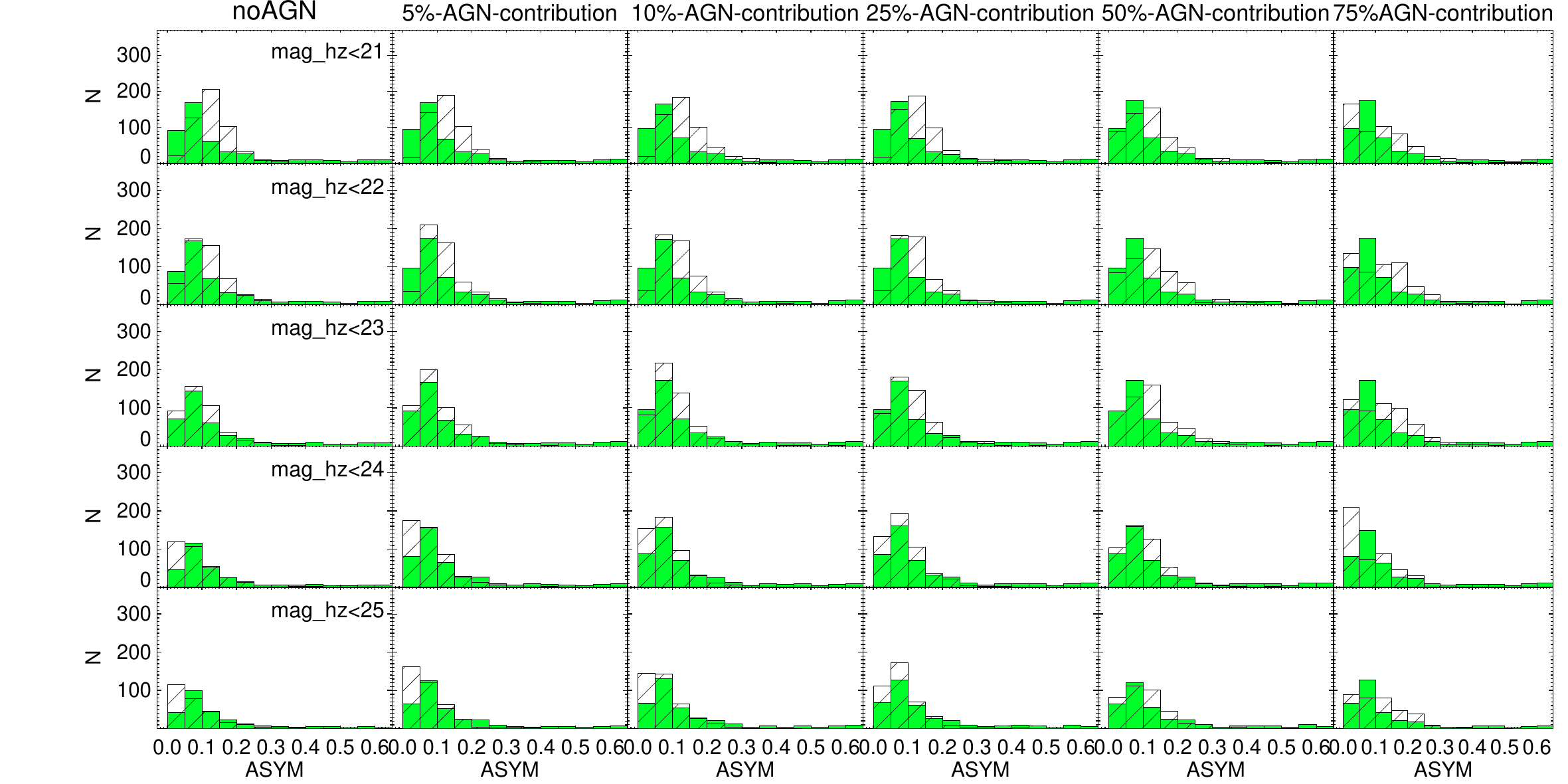}}
				\caption{Same as Figure \ref{fig4.1:part1}, but for the ASYM parameter.}
				\label{fig4.5:part1}
			\end{center}
		\end{figure*}
		
		\begin{figure*}
			\begin{center}
				{\includegraphics[height= 2.2in, width=5.54in]{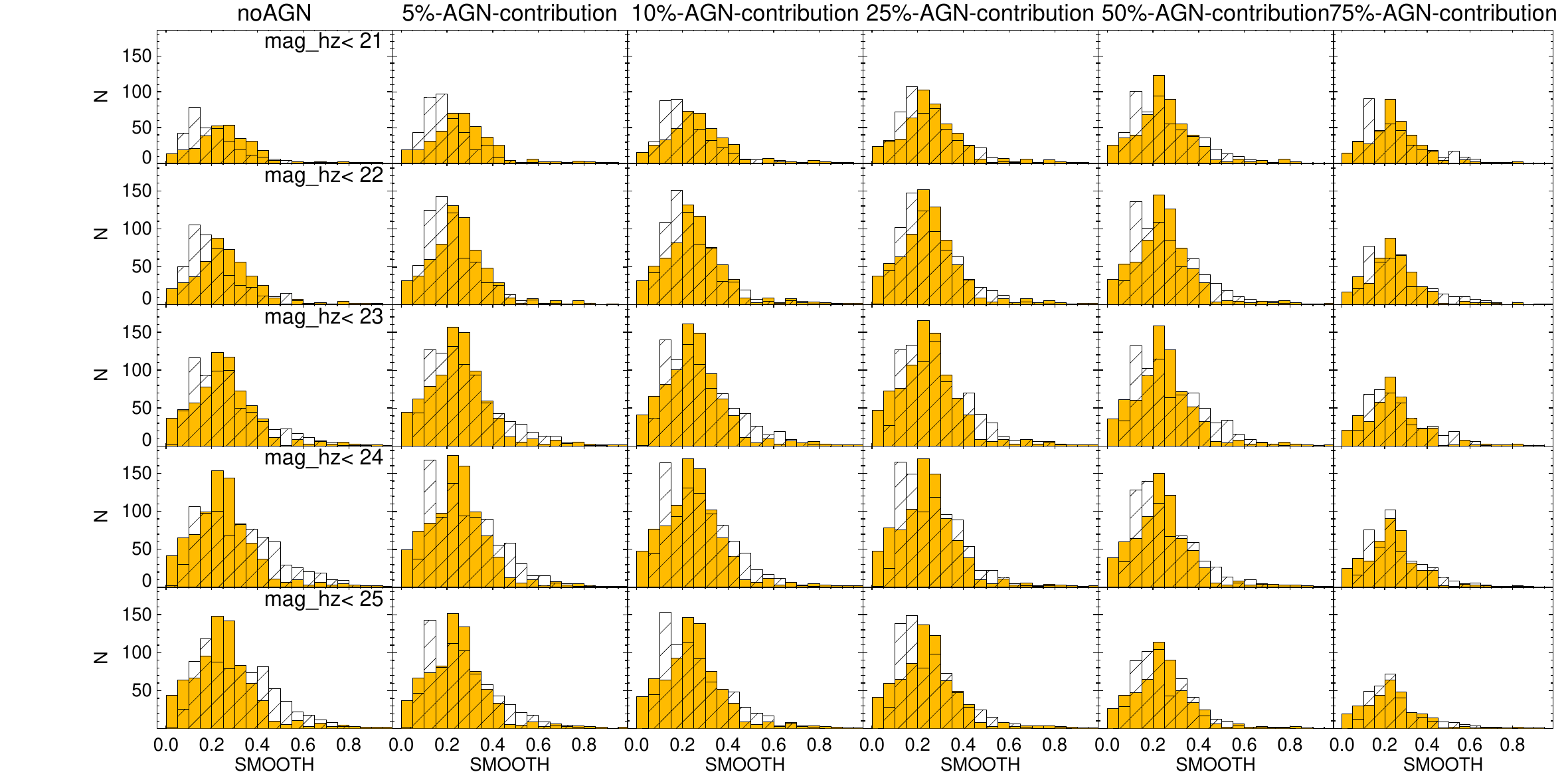}}
				{\includegraphics[height= 2.2in, width=5.54in]{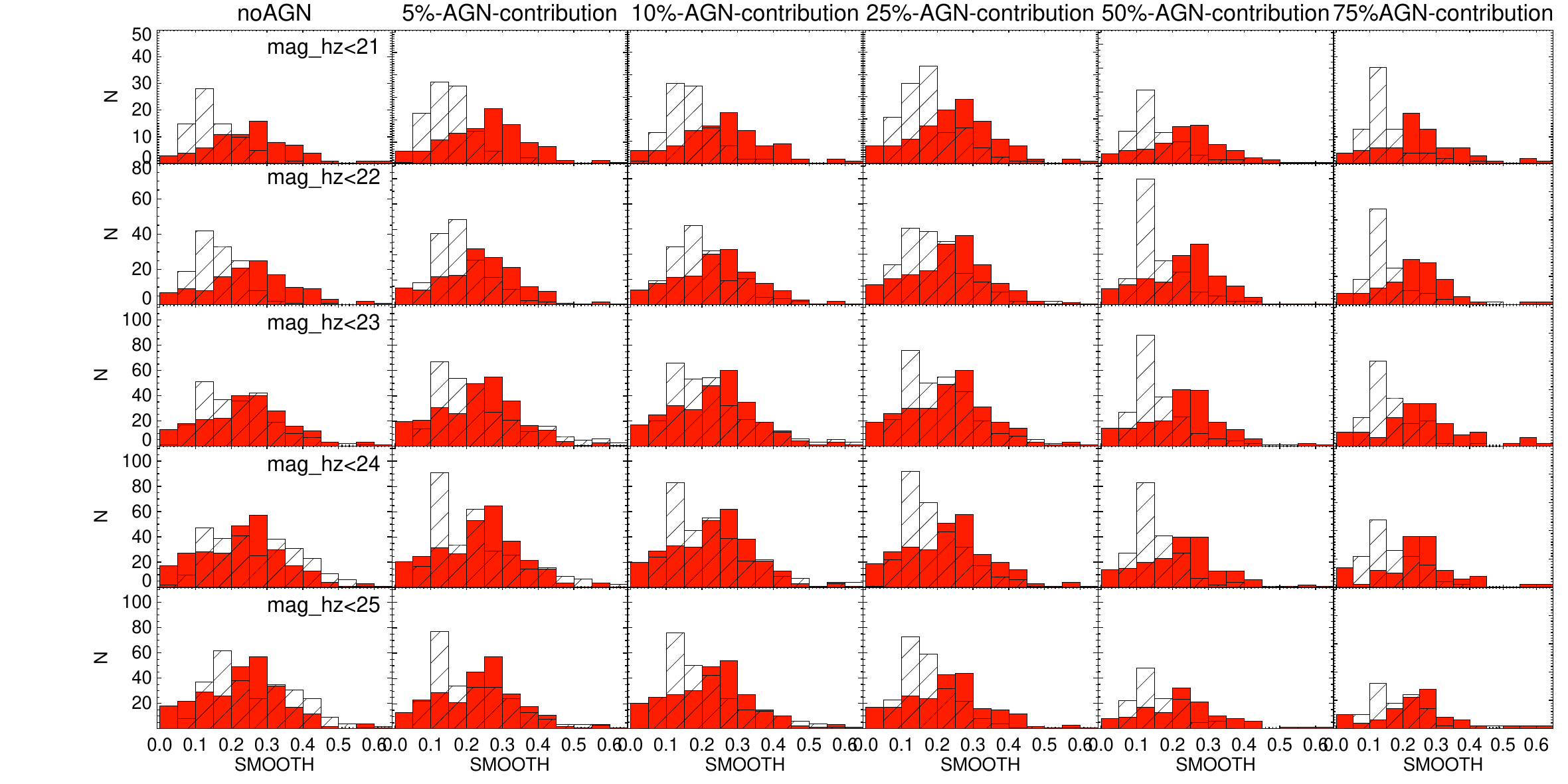}}
				{\includegraphics[height= 2.2in, width=5.54in]{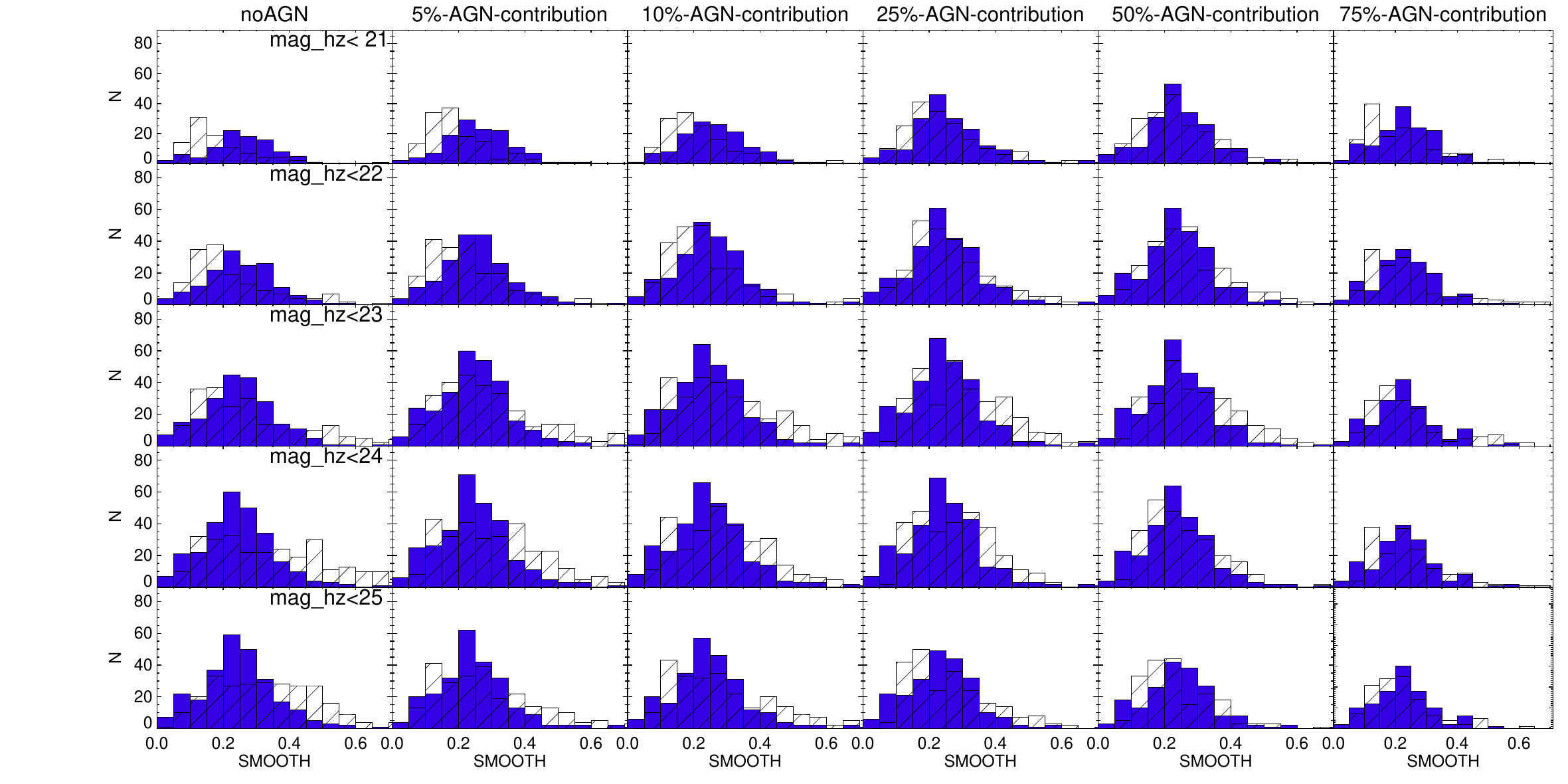}}
				{\includegraphics[height= 2.2in, width=5.54in]{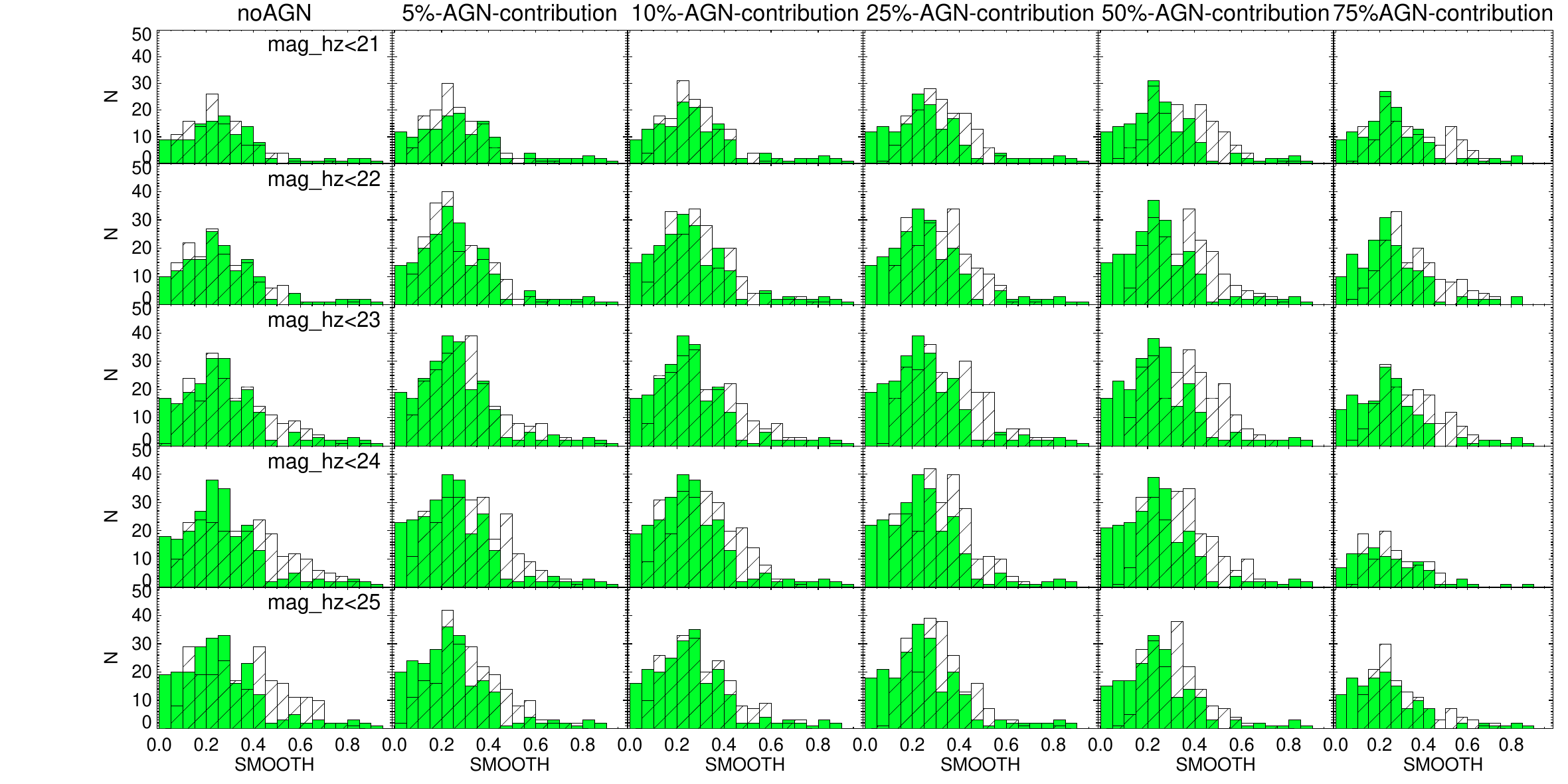}}
				\caption{Same as Figure \ref{fig4.1:part1}, but for the SMOOTHness parameter.}
				\label{fig4.6:part1}
			\end{center}
		\end{figure*}
		
		\newpage
		\onecolumn
		\section{: Comparing the local-like AGN sample with the COSMOS-like AGN sample}
		\label{appendixB}
		\begin{figure}[h!]
			\begin{center}
				{\includegraphics[height=1.9in, width=5.54in]{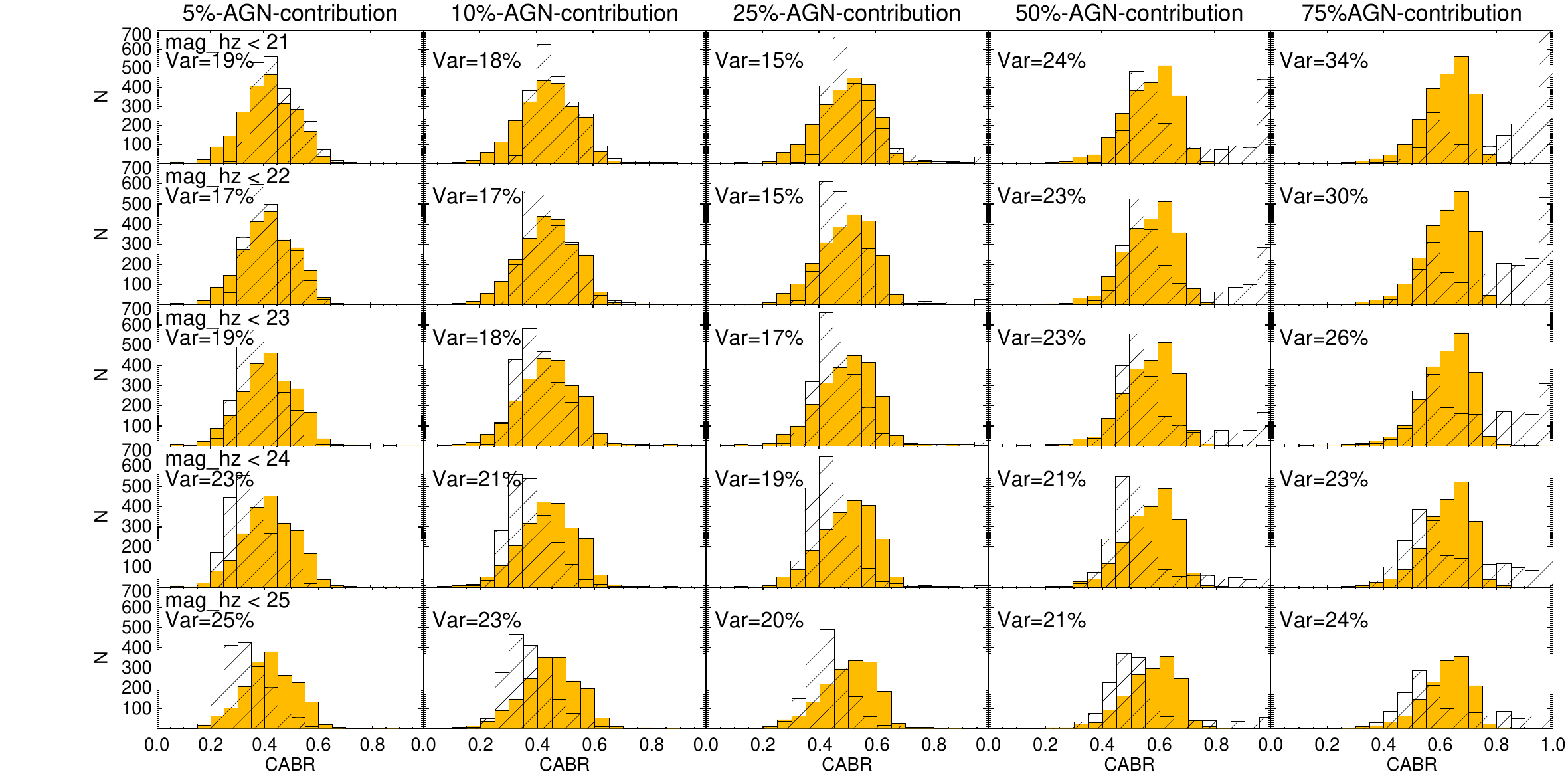}}
				{\includegraphics[height=1.9in, width=5.54in]{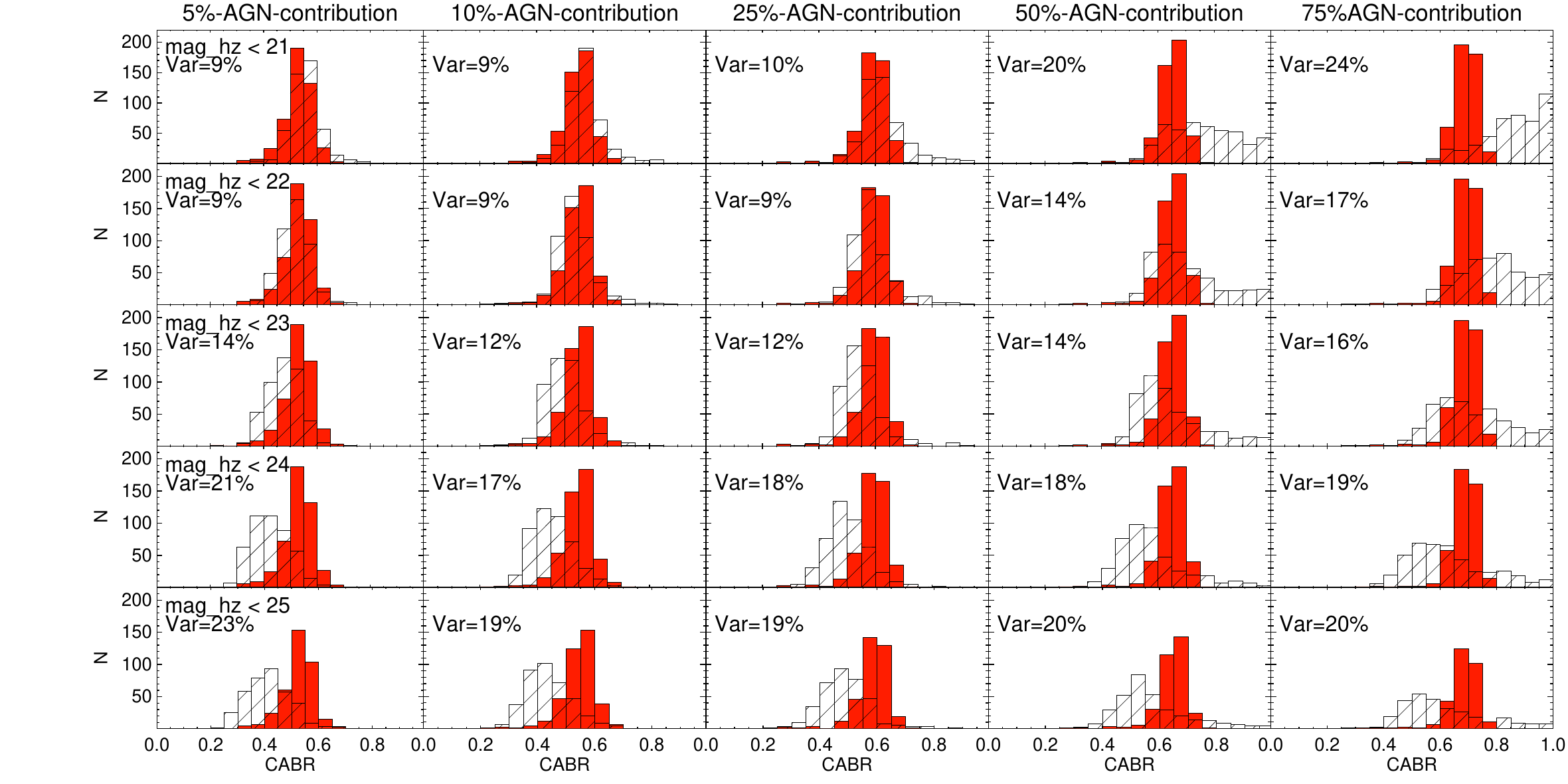}}
				{\includegraphics[height=1.9in, width=5.54in]{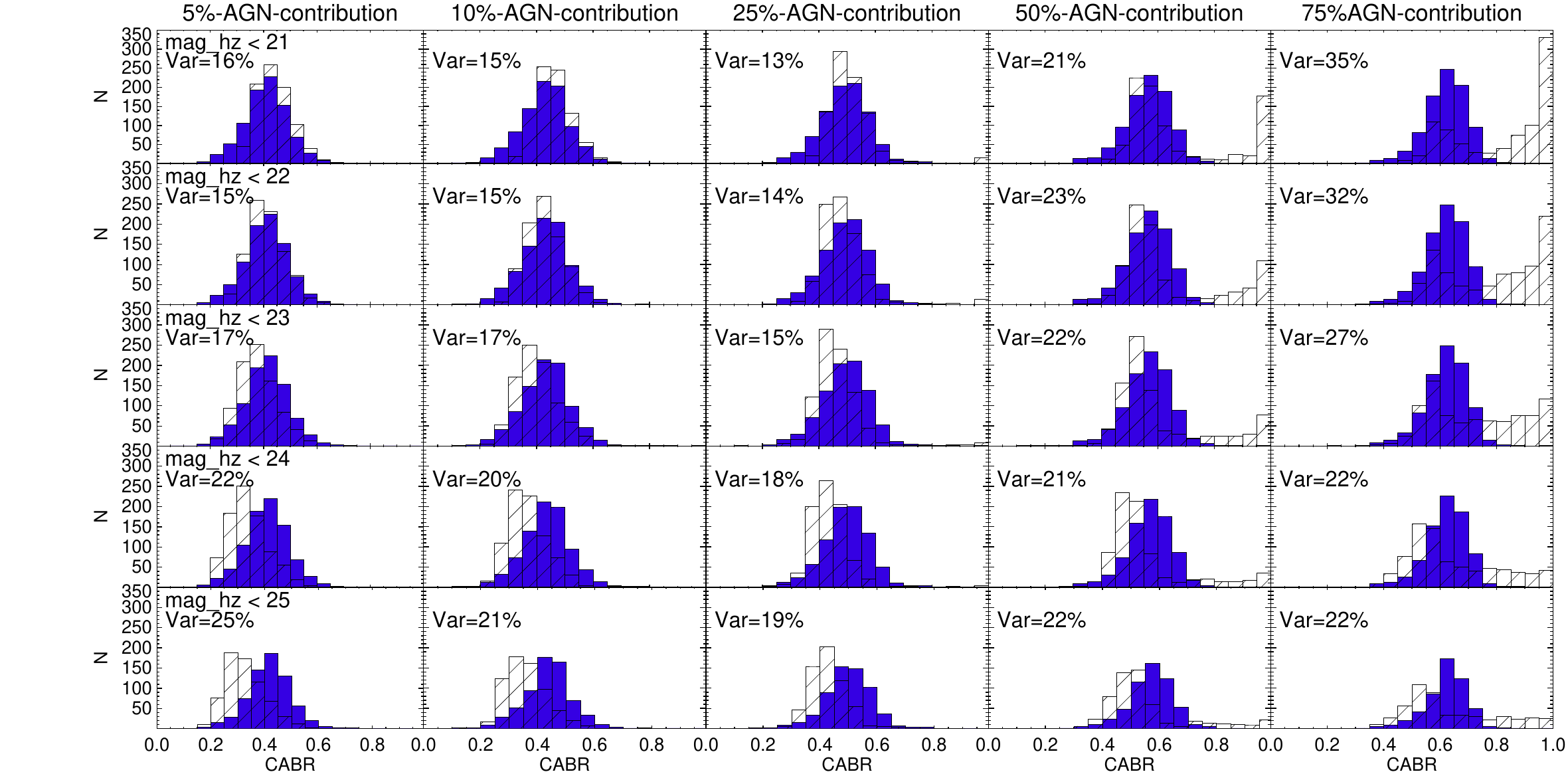}}
				{\includegraphics[height=1.9in, width=5.54in]{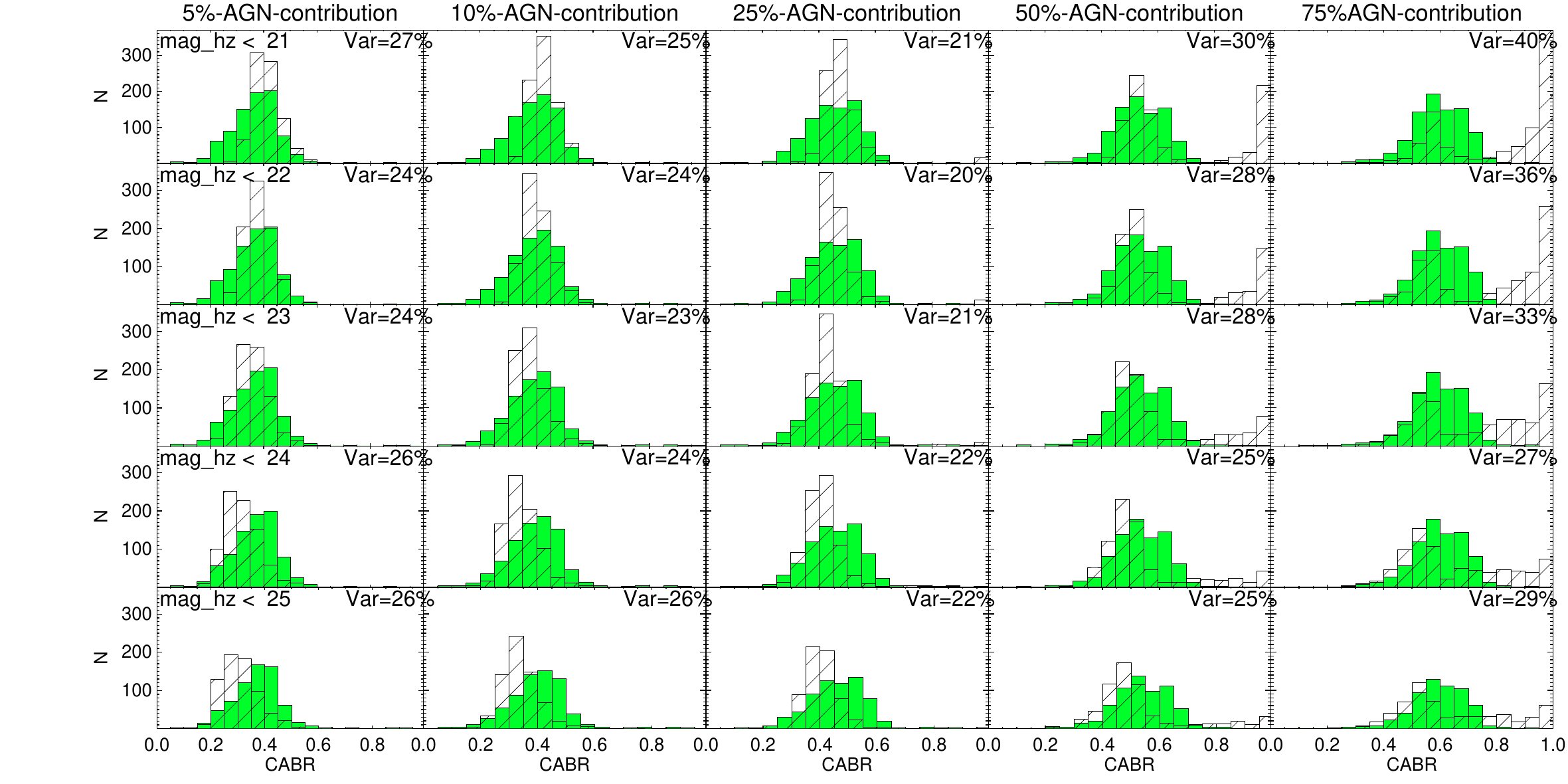}}
				\caption{The comparison of the CABR parameter measured after adding the AGN (from 5\%\,-\,75\%, columns 2 to 5, respectively) to the images of the original sample at z\,$\sim$ 0 (filled histograms), and to the sim ulated images (open histograms) after taking into account higher redshifts and 5 different magnitude cuts from 21 to 25 (first to the fifth row, respectively). From top to bottom, the distributions of the total sample (yellow), early-type galaxies (red), early-spirals (blue), and late-spirals (green) are shown, respectively.}
				\label{fig4.1:part2}
			\end{center}
		\end{figure}
		
		\begin{figure*}
			\begin{center}
				\includegraphics[height= 2.2in, width=5.54in]{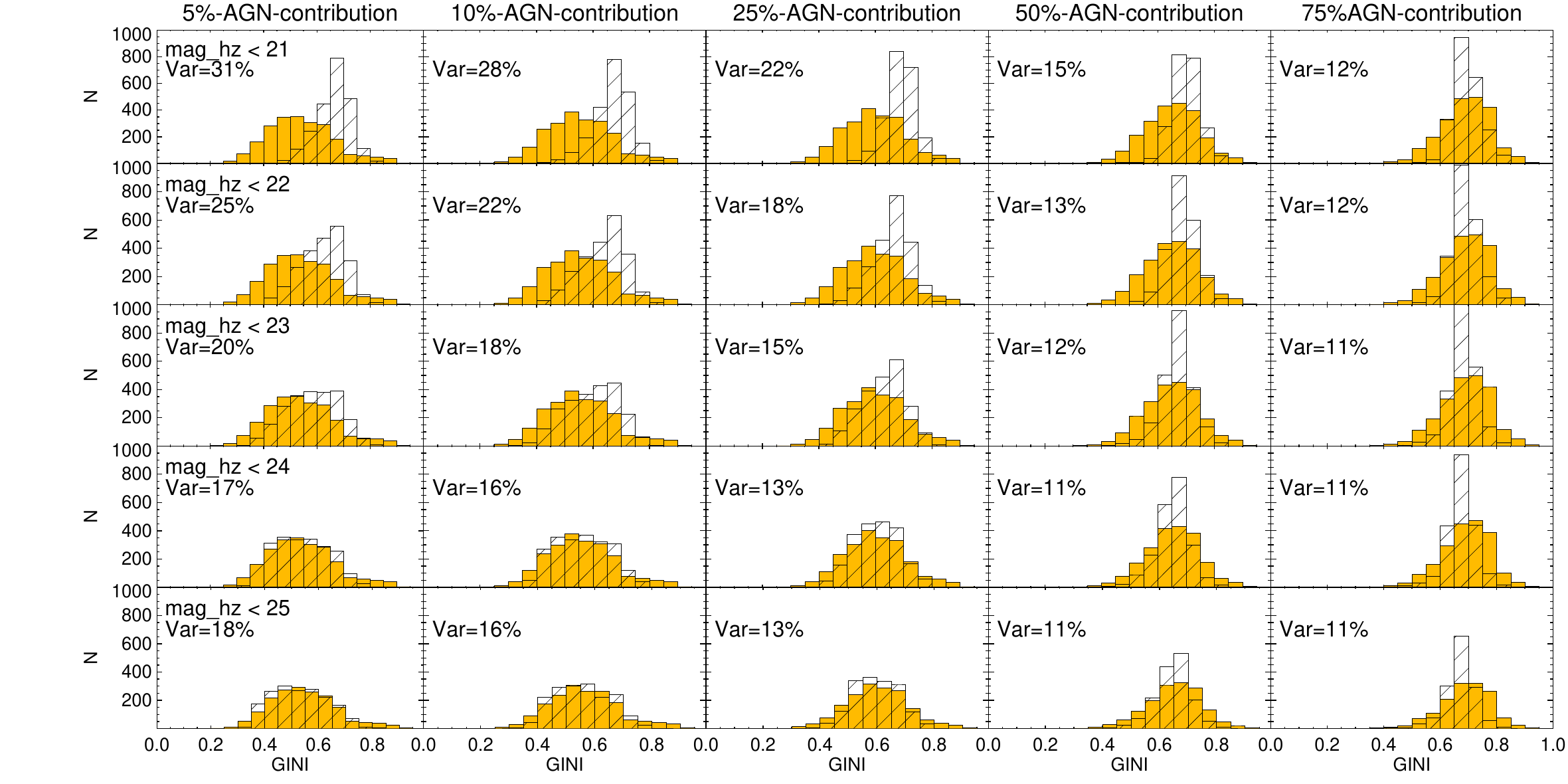}
				\includegraphics[height= 2.2in, width=5.54in]{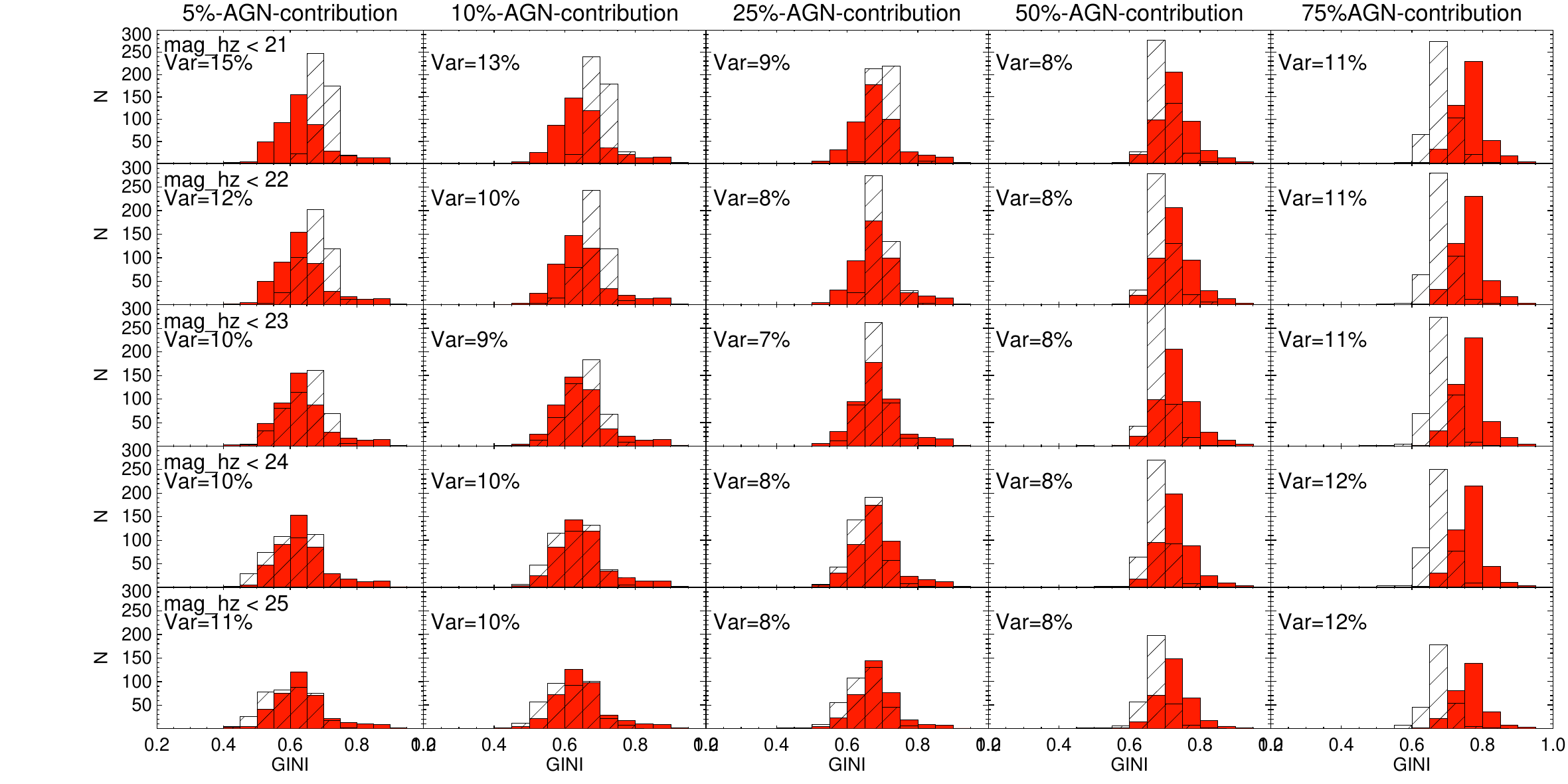}
				{\includegraphics[height= 2.2in, width=5.54in]{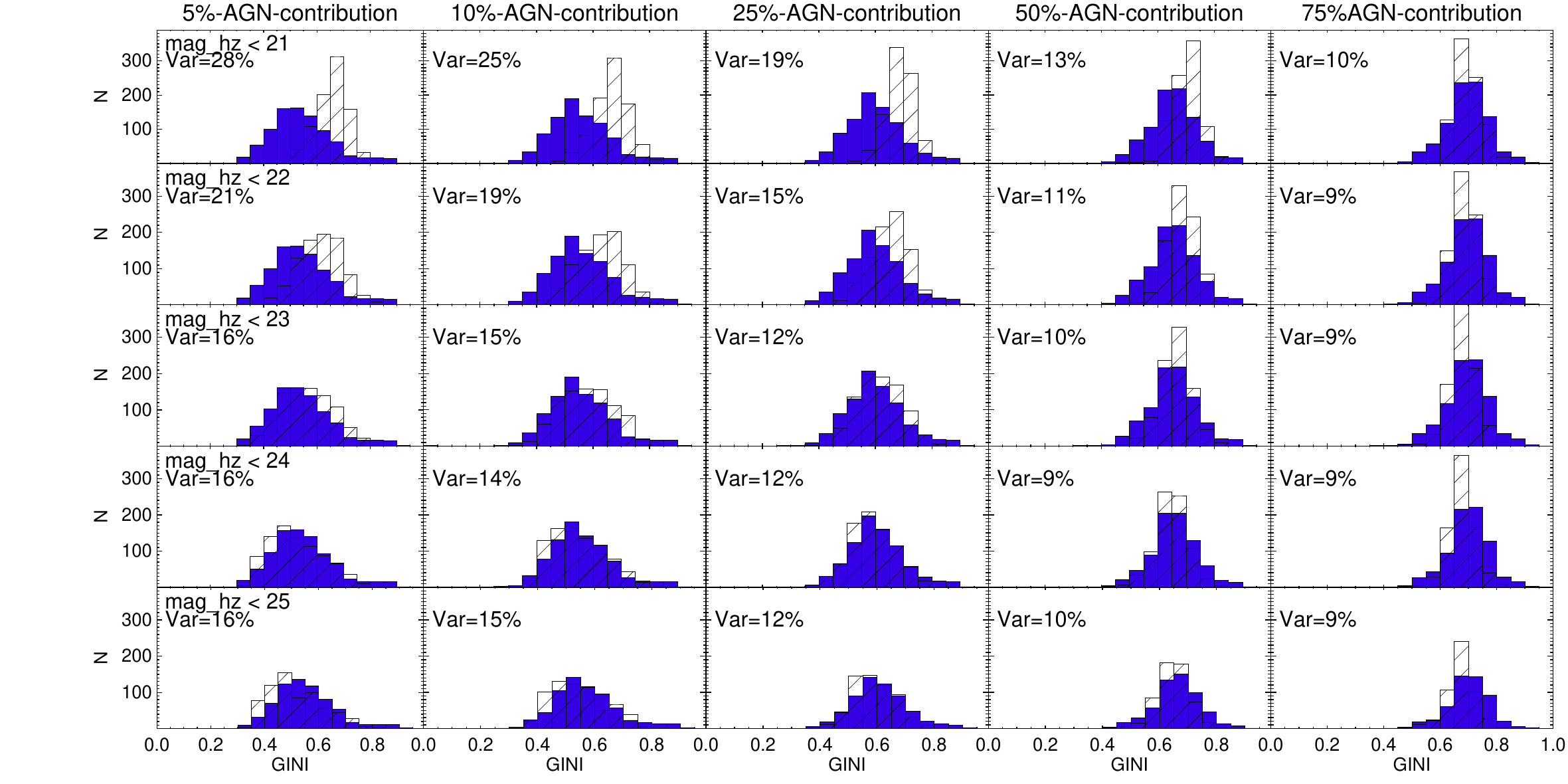}}
				{\includegraphics[height= 2.2in, width=5.54in]{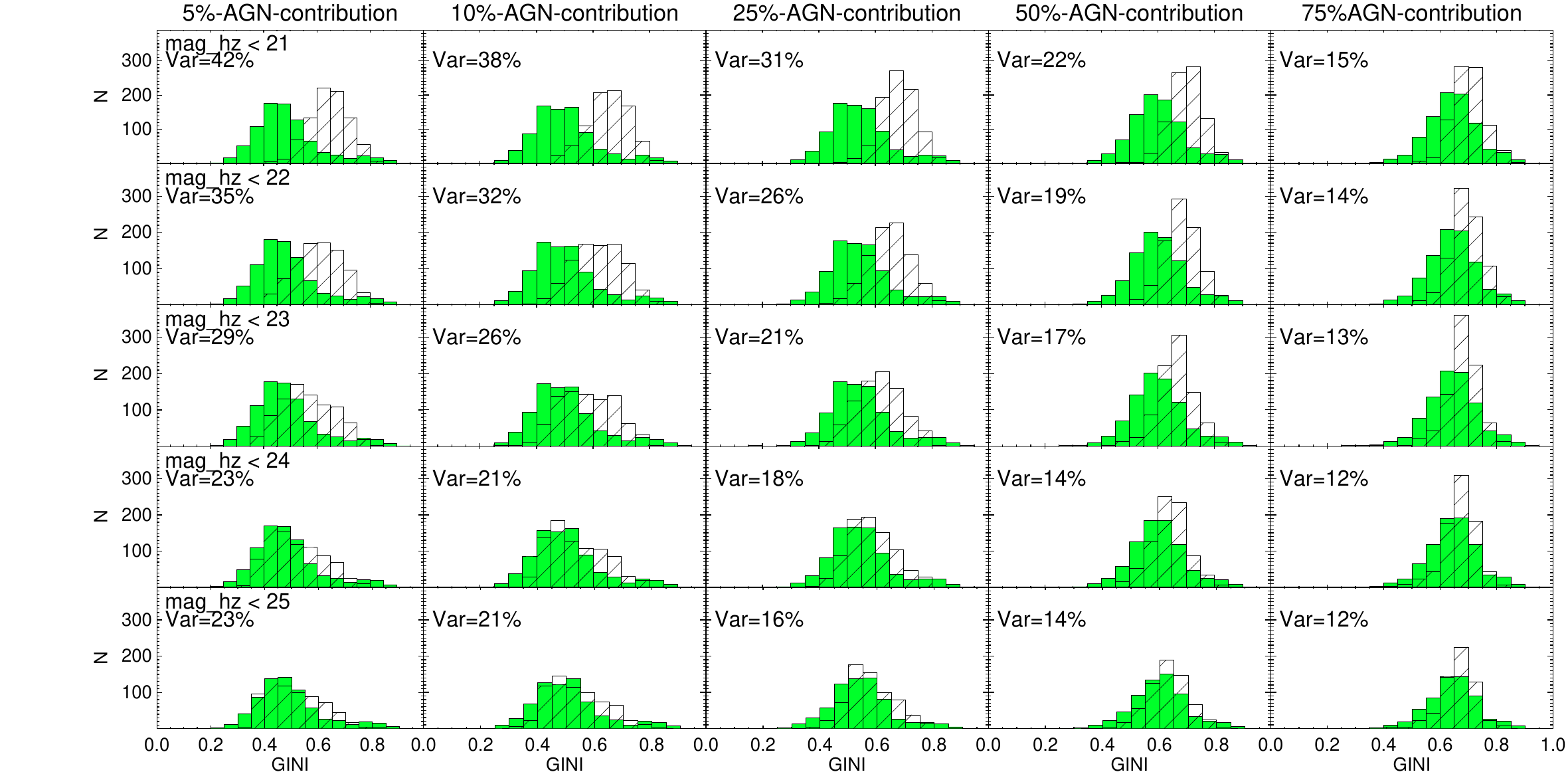}}
				\caption{Same as Figure \ref{fig4.1:part2}, but for the GINI parameter.}
				\label{fig4.2:part2}
			\end{center}
		\end{figure*}

		\begin{figure*}
			\begin{center}
				\includegraphics[height= 2.2in, width=5.54in]{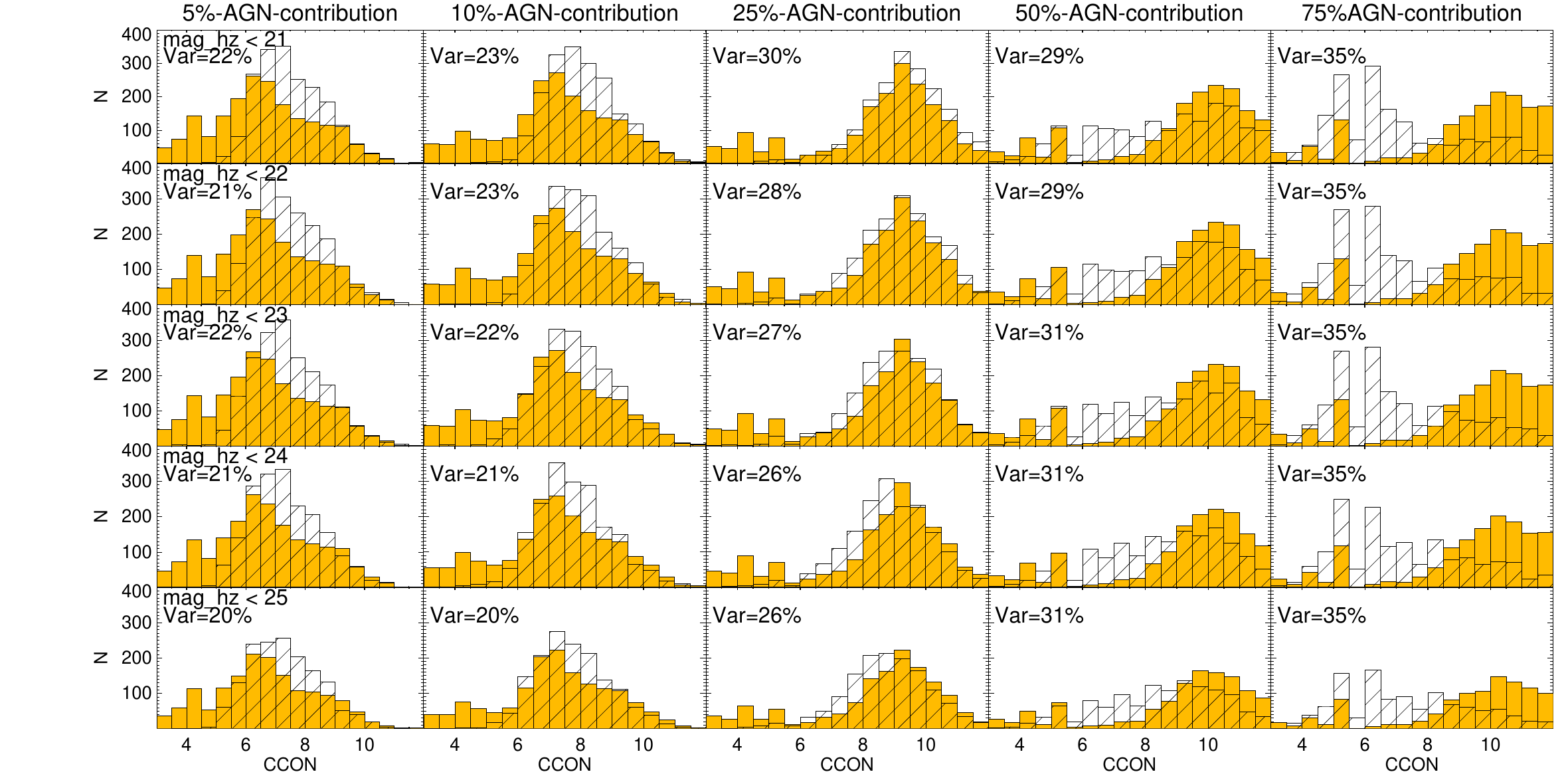}
				\includegraphics[height= 2.2in, width=5.54in]{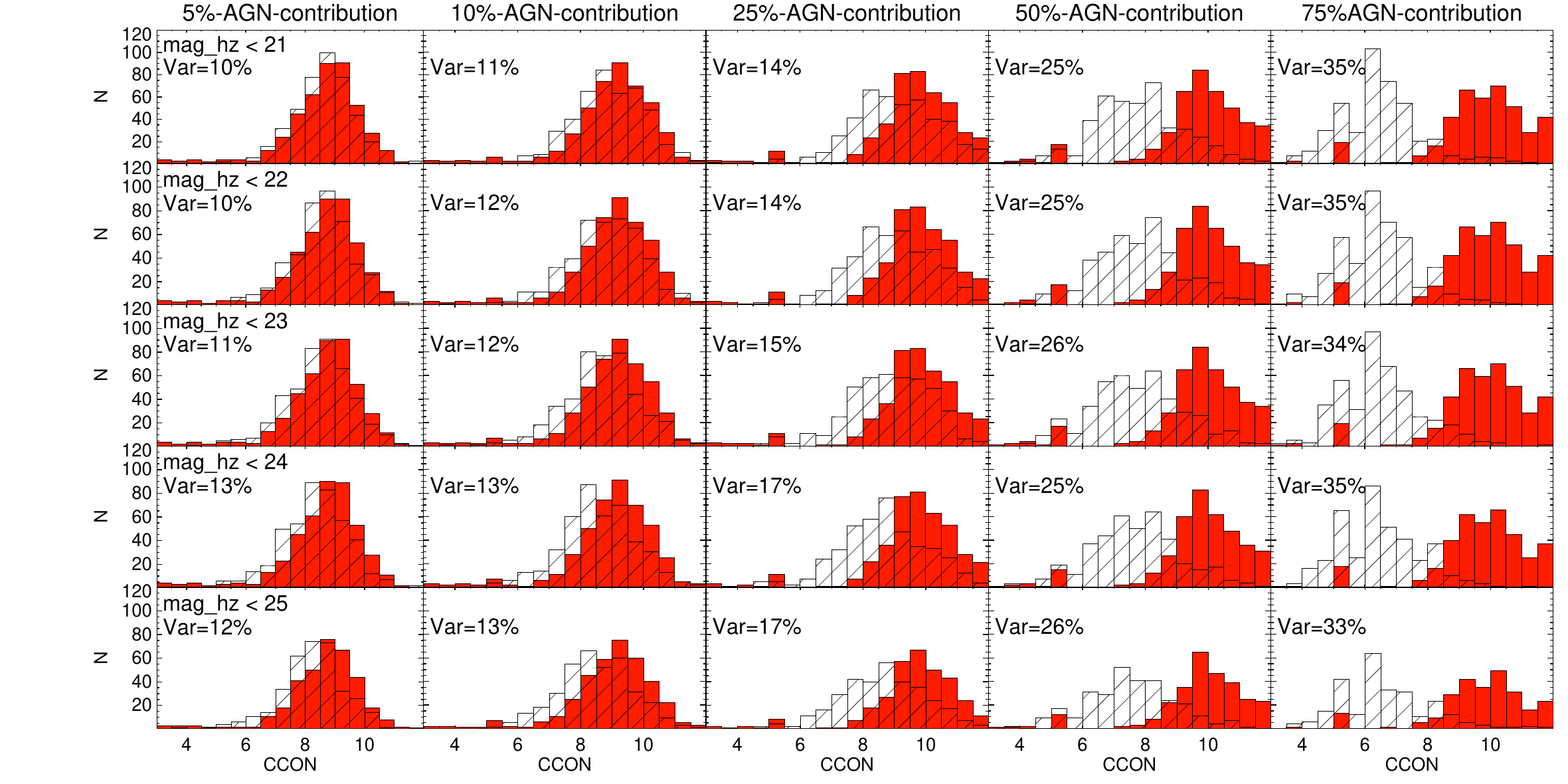}
				\includegraphics[height=  2.2in, width=5.54in]{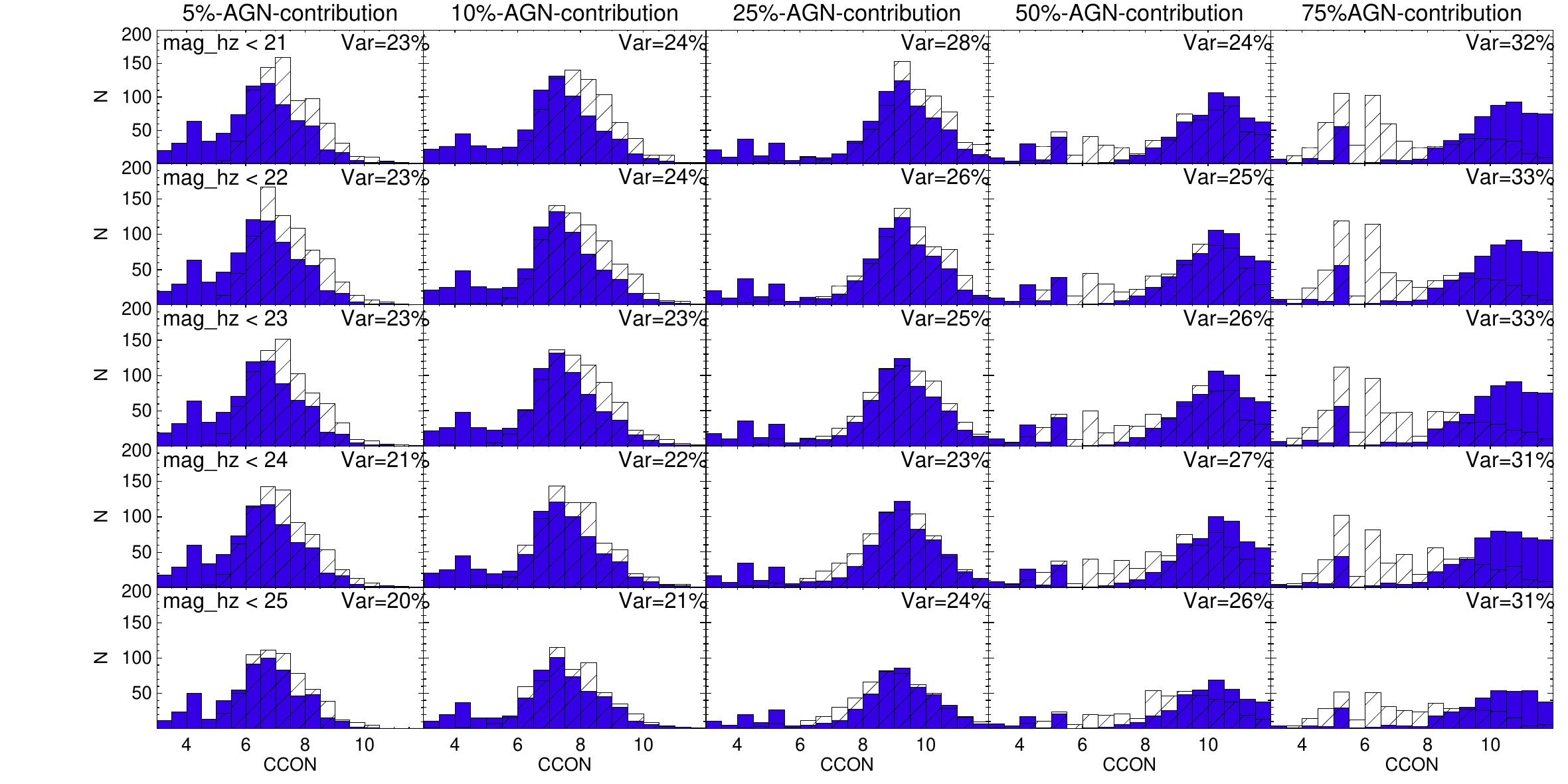}
				{\includegraphics[height= 2.2in, width=5.54in]{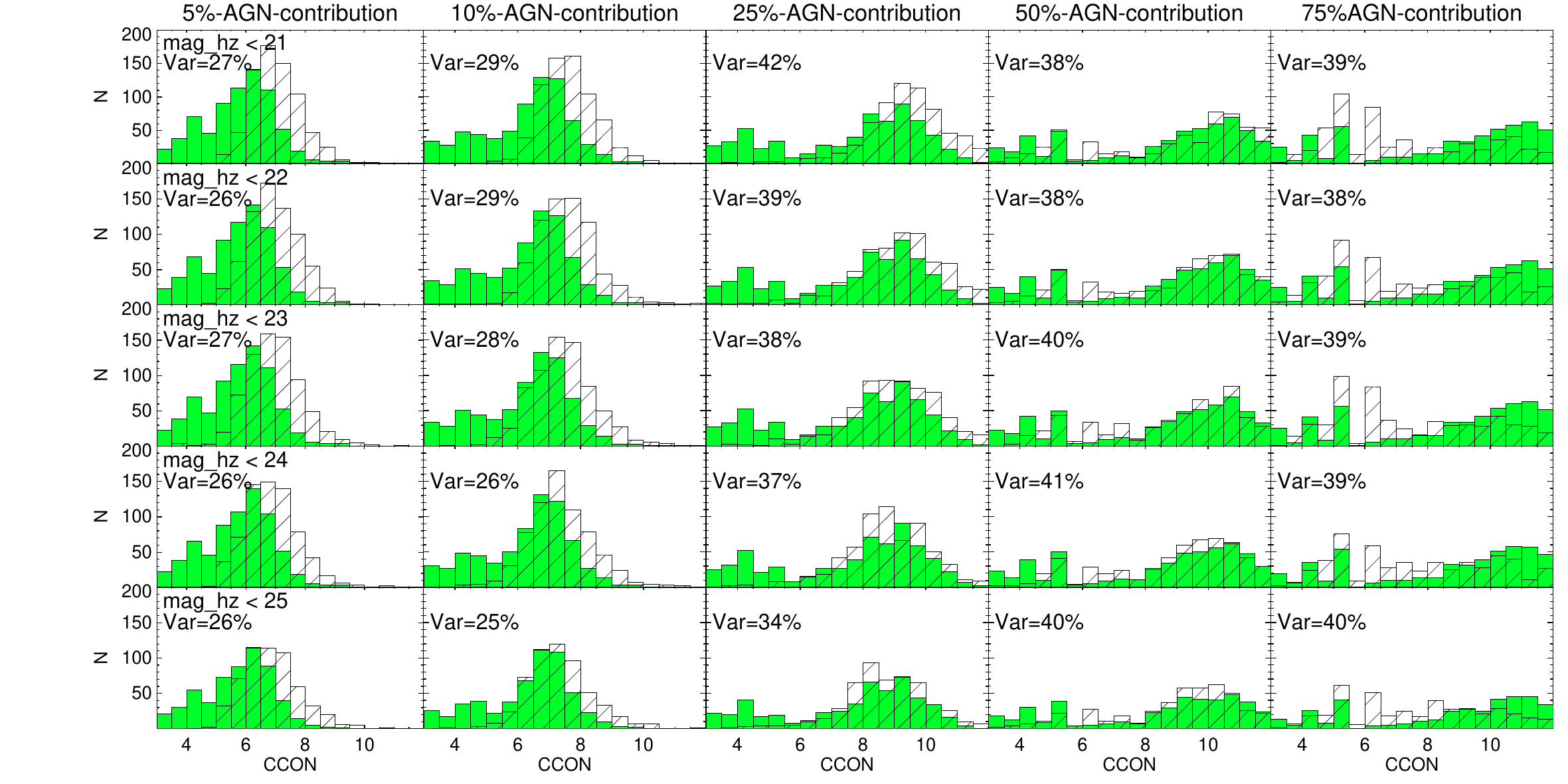}}
				\caption{Same as Figure \ref{fig4.1:part2}, but for the CCON parameter.}
				\label{fig4.3:part2}
			\end{center}
		\end{figure*}
		
		\begin{figure*}
			\begin{center}
				{\includegraphics[height= 2.2in, width=5.54in]{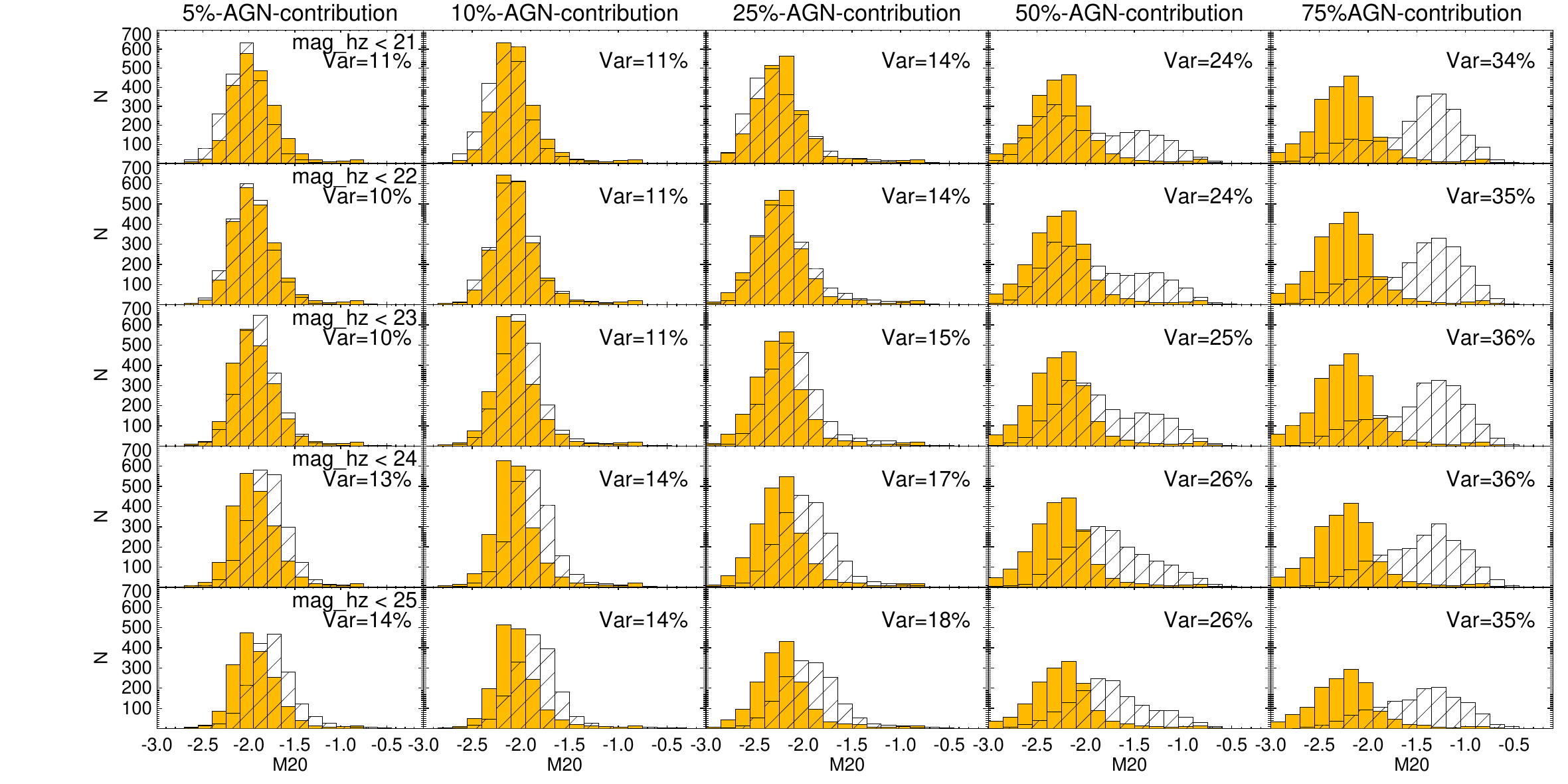}}
				{\includegraphics[height= 2.2in, width=5.54in]{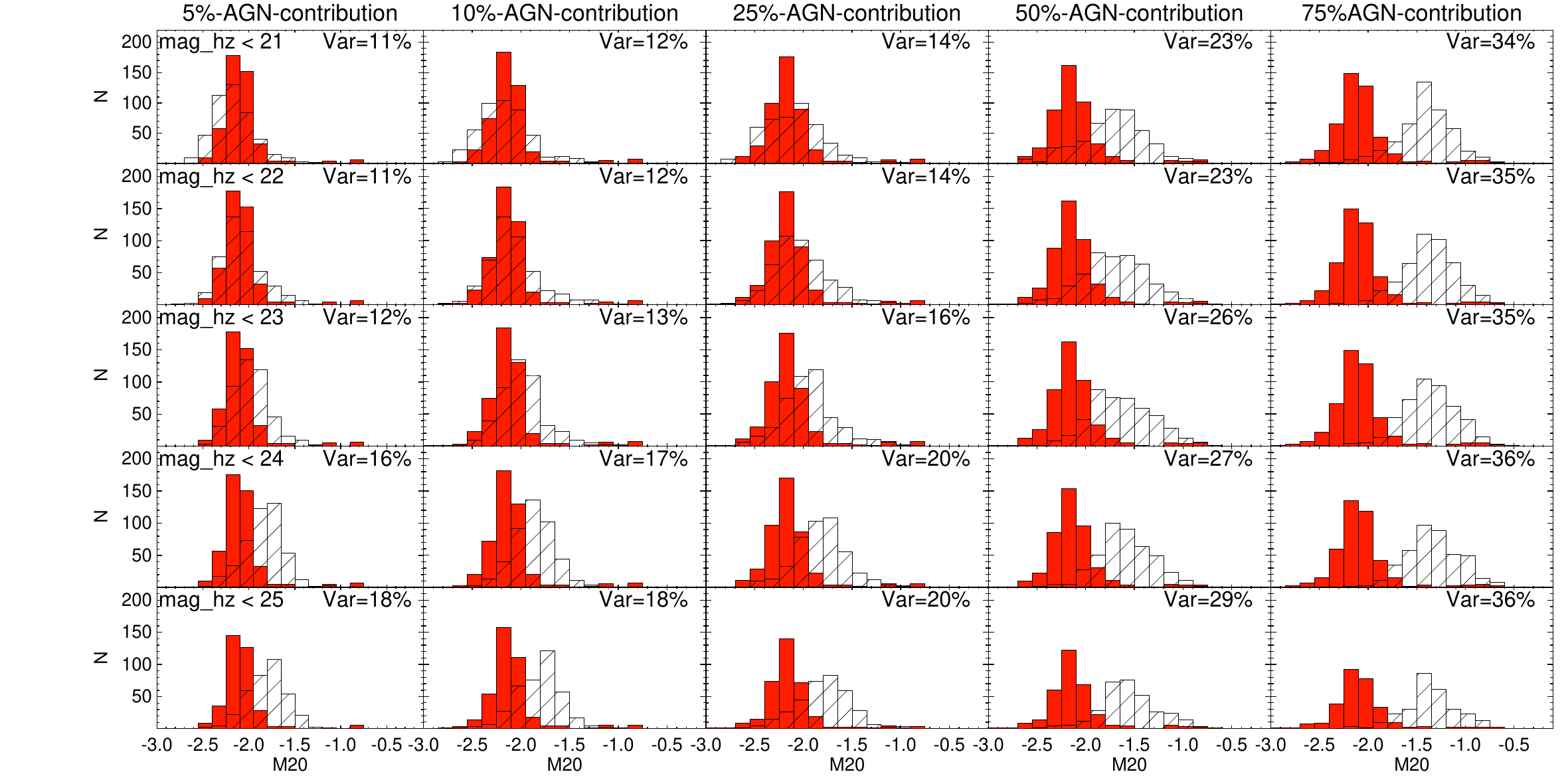}}
				{\includegraphics[height= 2.2in, width=5.54in]{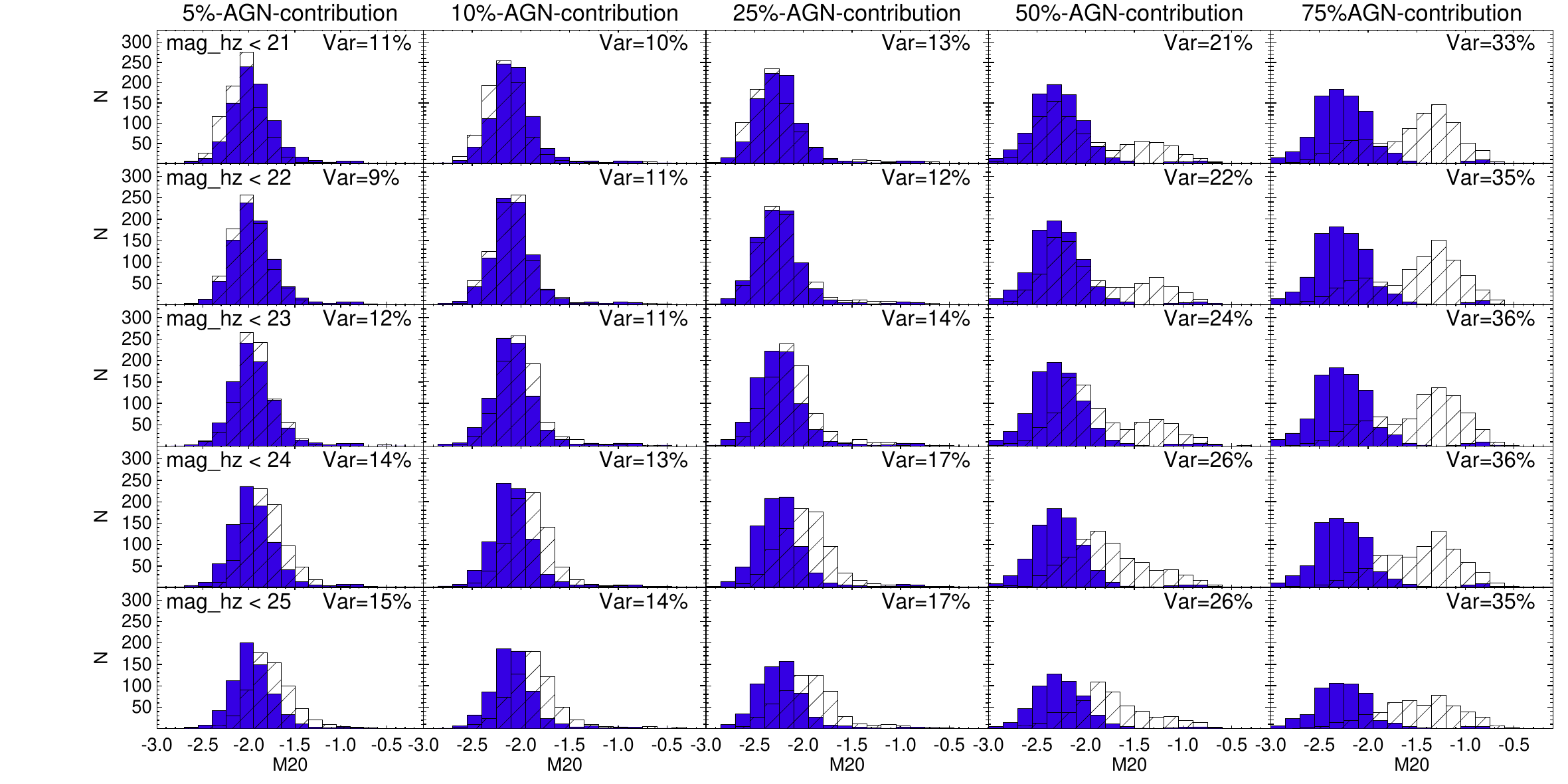}}
				{\includegraphics[height= 2.2in, width=5.54in]{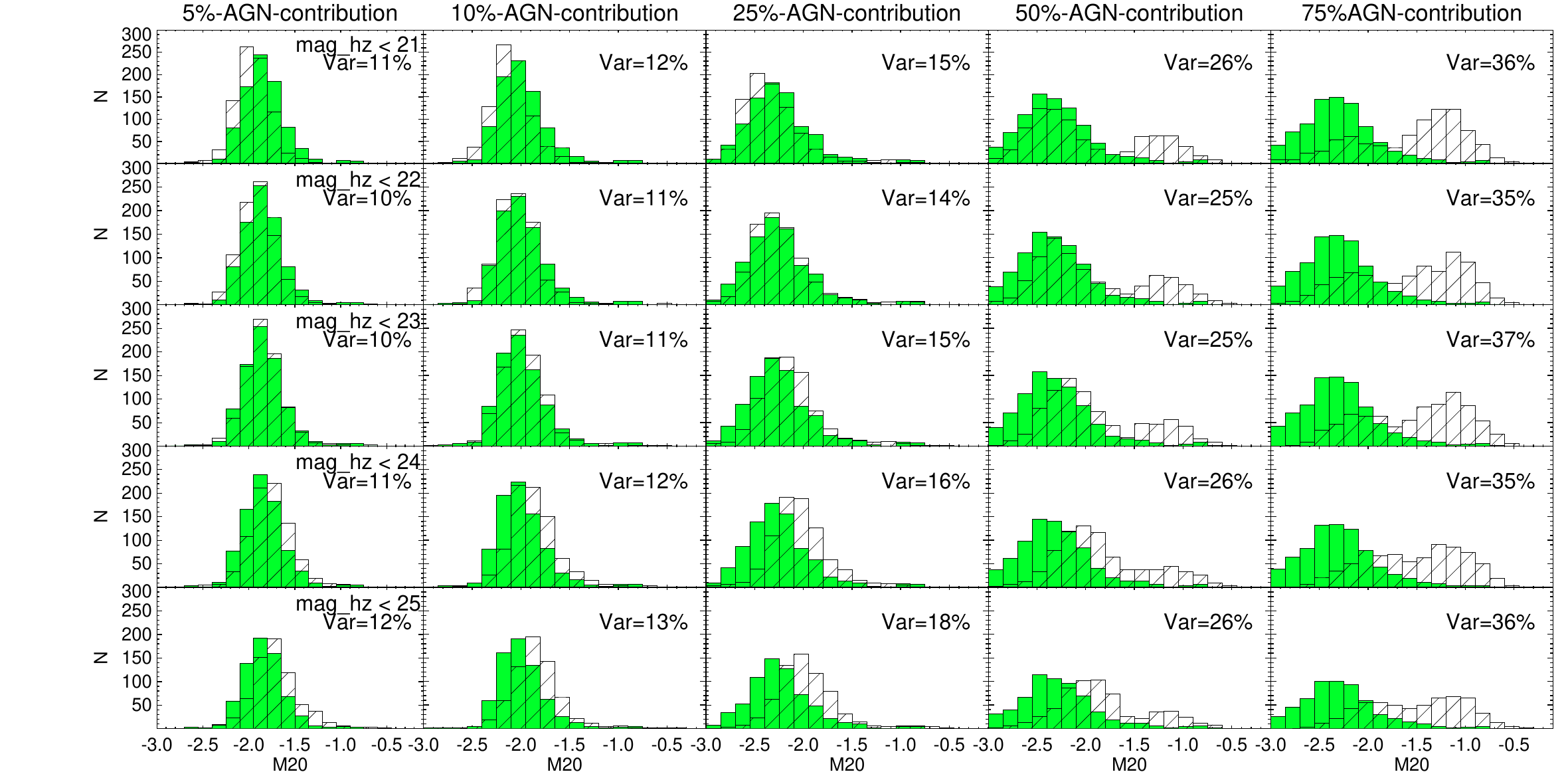}}
				\caption{Same as Figure \ref{fig4.1:part2}, but for the M20 moment of light.}
				\label{fig4.4:part2}
			\end{center}
		\end{figure*}
		
		\begin{figure*}
			\begin{center}
				{\includegraphics[height= 2.2in, width=5.54in]{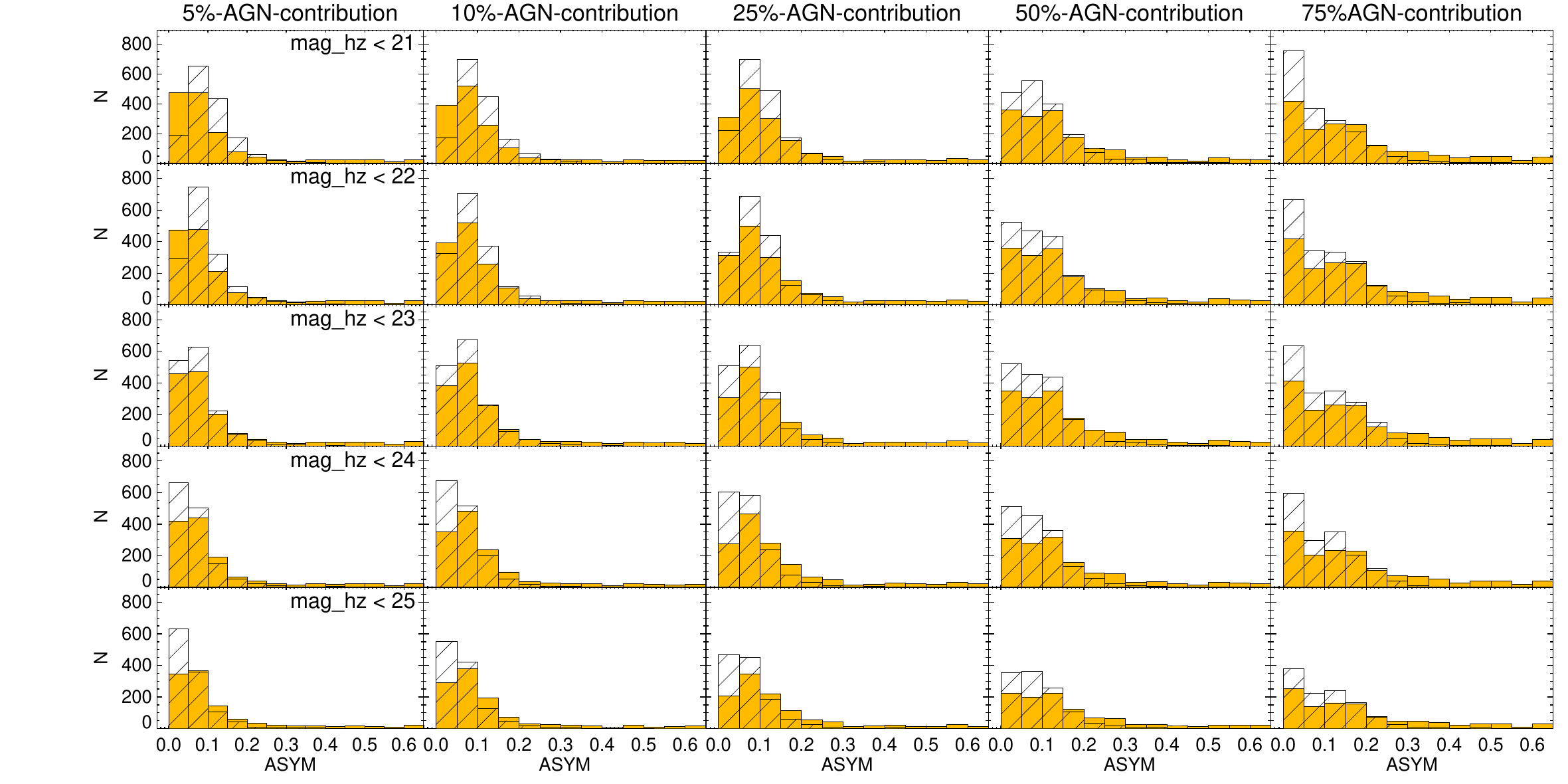}}
				{\includegraphics[height= 2.2in, width=5.54in]{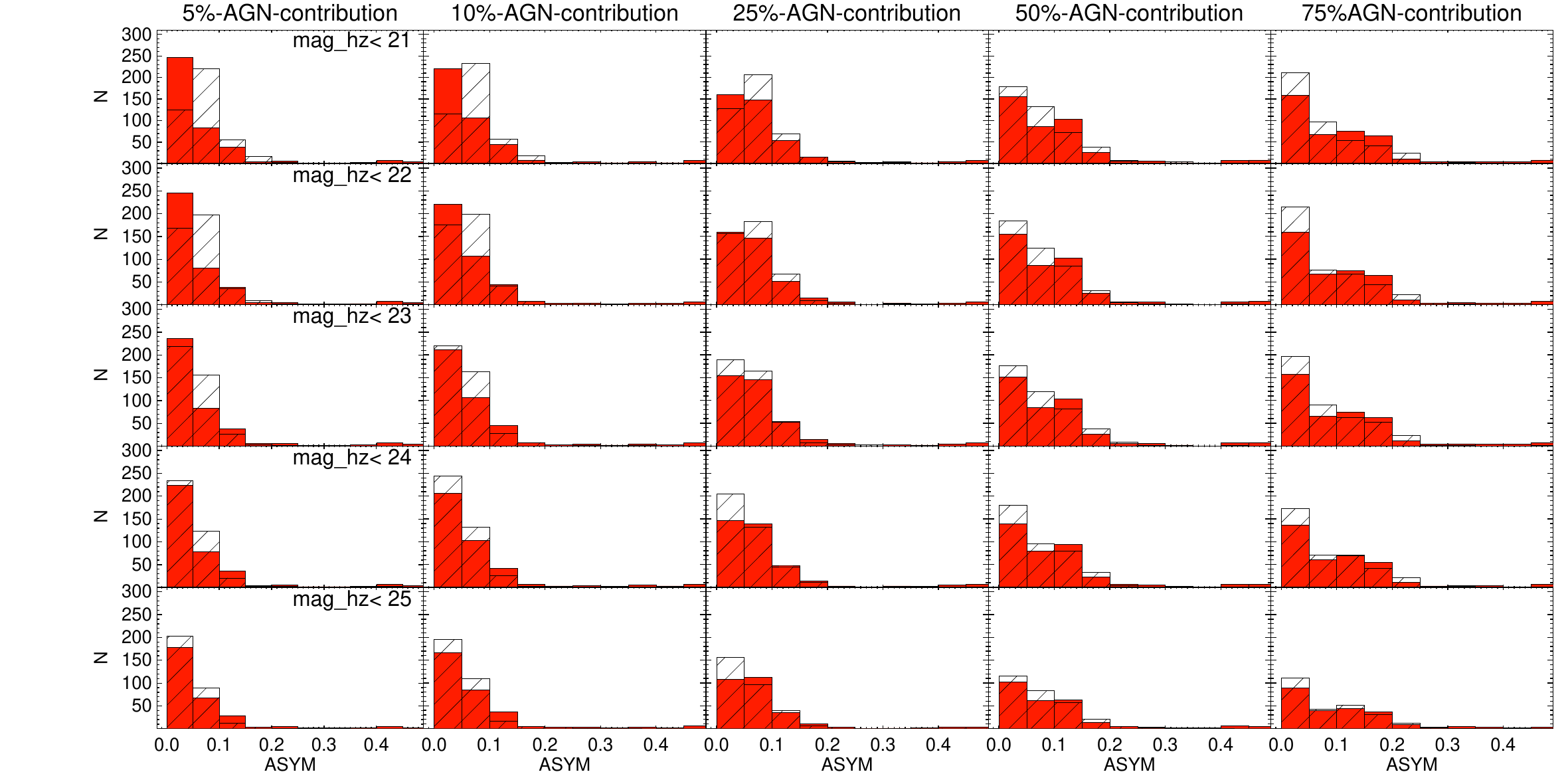}}
				{\includegraphics[height= 2.2in, width=5.54in]{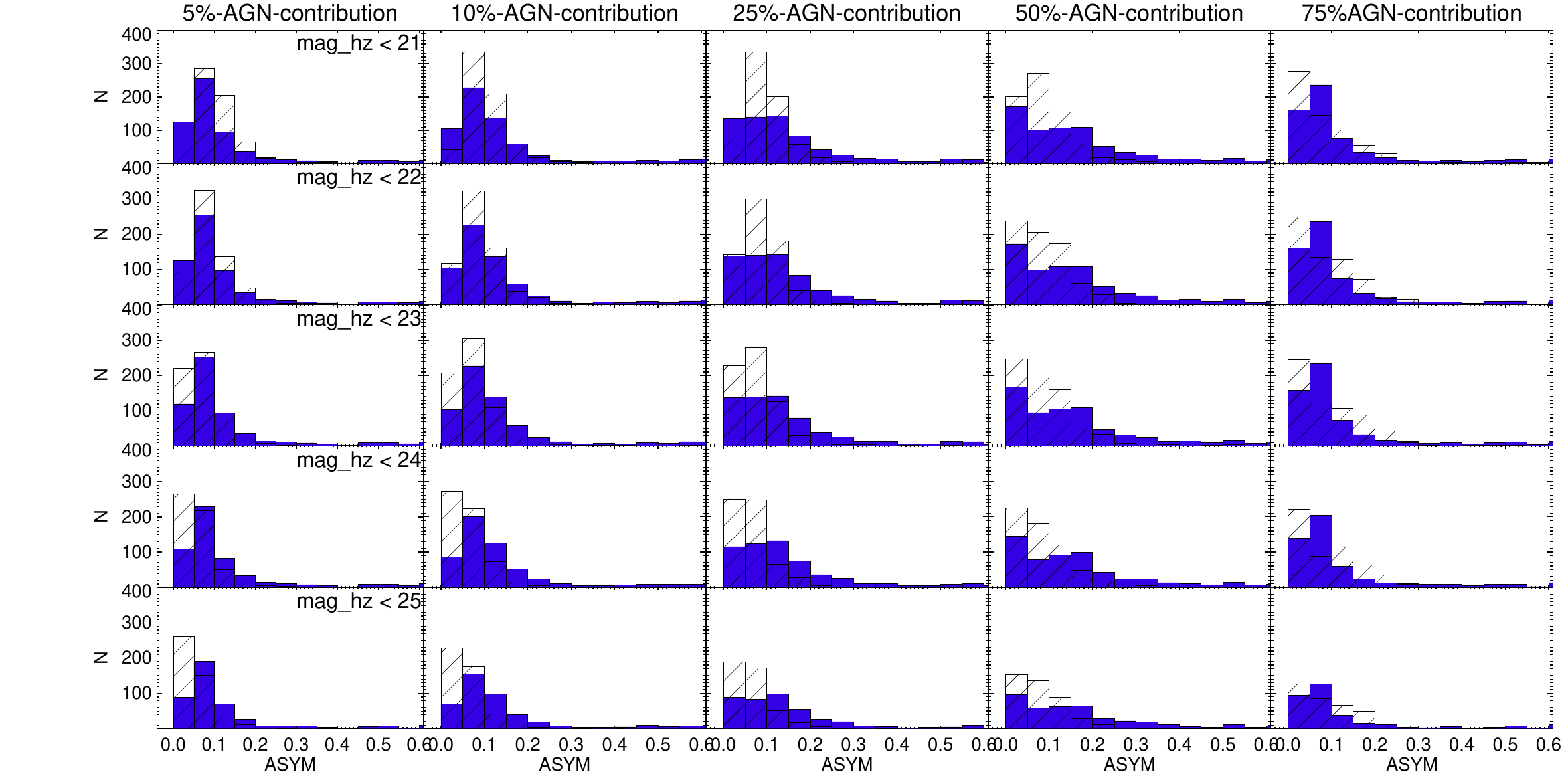}}
				{\includegraphics[height= 2.2in, width=5.54in]{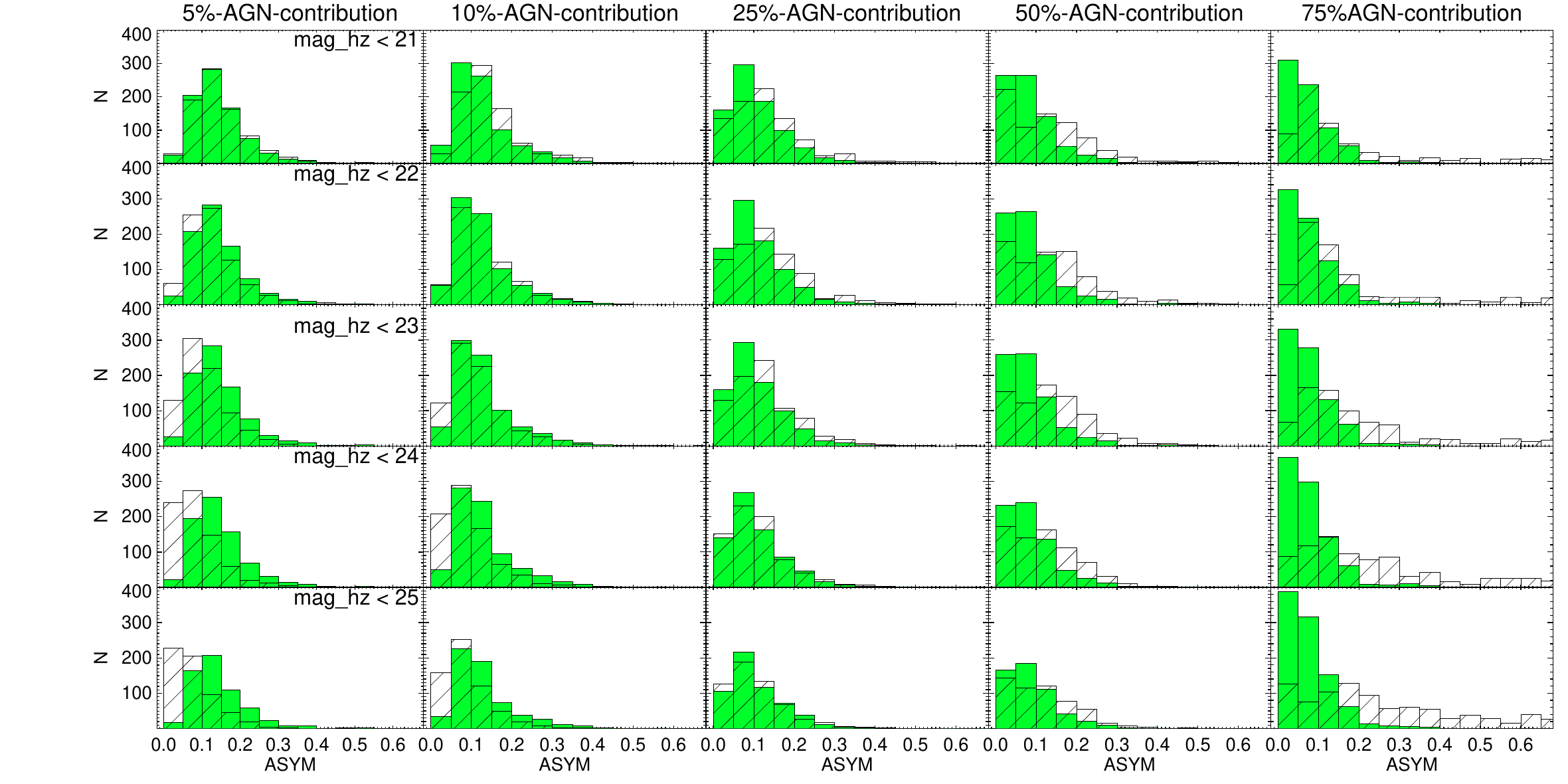}}
				\caption{Same as Figure \ref{fig4.1:part2}, but for the ASYM parameter.}
				\label{fig4.5:part2}
			\end{center}
		\end{figure*}
		
		\begin{figure*}
			\begin{center}
				{\includegraphics[height= 2.2in, width=5.54in]{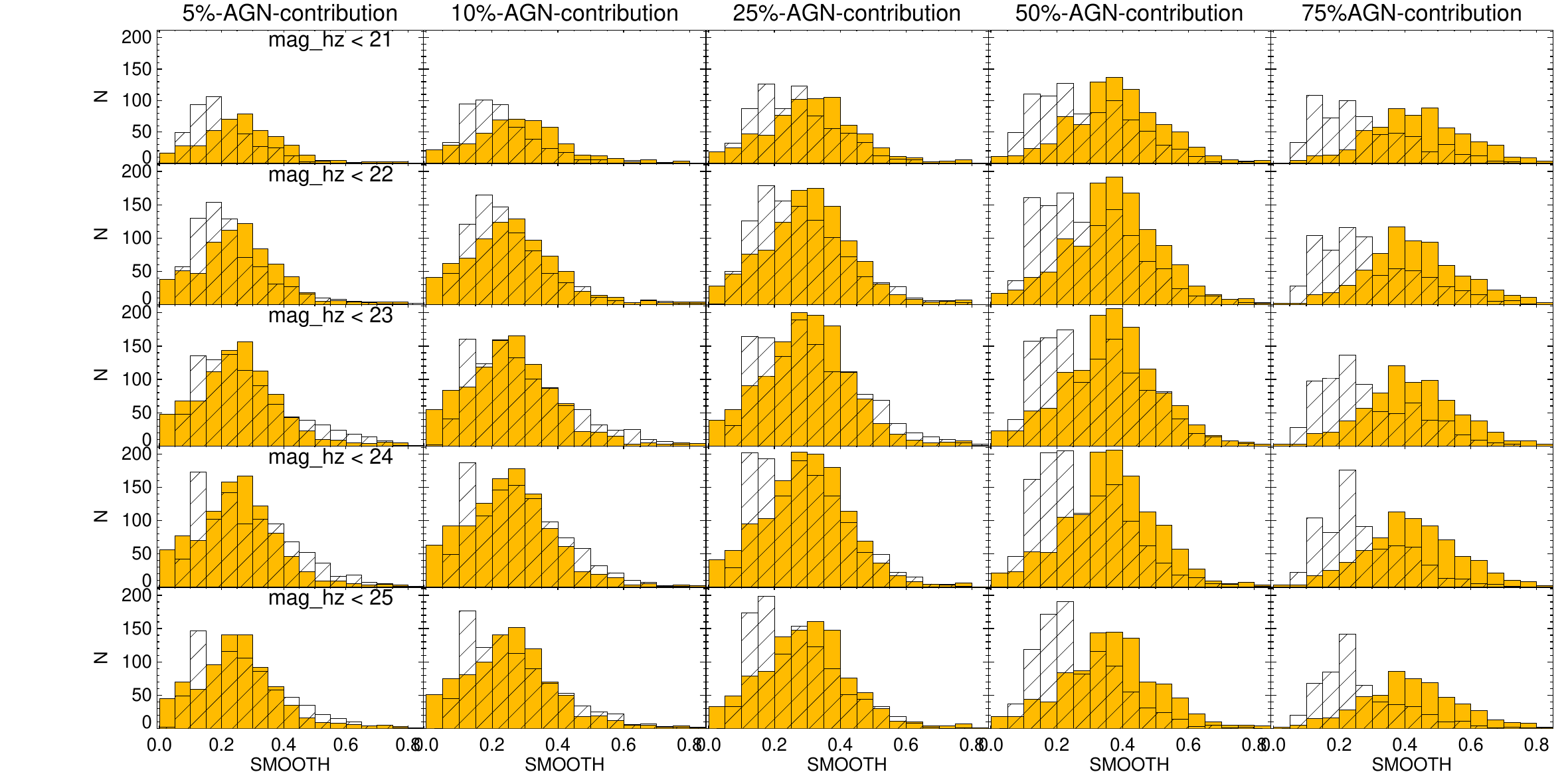}}
				{\includegraphics[height= 2.2in, width=5.54in]{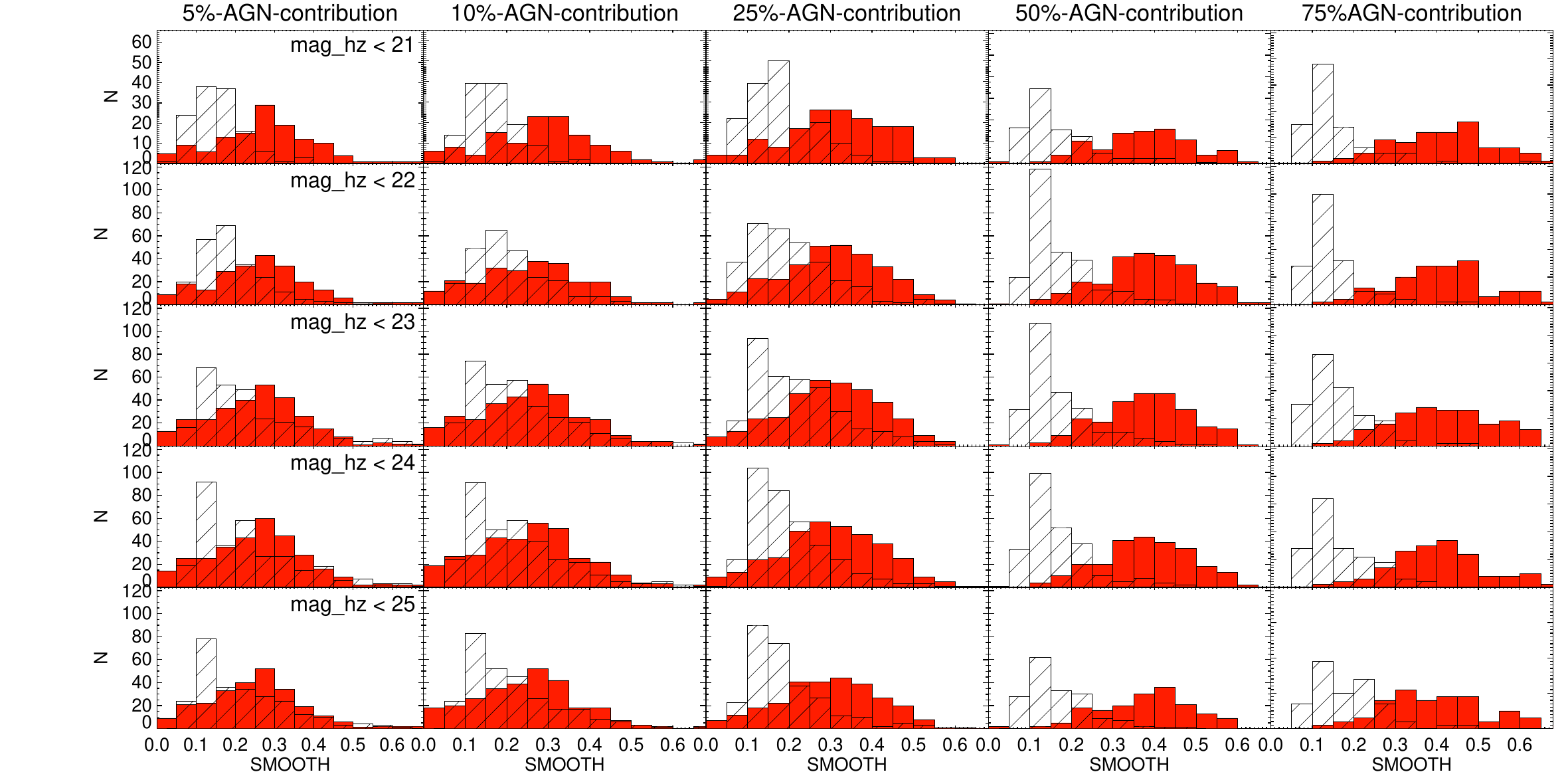}}
				{\includegraphics[height= 2.2in, width=5.54in]{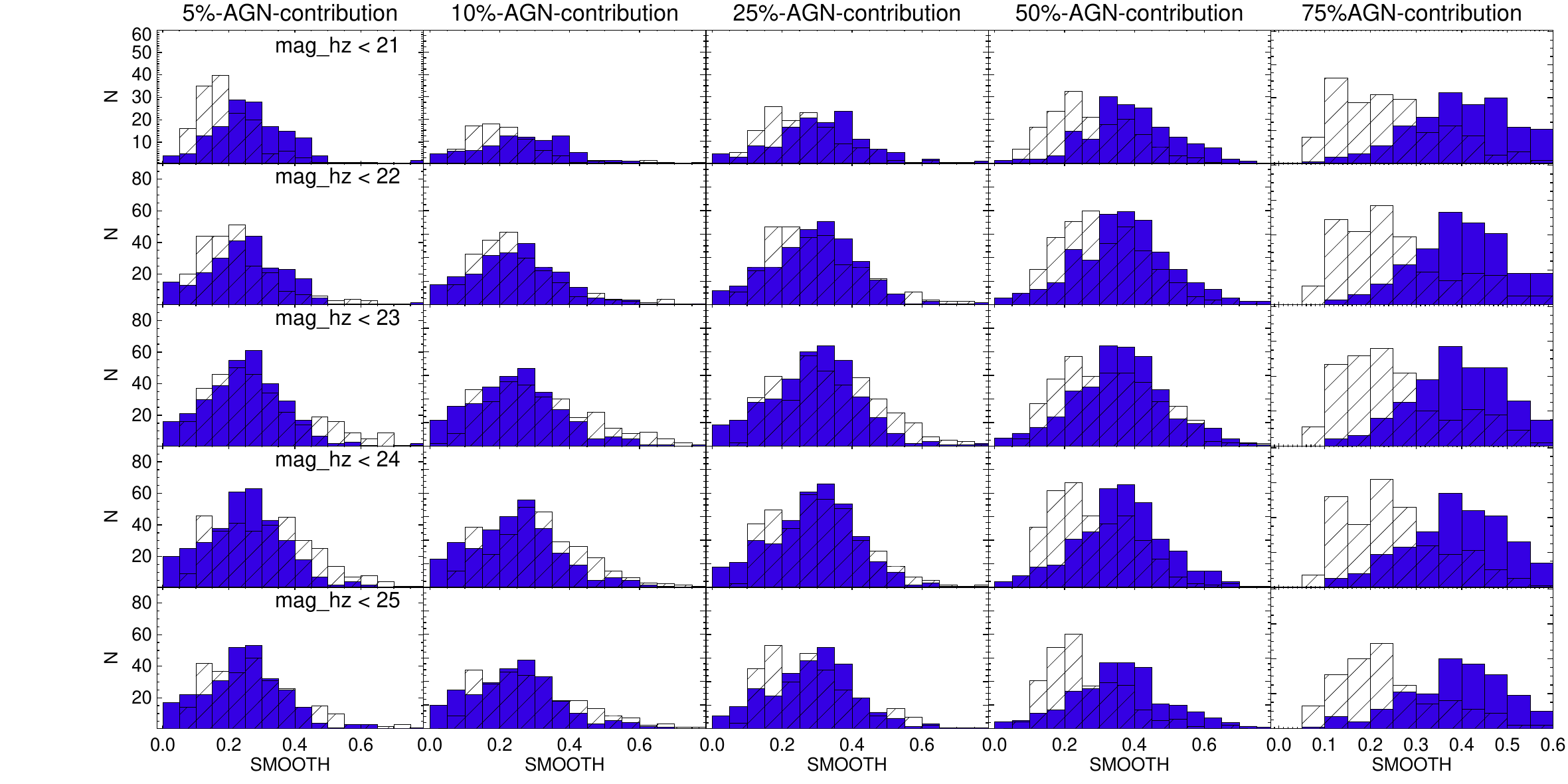}}
				{\includegraphics[height= 2.2in, width=5.54in]{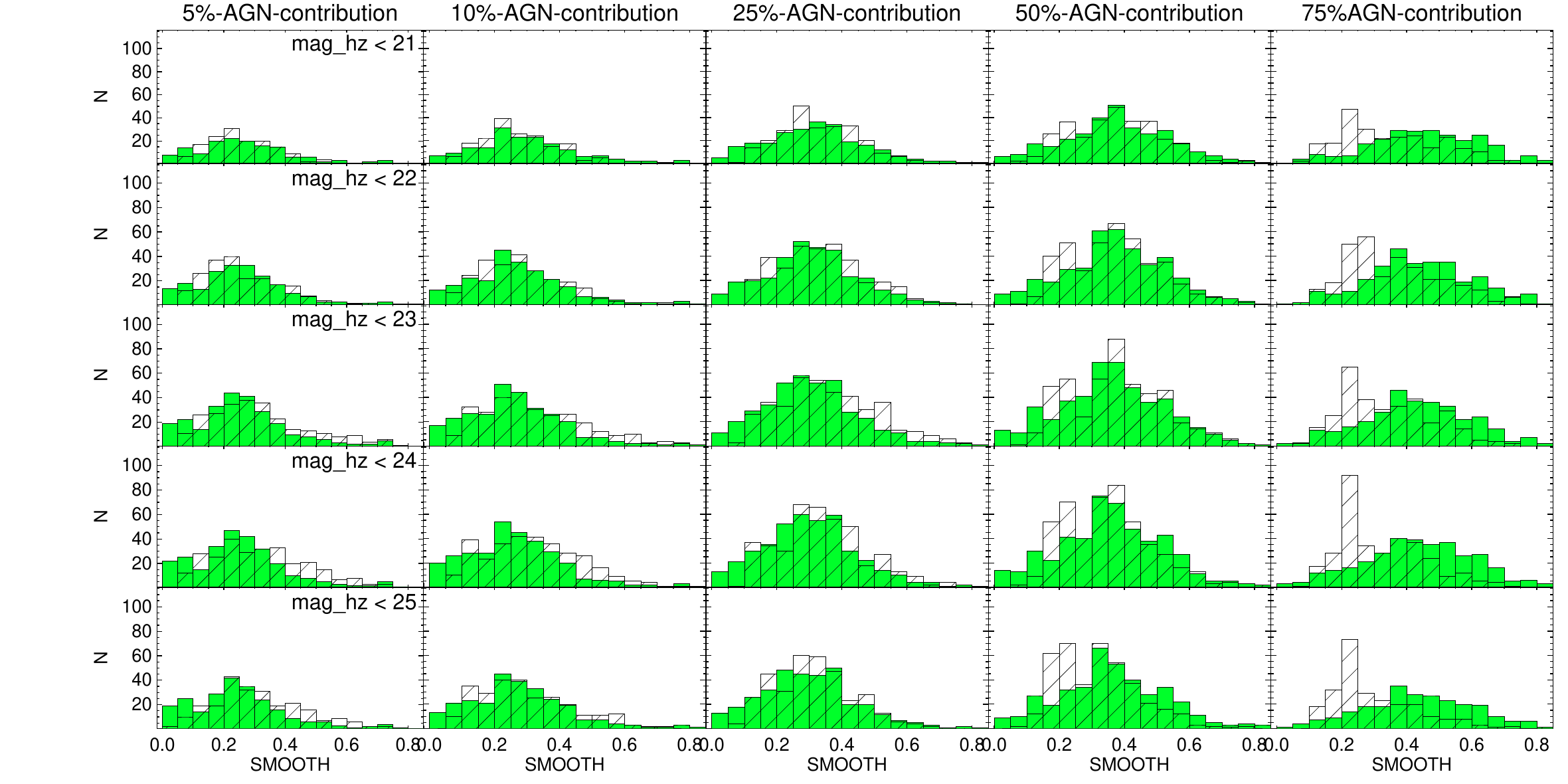}}
				\caption{Same as Figure \ref{fig4.1:part2}, but for the SMOOTHness parameter.}
				\label{fig4.6:part2}
			\end{center}
		\end{figure*}
		
		\clearpage
		\onecolumn
		\renewcommand{\thesection}{C}
		\section{: Comparing the COSMOS-like sample with and without AGN}
		\label{appendixC}
		\begin{figure}[h!]
			\begin{center}
				{\includegraphics[height=1.9in, width=5.54in]{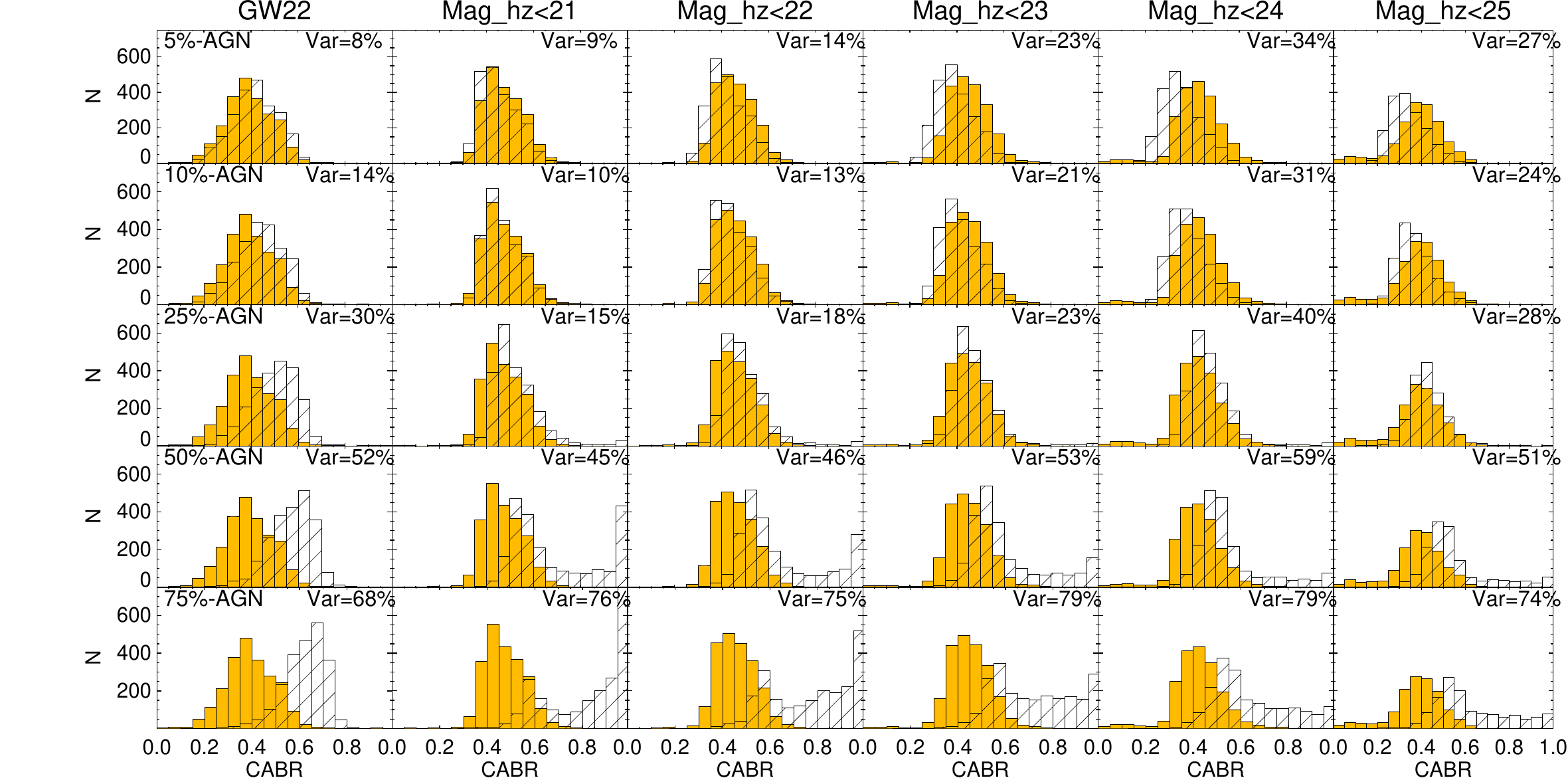}}
				{\includegraphics[height=1.9in, width=5.54in]{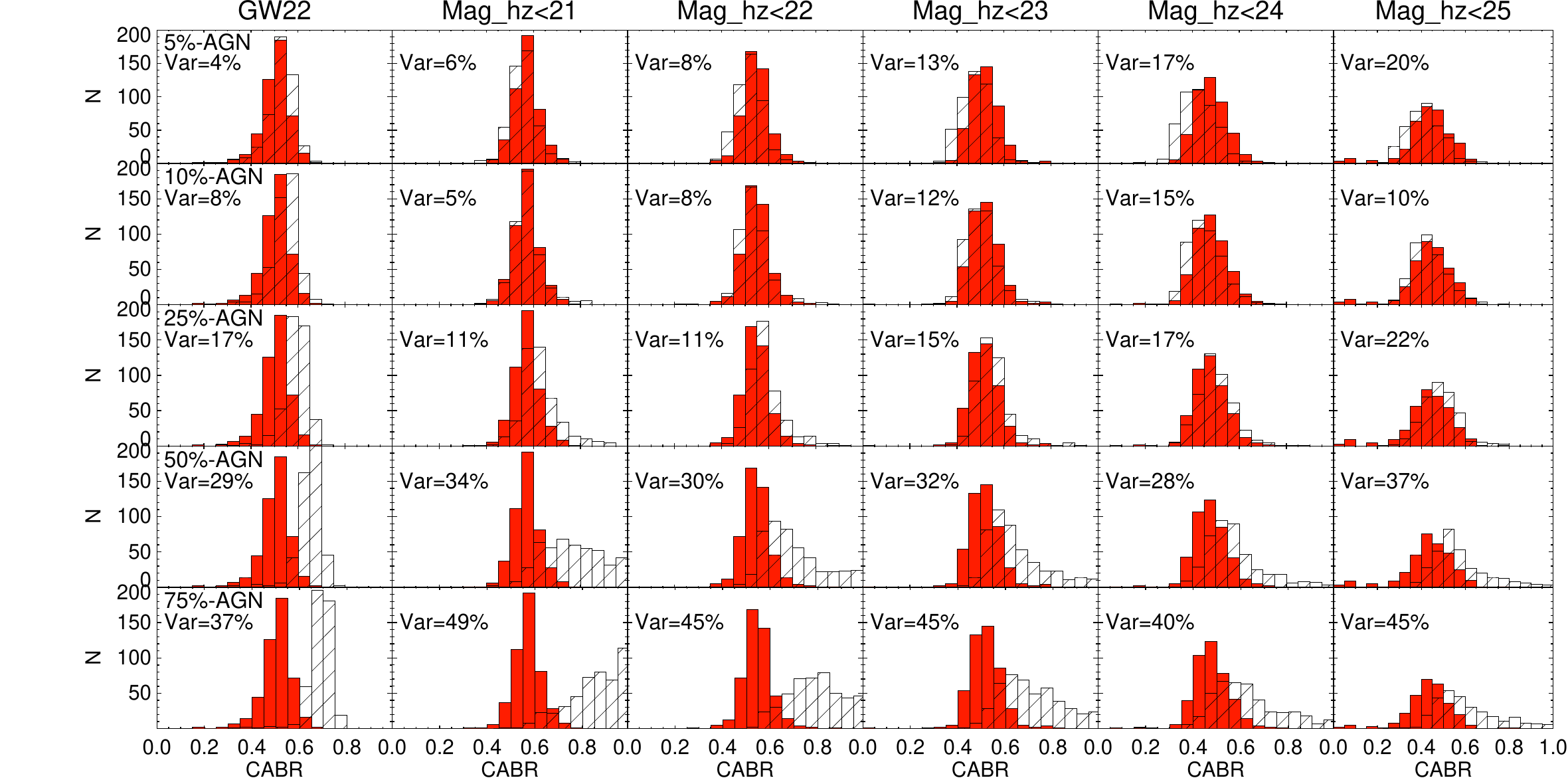}}
				{\includegraphics[height=1.9in, width=5.54in]{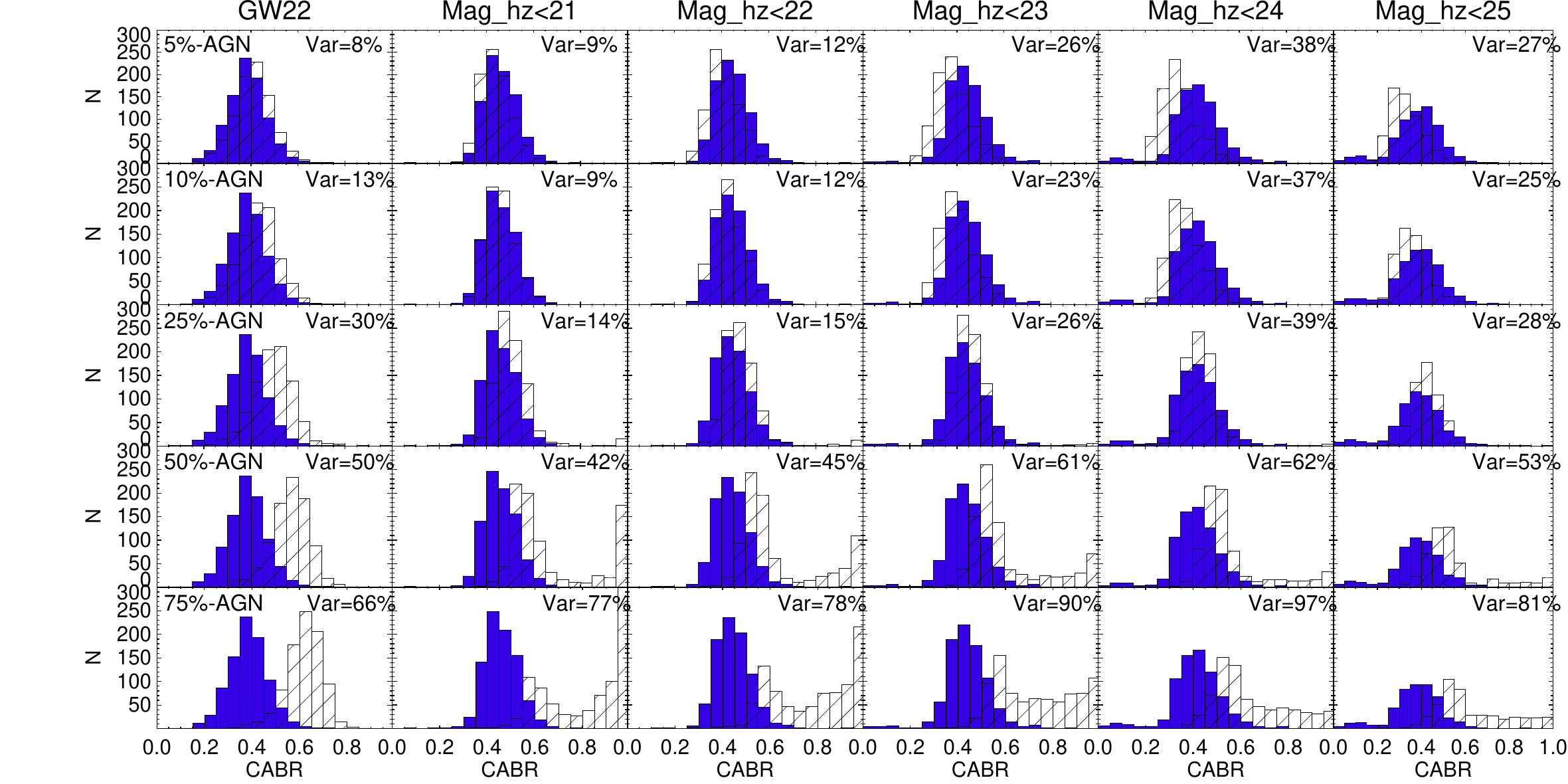}}
				{\includegraphics[height=1.9in, width=5.54in]{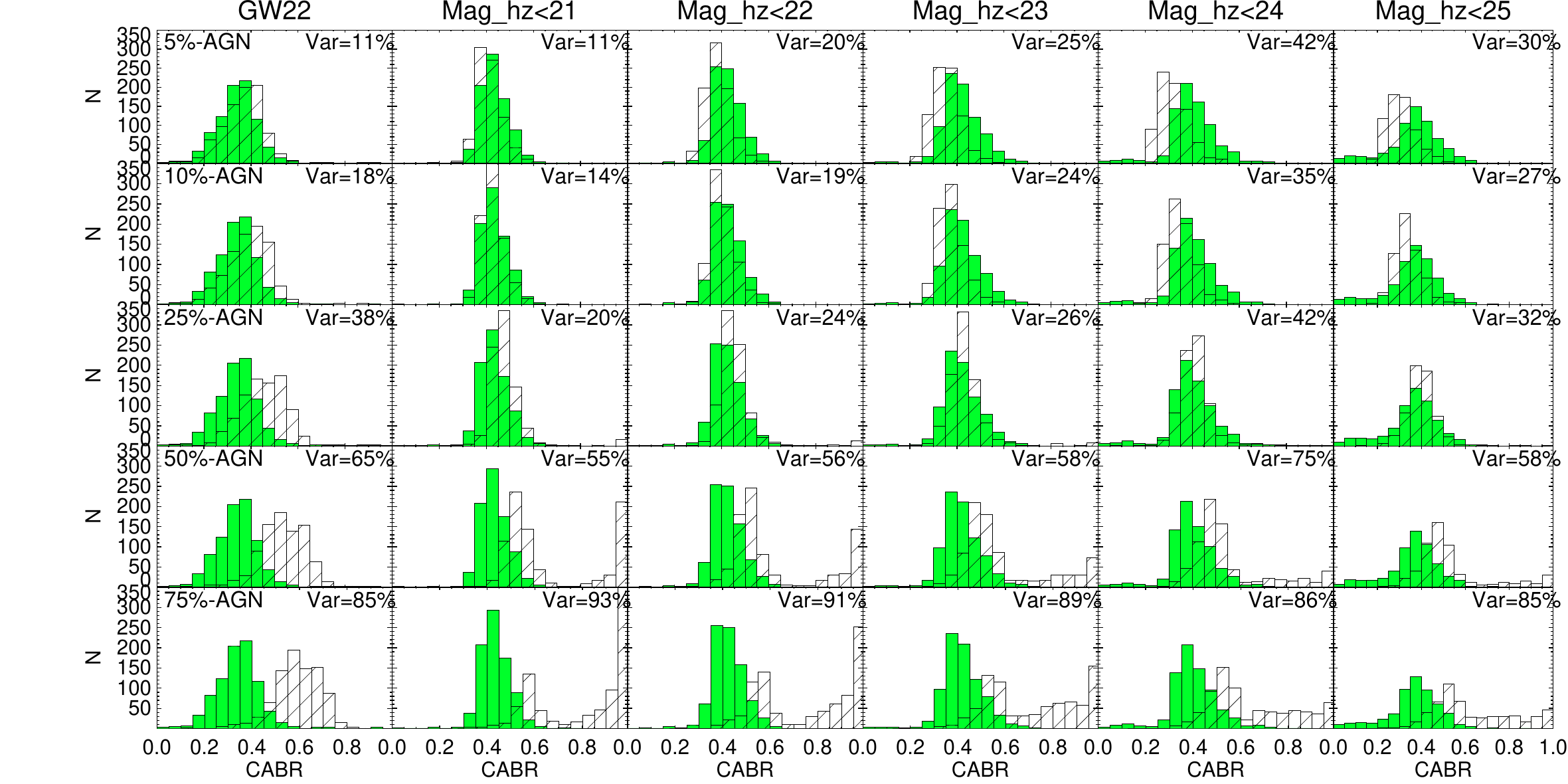}}
				\caption{The comparison of CABR between the original sample at z\,$\sim$\,0 without (filled histograms) and with (open histograms) added AGN contribution of 5\% to 75\% (top to bottom rows, respectively) using the original sample at z\,$\sim$\,0 (first columns) and the simulated samples moved to fainter magnitudes from 21 to 25 (columns 2 to 6, respectively). From top to bottom, the distributions of the total sample (yellow), early-type galaxies (red), early-spirals (blue), and late-spirals (green) are shown, respectively.}
				\label{fig4.1:part3}
			\end{center}
		\end{figure}
		
		\begin{figure*}
			\begin{center}
				{\includegraphics[height= 2.2in, width=5.54in]{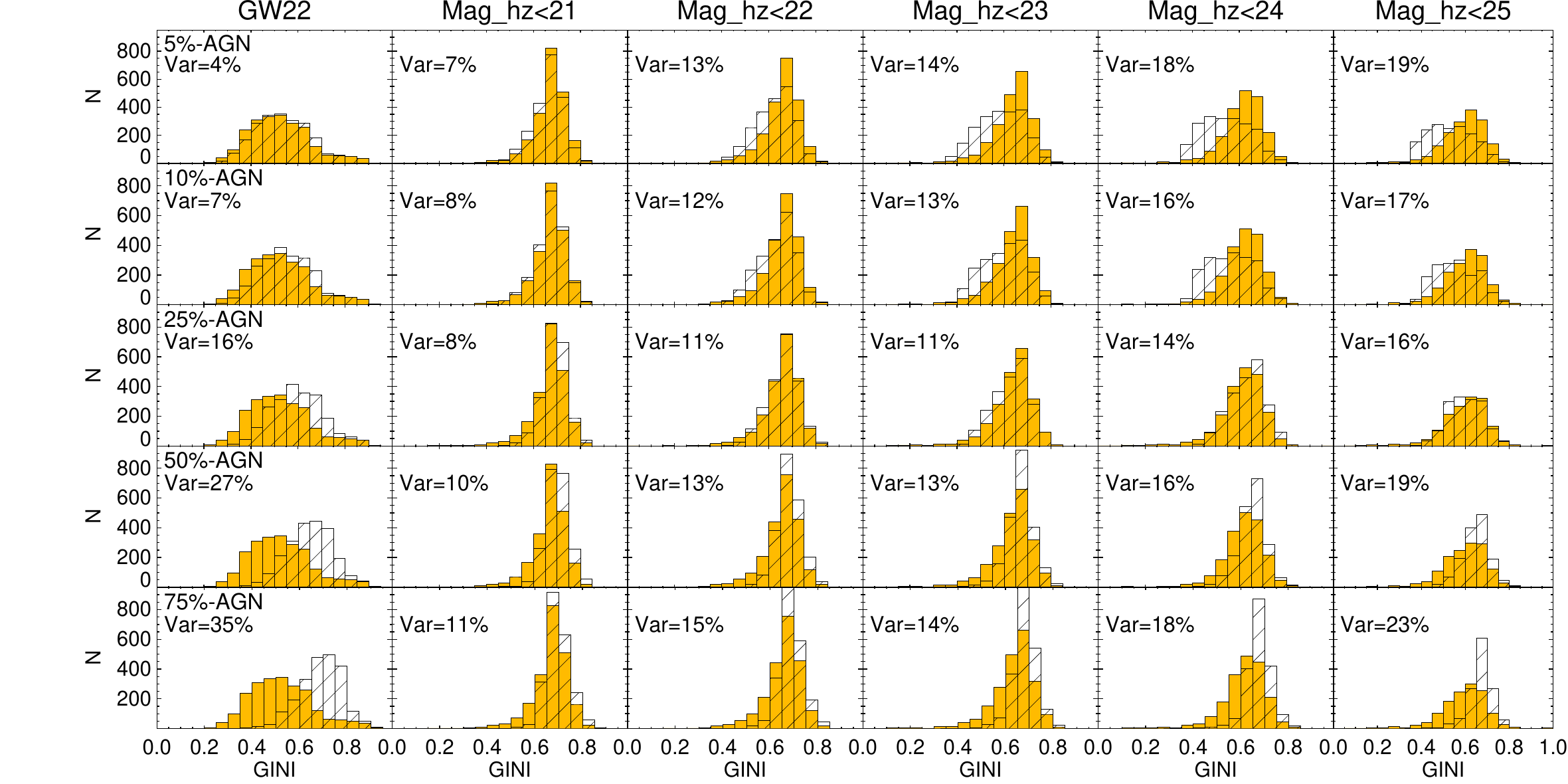}}
				{\includegraphics[height= 2.2in, width=5.54in]{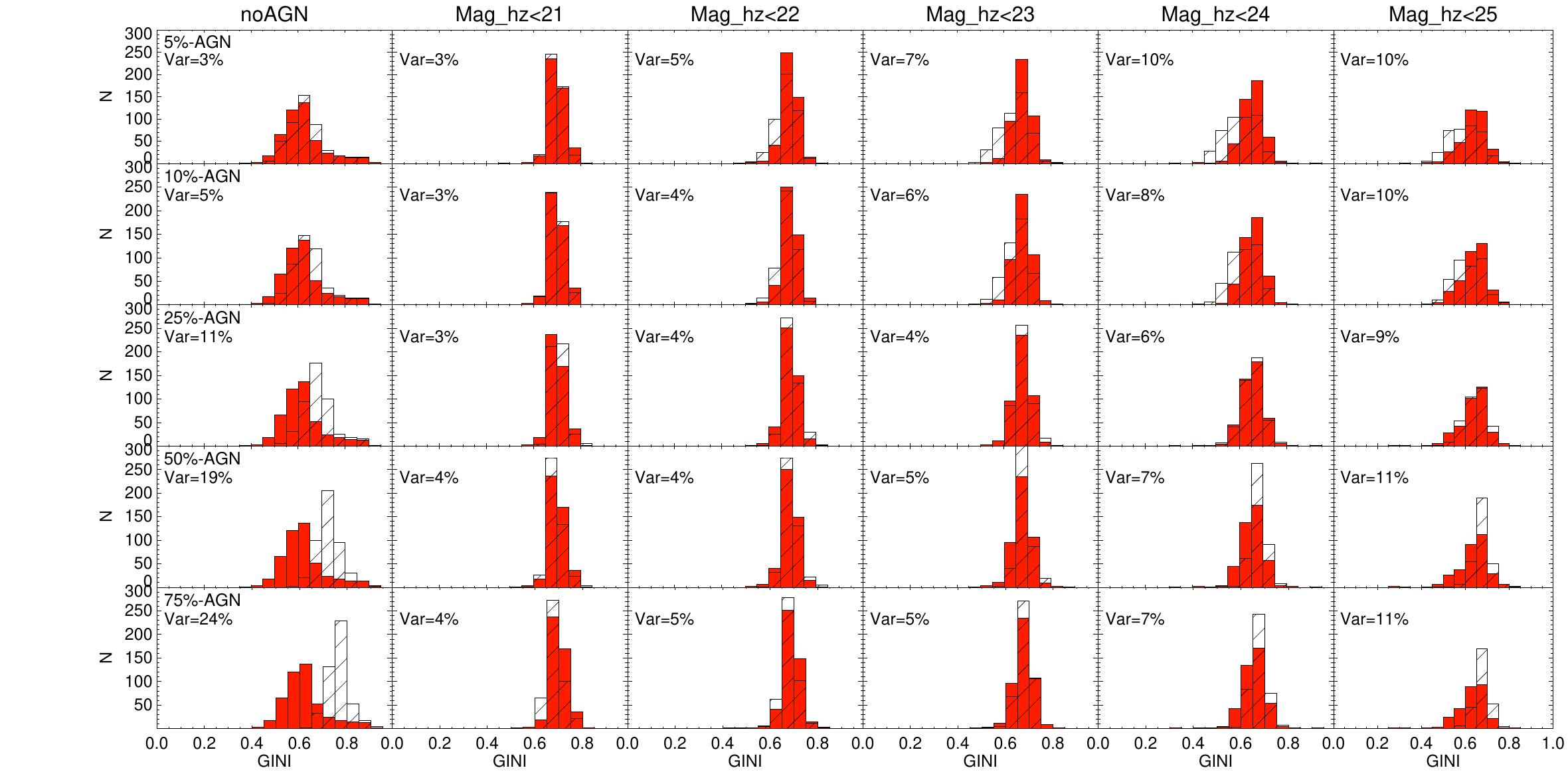}}
				{\includegraphics[height=2.2in, width=5.54in]{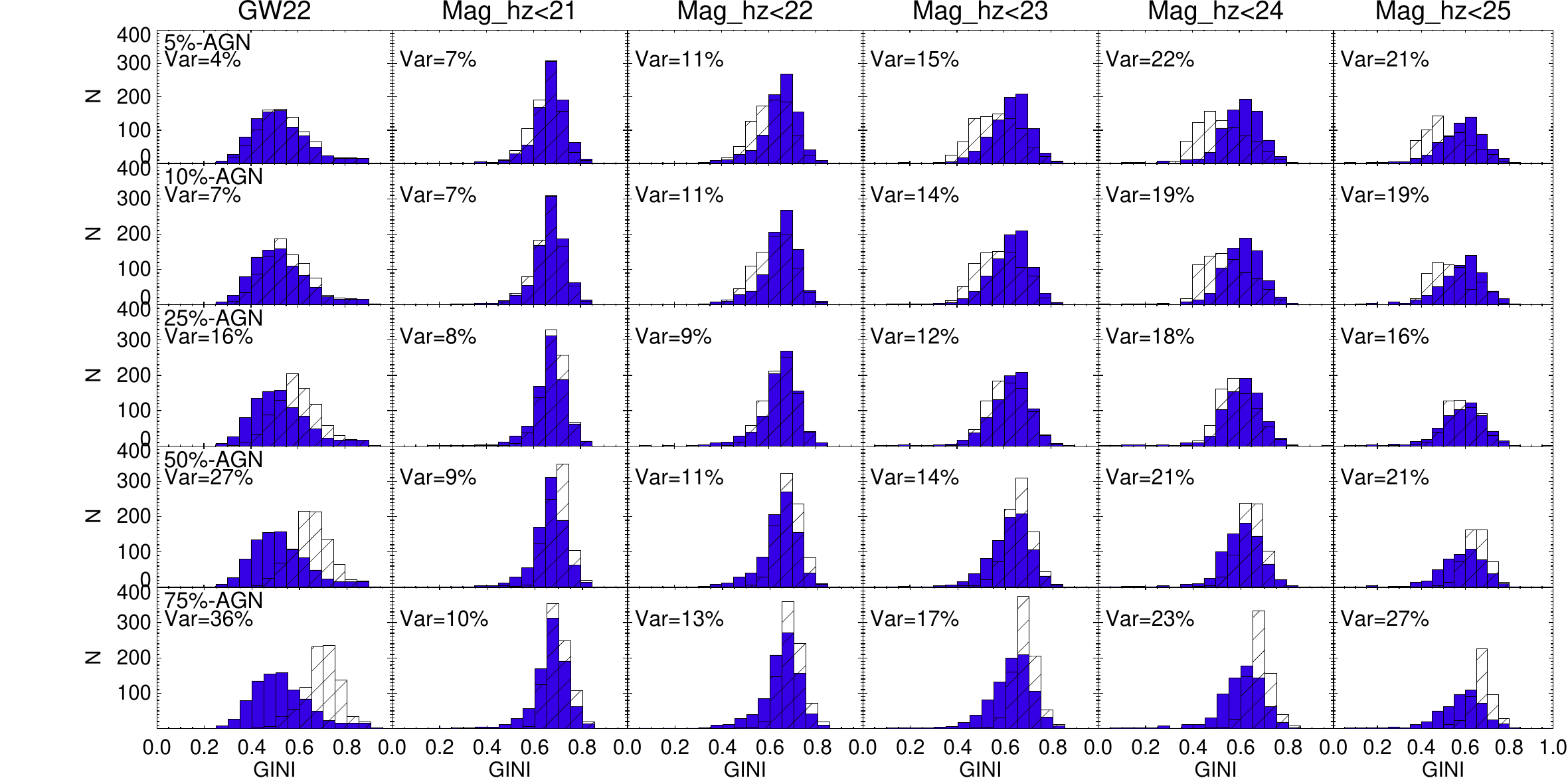}}
				{\includegraphics[height=2.2in, width=5.54in]{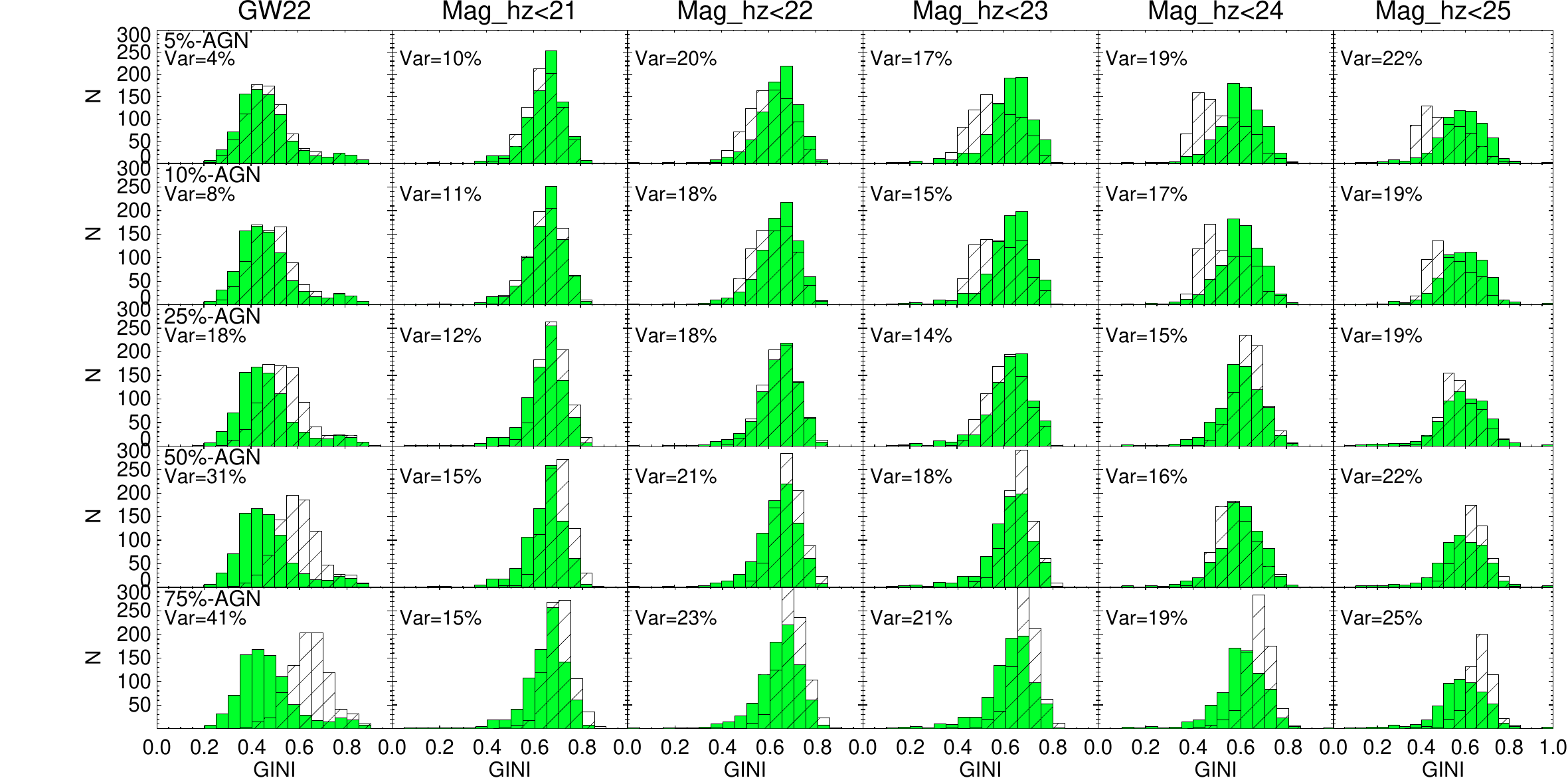}}
				\caption{Same as Figure \ref{fig4.1:part3}, but for the GINI parameter.}
				\label{fig4.2:part3}
			\end{center}
		\end{figure*}
		
		\begin{figure*}
			\begin{center}
				{\includegraphics[height= 2.2in, width=5.54in]{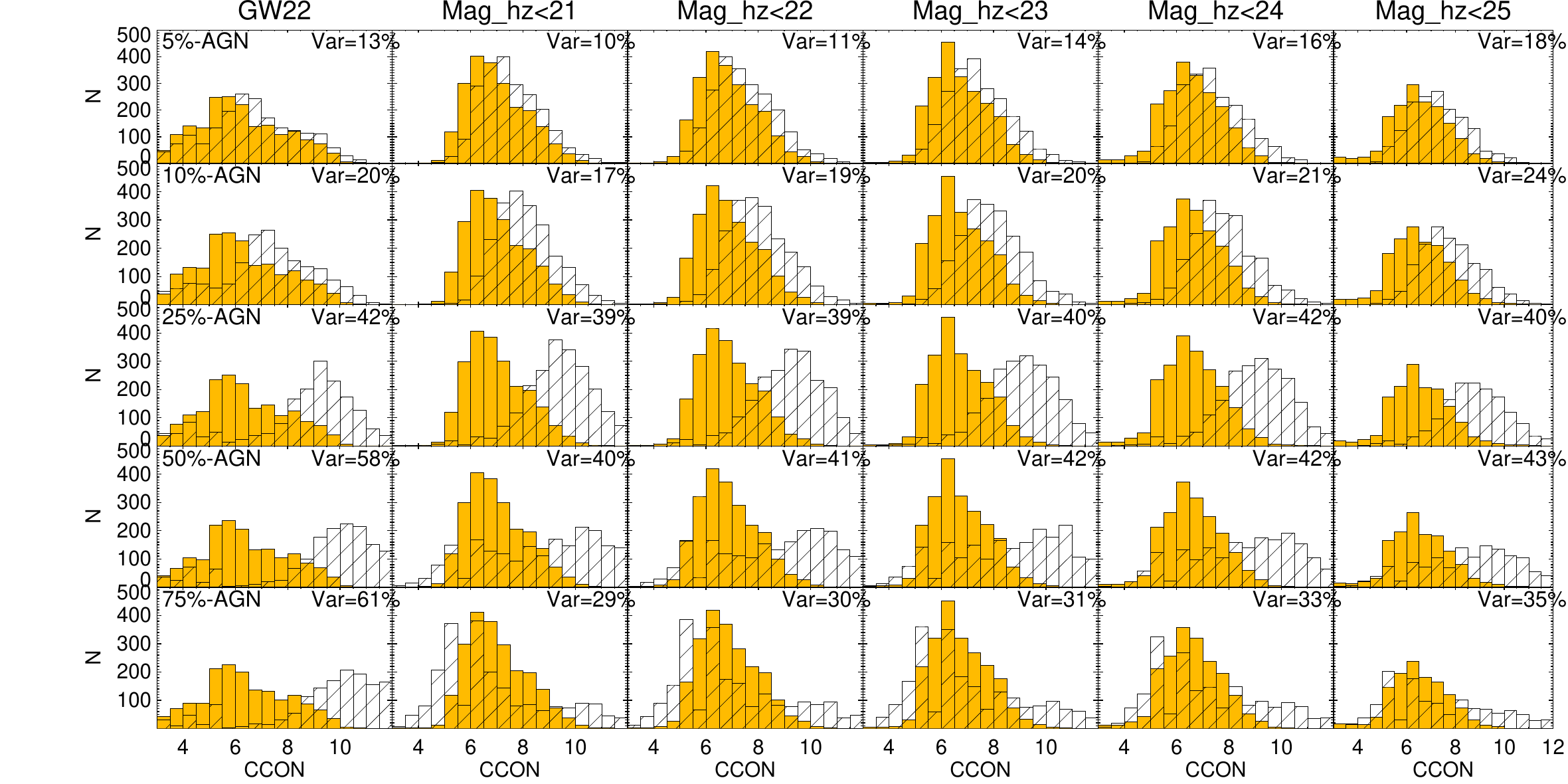}}
				{\includegraphics[height= 2.2in, width=5.54in]{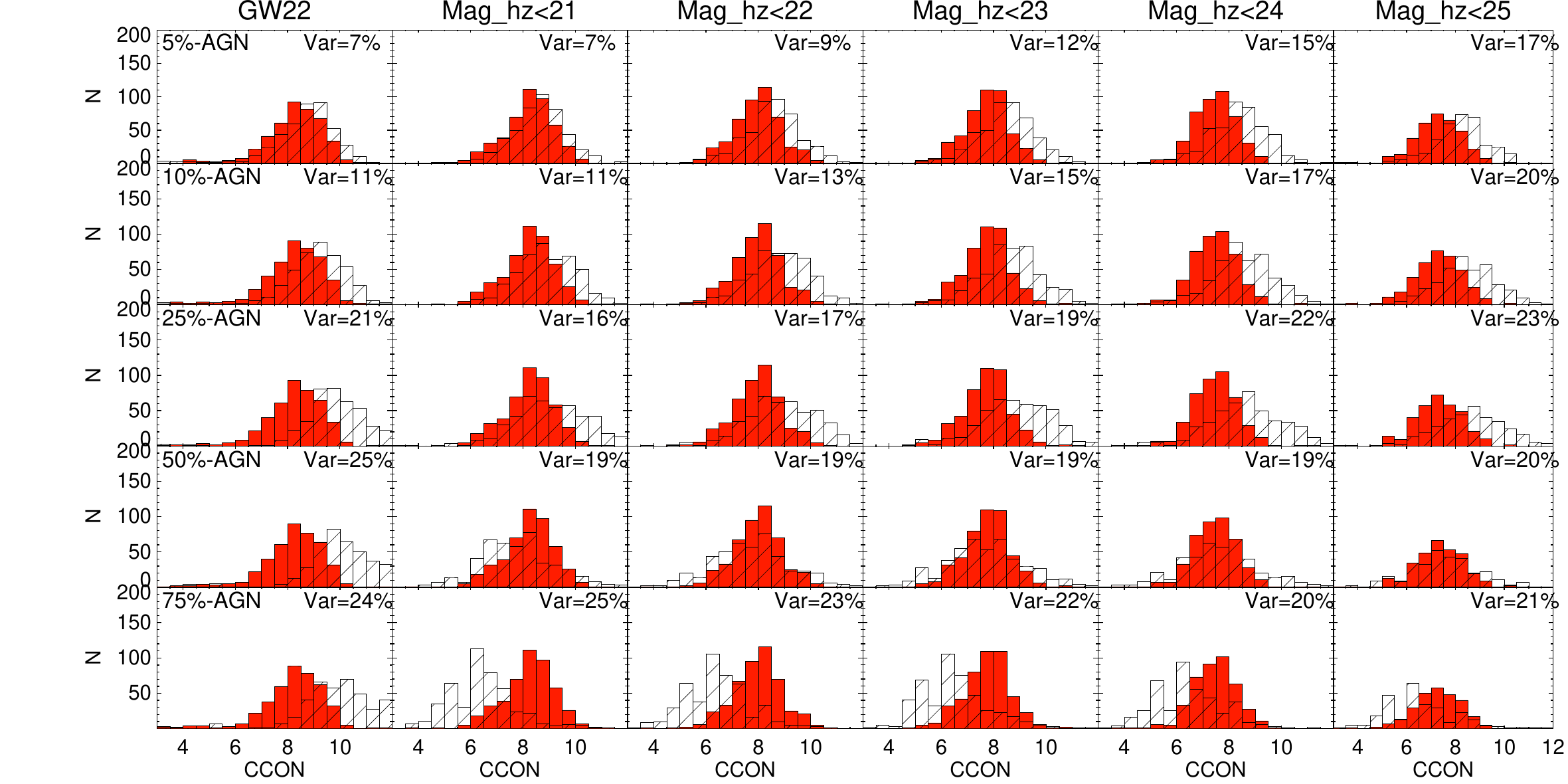}}
				{\includegraphics[height= 2.2in, width=5.54in]{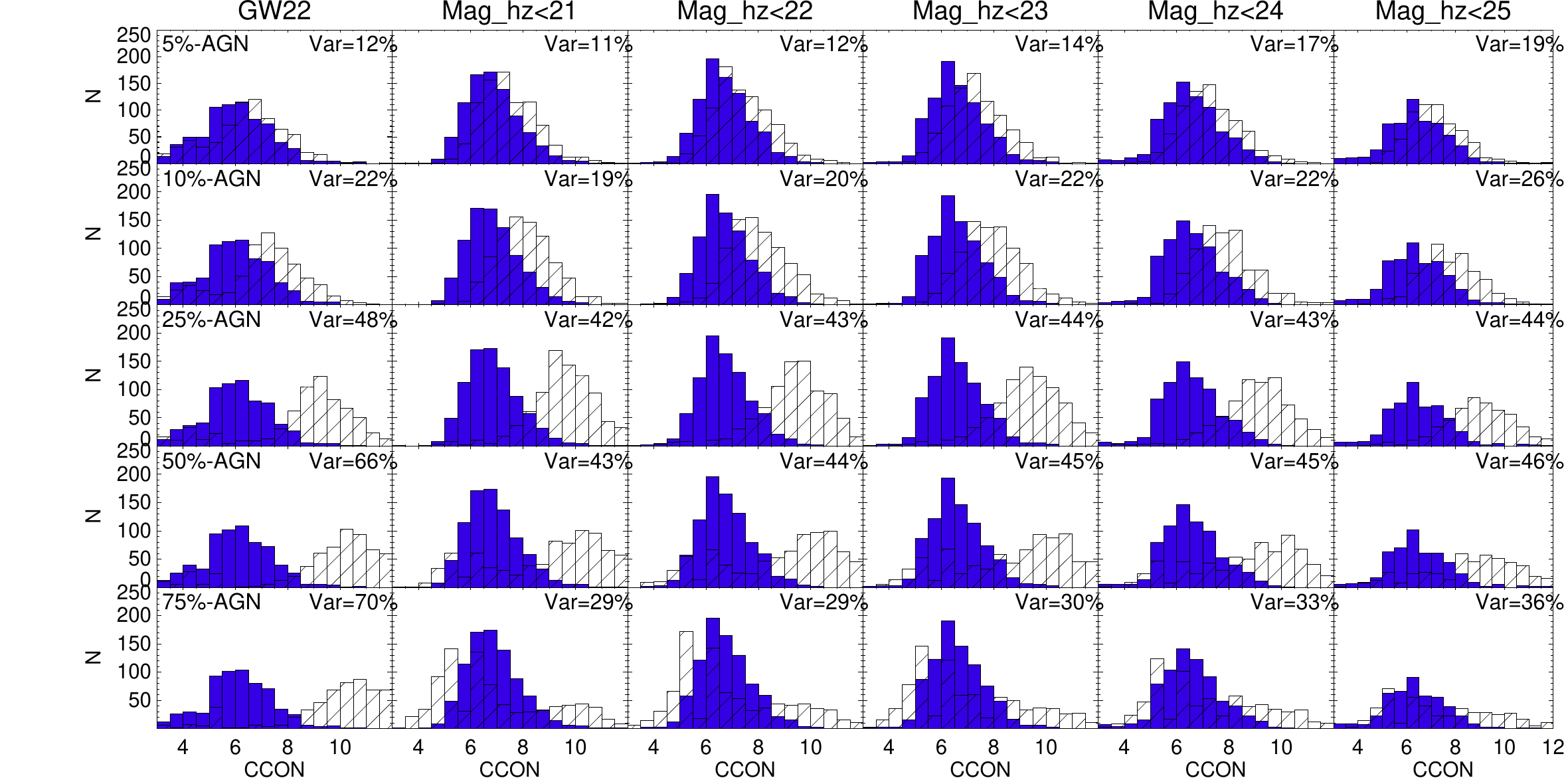}}
				{\includegraphics[height= 2.2in, width=5.54in]{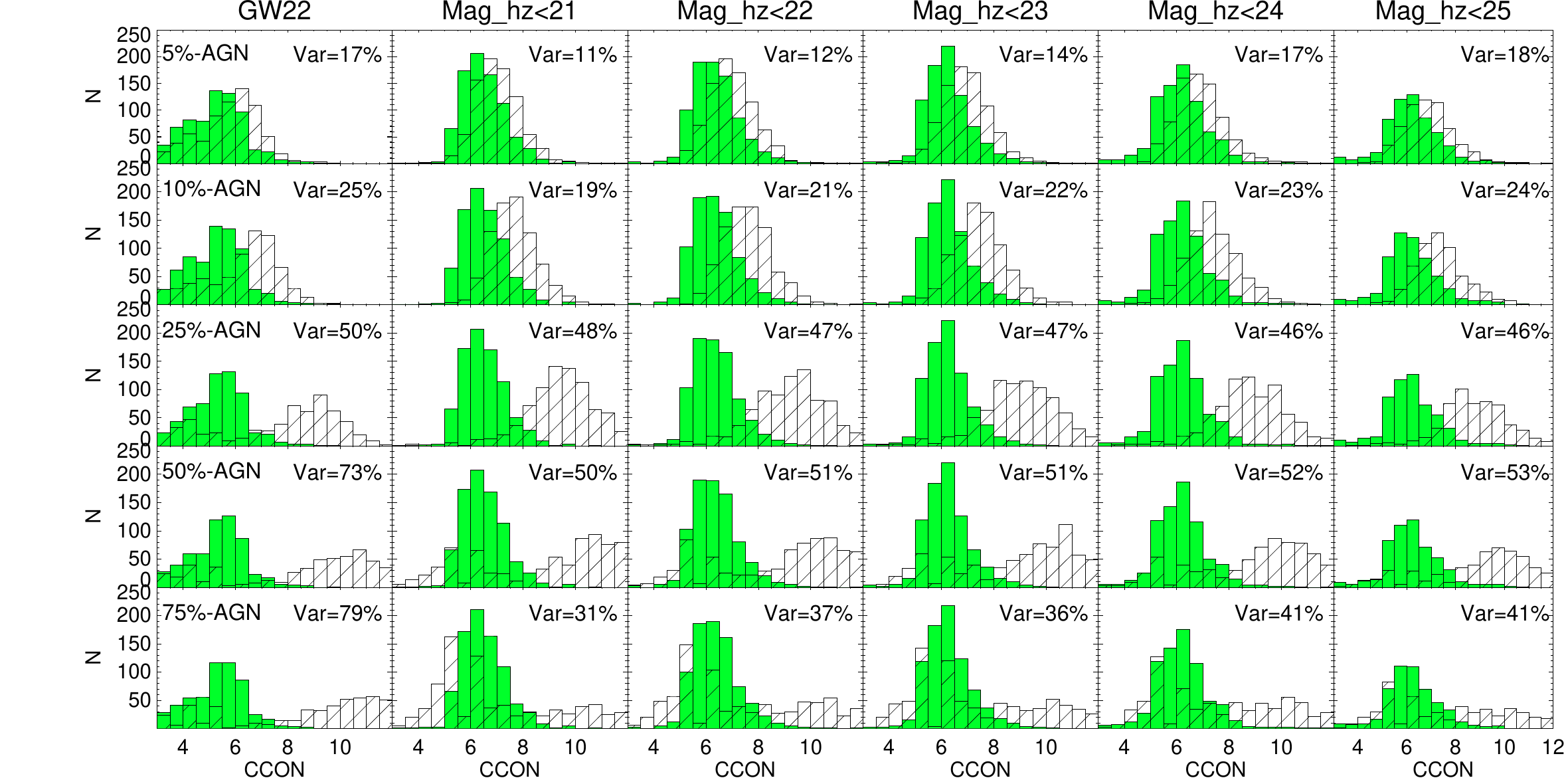}}
				\caption{Same as Figure \ref{fig4.1:part3}, but for the CCON parameter.}
				\label{fig4.3:part3}
			\end{center}
		\end{figure*}
		
		\begin{figure*}
			\begin{center}
				{\includegraphics[height= 2.2in, width=5.54in]{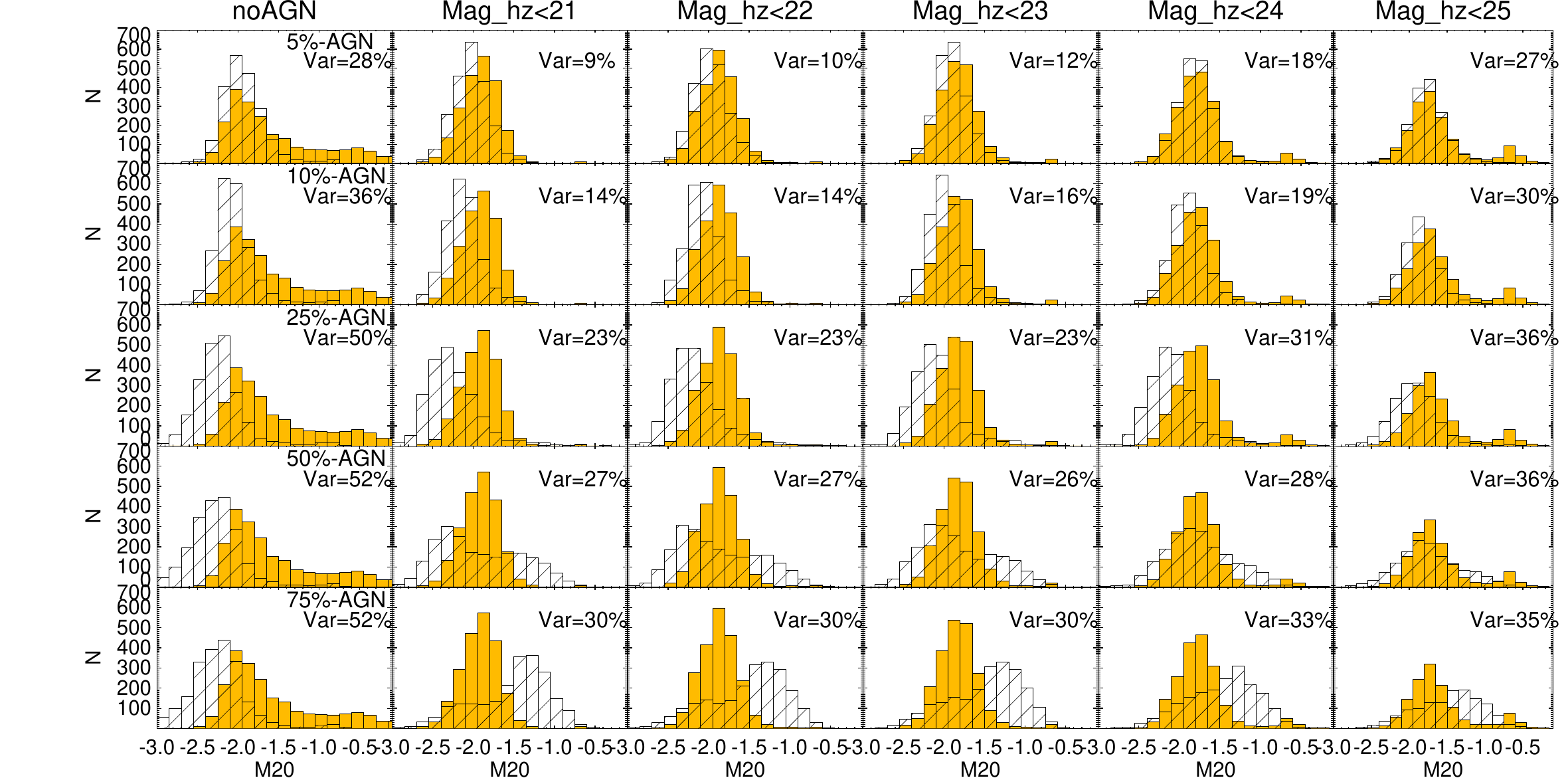}}
				{\includegraphics[height= 2.2in, width=5.54in]{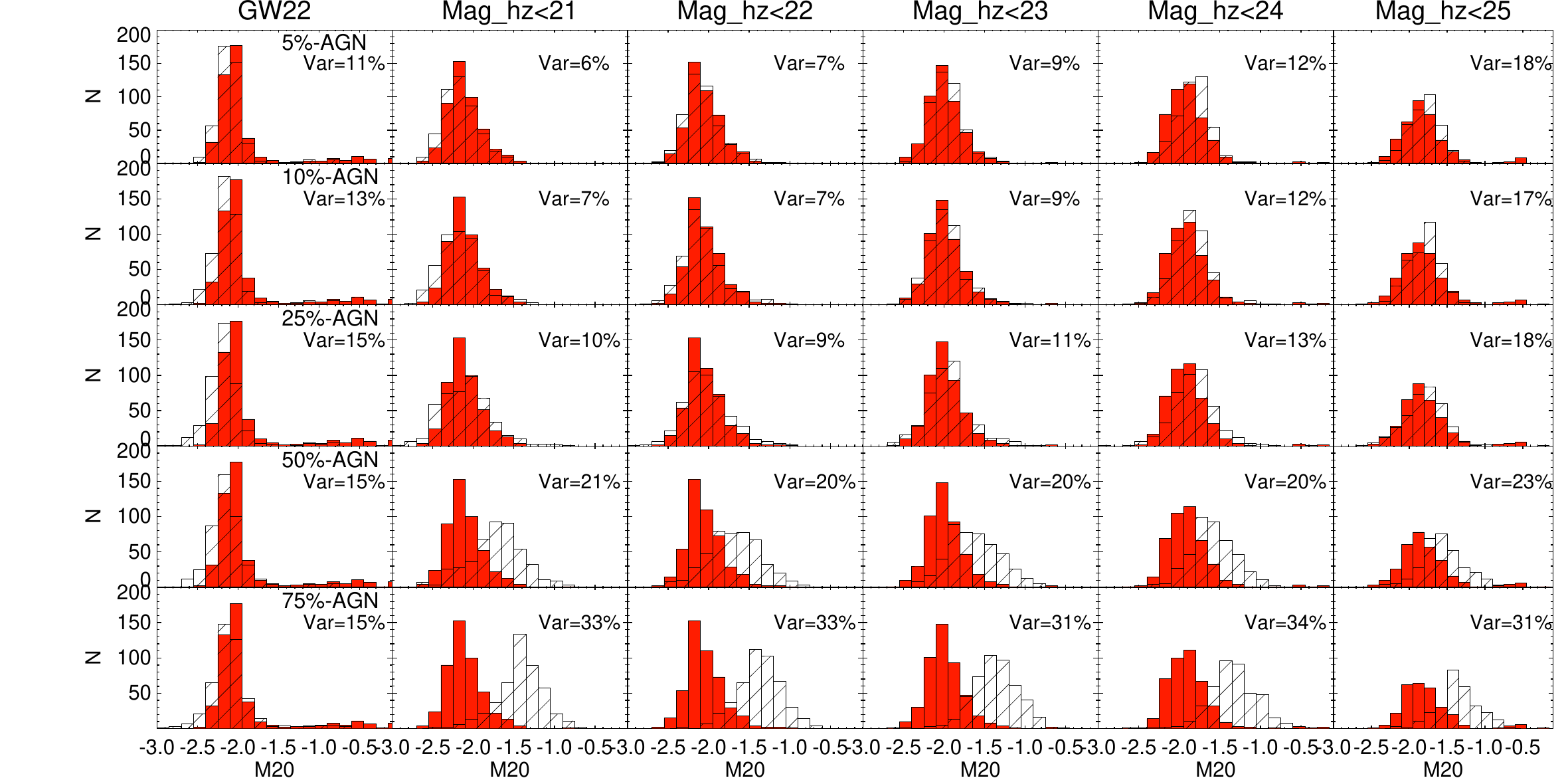}}
				{\includegraphics[height= 2.2in, width=5.54in]{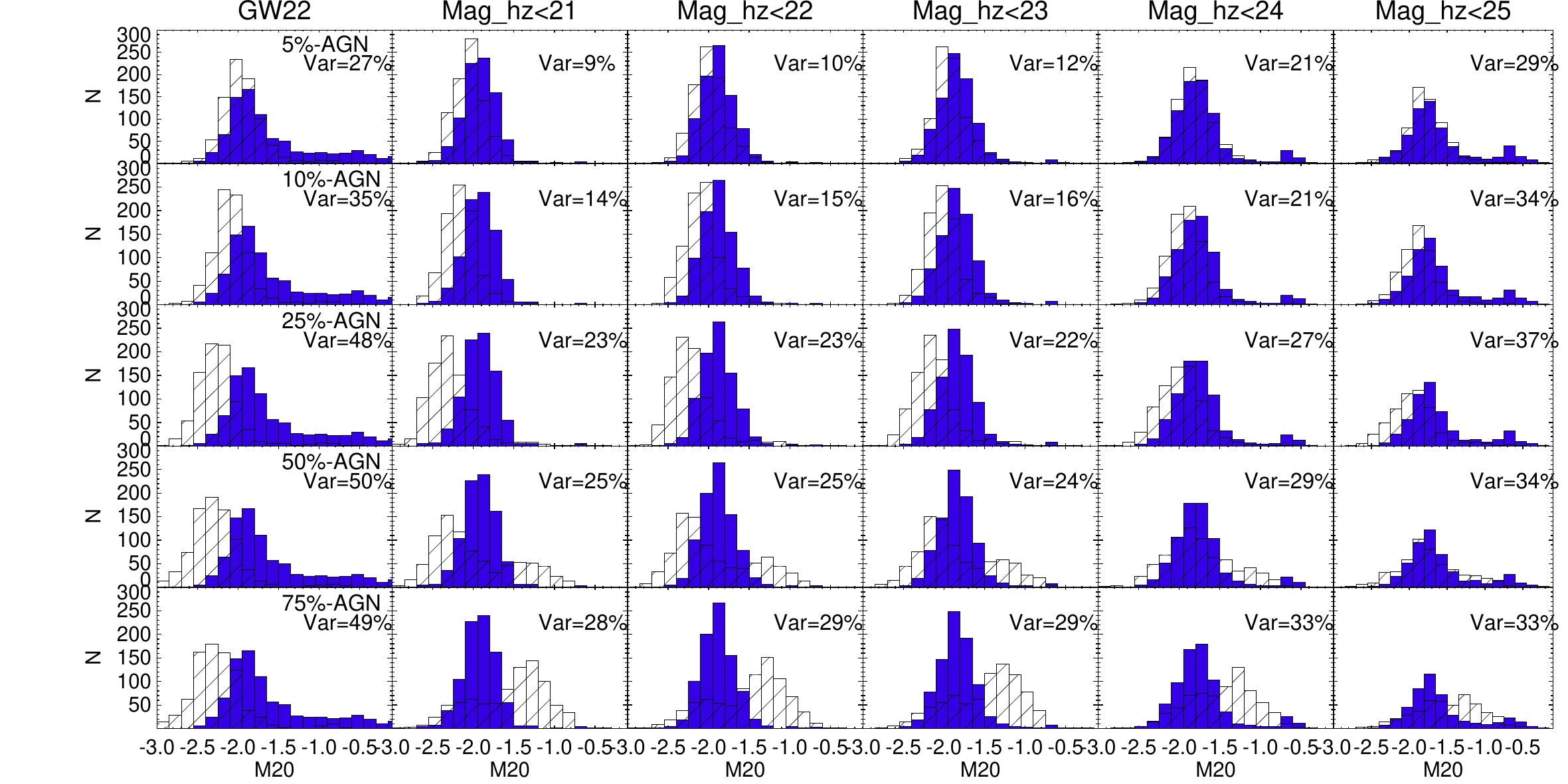}}
				{\includegraphics[height= 2.2in, width=5.54in]{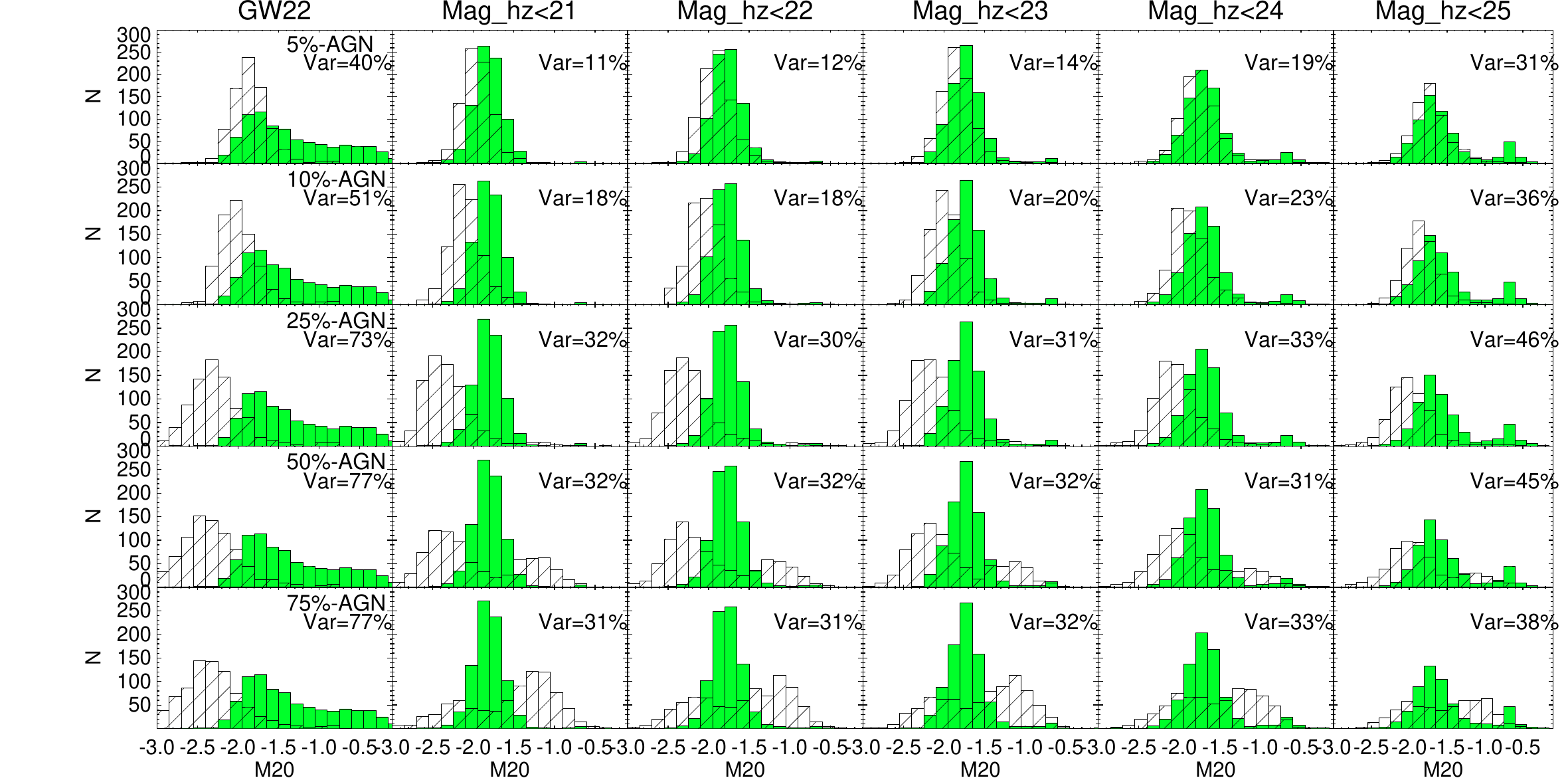}}
				\caption{Same as Figure \ref{fig4.1:part3}, but for the M20 moment of light.}
				\label{fig4.4:part3}
			\end{center}
		\end{figure*}
		
		\begin{figure*}
			\begin{center}
				{\includegraphics[height= 2.2in, width=5.54in]{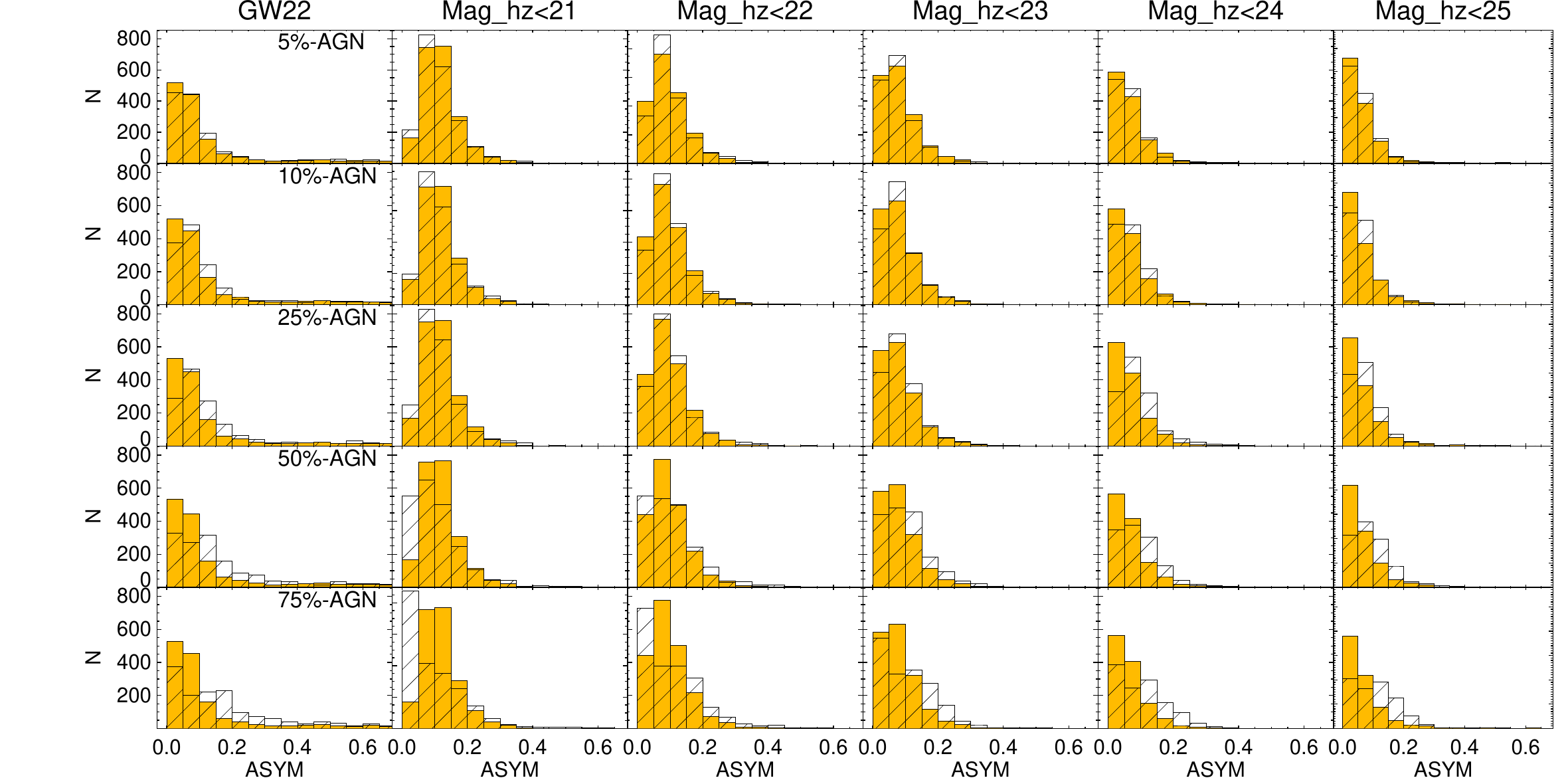}}
				{\includegraphics[height= 2.2in, width=5.54in]{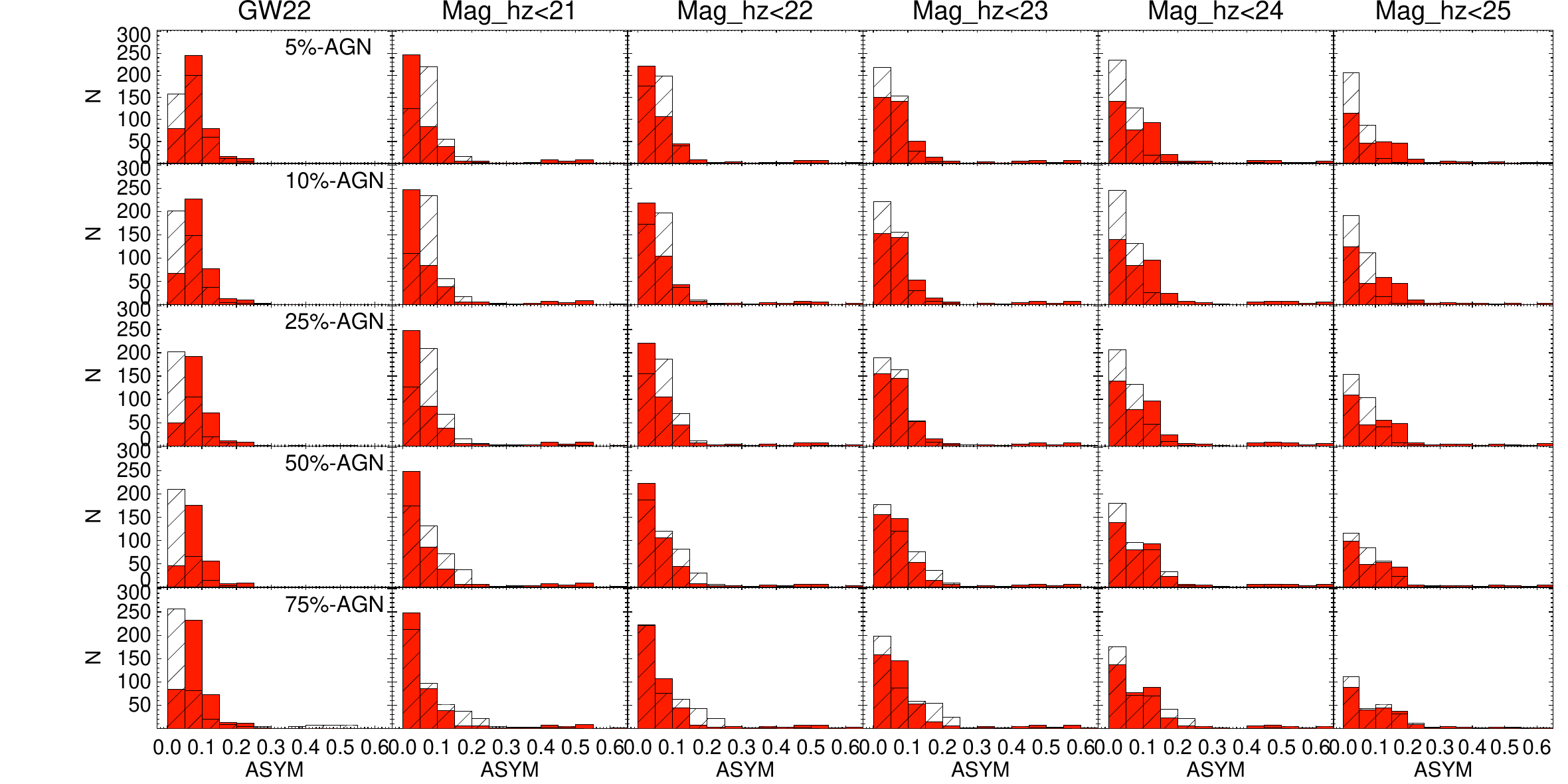}}
				{\includegraphics[height= 2.2in, width=5.54in]{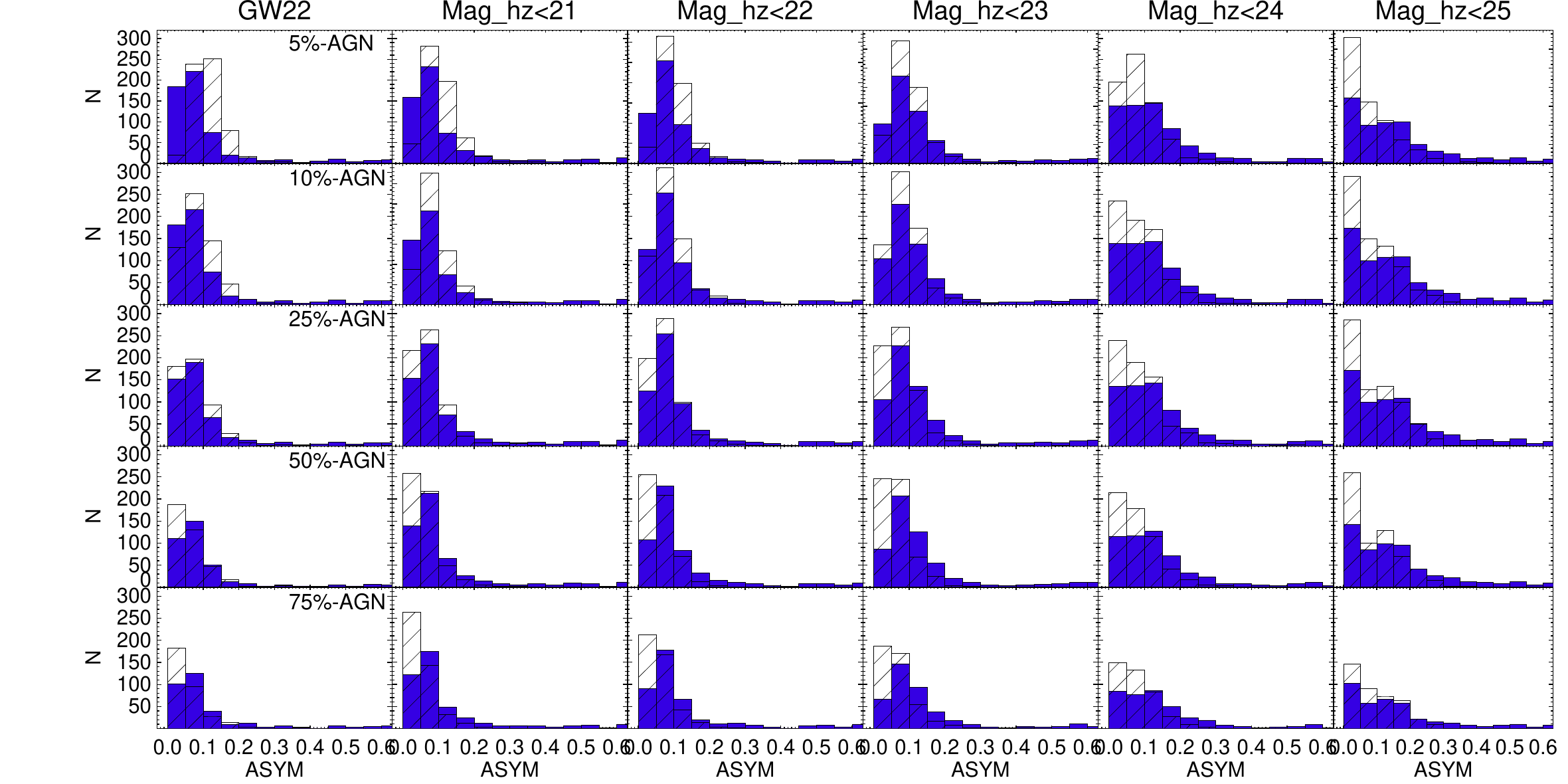}}
				{\includegraphics[height= 2.2in, width=5.54in]{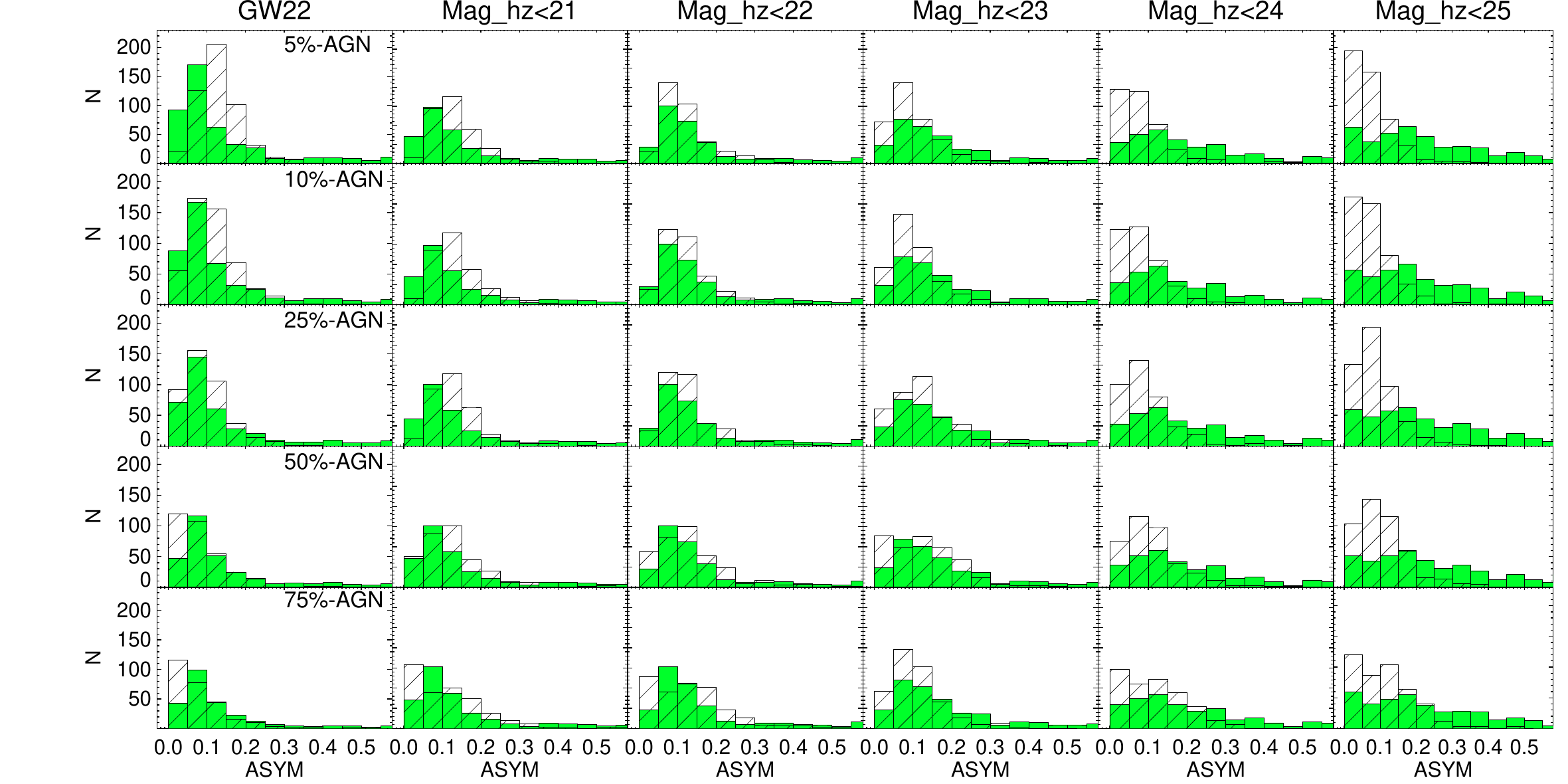}}
				\caption{Same as Figure \ref{fig4.1:part3}, but for the ASYM parameters.}
				\label{fig4.5:part3}
			\end{center}
		\end{figure*}
		
		\begin{figure*}
			\begin{center}
				{\includegraphics[height= 2.2in, width=5.54in]{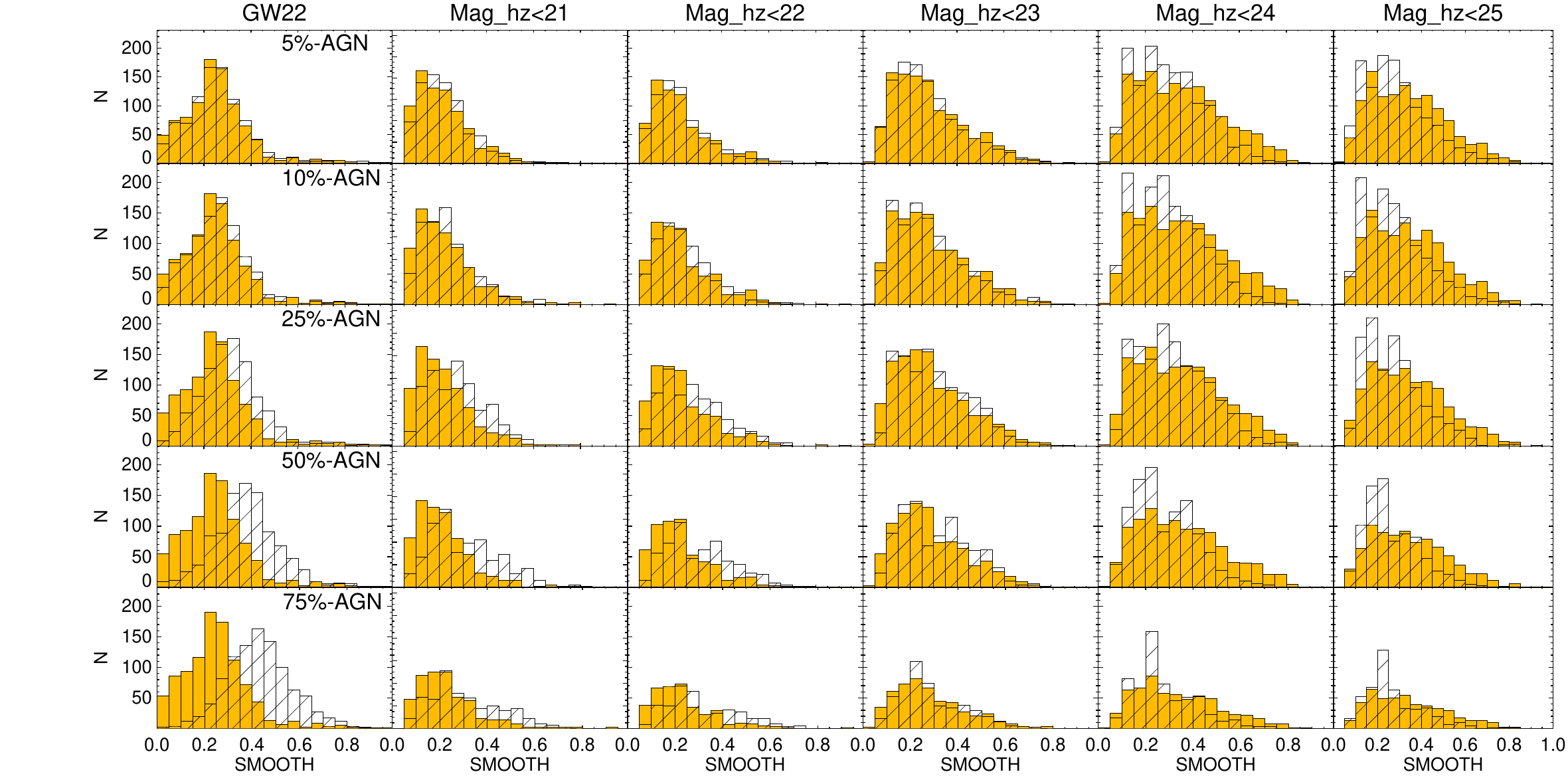}}
				{\includegraphics[height= 2.2in, width=5.54in]{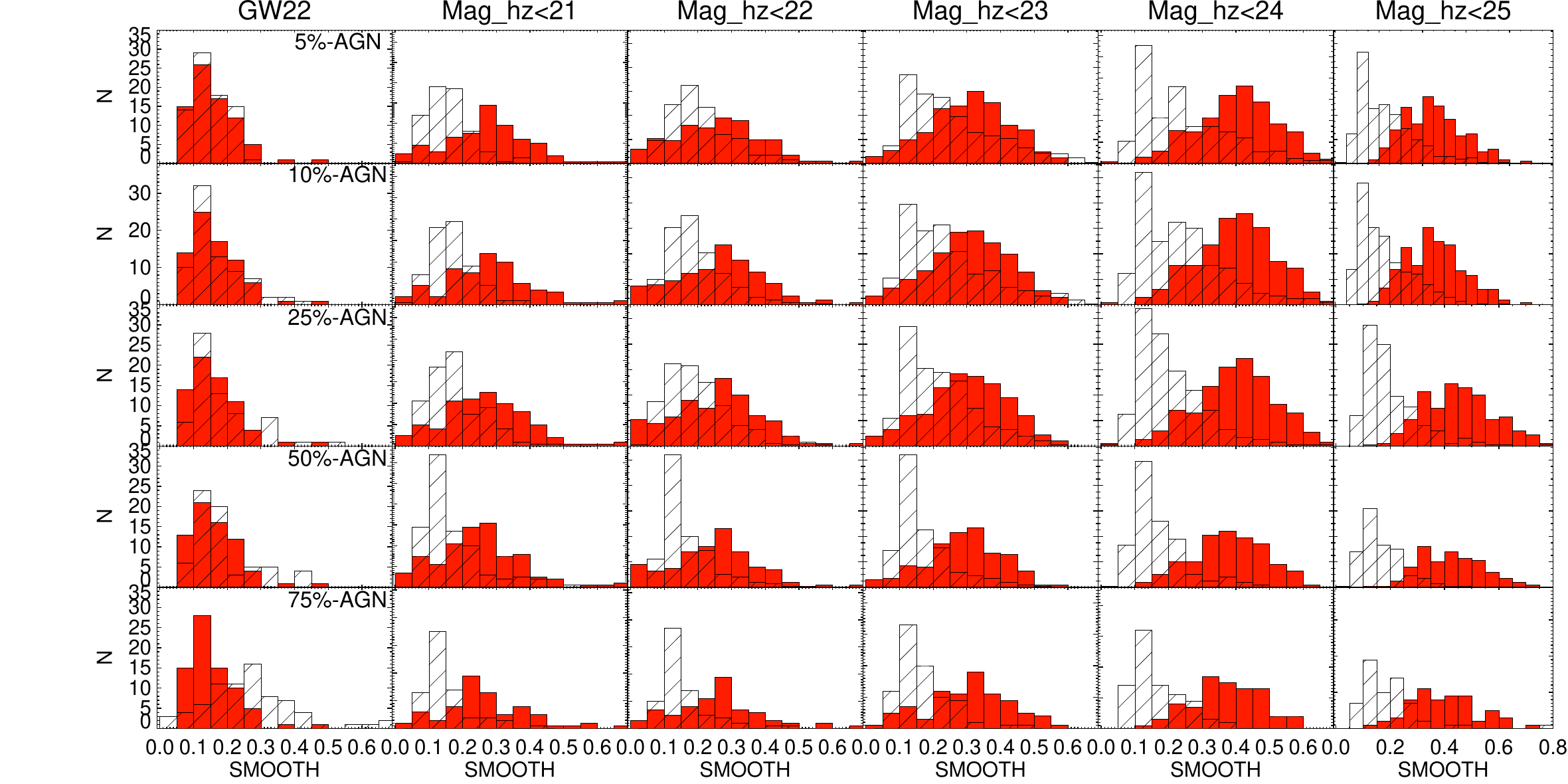}}
				{\includegraphics[height= 2.2in, width=5.54in]{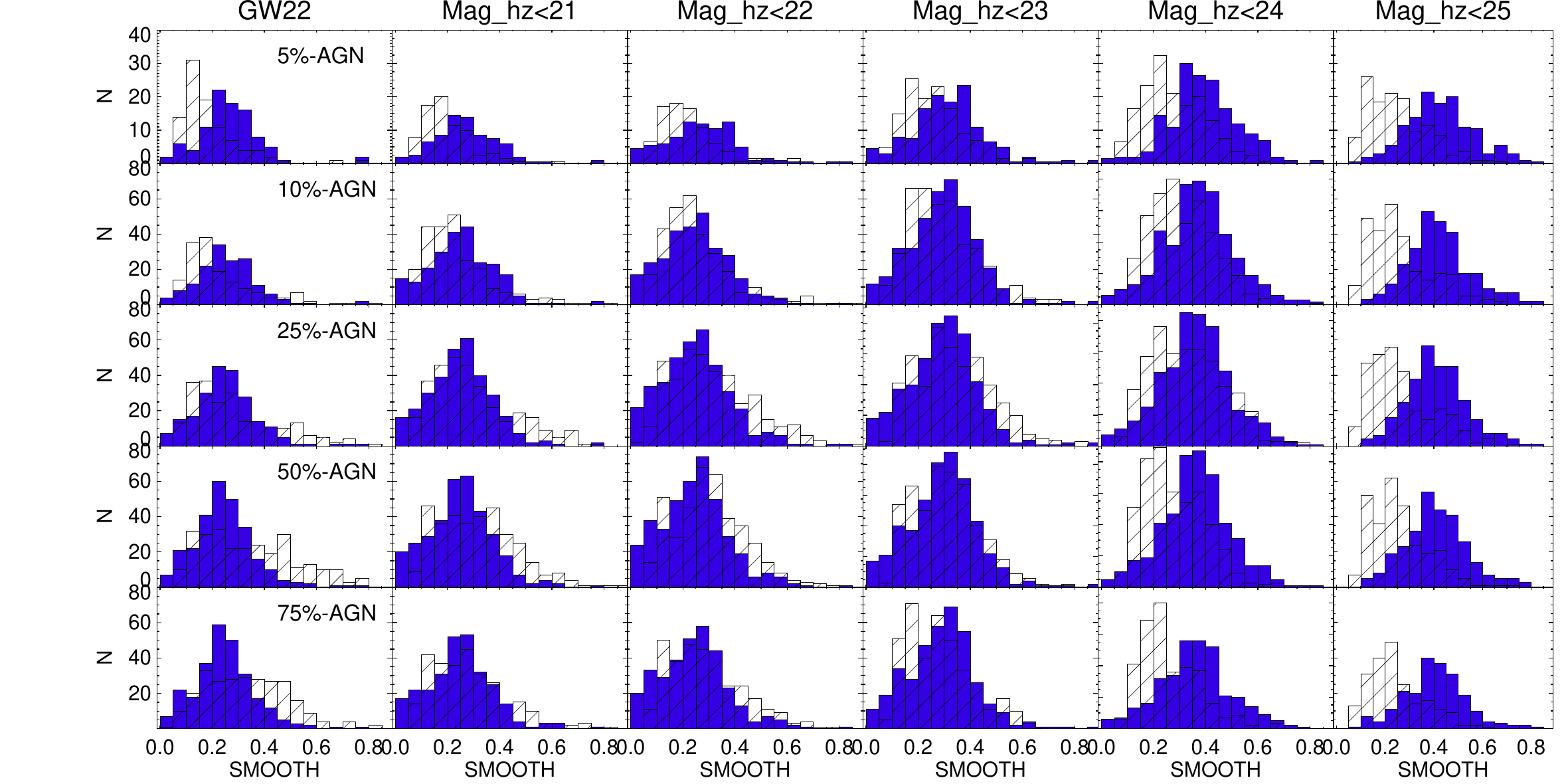}}
				{\includegraphics[height= 2.2in, width=5.54in]{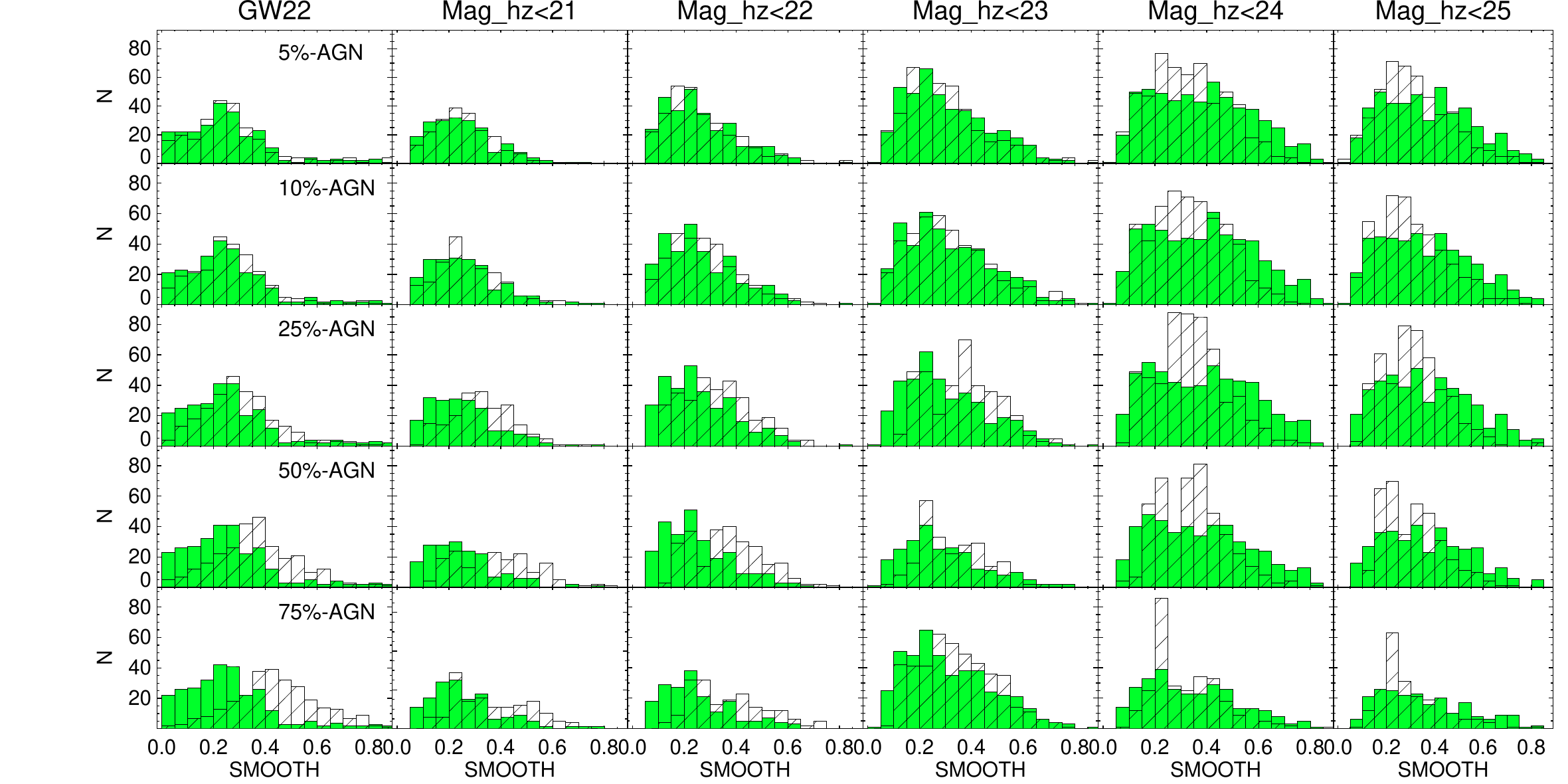}}
				\caption{Same as Figure \ref{fig4.1:part3}, but for the SMOOTH parameters.}
				\label{fig4.6:part3}
			\end{center}
		\end{figure*}
	\end{appendix}
	\section*{Notes}
	\begin{enumerate}
		\item \label{note1} \url{https://www.mpa.mpa-garching.mpg.de/SDSS/} (accessed on 31 October 2008)
	\end{enumerate}
		\begin{adjustwidth}{-\extralength}{0cm}
			\reftitle{References}
			\isAPAandChicago{}{
				}
			\PublishersNote{}
		\end{adjustwidth}
	\end{document}